\documentclass[structabstract,longauth]{aa}  
\usepackage{longtable,lscape,graphicx,float,amsmath,amssymb,mathrsfs,txfonts,color}
\usepackage{multirow,natbib,rotating,enumerate} 
\def\msol{$M_{\odot}$}
\def\lsol{$L_{\odot}$}
\def\rsol{$R_{\odot}$}
\def\HII{H\,{\sc ii}}

\def\arcsec{$^{\prime}$$^{\prime}$}
\def\arcmin{$^{\prime}$}
\def\deg{$^{\circ}$}
\def\micron{\,$\mu$m}

\def\kms{\,km\,s$^{-1}$}

\def\apjs{ApJS}
\def\apjl{ApJL}%
\def\aap{A\&A}%

\newcounter{ppnum4}
\newcounter{ppnum5}
\newcounter{ppnum6}
\begin{document}
\setlength{\tabcolsep}{3pt}
\title{Outflows, infall and evolution of a sample of embedded low-mass protostars}

\subtitle{The William Herschel Line Legacy (WILL) survey\thanks{\textit{Herschel} is an ESA space observatory with science instruments provided by European-led Principal Investigator consortia and with important participation from NASA.}}

\author{J.\,C.~Mottram\inst{1,2} 
          \and
          E.\,F.~van~Dishoeck\inst{1,3}
          \and
          L.\,E.~Kristensen\inst{4,5}
          \and
          A.~Karska\inst{3,6}
          \and
          I.~San~Jos\'{e}-Garc\'{i}a\inst{1}
          \and
          S.~Khanna\inst{1}
          \and
          G.\,J.~Herczeg\inst{7}
          \and
          Ph.~Andr\'{e}\inst{8}
          \and
          S.~Bontemps\inst{9,10}
          \and
          S.~Cabrit\inst{11,12}
          \and
          M.\,T.~Carney\inst{1}
          \and
          M.\,N.~Drozdovskaya\inst{1}
          \and
          M.\,M.~Dunham\inst{4}
          \and
          N.\,J.~Evans\inst{13}
          \and
          D.~Fedele\inst{2,14}
          \and
          J.\,D.~Green\inst{13,15}
          \and
          D.~Harsono\inst{1,16}
          \and
          D.~Johnstone\inst{17,18}
          \and
          J.\,K.~J\o rgensen\inst{5}
          \and
          V.~K\"onyves\inst{8}
          \and
          B.~Nisini\inst{19}
          \and
          M.\,V.~Persson\inst{1}
          \and
          M.~Tafalla\inst{20}
          \and
          R.~Visser\inst{21}
          \and
          U.\,A.~Y{\i}ld{\i}z\inst{22} 
}

\institute{Leiden Observatory, Leiden University, PO Box 9513, 2300 RA Leiden, The Netherlands
  \and
Max Planck Institute for Astronomy, K\"{o}nigstuhl 17, 69117 Heidelberg, Germany\\ \email{mottram@mpia.de}
  \and
Max Planck Institut f\"{u}r Extraterrestrische Physik, Giessenbachstrasse 1, 85748 Garching, Germany
\and
Harvard-Smithsonian Center for Astrophysics, 60 Garden Street, Cambridge, MA 02138, USA
\and
Centre for Star and Planet Formation, Niels Bohr Institute and Natural History Museum of Denmark, University of Copenhagen, {\O}ster Voldgade 5-7, DK-1350 Copenhagen K, Denmark
\and
Centre for Astronomy, Nicolaus Copernicus University, Faculty of Physics, Astronomy and Informatics, Grudziadzka 5, PL-87100 Torun, Poland
\and
Kavli Institut for Astronomy and Astrophysics, Yi He Yuan Lu 5, HaiDian Qu, Peking University, 100871 Beijing, PR China
\and
Laboratoire AIM, CEA/DSM-CNRS-Universit\'e Paris Diderot, IRFU/Service d'Astrophysique, CEA Saclay, 91191 Gif-sur-Yvette, France
\and
Univ. Bordeaux, LAB, UMR5804, 33270 Floirac, France
\and
CNRS, LAB, UMR5804, 33270 Floirac, France
\and
LERMA, Observatoire de Paris, UMR 8112 du CNRS, ENS, UPMC, UCP, 61 Av. de l’Observatoire, F-75014 Paris, France
\and
Institut de Plan\'{e}tologie et d’Astrophysique de Grenoble (IPAG) UMR 5274, Grenoble, 38041, France
\and
Department of Astronomy, University of Texas at Austin, Austin, TX 78712, USA
\and
INAF-Osservatorio Astrofisico di Arcetri, L.go E. Fermi 5, I-50125 Firenze, Italy,
\and
Space Telescope Science Institute, Baltimore, MD, USA
\and
Universit\"{a}t Heidelberg, Zentrum f\"{u}r Astronomie, Institut f\"{u}r Theoretische Astrophysik (ITA), Albert-Ueberle-Str. 2, 69120, Heidelberg, Germany
\and
NRC Herzberg Astronomy and Astrophysics, 5071 West Saanich Rd, Victoria, BC, V9E 2E7, Canada
\and
Department of Physics and Astronomy, University of Victoria, Victoria, BC, V8P 1A1, Canada 
\and
INAF-Osservatorio Astronomico di Roma, via Frascati 33, I-00040 Monte Porzio Catone, Italy
\and
Observatorio Astronomico Nacional (OAN-IGN), Alfonso XII 3, 28014, Madrid, Spain
\and
European Southern Observatory, Karl-Schwarzschild-Strasse 2, D-85748 Garching, Germany
\and
Jet Propulsion Laboratory, California Institute of Technology, 4800 Oak Grove Drive, Pasadena CA, 91109, USA
}

%
%

   \date{Received XXXX; accepted XXXX}

 
  \abstract
   {\textit{Herschel} observations of water and highly excited CO ($J>$9) have allowed the physical and chemical conditions in the more active parts of protostellar outflows to be quantified in detail for the first time. However, to date, the studied samples of Class 0/I protostars in nearby star-forming regions have been selected from bright, well-known sources and have not been large enough for statistically significant trends to be firmly established.}
   {We aim to explore the relationships between the outflow, envelope and physical properties of a flux-limited sample of embedded low-mass Class 0/I protostars.}
   {We present spectroscopic observations in H$_{2}$O, CO and related species with \textit{Herschel} HIFI and PACS, as well as ground-based follow-up with the JCMT and APEX in CO, HCO$^{+}$ and isotopologues, of a sample of 49 nearby ($d<$500\,pc) candidate protostars selected from \textit{Spitzer} and \textit{Herschel} photometric surveys of the Gould Belt. This more than doubles the sample of sources observed by the WISH and DIGIT surveys. These data are used to study the outflow and envelope properties of these sources. We also compile their continuum spectral energy distributions (SEDs) from the near-IR to mm wavelengths in order to constrain their physical properties (e.g. $L_{\mathrm{bol}}$, $T_{\mathrm{bol}}$ and $M_{\mathrm{env}}$).}
   {Water emission is dominated by shocks associated with the outflow, rather than the cooler, slower entrained outflowing gas probed by ground-based CO observations. These shocks become less energetic as sources evolve from Class 0 to Class I. Outflow force, measured from low-$J$ CO, also decreases with source evolutionary stage, while the fraction of mass in the outflow relative to the total envelope (i.e. $M_{\mathrm{out}}/M_{\mathrm{env}}$) remains broadly constant between Class 0 and I. The median value of $\sim$1$\%$ is consistent with a core to star formation efficiency on the order of 50$\%$ and an outflow duty cycle on the order of 5$\%$. Entrainment efficiency, as probed by $F_{\mathrm{CO}}/\dot{M}_{\mathrm{acc}}$, is also invariant with source properties and evolutionary stage. The median value implies a velocity at the wind launching radius of 6.3\kms{}, which in turn suggests an entrainment efficiency of between 30 and 60$\%$ if the wind is launched at $\sim$1AU, or close to 100$\%$ if launched further out. $L$[O\,{\sc i}] is strongly correlated with $L_{\mathrm{bol}}$ but not with $M_{\mathrm{env}}$, in contrast to low-$J$ CO, which is more closely correlated with the latter than the former. This suggests that [O\,{\sc i}] traces the present-day accretion activity of the source while CO traces time-averaged accretion over the dynamical timescale of the outflow. H$_{2}$O is more strongly correlated with $M_{\mathrm{env}}$ than $L_{\mathrm{bol}}$, but the difference is smaller than low-$J$ CO, consistent with water emission primarily tracing actively shocked material between the wind, traced by [O\,{\sc i}], and the entrained molecular outflow, traced by low-$J$ CO. $L$[O\,{\sc i}] does not vary from Class 0 to Class I, unlike CO and H$_{2}$O. This is likely due to the ratio of atomic to molecular gas in the wind increasing as the source evolves, balancing out the decrease in mass accretion rate. Infall signatures are detected in HCO$^{+}$ and H$_{2}$O in a few sources, but still remain surprisingly illusive in single-dish observations.}
   {}


   \maketitle

%

\section{Introduction}
\label{S:Intro}

The general, cartoon picture of how stars form has been agreed for some time: a dense core within a molecular cloud becomes gravitationally unstable, causing material to fall inwards towards the centre; a protostar forms and launches a bi-polar molecular outflow; over time the outflow and infall combine to remove the envelope, eventually starving the protostar, which then slowly settles to the main sequence \citep[e.g.][]{Shu1987}. However, a more detailed understanding is still required, particularly on infall and outflow, in order to quantitatively track the conversion of matter into stars and accurately predict the evolution and outcome of the star-formation process for individual sources, stellar clusters and even whole galaxies.

The first step is improved quantification of the basic physical properties (e.g. $L_{\mathrm{bol}}$, $M_{\mathrm{env}}$) and evolutionary state of low-mass protostars, on which considerable progress has been made. Improvements in detectors and telescopes have lead to full-wavelength coverage from optical to radio wavelengths at better sensitivity and resolution, while dedicated very long baseline interferometry (VLBI) campaigns in the radio are providing much more accurate distances for nearby star-forming regions \citep[e.g.][for a recent review]{Loinard2013}. 

A framework for defining the evolutionary status of protostars has also been developed, dividing protostellar sources into one of five categories (Class 0, Class I, Flat, Class II and Class III) using various ways of quantifying the shift in the spectral energy distribution (SED) to shorter wavelengths as the source evolves: the infrared spectral index \citep[$\alpha_{\mathrm{IR}}$, e.g.][]{Lada1984,Lada1987,Greene1994}; the submillimetre ($\lambda>350$\micron) to bolometric luminosity ratio \citep[$L_{\mathrm{submm}}/L_{\mathrm{bol}}$ used as a proxy for $M_{\mathrm{env}}/L_{\mathrm{bol}}$, e.g.][]{Andre1993}; and bolometric temperature \citep[$T_{\mathrm{bol}}$, e.g.][]{Myers1993,Chen1995}. For this latter measure, which is the intensity-weighted peak of the SED, these classifications are defined as: Class 0 ($T_{\mathrm{bol}}<70$\,K), Class I (70$\leq T_{\mathrm{bol}}<$650\,K), Class II (650$\leq T_{\mathrm{bol}}<$2800\,K) and Class III ($T_{\mathrm{bol}}\geq$2800\,K). Flat-SED sources have $T_{\mathrm{bol}}$ values in the 350$-$950\,K range with a mean around 650\,K \citep{Evans2009}.

The \textit{Spitzer} Space Telescope \citep{Gallagher2003} and more recently the \textit{Herschel} Space Observatory \citep{Pilbratt2010} have allowed the full potential of this evolutionary framework to be exploited in constraining how the properties of protostars change as the source evolves through large-area, high spatial resolution, uniform photometric surveys of many nearby star-forming regions \citep[e.g.][]{Evans2003,Evans2009,Andre2010,Rebull2010,Megeath2012,Dunham2014a,Furlan2016}. Furthermore, the statistics available from such large surveys have enabled estimates of the relative lifetimes of the different Classes to be obtained, showing in particular that the combined Class 0 and I phases, where the majority of the protostellar mass is accreted and the final properties of the star and its accompanying disk are imprinted, last approximately 0.4$-$0.7\,Myr \citep{Dunham2015,Heiderman2015,Carney2016}. 

For a 1\msol{} star, such lifetimes imply typical time-averaged mass-accretion rates onto the protostar of approximately 10$^{-6}$\msol{}\,yr$^{-1}$. Since not all material in the core will end up on the star, the infall rate in the envelope must presumably be higher than this by at least a factor of 2 or 3. Searches to quantify the infall in protostars have presented candidates using molecular line observations \citep[e.g.][]{Gregersen1997,Mardones1997} based on the doppler-shift of infalling material causing asymmetries in the line profile \citep{Myers2000}. However, confirming and quantifying infall in protostellar envelopes remains extremely challenging, limiting our understanding of the rate at which, and route by which, material reaches the disk and protostar, as well as how this changes with time and depends on the mass of the core/star.

Bipolar molecular outflows also play an important role in the evolution and outcome of star formation, as they remove mass from and inject energy into the envelope and surrounding material. However, the driving mechanism for protostellar outflows is still uncertain \citep[e.g.][]{Arce2007,Frank2014}. A decrease in the driving force was measured between Class 0 and I sources, in addition to relations with $L_{\mathrm{bol}}$ and M$_{\mathrm{env}}$, by \citet{Bontemps1996} using ground-based observations of CO. They attributed the decrease in outflow driving force with Class to a decrease in the accretion/infall rate as the source evolves. However, their study only included ten Class 0 sources, as few were known at the time. 

Recent observations of H$_{2}$O and highly-excited CO using the Heterodyne Instrument for the Far-Infrared \citep[HIFI;][]{deGraauw2010} and Photodetector Array Camera and Spectrometer \citep[PACS;][]{Poglitsch2010} with \textit{Herschel} have shown that these primarily trace active shocks related to the outflow and/or warm disk winds heated by ambipolar diffusion, rather than the entrained outflow as is accessible with ground-based CO observations \citep{Nisini2010,Kristensen2013,Tafalla2013,Santangelo2013,Santangelo2014,Mottram2014,Yvart2016}. The line-width and intensity in these tracers decreases between Class 0 and I while the excitation conditions ($T$,$N$,$n$) remain the same \citep{Mottram2014,Manoj2013,Karska2013,Green2013a,Karska2014a,Matuszak2015}. However, these studies have typically considered relatively small samples ($N\lesssim$30) of bright, well-known sources and so the statistical significance of trends with evolution and other source parameters has, in some cases, been low.

Two of the main surveys studying nearby Class 0/I protostars with \textit{Herschel} were the ``Water in star-forming regions with \textit{Herschel}'' (WISH) guaranteed time key program \citep{vanDishoeck2011}, which observed 29 Class 0/I protostars with HIFI and PACS plus ground-based follow-up, and the ``Dust, Ice, and Gas in Time'' (DIGIT) \textit{Herschel} key program \citep{Green2013a,Green2016}, which observed a further 13 Class 0/I protostars, primarily with full-scan PACS spectroscopy. Both the WISH and DIGIT surveys selected their samples to target well known, archetypal sources, ensuring success in detecting water, CO and other species and the availability of complementary data. As a result, these samples favoured luminous sources with particularly prominent and extended outflows, which may not be representative of the general population of protostars. In addition, both programs together only included a total of 42 low-mass sources split between Classes 0 and I, limiting the statistical significance of trends with evolution that might otherwise have been expected, for example between integrated intensity in water emission and $T_{\mathrm{bol}}$.

The motivation of the ``William Herschel Line Legacy'' (WILL) survey was therefore to further explore the physics (primarily infall and outflow) and chemistry of water, CO and other complementary species in Class 0/I protostars in nearby low-mass star forming regions using a combination of \textit{Herschel} and ground-based observations, building on WISH and DIGIT. The aim was to increase the number of Class 0/I protostars observed, thus improving the statistical significance of the existing correlations found by for example \citet{Kristensen2012}, and allowing shallower correlations to be tested, as well as improving the sampling of fainter and colder sources.

This paper is structured as follows. Section~\ref{S:sample} discusses the selection of the WILL sample, the basic physical properties of the sources and evaluates the properties of the combined WISH+DIGIT+WILL sample. Section~\ref{S:observations} gives the details and basic results of both the \textit{Herschel} observations and a complementary ground-based follow-up campaign. More detailed results and analysis are then presented thematically, centred around outflows (Sect.~\ref{S:outflow}) and envelope emission (Sect.~\ref{S:envelope}), followed by a discussion on the variation of water with evolution (Sect.~\ref{S:evolution}). Finally, we summarise our main conclusions in Sect.~\ref{S:conclusions}.

\section{Sample}
\label{S:sample}

\subsection{Selection}
\label{S:sample_selection}

The starting point for selecting a flux-limited sample of low-mass protostars was the catalogue of Class 0/I protostars identified as part of photometric surveys with the \textit{Spitzer} Space Telescope of the closest major star-forming clouds that make up the Gould Belt \citep{Gould1879}. In particular, these were drawn from the \textit{Spitzer} c2d \citep[][]{Evans2009}, \textit{Spitzer} Gould Belt \citep[][]{Dunham2015} and Taurus \textit{Spitzer} \citep[][]{Rebull2010} surveys. 

The initial catalogue was compiled from individual cloud catalogues for the Perseus, Taurus, Ophiuchus, Scorpius (also known as Ophiuchus North), Corona Australis and Chameleon star-forming regions \citep[for more details, see][]{Jorgensen2007,Rebull2007,Rebull2010,Padgett2008,Jorgensen2008,Hatchell2012,Peterson2011,Alcala2008}. At the time of selection in 2011, the \textit{Herschel} Gould Belt \citep[][]{Andre2010} survey had also produced catalogues of protostellar candidates in the Aquila Rift region \citep{Maury2011}, so these were also considered in an attempt to extend the coverage of the WILL survey to particularly young (cold) embedded young stellar objects (YSOs).

From this master catalogue of protostars in major star-forming regions within 500\,pc, the following criteria were used to select the final WILL sample:

\begin{enumerate}[(i)]

\item infrared slope (2$-$24\micron{}) $\alpha_{\mathrm{IR}} >$ 0.3 or non-detection,

\item $T_{\mathrm{bol}} <$ 350~K,

\item $L_{\mathrm{bol}}$ $>$ 0.4\,\lsol{} for Class 0 ($T_{\mathrm{bol}} <$ 70\,K),\\ $L_{\mathrm{bol}} \geq$ 1\,\lsol{} for Class I (70 $\leq T_{\mathrm{bol}} <$ 350\,K),

\item $\delta$ $<$ 35\deg{}.

\end{enumerate}

\noindent The distinction between Class I and II sources is normally made at $T_{\mathrm{bol}}$=650\,K \citep{Chen1995}, however \citet{Evans2009} found that Flat SED sources cover the range 350$-$950\,K with a mean around 650\,K and therefore likely consist of more evolved Class I or younger Class II sources. An upper limit of 350\,K was therefore imposed in order to exclude more evolved Class I sources from the sample. Water emission is typically weaker for Class I sources than Class 0s and is generally higher for more luminous sources \citep[e.g.][]{Kristensen2012}, so a higher $L_{\mathrm{bol}}$ cut was used for Class I sources in an attempt to ensure detections. Criteria \rm{i}$-$\rm{iii} were therefore designed to ensure that the sample includes only young, deeply embedded protostars that are bright enough to be detected in H$_{2}$O and related species based on the experience of the WISH and DIGIT surveys. Criterion \rm{iv} ensures that all WILL sources can be observed with ALMA to allow high spectral and spatial resolution ground-based interferometric follow-up of interesting sources.

Unfortunately, edge-on disks, reddened background sources and evolved asymptotic giant-branch (AGB) stars all have the potential to present similar infrared colours and thus contaminate any sample selected purely based on continuum properties. As first highlighted by \citet{vanKempen2009} for a sample of sources in Ophiuchus, molecular emission tracing dense gas can help to break this degeneracy. More specifically, the high critical density of HCO$^{+}$ $J$=4$-$3 or $J$=3$-$2 means it will not be strong in foreground cloud material, while the rarity of C$^{18}$O similarly means that the $J$=3$-$2 transition is only bright and concentrated in protostellar sources. In addition, more evolved disk sources will not present strong emission in single-dish HCO$^{+}$ spectra due to beam-dilution. Such data, particularly for HCO$^{+}$, have been collected and used to remove contaminants in a number of Gould Belt samples by \citet[][]{Heiderman2010}, \citet{Heiderman2015} and \citet{Carney2016}, which have some overlap with the initial candidate sample. Therefore, following the cuts detailed above, non-detection in HCO$^{+}$ $J$=4$-$3 or 3$-$2 was used, where data were available, to exclude contaminant sources.  

Most of the sources observed by the WISH and DIGIT surveys also conform to the above criteria, so any initial candidates within 5\arcsec{} of a WISH or DIGIT source were also excluded to avoid repeat observations. However, two sources, PER\,03 and PER\,11, have enough overlap with the WISH observations of L1448-MM (offset by 7.7\arcsec{}) and NGC1333-IRAS4B (offset by 6.4\arcsec{}), respectively, particularly in the H$_{2}$O 1$_{10}-$1$_{01}$ (557\,GHz) ground-state line obtained in a 39\arcsec{} beam, that they are removed from the WILL sample as presented here. Finally, source TAU\,05 was removed as it is the young and active Class II source DG~Tau B, which has an edge-on disk \citep{Podio2013}.

\begin{figure*}[htb!]
\begin{center}
\includegraphics[width=0.70\textwidth]{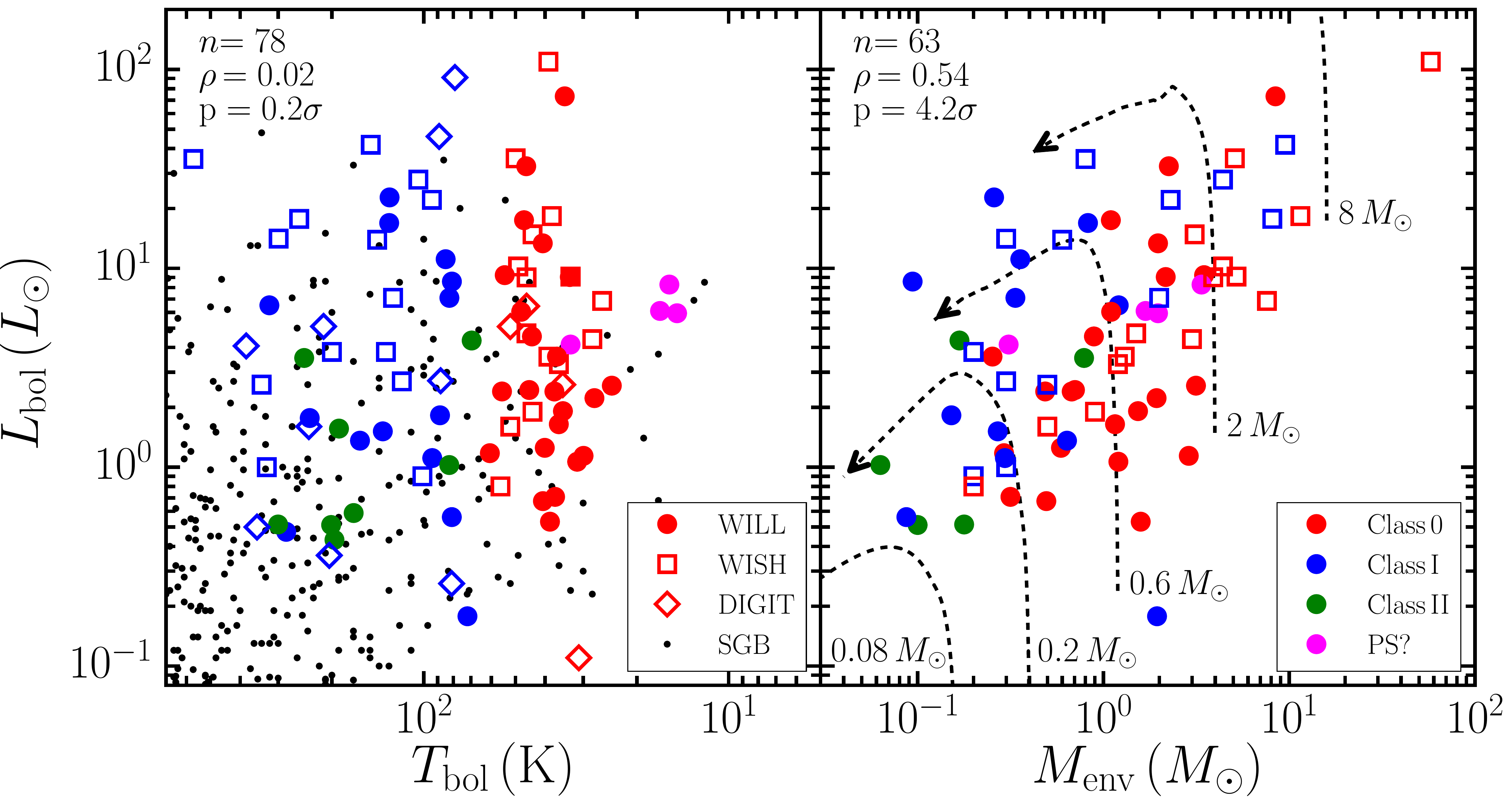}
\includegraphics[width=0.70\textwidth]{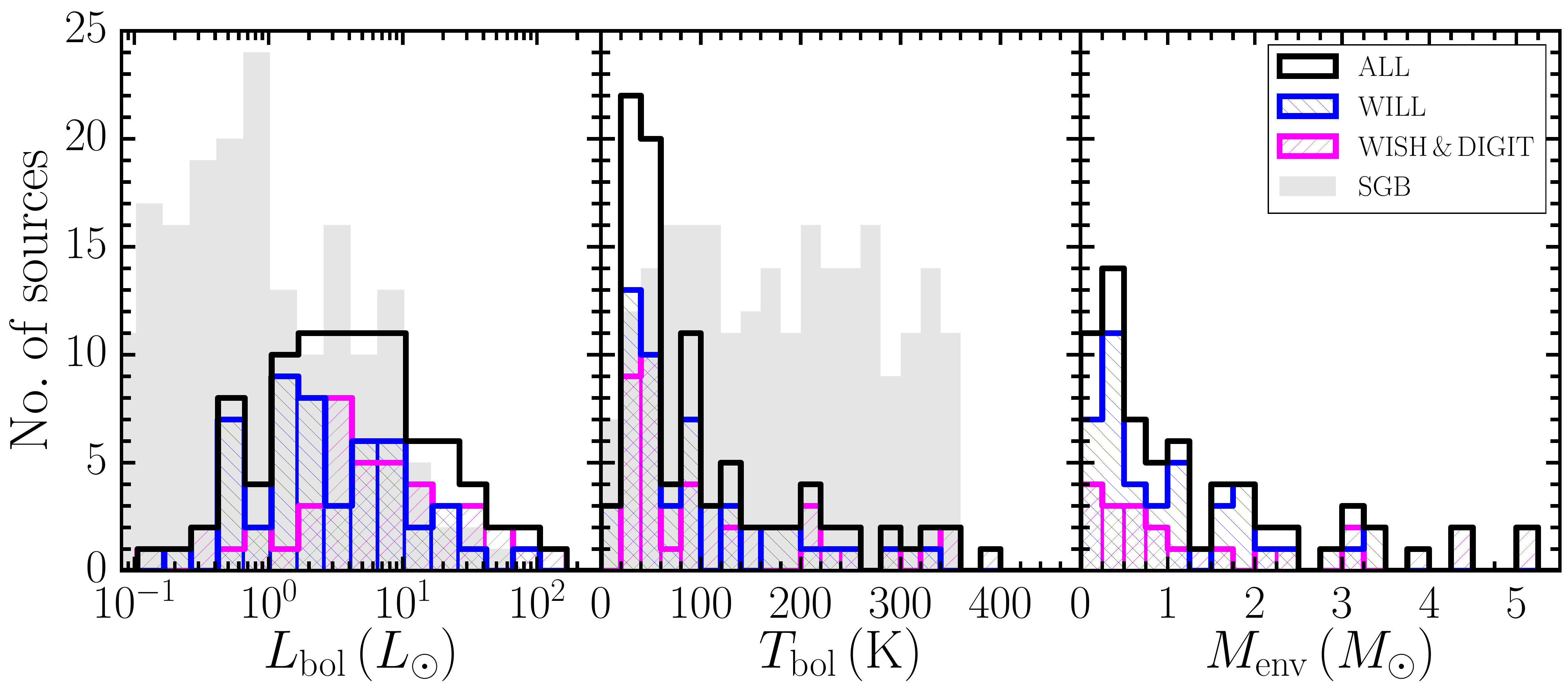}
\caption{Top: The distribution of $L_{\mathrm{bol}}$ vs. $T_{\mathrm{bol}}$ and $M_{\mathrm{env}}$ for the WILL (filled circles), WISH (open squares) and DIGIT (open diamonds) surveys. In the left-hand panel, the \textit{Spitzer} Gould Belt (SGB) determinations from \citet{Dunham2015} are shown for comparison (black dots). The different colours are used to distinguish between different source classifications: Class 0 (red), Class I (blue) Class II (green) and pre-stellar (PS, magenta). The number of sources ($n$), Pearson correlation coefficient ($\rho$), and the probability ($p$) that the correlation is not just due to random distributions in the variables are shown in the upper-left of each panel including only Class 0/I sources. Evolutionary tracks between $L_{\mathrm{bol}}$ and $M_{\mathrm{env}}$ from \citet{Duarte-Cabral2013} are shown in the right-hand panel (see text for details), with the final stellar mass indicated for each track. Bottom: histograms showing the distribution of $L_{\mathrm{bol}}$, $T_{\mathrm{bol}}$ and $M_{\mathrm{env}}$ for the WILL (blue), combined WISH and DIGIT (magenta hatched), and total WILL, WISH and DIGIT (black) samples. The grey shaded region indicates the distribution of the \textit{Spitzer} Gould Belt determinations for sources with $T_{\mathrm{bol}}\leq$350\,K.}
\label{F:lboltbol}
\end{center}
\end{figure*}

\subsection{Properties and evaluation}
\label{S:sample_evaluation}

The properties of the final sample of 49 sources that make up the WILL sample are presented in Table~\ref{T:properties}. For simplicity, we give each a name based on the region and a number ordered by right ascension, but many are already well known and therefore the table also gives details of common names used by previous studies for the same sources. 

The following distances are used for the various regions covered by our sample: 235\,pc for Perseus \citep{Hirota2008}, 140\,pc for Taurus \citep{Kenyon2008}, 125\,pc for Ophiuchus and Scorpius \citep{deGeus1989}, 130\,pc for Corona Australis \citep{Knude1998}, 150\,pc for Chameleon I and 178\,pc for Chameleon II \citep{Whittet1997}. For Aquila, W40 and Serpens South, \citet{Ortiz-Leon2016} recently found that these regions, as well as Serpens Main, are at a common distance of 436\,pc.

The determination of the source properties and evolutionary classification is discussed in detail in Appendix~\ref{S:properties}. To summarise briefly, the SED for each source is constructed from the near-IR to (sub-)mm and used to calculate $L_{\mathrm{bol}}$, $L_{\mathrm{submm}}$/$L_{\mathrm{bol}}$, $T_{\mathrm{bol}}$ and $\alpha_{\mathrm{IR}}$. $M_{\mathrm{env}}$ is obtained from sub-mm or mm photometry assuming that the dust is optically thin, while $\varv_{\mathrm{LSR}}$ is calculated from molecular line observations. Finally, the classification of each source is reached by considering the spatial and spectral properties of both the gas and dust associated with each source (see Appendix~\ref{S:properties_evolution} for more details). 

The sample comprises 23 Class 0, 14 Class I, 8 Class II and 4 uncertain, potentially pre-stellar sources. In the case of this last group of sources, all in W40, they are faint or not detected at $<$160\micron{}, show few detections in PACS and have no indications of outflow activity, but the presence of the W40 PDR, detected in some of the HIFI and ground-based lines, leaves some ambiguity. These and other cases of note are discussed in more detail in Appendix~\ref{S:cases}.

\begin{table*}
\caption{The WILL survey source sample.}
\begin{center}
\vspace{-2mm}
\begin{tabular}{l@{\,}ccccccccccl}
\hline \noalign {\smallskip}
Name & RA ($J$2000) & Dec ($J$2000) & $d$ & $\varv_{\mathrm{LSR}}$\tablefootmark{a} & $L_{\mathrm{bol}}$\tablefootmark{b} & $\frac{L_{\mathrm{submm}}}{L_{\mathrm{bol}}}$\tablefootmark{b} & $T_{\mathrm{bol}}$\tablefootmark{b} & $\alpha_{\mathrm{IR}}$\tablefootmark{b} & $M_{\mathrm{env}}$\tablefootmark{b} & Class\tablefootmark{c} & Other names\tablefootmark{d} \\
       & (h m s) & (\degr\ \arcmin\ \arcsec) & (pc) & (km\,s$^{-1}$) & (\lsol) & ($\%$) & (K) & & (\msol) & & \\ 
\hline\noalign {\smallskip}
AQU\,01\tablefootmark{e}&18:29:03.82&$-$01:39:01.5&436&$+$7.4&\phantom{0}2.6&11.8&\phantom{0}24&$-$&3.15&0&Aqu-MM2\\
AQU\,02\tablefootmark{e}&18:29:08.60&$-$01:30:42.8&436&$+$7.5&\phantom{0}9.0&\phantom{0}7.8&\phantom{0}33&$-$&2.17&0&Aqu-MM4, IRAS 18265-0132\\
AQU\,03\tablefootmark{e}&18:30:25.10&$-$01:54:13.4&436&$+$7.1&\phantom{0}3.5&\phantom{0}5.3&246&0.7&0.79&II&Aqu-MM6, IRAS 18278-0156\\
AQU\,04\tablefootmark{e}&18:30:28.63&$-$01:56:47.7&436&$+$7.6&\phantom{0}6.5&\phantom{0}4.5&320&0.5&1.21&I&Aqu-MM7, IRAS 18278-0158\\
AQU\,05\phantom{\tablefootmark{e}}&18:30:29.03&$-$01:56:05.4&436&$+$7.3&\phantom{0}2.4&\phantom{0}9.2&\phantom{0}37&1.4&0.68&0&Aqu-MM10\\
AQU\,06\phantom{\tablefootmark{e}}&18:30:49.94&$-$01:56:06.1&436&$+$8.3&\phantom{0}1.3&\phantom{0}8.2&\phantom{0}40&1.9&0.59&0&Aqu-MM14\\
CHA\,01\phantom{\tablefootmark{e}}&11:09:28.51&$-$76:33:28.4&150&$+$4.9&\phantom{0}1.6&$-$&189&1.6&$-$&II&GM Cha, ISO-ChaI 192, CaINa2\\
CHA\,02\phantom{\tablefootmark{e}}&12:59:06.58&$-$77:07:39.9&178&$+$3.0&\phantom{0}1.8&\phantom{0}0.6&236&1.3&$-$&I&ISO-ChaII 28, IRAS 12553-7651\\
CRA\,01\phantom{\tablefootmark{e}}&19:02:58.67&$-$37:07:35.9&130&$+$5.6&\phantom{0}2.4&\phantom{0}2.2&\phantom{0}55&1.7&0.49&0&ISO-CrA 182, IRAS 18595-3712\\
OPH\,01\phantom{\tablefootmark{e}}&16:26:59.10&$-$24:35:03.3&125&$+$3.8&\phantom{0}4.3&$-$&\phantom{0}69&2.0&0.17&II+PDR?&ISO-Oph 90, WL 22\\
OPH\,02\phantom{\tablefootmark{e}}&16:32:00.99&$-$24:56:42.6&125&$+$4.2&\phantom{0}8.6&\phantom{0}0.1&\phantom{0}80&1.8&0.09&I&ISO-Oph 209, Oph-emb 10\\
PER\,01\phantom{\tablefootmark{e}}&03:25:22.32&+30:45:13.9&235&$+$4.1&\phantom{0}4.5&\phantom{0}2.7&\phantom{0}44&2.3&0.89&0&L1448 IRS2, Per-emb 22\\
PER\,02\phantom{\tablefootmark{e}}&03:25:36.49&+30:45:22.2&235&$+$4.5&\phantom{0}9.2&\phantom{0}1.7&\phantom{0}54&2.6&3.48&0&L1448 N(A), L1448 IRS3, Per-emb 33\\
PER\,04\phantom{\tablefootmark{e}}&03:26:37.47&+30:15:28.1&235&$+$5.2&\phantom{0}1.2&\phantom{0}4.2&\phantom{0}60&1.2&0.29&0&IRAS 03235+3004, Per-emb 25\\
PER\,05\phantom{\tablefootmark{e}}&03:28:37.09&+31:13:30.8&235&$+$7.3&11.1&\phantom{0}0.6&\phantom{0}84&2.2&0.36&I&NGC1333 IRAS1, Per-emb 35\\
PER\,06\phantom{\tablefootmark{e}}&03:28:57.36&+31:14:15.9&235&$+$7.3&\phantom{0}7.1&$-$&\phantom{0}82&1.5&0.34&I&NGC1333 IRAS2B, Per-emb 36\\
PER\,07\phantom{\tablefootmark{e}}&03:29:00.55&+31:12:00.8&235&$+$7.4&\phantom{0}0.7&\phantom{0}3.9&\phantom{0}37&2.1&0.32&0&Per-emb 3\\
PER\,08\phantom{\tablefootmark{e}}&03:29:01.56&+31:20:20.6&235&$+$7.7&16.9&\phantom{0}1.3&129&2.5&0.83&I&Per-emb 54, NGC1333 IRAS6\\
PER\,09\phantom{\tablefootmark{e}}&03:29:07.78&+31:21:57.3&235&$+$7.5&22.7&$-$&129&2.6&0.26&I&IRAS 03260+3111(W), Per-emb 50\\
PER\,10\phantom{\tablefootmark{e}}&03:29:10.68&+31:18:20.6&235&$+$8.7&\phantom{0}6.0&\phantom{0}2.2&\phantom{0}47&1.9&1.10&0&NGC1333 IRAS7, Per-emb 21\\
PER\,12\phantom{\tablefootmark{e}}&03:29:13.54&+31:13:58.2&235&$+$7.8&\phantom{0}1.1&\phantom{0}8.7&\phantom{0}31&2.4&1.20&0&NGC1333 IRAS4C, Per-emb 14\\
PER\,13\phantom{\tablefootmark{e}}&03:29:51.82&+31:39:06.0&235&$+$8.0&\phantom{0}0.7&\phantom{0}5.0&\phantom{0}40&3.5&0.49&0&IRAS 03267+3128, Per-emb 9\\
PER\,14\phantom{\tablefootmark{e}}&03:30:15.14&+30:23:49.4&235&$+$6.2&\phantom{0}1.8&\phantom{0}1.6&\phantom{0}88&1.8&0.15&I&IRAS 03271+3013, Per-emb 34\\
PER\,15\phantom{\tablefootmark{e}}&03:31:20.98&+30:45:30.1&235&$+$6.9&\phantom{0}1.6&\phantom{0}5.8&\phantom{0}36&1.2&1.16&0&IRAS 03282+3035, Per-emb 5\\
PER\,16\phantom{\tablefootmark{e}}&03:32:17.96&+30:49:47.5&235&$+$7.0&\phantom{0}1.1&13.3&\phantom{0}29&1.0&2.88&0&IRAS 03292+3039, Per-emb 2\\
PER\,17\phantom{\tablefootmark{e}}&03:33:14.38&+31:07:10.9&235&$+$6.6&\phantom{0}0.2&$-$&\phantom{0}71&2.4&1.94&I&B1 SMM3, Per-emb 6\\
PER\,18\phantom{\tablefootmark{e}}&03:33:16.44&+31:06:52.5&235&$+$6.6&\phantom{0}0.5&$-$&\phantom{0}38&1.6&1.59&0&B1d, Per-emb 10\\
PER\,19\phantom{\tablefootmark{e}}&03:33:27.29&+31:07:10.2&235&$+$6.8&\phantom{0}1.1&\phantom{0}1.7&\phantom{0}93&1.9&0.29&I&B1 SMM11, Per-emb 30\\
PER\,20\phantom{\tablefootmark{e}}&03:43:56.52&+32:00:52.8&235&$+$8.9&\phantom{0}2.2&\phantom{0}6.3&\phantom{0}27&0.7&1.93&0&IRAS 03407+3152, HH 211, Per-emb 1\\
PER\,21\phantom{\tablefootmark{e}}&03:43:56.84&+32:03:04.7&235&$+$8.8&\phantom{0}1.9&\phantom{0}3.8&\phantom{0}35&1.5&1.54&0&IC348 MMS, Per-emb 11\\
PER\,22\phantom{\tablefootmark{e}}&03:44:43.96&+32:01:36.2&235&$+$9.8&\phantom{0}2.4&\phantom{0}3.4&\phantom{0}45&0.9&0.70&0&IRAS 03415+3152, Per-emb 8\\
SCO\,01\phantom{\tablefootmark{e}}&16:46:58.27&$-$09:35:19.8&125&\phantom{\tablefootmark{f}}$+$3.6\tablefootmark{f}&\phantom{0}0.5&\phantom{0}0.6&201&0.9&0.10&II&IRAS 16442-0930, L260 SMM1\\
SERS\,01\phantom{\tablefootmark{e}}&18:29:37.70&$-$01:50:57.8&436&$+$8.2&17.4&\phantom{0}3.9&\phantom{0}46&1.3&1.10&0&IRAS 18270-0153, SerpS-MM1\\
SERS\,02\phantom{\tablefootmark{e}}&18:30:04.13&$-$02:03:02.1&436&$+$7.8&73.2&\phantom{0}4.6&\phantom{0}34&2.5&8.44&0&SerpS-MM18\\
TAU\,01\phantom{\tablefootmark{e}}&04:19:58.40&+27:09:57.0&140&$+$6.8&\phantom{0}1.5&\phantom{0}3.3&136&1.4&0.27&I&IRAS 04169+2702\\
TAU\,02\phantom{\tablefootmark{e}}&04:21:11.40&+27:01:09.0&140&$+$6.6&\phantom{0}0.5&\phantom{0}0.8&282&0.5&$-$&I&IRAS 04181+2654A\\
TAU\,03\phantom{\tablefootmark{e}}&04:22:00.60&+26:57:32.0&140&\phantom{\tablefootmark{f}}$+$7.4\tablefootmark{f}&\phantom{0}0.4&\phantom{0}0.2&196&1.0&$-$&II&IRAS 04189+2650(W)\\
TAU\,04\phantom{\tablefootmark{e}}&04:27:02.60&+26:05:30.0&140&$+$6.3&\phantom{0}1.4&\phantom{0}1.5&161&0.8&0.64&I&DG TAU B\\
TAU\,06\phantom{\tablefootmark{e}}&04:27:57.30&+26:19:18.0&140&$+$7.2&\phantom{0}0.6&\phantom{0}2.7&\phantom{0}80&0.8&0.09&I&HH31 IRS 2, IRAS 04248+2612\\
TAU\,07\phantom{\tablefootmark{e}}&04:29:30.00&+24:39:55.0&140&\phantom{\tablefootmark{f}}$+$6.3\tablefootmark{f}&\phantom{0}0.6&\phantom{0}0.2&169&0.9&$-$&II&HH 414, IRAS 04264+2433\\
TAU\,08\phantom{\tablefootmark{e}}&04:32:32.00&+22:57:26.0&140&\phantom{\tablefootmark{g}}$+$5.5\tablefootmark{g}&\phantom{0}0.5&\phantom{0}1.2&300&0.5&0.18&II&L1536 IRS, IRAS 04295+2251\\
TAU\,09\phantom{\tablefootmark{e}}&04:35:35.30&+24:08:19.0&140&$+$5.5&\phantom{0}1.0&\phantom{0}1.7&\phantom{0}82&1.4&0.06&II&L1535 IRS, IRAS 04325+2402\\
W40\,01\phantom{\tablefootmark{e}}&18:31:09.42&$-$02:06:24.5&436&$+$4.9&13.3&\phantom{0}7.4&\phantom{0}40&2.3&1.97&0+PDR&W40-MM3\\
W40\,02\phantom{\tablefootmark{e}}&18:31:10.36&$-$02:03:50.4&436&$+$4.8&32.6&\phantom{0}3.7&\phantom{0}46&4.6&2.25&0&W40-MM5\\
W40\,03\phantom{\tablefootmark{e}}&18:31:46.54&$-$02:04:22.5&436&$+$6.4&\phantom{0}8.3&20.6&\phantom{0}15&$-$&3.37&PS?+PDR&W40-MM26\\
W40\,04\phantom{\tablefootmark{e}}&18:31:46.78&$-$02:02:19.9&436&$+$6.7&\phantom{0}6.1&\phantom{0}9.4&\phantom{0}16&$-$&1.69&PS?+PDR&W40-MM27\\
W40\,05\phantom{\tablefootmark{e}}&18:31:47.90&$-$02:01:37.2&436&$+$6.5&\phantom{0}5.9&27.3&\phantom{0}14&$-$&1.97&PS?+PDR&W40-MM28\\
W40\,06\phantom{\tablefootmark{e}}&18:31:57.24&$-$02:00:27.7&436&$+$6.6&\phantom{0}4.1&\phantom{0}2.2&\phantom{0}33&$-$&0.31&PS?+PDR&W40-MM34\\
W40\,07\phantom{\tablefootmark{e}}&18:32:13.36&$-$01:57:29.6&436&$+$7.4&\phantom{0}3.6&\phantom{0}3.3&\phantom{0}36&0.9&0.25&0&W40-MM36\\
\hline
\end{tabular}
\vspace{-2mm}
\tablefoot{\tablefoottext{a}{From Gaussian fits to the C$^{18}$O $J$=3$-$2 observations (see Table~\ref{T:ground_spectra}).} \tablefoottext{b}{Calculated as discussed in Sect.~\ref{S:properties_evolution}.} \tablefoottext{c}{Evolutionary classification, see Sect.~\ref{S:properties_evolution} for details of the determination. PS=pre-stellar, PDR=narrow, bright $^{12}$CO $J$=10$-$9 emission consistent with a photon-dominated region.} \tablefoottext{d}{First additional names for Aquila, Serpens South and W40 are from \citet{Maury2011}, `-emb' names from \citet{Enoch2009}.} \tablefoottext{e}{Sources off-centre in beam. Peak coordinates in PACS maps used for extraction of ground-based data: AQU\,01 18:29:03.61 $-$01:39:05.6; AQU\,02 18:29:08.20 $-$01:30:46.6; AQU\,03 18:30:24.69 $-$01:54:11.0; AQU\,04 18:30:29.32 $-$01:56:42.4.} \tablefoottext{f}{From Gaussian fits to the $^{13}$CO ($J$=3$-$2) observations as C$^{18}$O is not detected.} \tablefoottext{g}{Taken from \cite{Caselli2002}.}}
\end{center}
\label{T:properties}
\end{table*}

Figure~\ref{F:lboltbol} shows the $L_{\mathrm{bol}}$, $T_{\mathrm{bol}}$ and $M_{\mathrm{env}}$ distribution of the WILL sample, along with the WISH and DIGIT samples for comparison. The properties of the WISH sample are taken from \citet{Kristensen2012} while those for the DIGIT sample are taken from \citet{Green2013a} and \citet{Lindberg2014}. These are corrected to the distances for the various regions discussed above where needed. It should be noted that $M_{\mathrm{env}}$ values are not available for the DIGIT sample, leading to the difference in the number of sources between the upper-left and upper-right panels. 

The probability (p) that a given value of the Pearson correlation coefficient ($\rho$) for sample size $n$ represents a real correlation (i.e. the likelihood that a two-tailed test can reject the null-hypothesis that the two variables are uncorrelated with $\rho$=0) can be expressed in terms of the standard deviation of a normal distribution, $\sigma$, as:

\begin{equation}
p=\mid\rho\mid\sqrt{n-1}\sigma,
\label{E:sigma}
\end{equation}

\noindent following \citep{Marseille2010}. We consider $p=$3$\sigma$ (i.e. 99.7$\%$) to be the threshold for statistical significance. Thus, for a sample size of 30, values of $\mid\rho\mid>0.56$ indicate real, statistically significant correlations while for a sample size of 50, this is true for $\mid\rho\mid>0.43$. While one might expect correlations between some of the observed properties of embedded protostars due to the related nature of their different components (e.g. envelope, outflow and driving source), such tests are a simple way of ascertaining whether or not the data are able to support such links. As mentioned above, the extension of the sample of sources studied in spectral lines with PACS and HIFI enabled by the WILL survey and presented here allows us to study these more completely for the first time. 

The evolutionary tracks between $L_{\mathrm{bol}}$ and M$_{\mathrm{env}}$ shown in the top-right panel of Figure~\ref{F:lboltbol} are taken from \citet{Duarte-Cabral2013}. They assume an exponential decrease of $M_{\mathrm{env}}$ and a core-to-star formation efficiency of 50$\%$, such that the net accretion rate is given by:

\begin{equation}
\dot{M}_{\mathrm{acc}}(t)~=~0.5\,\frac{M_{\mathrm{env}}(t)}{\tau}\,,
\label{E:dcmdotacc}
\end{equation}

\noindent where $\tau$ is the e-folding time, which is assumed to be 3$\times$10$^{5}$\,yrs.

The WILL sample doubles the number of low-mass YSOs observed, which have slightly lower values of $L_{\mathrm{bol}}$ and $M_{\mathrm{env}}$, as well as lower $T_{\mathrm{bol}}$ for Class 0 sources, than the WISH and DIGIT samples. Comparing to \textit{Spitzer} Gould Belt (SGB) sources with $T_{\mathrm{bol}}\leq$350\,K, taken from \citet{Dunham2015}, it can be seen in Fig.~\ref{F:lboltbol} that the combined WILL+DIGIT+WISH sample is representative of the overall Class 0/I population and contains most sources above $\sim$1\lsol. Below this luminosity, the sample rapidly becomes incomplete, and thus the combined sample is still biased towards higher mean $L_{\mathrm{bol}}$ compared with the general distribution, but the addition of the WILL sources shifts the completeness limit approximately a factor of three lower. In terms of $T_{\mathrm{bol}}$, the sample is biased towards lower values, but judging from upper-left panel of Fig.~\ref{F:lboltbol}, the higher $T_{\mathrm{bol}}$ sources in the SGB data are primarily those below our $L_{\mathrm{bol}}$ limit, that is, the mean $L_{\mathrm{bol}}$ decreases as $T_{\mathrm{bol}}$ increases for SGB sources. The differences between the values of \citet{Dunham2015} and those given here for individual sources are likely due to our inclusion of far-IR data in these determinations.

It is worth mentioning a couple of caveats. Firstly, the sample of Class 0 sources is dominated by sources in the Perseus molecular cloud, while the Class I sources are drawn from a number of regions that vary in the concentration and activity of their star formation (e.g. Taurus vs. Ophiuchus). There may well be regional differences due to environmental effects, which we cannot test due to the overall small sample size for a given region. Secondly, by excluding older Class I and flat-spectrum sources, we introduce a bias towards younger Class I sources, so the properties of an average Class I source may well be slightly different from those determined with this sample. However, in general for the part of parameter space that WILL, WISH and DIGIT are designed to probe, the addition of the WILL survey leaves the combined sample broadly complete.

\section{Observations and results}
\label{S:observations}

The primary observations for the WILL survey were taken with \textit{Herschel}\footnote{\textit{Herschel} is an ESA space observatory with science instruments provided by European-led Principal Investigator consortia and with important participation from NASA.} using the Heterodyne Instrument for the Far-Infrared \citep[HIFI,][]{deGraauw2010} and Photodetector Array Camera and Spectrometer \citep[PACS,][]{Poglitsch2010} detectors between the 31$^{st}$ October 2012 and 27$^{th}$ March 2013. The observing modes, observational properties, data reduction and detection statistics are described for each instrument separately in Sections~\ref{S:observations_hifi} and \ref{S:observations_pacs}. Complementary spectroscopic maps obtained through follow-up observations of the sample with ground-based facilities are then described in Section~\ref{S:observations_jcmtapex}.

\subsection{HIFI}
\label{S:observations_hifi}

\subsubsection{Observational details}
\label{S:observations_hifi_obs}

HIFI was a set of seven single-pixel dual-sideband heterodyne receivers that combined to cover the frequency ranges 480$-$1250\,GHz and 1410$-$1910\,GHz with a sideband ratio of approximately unity. Spectra were simultaneously observed in two polarisations, $H$ and $V$, which pointed at slightly different positions on the sky ($\sim$6.5\arcsec{} apart at 557\,GHz decreasing to $\sim$2.8\arcsec{} at 1153\,GHz), with two spectrometers simultaneously providing both wideband (WBS, 4\,GHz bandwidth at 1.1\,MHz resolution) and high-resolution (HRS, typically 230\,MHz bandwidth at 250\,kHz resolution) frequency coverage.

The HIFI component of the WILL \textit{Herschel} observations consists of single pointed spectra at four frequency settings, principally targeting the H$_{2}$O 1$_{10}-$1$_{01}$, 1$_{11}-$0$_{00}$ and 2$_{02}-$1$_{11}$ transitions at 557, 1113 and 988\,GHz respectively and the $^{12}$CO $J$=10$-$9 transition at 1152\,GHz, which also includes the H$_{2}$O 3$_{12}-$2$_{21}$ transition. All observations were carried out in dual-beam-switch mode with a nod of 3\arcmin{} using fast chopping. The specific central frequencies of the settings were chosen to maximise the number of observable H$_{2}$O, CO and H$_{2}^{18}$O transitions, the details of which are given in Table~\ref{T:observations_hifi_lines} along with the corresponding instrumental properties, spectral and spatial resolution, and observing time. The main difference compared to the WISH HIFI observations of low-mass sources \citep[see][]{Kristensen2012,Mottram2014} was that the frequency of the WILL observations for the H$_{2}$O 1$_{10}-$1$_{01}$ and 1$_{11}-$0$_{00}$ settings was set so that the corresponding H$_{2}^{18}$O transition was observed simultaneously, and longer observing times were used for the H$_{2}$O 1$_{10}-$1$_{01}$ setting. The observation ID numbers for all WILL HIFI observations are given in Table~\ref{T:obsids_will}. 

\begin{table*}
\begin{center}
\caption[]{Principle lines observed with HIFI.}
\vspace{-2mm}
\begin{tabular}{lccccccccccc}
\hline \noalign {\smallskip}
Species & Transition & Rest Frequency\tablefootmark{a}& $E_{\mathrm{u}}$/$k_{\mathrm{b}}$ & $A_{\mathrm{ul}}$\tablefootmark{b} & $n_{\mathrm{cr}}$\tablefootmark{c} & $\eta_{mb}$\tablefootmark{d} & $\theta_{mb}$\tablefootmark{e} & WBS resolution & HRS resolution & Obs. Time\tablefootmark{f} & Det.\tablefootmark{g} \\
& & (GHz) & (K) & (s$^{-1}$) & (cm$^{-3}$) & (H/V) & (\arcsec{}) & (\kms{}) & (\kms{}) & (min) & \\
\hline\noalign {\smallskip}
o-H$_{2}$O & 1$_{10}$-1$_{01}$ & \phantom{0}556.93599 & \phantom{0}61.0 & 3.46$\times$10$^{-3}$ & 1$\times$10$^{7}$ & 0.62/0.62 & 38.1 & 0.27 & 0.03 & 38 & 39/46 \\
 & 3$_{12}$-2$_{21}$ & 1153.12682 & 249.4 & 2.63$\times$10$^{-3}$ & 8$\times$10$^{6}$ & 0.59/0.59 & 18.4 & 0.13 & 0.06 & 13 & \phantom{0}7/46 \\
\hline\noalign {\smallskip}
p-H$_{2}$O & 1$_{11}$-0$_{00}$ & 1113.34301 & \phantom{0}53.4 & 1.84$\times$10$^{-2}$ & 1$\times$10$^{8}$ & 0.63/0.64 & 19.0 & 0.13 & 0.06 & 28 & 28/46 \\
 & 2$_{02}$-1$_{11}$ & \phantom{0}987.92676 & 100.8 & 5.84$\times$10$^{-3}$ & 4$\times$10$^{7}$ & 0.63/0.64 & 21.5 & 0.15 & 0.07 & 36 & 25/46 \\
\hline\noalign {\smallskip}
o-H$_{2}^{18}$O & 1$_{10}$-1$_{01}$ & \phantom{0}547.67644 & \phantom{0}60.5 & 3.29$\times$10$^{-3}$& 1$\times$10$^{7}$ & 0.62/0.62 & 38.7 & 0.27 & 0.07 & 38 & \phantom{0}1/46\\
\hline\noalign {\smallskip}
p-H$_{2}^{18}$O & 1$_{11}$-0$_{00}$ & 1101.69826 & \phantom{0}52.9 & 1.79$\times$10$^{-2}$ & 1$\times$10$^{8}$ & 0.63/0.64 & 19.0 & 0.13 & 0.06 & 28 & \phantom{0}0/46\\
\hline\noalign {\smallskip}
C$^{18}$O & 9$-$8 & \phantom{0}987.56038 & 237.0 & 6.38$\times$10$^{-5}$ & 2$\times$10$^{5}$ & 0.63/0.64 & 21.5 & 0.15 & 0.07 & 36 & \phantom{0}4/46\\
CO & 10$-$9 & 1151.98545 & 304.2 & 1.01$\times$10$^{-4}$ & 3$\times$10$^{5}$ & 0.59/0.59 & 18.4 & 0.13 & 0.06 & 13 & 40/46 \\
$^{13}$CO & 10$-$9 & 1101.34966 & 290.8 & 8.86$\times$10$^{-5}$ & 3$\times$10$^{5}$ & 0.63/0.64 & 19.3 & 0.13 & 0.06 & 28 & 20/46 \\
\hline\noalign {\smallskip}
\label{T:observations_hifi_lines}
\end{tabular}
\vspace{-5mm}
\tablefoot{\tablefoottext{a}{Taken from the JPL database \citep{Pickett2010}.} \tablefoottext{b}{Taken from \citet{Daniel2011} and \citet{Dubernet2009} for H$_{2}$O, the JPL database \citep{Pickett2010} for H$_{2}^{18}$O and CO isotopologues.}  \tablefoottext{c}{Calculated for T=300\,K.} \tablefoottext{d}{Taken from the latest HIFI calibration document at http://herschel.esac.esa.int/twiki/pub/Public/HifiCalibrationWeb/HifiBeamReleaseNote\_Sep2014.pdf .} \tablefoottext{e}{Calculated using equation 3 from \citet{Roelfsema2012}.} \tablefoottext{f}{Total time including on+off source and overheads.}  \tablefoottext{g}{Number of detections. Due to contamination of the reference positions, the status for observations of W40 sources 01, 03 and 06 cannot be determined.}}
\end{center}
\vspace{-2mm}
\end{table*}

Initial data reduction was conducted using the Herschel Interactive Processing Environment \citep[\textsc{hipe} v. 10.0,][]{Ott2010}. After initial spectrum formation, any instrumental standing waves were removed. Next, a low-order ($\leq$2) polynomial baseline was subtracted from each sub-band. The fit to the baseline was then used to calculate the continuum level, compensating for the dual-sideband nature of the HIFI detectors (the initial continuum level is the combination of emission from both the upper and lower sideband, which we assume to be equal). Following this the WBS sub-bands were stitched into a continuous spectrum and all data were converted to the $T_{\mathrm{MB}}$ scale using the latest beam efficiencies (see Table~\ref{T:observations_hifi_lines}). Finally, for ease of analysis, all data were converted to FITS format and resampled to 0.3\kms{} spectral resolution on the same velocity grid using bespoke \textsc{python} routines. 

Few differences have been found in line-shape or gain between the $H$ and $V$ polarisations \citep[e.g.][]{Kristensen2012,Yildiz2013,Mottram2014}, so after visual inspection the two polarisations were co-added to improve signal-to-noise. The velocity calibration is better than 100\,kHz, while the pointing uncertainty is better than 2\arcsec{} and the intensity calibration uncertainty is $\lesssim$10$\%$ \citep[][]{Mottram2014}.

\subsubsection{Results}

\begin{figure*}
\begin{center}
\includegraphics[width=0.88\textwidth]{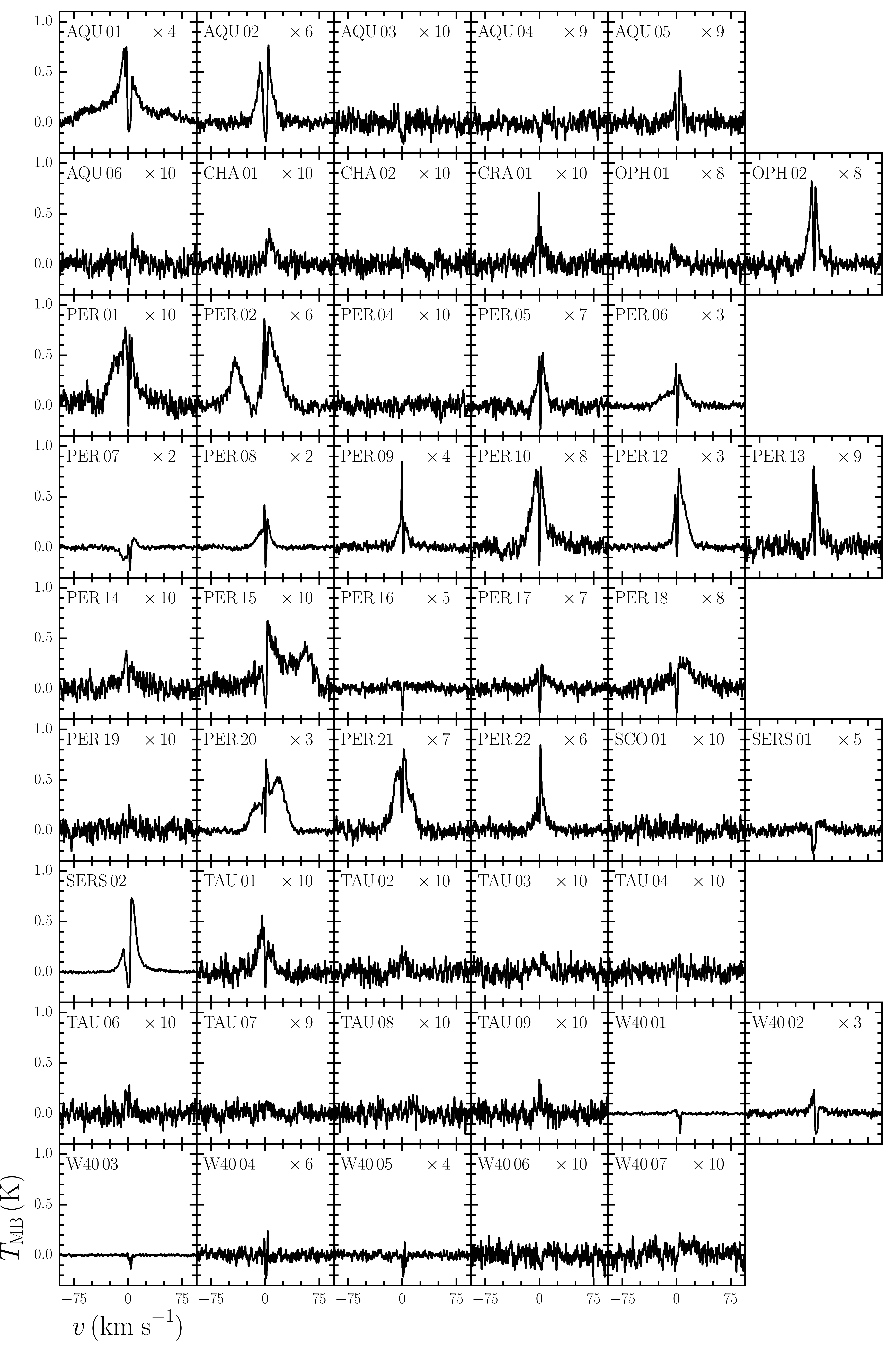}
\caption{H$_{2}$O 1$_{10}-$1$_{01}$ (557\,GHz) continuum-subtracted spectra for the final WILL sample. All have been recentred so that the source velocity is at zero.  The number in the upper-right corner of each panel indicates what factor the spectra have been multiplied by in order to show them on a common scale.}
\label{F:hifi_557}
\end{center}
\end{figure*}

\begin{figure*}
\begin{center}
\includegraphics[width=0.88\textwidth]{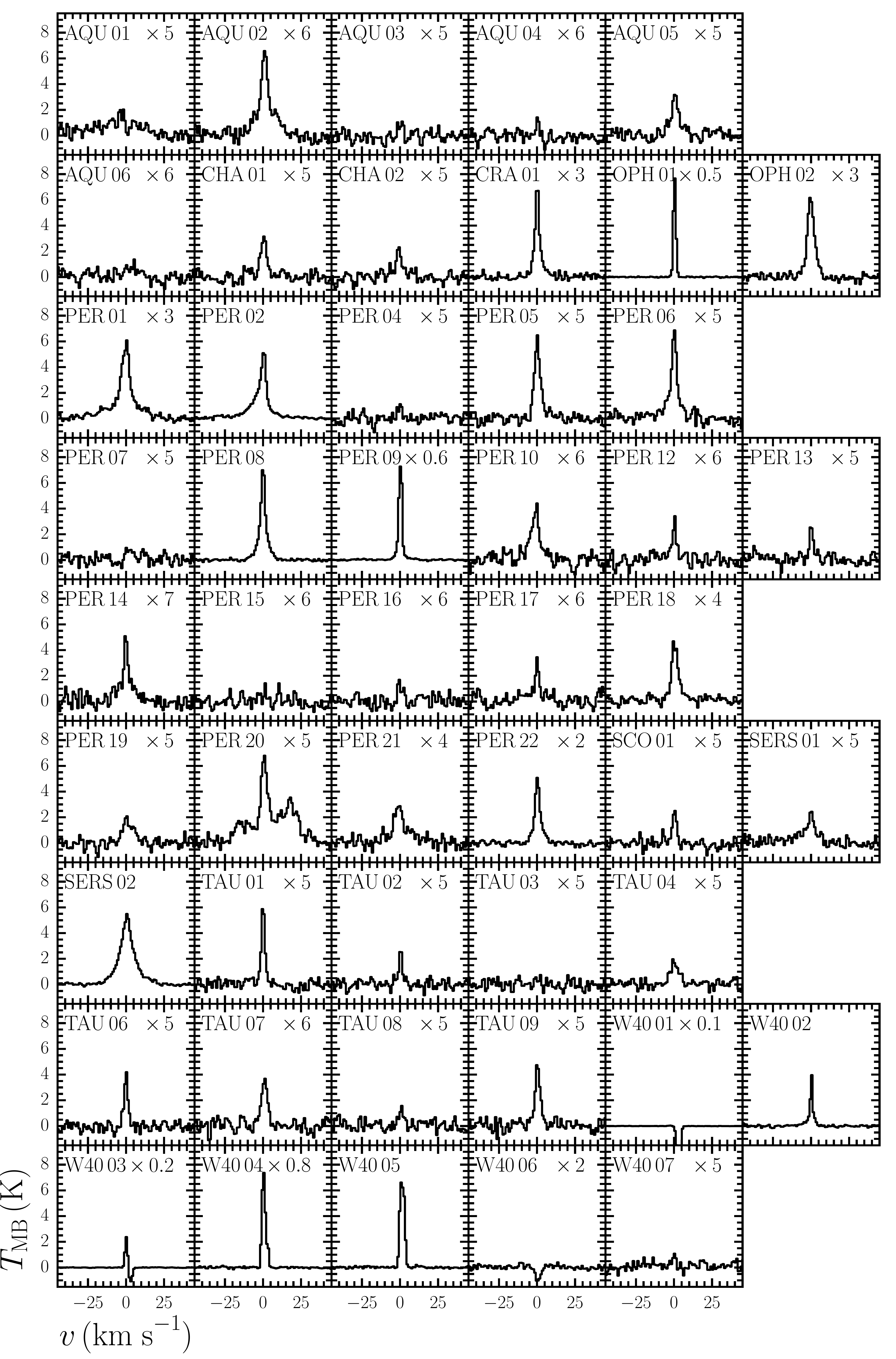}
\caption{CO $J$=10$-$9 continuum-subtracted spectra for the final WILL sample. All have been recentred so that the source velocity is at zero.  The number in the upper-right corner of each panel indicates what factor the spectra have been multiplied by in order to show them on a common scale.}
\label{F:hifi_co10-9}
\end{center}
\end{figure*}

Figures~\ref{F:hifi_557} and \ref{F:hifi_co10-9} present the observed HIFI ortho-H$_{2}$O 1$_{10}-$1$_{01}$ (557\,GHz) ground-state transition and $^{12}$CO $J$=10$-$9, respectively, for all WILL sources. The water spectra are complex, containing multiple components, some absorption, which is usually narrow, and emission up to $\pm\sim$100\kms{} from the source velocity, similar to other \textit{Herschel} HIFI observations of water towards Class 0/I sources \citep[e.g.][]{Kristensen2012}. $^{12}$CO $J$=10$-$9 typically shows two gaussian emission components with a lower total velocity extent than H$_{2}$O. Strong, narrow absorption in $^{12}$CO $J$=10$-$9 for W40 sources 01, 03 and 06 (see Fig.~\ref{F:hifi_co10-9}) indicates that contamination in at least one of the reference positions affects these spectra and also likely affects most of the H$_{2}$O transitions for these sources as well. The narrow yet bright nature of the $^{12}$CO $J$=10$-$9 seen in six sources (OPH\,01, W40\,01 and W40\,03$-$06, see Fig.~\ref{F:hifi_co10-9}), combined with the narrow and low-intensity nature of the H$_{2}$O emission, suggests that they are related to photon-dominated regions \citep[PDRs, c.f. for example CO observations of the Orion Bar PDR,][]{Hogerheijde1995,Jansen1996,Nagy2013}.

The detection statistics for all transitions are given in the last column of Table~\ref{T:observations_hifi_lines}, excluding W40 sources 01, 03 and 06 due to the contamination of these spectra. The H$_{2}$O 1$_{10}-$1$_{01}$ transition is detected towards 39/46 sources in total, including 33/36 confirmed Class 0/I sources (not detected in CHA\,02, PER\,04 and W40\,07, see Fig.~\ref{F:hifi_557}), while $^{12}$CO $J$=10$-$9 is detected towards 40/46 sources in total including 32/36 Class 0/Is (not detected in CHA\,02, PER\,07, PER\,15 and W40\,07). H$_{2}^{18}$O 1$_{10}-$1$_{01}$ is only detected towards the source with the strongest H$_{2}$O emission, SERS\,02, while C$^{18}$O $J$=9$-$8 is only detected towards four sources (PER\,02, SERS\,02, W40\,04 and \,05).

A more detailed analysis of the kinematics of the HIFI lines is presented and discussed in \citet{SanJoseGarciaThesis}, including the results of Gaussian decomposition of the lines using the methods outlined for the WISH sample by \citet{Mottram2014} and \citet{SanJoseGarcia2013} for H$_{2}$O and CO, respectively. In summary, the minimum number of Gaussian components is found that results in no residuals above 3$\sigma$, with these components then categorised between the envelope and C or J-type outflow-related shocks depending on their width and offset from the source velocity. A global fit is used for the H$_{2}$O transitions with the component peak velocity and line-widths constrained by all lines and the intensity allowed to vary between transitions because the lines all have a consistent shape. The different CO transitions are fit independently as their line profile shapes vary between different transitions.

\subsection{PACS}
\label{S:observations_pacs}

\subsubsection{Observational details}

PACS consisted of four detectors, two photoconductor arrays with 16$\times$25 pixels for integral field unit (IFU) spectroscopy and two bolometer arrays with 16$\times$32 and 32$\times$64 pixels for broad-band imaging photometry. In IFU spectroscopy mode, observations were taken simultaneously in the red 1st order grating (102$-$210\micron{}) and one of the 2nd or 3rd order blue gratings (51$-$73\micron{} or 71$-$105\micron{}) over 5$\times$5 spatial pixels (spaxels), which covered a 47\arcsec{}$\times$47\arcsec{} field of view. For details of the 70, 100 and 160\micron{} PACS \citep[and 250, 350 and 500\micron{} SPIRE,][]{Griffin2010} photometric maps used to determine the continuum flux densities for the SEDs (discussed in Section~\ref{S:properties_seds}) see \citet{Andre2010}.

\begin{table}
\begin{center}
\caption[]{Wavelength ranges covered by WILL PACS line-scan settings.}
\vspace{-2mm}
\begin{tabular}{lcc}
\hline \noalign {\smallskip}
Setting & Wavelengths& Primary Transitions \\
& (\micron{}) & \\
\hline\noalign {\smallskip}
\multirow{8}{*}{1} & \phantom{0}78.6 $-$ \phantom{0}79.5 & H$_{2}$O 4$_{23}-$3$_{12}$, 6$_{15}-$5$_{24}$, CO 33$-$32, OH  \\
 & \phantom{0}81.3 $-$ \phantom{0}82.2 & H$_{2}$O 6$_{16}-$5$_{05}$, CO 32$-$31 \\
 & \phantom{0}84.2 $-$ \phantom{0}85.0 & H$_{2}$O 7$_{16}-$7$_{07}$, CO 31$-$30, OH \\
 & \phantom{0}89.5 $-$ \phantom{0}90.4 & H$_{2}$O 3$_{22}-$2$_{11}$, CO 29$-$28 \\
 & 123.7 $-$ 126.1 & H$_{2}$O 4$_{04}-$3$_{13}$, CO 21$-$20\\
 & 157.0 $-$ 158.0 & [C\,{\sc ii}] \\
 & 162.5 $-$ 164.5 & CO 16$-$15, OH\\
 & 168.3 $-$ 170.0 & \\
 & 179.0 $-$ 180.8 & H$_{2}$O 2$_{12}-$1$_{01}$, 2$_{21}-$2$_{12}$\\
\hline\noalign {\smallskip}
\multirow{4}{*}{2} & \phantom{0}53.6 $-$ \phantom{0}55.0 & \\
 & \phantom{0}63.0 $-$ \phantom{0}63.5 & H$_{2}$O 8$_{18}-$7$_{07}$, [O\,{\sc i}] \\
 & 107.3 $-$ 109.7 & H$_{2}$O 2$_{21}-$1$_{10}$, CO 24$-$23\\
 & 189.0 $-$ 190.5 & \\
\hline\noalign {\smallskip}
\label{T:observations_pacs_settings}
\end{tabular}
\vspace{-5mm}
\end{center}
\vspace{-2mm}
\end{table}

WILL PACS observations were carried out using the IFU in line-scan mode where deep observations were obtained for targeted wavelength regions (bandwidth $\Delta\lambda/\lambda$=0.01) around selected transitions. Two wavelength settings were used, each including observations in both the blue and red gratings, as summarised in Table~\ref{T:observations_pacs_settings}. The principle transitions within these regions are from H$_{2}$O, OH, [O\,\textsc{i}], CO and [C\,\textsc{ii}], the properties of which are given in Table~\ref{T:observations_pacs_lines}. While WILL targeted the key lines observed by WISH, some of the wavelength ranges were shifted slightly in order to allow for better baseline subtraction or additional line detections (e.g. those around 82 and 90\micron{} were shifted to slightly longer wavelengths) while others were omitted to save time (e.g. around CO $J$=14-13). The velocity resolution of PACS ranges from 75\kms{} at the shortest wavelength to 300\kms{}, with only [O\,\textsc{i}] sometimes showing velocity resolved line profiles in a few sources. All observations used a chopping/nodding observing mode with off-positions within 6\arcmin{} of the target coordinates. The obsids for WILL PACS observations are given in Table~\ref{T:obsids_will}. For one source, TAU~08, PACS data were not obtained because the coolant on \textit{Herschel} ran out before they could be successfully observed.

Data reduction was performed with \textsc{hipe} v.10 with Calibration Tree 45, including spectral flat-fielding \citep[see][for more details]{Herczeg2012,Green2013a}. The flux density was normalised to the telescopic background and calibrated using observations of Neptune, resulting in an overall calibration uncertainty in flux densities of approximately 20$\%$ \citep{Karska2014b}. 1D spectra were obtained by summing over a number of spaxels chosen after inspection of the 2D spectral maps \citep{Karska2013}, with only the central spaxel used for point-like emission multiplied by the wavelength-dependent instrumental correction factors to account for the PSF (see PACS Observers Manual\footnote{http://herschel.esac.esa.int/Docs/PACS/html/pacs$\_$om.html}).

\subsubsection{Results}


\begin{figure*}
\begin{center}
\includegraphics[width=0.85\textwidth]{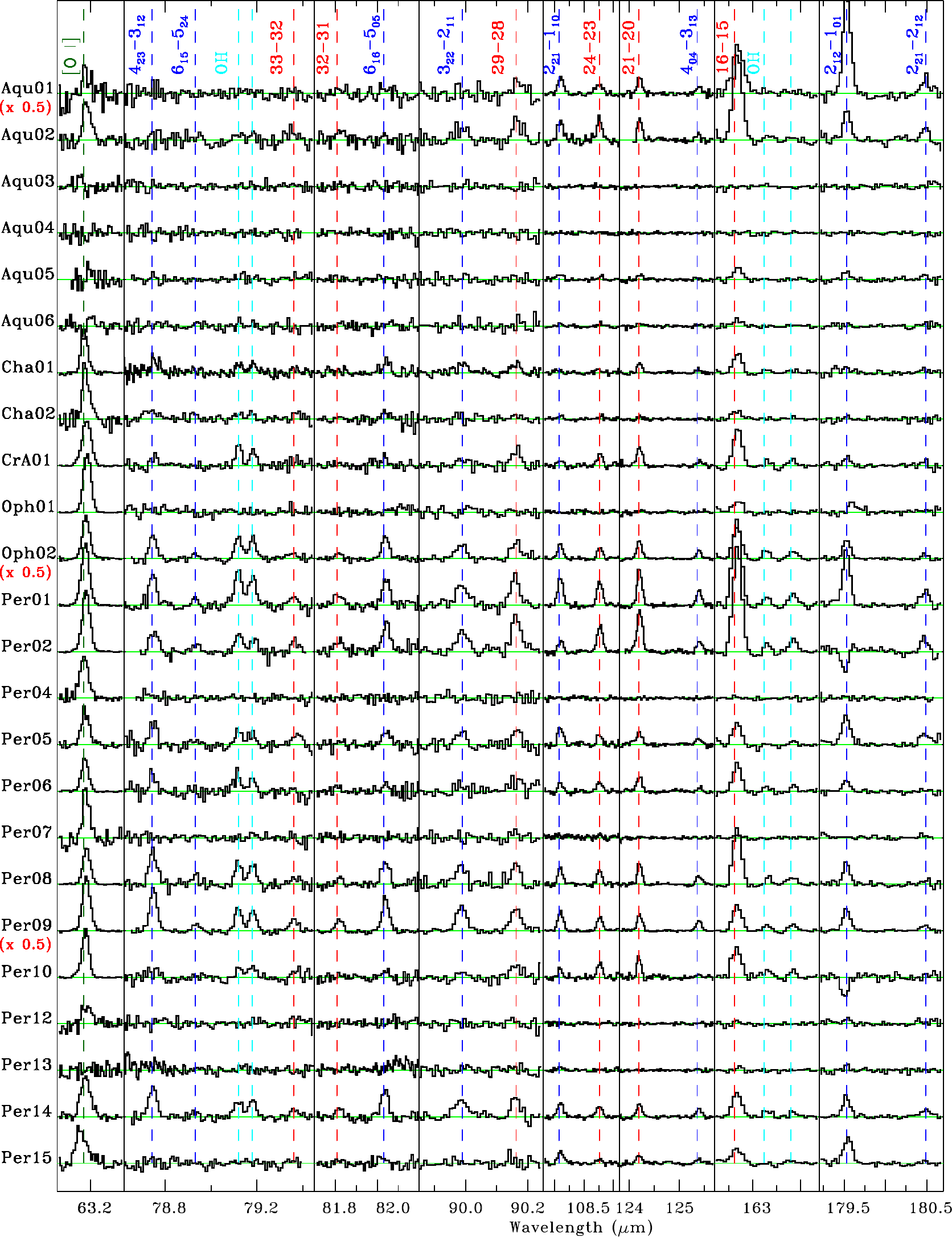}
\caption{Overview of continuum-subtracted PACS spectra for selected lines. These are not corrected for the PSF. H$_{2}$O, CO and OH lines are marked in blue, red and cyan, respectively, with the [O\,{\sc i}] marked in green. The y-axis of each spectrum for all lines except [O\,{\sc i}] goes from 0 to 5\,Jy, with the brightest sources scaled down by the factor indicated in red below the source name. The [O\,{\sc i}] spectra are scaled separately by a factor between 0.05 and 1.}
\vspace{-5mm}
\label{F:pacs_overview}
\end{center}
\end{figure*}

\renewcommand{\thefigure}{\arabic{figure} (Cont.)}
\addtocounter{figure}{-1}

\begin{figure*}
\begin{center}
\includegraphics[width=0.85\textwidth]{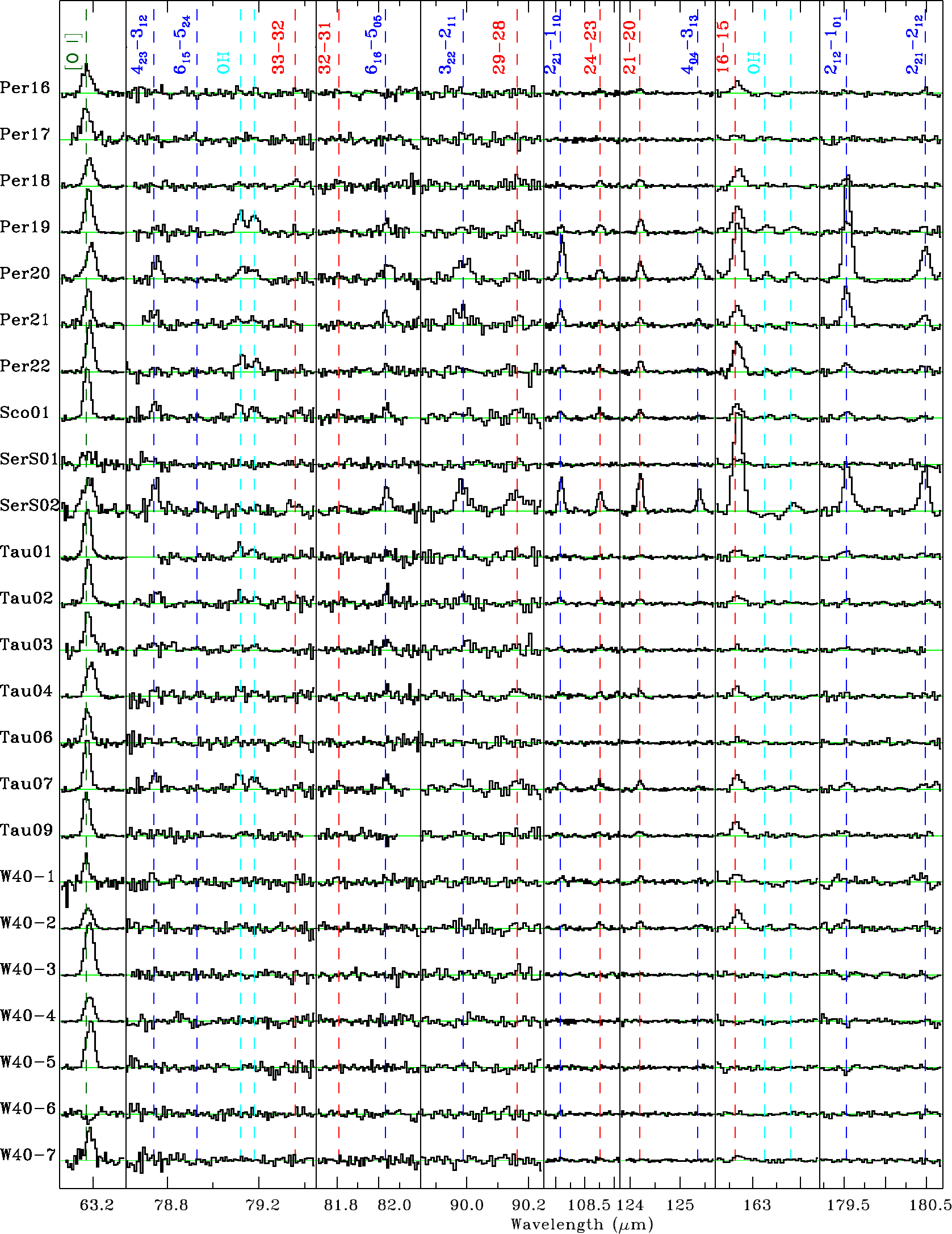}
\caption{Overview of continuum-subtracted PACS spectra for selected lines. These are not corrected for the PSF. H$_{2}$O, CO and OH lines are marked in blue, red and cyan, respectively, with the [O\,{\sc i}] marked in green. The y-axis of each spectrum for all lines except [O\,{\sc i}] goes from 0 to 5\,Jy, with the brightest sources scaled down by the factor indicated in red below the source name. The [O\,{\sc i}] spectra are scaled separately by a factor between 0.05 and 1.}
\vspace{-5mm}
\end{center}
\end{figure*}

\renewcommand{\thefigure}{\arabic{figure}}

An overview of the PACS spectra for all sources is shown in Fig.~\ref{F:pacs_overview}, while an overview of the detection of all transitions is given in Table~\ref{T:pacs_detections}. An extensive analysis of the PACS data for WILL sources in the Perseus molecular cloud was published in \citet{Karska2014b}, while a global study of PACS spectroscopy towards all WILL, DIGIT and WISH sources will be presented in Karska et al. (in prep.). Line flux densities were extracted from the PACS data as described in \citet{Karska2013}. 

The detection statistics for the main transitions are also given in Table~\ref{T:pacs_detections}. The most frequently detected line is [O\,{\sc i}], which is detected in 42 out of 48 sources. Those sources not showing [O\,{\sc i}] detections (AQU\,03$-$06, PER\,13 and W40\,06) are generally not detected in other PACS lines. These sources have weak and/or narrow lines, where detected, in the HIFI observations (c.f. Fig.~\ref{F:hifi_557}). There are 30 sources detected in at least one PACS water transition, while 27 are detected in at least one OH line and 32 in at least one CO line, with a detection more likely in the lower-energy transitions.

\subsection{Ground-based follow-up}
\label{S:observations_jcmtapex}

Follow-up ground-based observations were conducted towards the WILL sample, where not already available, to complement the \textit{Herschel} spectral line information. Approximately half of the sources in the final catalogue were not part of the samples and regions already observed in HCO$^{+}$ $J$=4$-$3 by \citet{Carney2016}, so such observations were undertaken to confirm the embedded protostellar nature of the sample (see Appendix \ref{S:properties_evolution}). The follow-up observations also included maps of $^{12}$CO $J$=3$-$2 to characterise the entrained molecular outflow and C$^{18}$O $J$=3$-$2 to obtain the source velocity and turbulent line-width in the cold envelope.

All but the two WILL sources in Chameleon are observable from the James Clerk Maxwell Telescope (JCMT\footnote{The James Clerk Maxwell Telescope has historically been operated by the Joint Astronomy Centre on behalf of the Science and Technology Facilities Council of the United Kingdom, the National Research Council of Canada and the Netherlands Organisation for Scientific Research.}) on Mauna Kea, Hawaii. Observations of C$^{18}$O $J$=3$-$2 and HCO$^{+}$ $J$=4$-$3 were obtained using HARP \citep{Buckle2009} and the ACSIS autocorrelator at the JCMT as 2\arcmin{}$\times$2\arcmin{} jiggle maps either as part of observing programs M12AN08 and M12BN07 or from the archive where these were already taken as part of other programs. These also included 2\arcmin{}$\times$2\arcmin{} jiggle map observations of all sources in $^{12}$CO and H$^{13}$CO$^{+}$ $J$=4$-$3, while $^{13}$CO  $J$=3$-$2 was obtained simultaneously with C$^{18}$O $J$=3$-$2 for those sources that had not been previously observed. In a few cases, the $^{12}$CO and$^{13}$CO observations were supplemented with cut-outs from the large basket-woven raster maps taken as part of the JCMT Gould Belt survey observations of Perseus, Taurus and Ophiuchus \citep[][]{Curtis2010,Davis2010,White2015}.

For the two Chameleon sources, a series of lines were observed with the Atacama Pathfinder EXperiment (APEX\footnote{APEX is a collaboration between the Max-Planck-Institut f{\"u}r Radioastronomie, the European Southern Observatory, and the Onsala Space Observatory.}) telescope at Llano de Chajnantor, Chile as part of project M0002\_90. These consisted of 2\arcmin{}$\times$2\arcmin{} on-the-fly maps of $^{12}$CO $J$=3$-$2 and 2$-$1, as well as single pointings of $^{12}$CO $J$=4$-$3, $^{13}$CO, C$^{18}$O and C$^{17}$O $J$=3$-$2 and 2$-$1, and HCO$^{+}$ and H$^{13}$CO$^{+}$ $J$=4$-$3, using the FLASH$^{+}$ (for the 300\,GHz and 450\,GHz bands) and APEX1 \citep[for the 225\,GHz band,][]{Vassilev2008} receivers.

The initial reduction of the JCMT jiggle maps was performed using the most up-to-date version of the \textsc{starlink}\footnote{http://starlink.eao.hawaii.edu/starlink} reduction package \textsc{orac-dr} \citep{Jenness2015}. Similar initial reduction was performed for the APEX data using \textsc{gildas-class}\footnote{http://www.iram.fr/IRAMFR/GILDAS}. Following this, all data were (re-)baselined, corrected to the $T_{\mathrm{mb}}$ scale, and re-sampled to a common velocity scale with 0.2\kms{} resolution using customised \textsc{python} scripts. A summary of the observed lines, adopted beam efficiencies and typical $\sigma_{\mathrm{rms}}$ values obtained is presented in Table~\ref{T:observations_jcmtapex}. $^{12}$CO emission is detected towards all sources but not all show evidence of outflows (see Sect.~\ref{S:outflow_jcmt} and \ref{S:properties_outflow} for more details). More details of detections and non-detections in the $^{13}$CO, C$^{18}$O, C$^{17}$O, HCO$^{+}$ and H$^{13}$CO$^{+}$ spectra can be found in Sect.~\ref{S:properties_ground}.

\section{Outflow characteristics and energetics}
\label{S:outflow}

In this section we present selected characterisation and comparative analysis of the \textit{Herschel} and ground-based spectral line observations, focusing on the entrained outflow as probed by $^{12}$CO $J$=3$-$2 and outflow/wind/jet-related shocks traced by PACS [O\,{\sc i}] observations and the broader components of the HIFI H$_{2}$O and $^{12}$CO $J$=10$-$9 lines. In this and the following section, the pre-stellar and Class II sources are excluded from all analyses as they do not show strong outflow or envelope signatures (see Sect.~\ref{S:properties_evolution} for characterisation of sources).

Details of how the various entrained outflow-related properties (i.e. mass, momentum, energy, force and mass-loss rate, maximum velocity, dynamical time, inclination and radius) were measured are given in Appendix~\ref{S:properties_outflow}, along with a table of their values for all WILL sources with detected outflows (Table~\ref{T:outflow_properties}). For consistency, the calculations are performed following the same method as that used by \citet{Yildiz2015} for the WISH sources, thus ensuring consistency between the WISH and WILL measurements. 

\subsection{Low-$J$ CO emission}
\label{S:outflow_jcmt}

\begin{figure*}
\begin{center}
\includegraphics[width=0.8\textwidth]{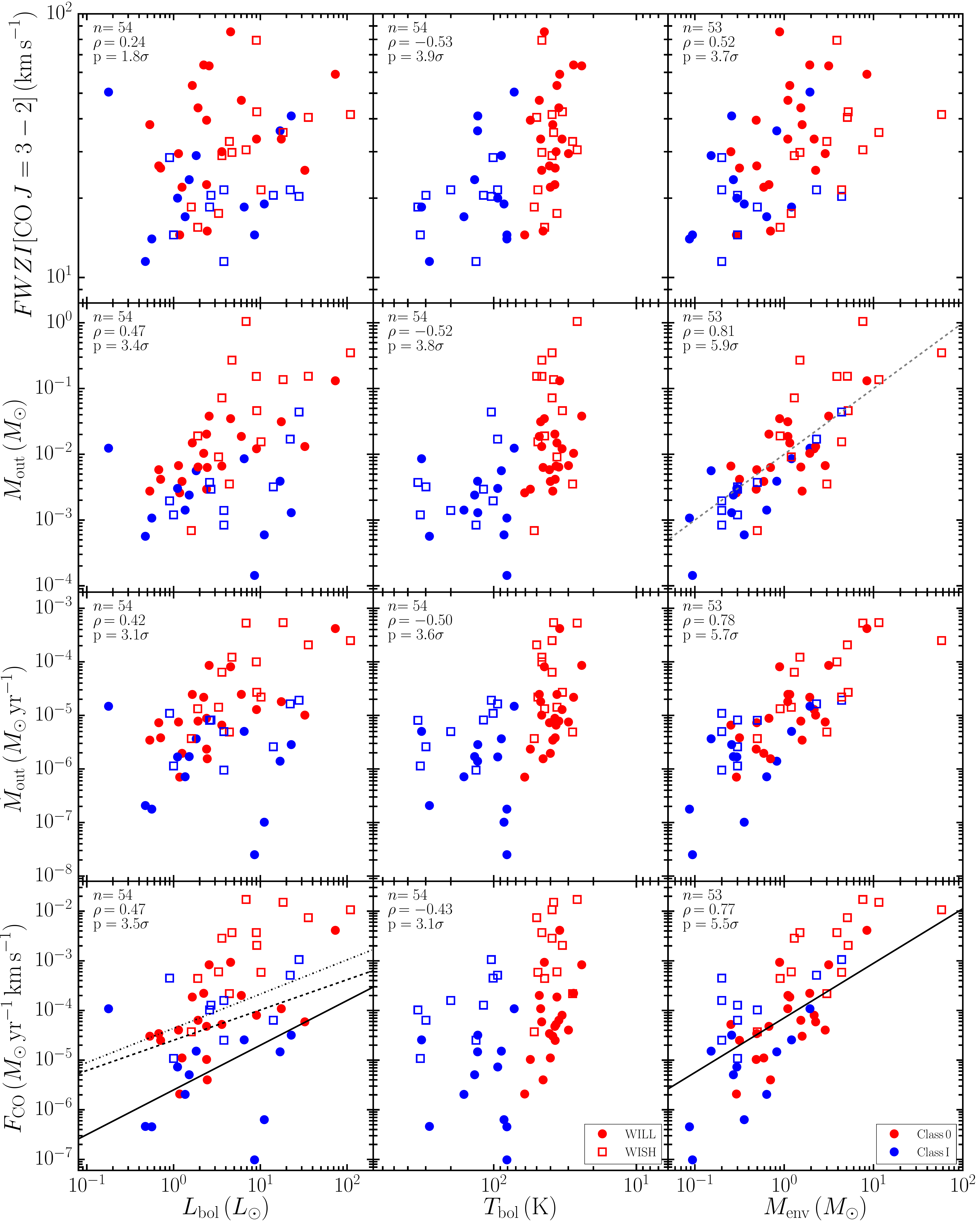}
\caption{Comparison of FWZI, $M_{\mathrm{out}}$, $\dot{M}_{\mathrm{out}}$ and $F_{\mathrm{CO}}$ obtained from CO $J$=3$-$2 maps with $L_{\mathrm{bol}}$, $T_{\mathrm{bol}}$ and $M_{\mathrm{env}}$ for the WILL (filled symbols) and WISH (open symbols) sources. The number of sources, correlation coefficient and probability that the correlation is not simply due to random distributions in the variables are shown in the upper-left of each panel. The grey dashed line in the panel for $M_{\mathrm{out}}$ vs. $M_{\mathrm{env}}$ indicates where $M_{\mathrm{out}}/M_{\mathrm{env}}$=1$\%$. The solid black lines show the relations found by \citet{Bontemps1996} between $F_{\mathrm{CO}}$, $L_{\mathrm{bol}}$ and $M_{\mathrm{env}}$ for a sample of Class I sources. The dot-dashed black line shows the best-fit found between $F_{\mathrm{CO}}$ and $L_{\mathrm{bol}}$ by \citet{Cabrit1992} for a sample of Class 0 sources, while the dashed black line shows an extension to the low-mass regime of the fit to a sample of massive young sources from \citet{Maud2015}.}
\label{F:outflow_fout}
\end{center}
\end{figure*}

One simple, initial question to ask is whether or not the observations are consistent with the common assumption that all embedded protostars have outflows. Overall 34/37 (92$\%$) of the Class 0/I sources in the WILL sample show outflow emission associated with the source in CO $J$=3$-$2 (shown in Fig.~\ref{F:outflow_maps}).  Two of the Class II sources (CHA\,01 and TAU\,03) also show outflow activity, which is discussed further in Appendix~\ref{S:cases}. Of the three Class 0/I sources without detections, Per\,12, a Class 0 source, shows indications in \textit{Spitzer} images that the outflow is in the plane of the sky \citep[see Fig.~19 in][and associated discussion]{Tobin2015}. Cha\,02, a Class I, is faint or not detected in most tracers, but is detected in [O\,{\sc i}] with PACS. W40\,01, a Class 0, is also not detected in most PACS lines but shows a faint broad blue-shifted line-wing in the HIFI spectra. Thus the lack of detection for these three sources is likely due to sensitivity, meaning our observations are consistent with the hypothesis that all Class 0/I sources drive a molecular outflow.

An important next step in understanding the mechanism and impact of outflows on star formation is to constrain how the properties of outflows are related to those of the protostar. Figure~\ref{F:outflow_fout} shows comparisons of the full-width at zero intensity (FWZI), which is the sum of the maximum velocities in the red and blue outflow lobes, mass in the entrained outflow ($M_{\mathrm{out}}$), the mass entrainment rate in the outflow ($\dot{M}_{\mathrm{out}}$) and the time-averaged momentum or outflow force ($F_{\mathrm{CO}}$) measured from CO $J$=3$-$2 with $L_{\mathrm{bol}}$, $T_{\mathrm{bol}}$ and $M_{\mathrm{env}}$ for all WISH and WILL Class 0/I sources. $\dot{M}_{\mathrm{out}}$ and $F_{\mathrm{CO}}$, as calculated quantities, are corrected for the inclination (see Appendix~\ref{S:properties_outflow} for details), while we do not correct for inclination for directly measured quantities, such as FWZI and $M_{\mathrm{out}}$.

The strongest correlation (5.9$\sigma$) is between the outflow mass and envelope mass, perhaps unsurprisingly given that the outflow is entrained from the envelope, with $M_{\mathrm{out}}/M_{\mathrm{env}}$ centred around 1$\%$, as shown by the grey dashed line. A significant, though weaker, correlation is also found between outflow mass and luminosity (3.4$\sigma$), possibly due to the correlation between $L_{\mathrm{bol}}$ and $M_{\mathrm{env}}$ (see Fig.~\ref{F:lboltbol}). Since the mass is also an important factor in the calculation of $\dot{M}_{\mathrm{out}}$ and $F_{\mathrm{CO}}$, it is not surprising that both also show significant (i.e. $\geq$3$\sigma$) correlations with $M_{\mathrm{env}}$ and $L_{\mathrm{bol}}$. In general, all parameters decrease between Class 0 and Class I, as reflected in the significant negative correlations with $T_{\mathrm{bol}}$ seen for FWZI, $M_{\mathrm{out}}$, $\dot{M}_{\mathrm{out}}$ and $F_{\mathrm{CO}}$ (3.9, 3.8, 3.6 and 3.1$\sigma$ respectively).

Correlations of $F_{\mathrm{CO}}$ with $L_{\mathrm{bol}}$ and $M_{\mathrm{env}}$ have been known for some time, including \citet{Cabrit1992} who found the relationship between $F_{\mathrm{CO}}$ and $L_{\mathrm{bol}}$ indicated by the dot-dashed line in Fig.~\ref{F:outflow_fout} for a sample of Class 0 sources, and \citet{Bontemps1996} who found the relationships indicated by the solid lines for a sample of primarily Class I sources. These have subsequently been confirmed to hold when extended to the high-mass regime for a sample of young protostars in Cygnus-X by \citet{Duarte-Cabral2013} and for a sample of massive young stellar objects (MYSOs) and young \HII{} regions by \citet{Maud2015}. The relationships seen between these variables in the combined WISH and WILL sample are steeper than that found by \citet{Cabrit1992} and \citet{Bontemps1996}. This may be due to differences in the calculation method \citep{vanderMarel2013}, or to the fact that the luminosities were likely overestimated and the $F_{\mathrm{CO}}$ values underestimated in these previous studies due to the larger beam and lower sensitivity of older observations.

At first glance, the lower right panel of Fig.~\ref{F:outflow_fout} would seem to show a slight offset in the $F_{\mathrm{CO}}$ measurements between the WISH and WILL samples, suggesting that either there is a difference in the measurements or that they come from distinct populations. However, there is no distinct break between the WISH and WILL sources, or between Class 0 and I when considering $M_{\mathrm{env}}$ vs. $M_{\mathrm{out}}$ and $\dot{M}_{\mathrm{out}}$, so the WISH sources are merely the extreme upper end of a continuous distribution. The WISH Class 0 sources were all chosen to be strong outflow sources, and the $M_{\mathrm{out}}$ and $\dot{M}_{\mathrm{out}}$ panels of Fig.~\ref{F:outflow_fout} suggest that they are more prominent due to a larger reservoir of material (i.e. larger $M_{\mathrm{env}}$), rather than faster outflows as they have similar or even lower FWZI than the WILL sources.

For the Class I sources, there is little difference between the WISH and WILL sources in FWZI or $M_{\mathrm{out}}$, but the WISH Class I sources tend to have smaller outflows \citep[see Table~\ref{T:outflow_properties} for WILL sources and Table~3 in][for WISH sources]{Yildiz2015} resulting in larger values for the WISH sources of $\dot{M}_{\mathrm{out}}$ and $F_{\mathrm{CO}}$. This could be because the WISH Class I sources are typically in smaller, more isolated clouds with shorter distances from the protostar to the cloud edge than the Class I sources in WILL.

Let us now consider the physical implications of the main correlations between outflow and source properties. The correlation between $F_{\mathrm{CO}}$ and (current) $M_{\mathrm{env}}$ is often interpreted as the result of an underlying link between envelope mass and the mass accretion rate ($\dot{M}_{\mathrm{acc}}$), which is itself related to the driving of the outflow \citep{Bontemps1996,Duarte-Cabral2013}. As the central source evolves, $M_{\mathrm{env}}$ and $\dot{M}_{\mathrm{acc}}$ decrease, leading naturally to the decrease in $F_{\mathrm{CO}}$ and other outflow-related properties between Class 0 and I sources. The comparatively tight relationship between $M_{\mathrm{out}}$ and $M_{\mathrm{env}}$ further supports this interpretation.

\begin{figure*}
\begin{center}
\includegraphics[width=0.9\textwidth]{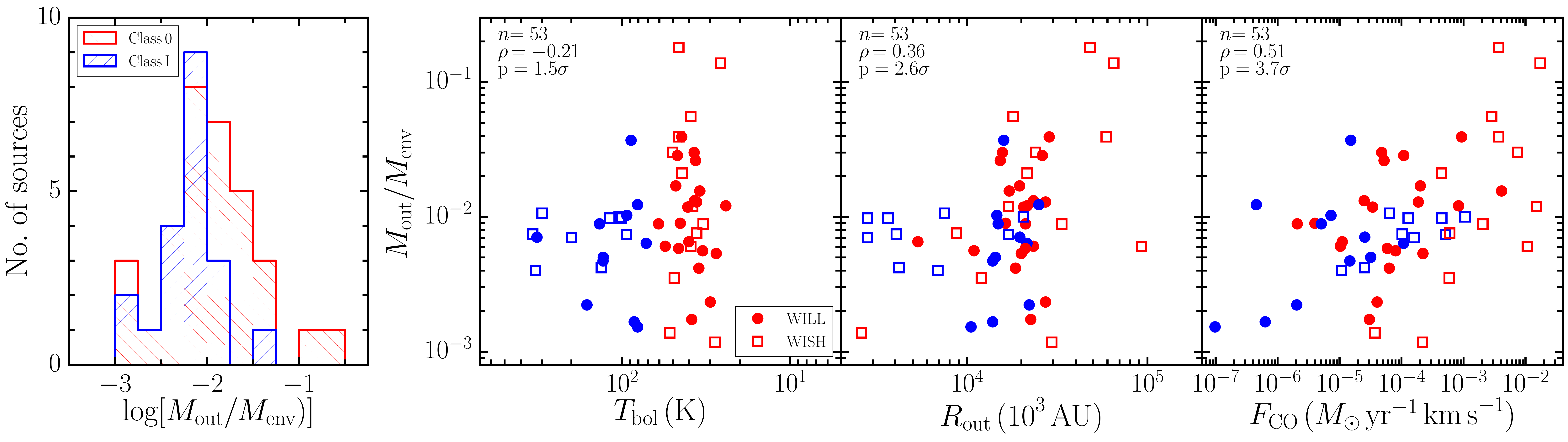}
\caption{Histogram of the ratio of outflow to envelope mass (i.e. $M_{\mathrm{out}}$/$M_{\mathrm{env}}$) for Class 0 (red) and I (blue) sources (left), as well as how this varies with $T_{\mathrm{bol}}$ (middle-left), the mean length of the outflow lobes ($R_{\mathrm{out}}$, middle-right) and outflow force ($F_{\mathrm{CO}}$, right). The colours and symbols have the same meaning as in Fig.~\ref{F:outflow_fout}.}
\label{F:outflow_moutmenv}
\end{center}
\end{figure*}

Indeed, the relation between $M_{\mathrm{out}}$ and $M_{\mathrm{env}}$ requires more investigation in its own right. Figure~\ref{F:outflow_moutmenv} shows a histogram of the fraction of mass in the outflow compared to the envelope (i.e. $M_{\mathrm{out}}$/$M_{\mathrm{env}}$), as well as how this varies with $T_{\mathrm{bol}}$ (as a more continuous proxy for source evolution), the mean length of the outflow lobes ($R_{\mathrm{out}}$), and the strength of the outflow as measured by $F_{\mathrm{CO}}$. The values of $M_{\mathrm{out}}$/$M_{\mathrm{env}}$ vary between $\sim$0.1 and 10$\%$, peaking around 1$\%$. The peak is similar between Class 0 and I with no significant trend with $T_{\mathrm{bol}}$, except that the Class 0 sources extend to larger values. This seems to be related to some Class 0 sources having longer outflows (i.e. larger $R_{\mathrm{out}}$) and thus have likely entrained additional material from the clump/cloud outside their original envelope. A statistically significant (3.7$\sigma$) correlation with $F_{\mathrm{CO}}$ is found, though with more than an order of magnitude spread.

The first impression of the peak value of $M_{\mathrm{out}}$/$M_{\mathrm{env}}$ being $\sim$10$^{-2}$ in the histogram shown in Figure~\ref{F:outflow_moutmenv} might be that this is rather low compared to a `typical' star formation efficiency of 30$-$50$\%$ \citep[e.g.][and references therein]{Myers2008,Offner2014,Frank2014}. In order to understand whether or not this value is actually reasonable, let us first assume that the outflow is responsible for removing all of the envelope material that does not end up on the star. In this case, the average mass entrainment rate in the outflow over the Class 0+I lifetime ($\tau_{\mathrm{Class\,0/I}}$) is given by:

\begin{equation}
\dot{M}_{\mathrm{out}}~=~(1-\epsilon_{sf})\,\frac{M_{\mathrm{env}}}{\tau_{\mathrm{Class\,0+I}}}\,,
\label{E:mdotout_class01}
\end{equation}

\noindent where $\epsilon_{sf}$ is the core-to-star formation efficiency, that is, the fraction of the envelope that will end up on the star. The observed mass-loss rate in the outflow, averaged along the flow, is given by:

\begin{equation}
M_{\mathrm{out}}~=~\dot{M}_{\mathrm{out}}\,t_{\mathrm{dyn}}\,,
\label{E:mdotout_tdyn}
\end{equation}

\noindent where $t_{\mathrm{dyn}}$ is the dynamical time of the flow.

It is worth pointing out that $t_{\mathrm{dyn}}$ is not necessarily the age of the source, particularly if outflow activity is time-variable. Indeed, if ejection stops then, after some time, radiative losses and mixing with the ambient cloud material will dissipate all the angular momentum and energy from the flow, meaning that the observed $t_{\mathrm{dyn}}$ is likely a lower limit to the true `age' of total accretion/outflow activity in a given source. However, if we assume that the overall mass outflow rate for a given burst is not significantly different from the average over the lifetime of the main accretion (i.e. Class 0+I) phase, or equally that protostellar outflows have an approximately constant entrainment efficiency per unit length, then we can combine and re-arrange Eqns.~\ref{E:mdotout_class01} and \ref{E:mdotout_tdyn} to get: 

\begin{equation}
\frac{M_{\mathrm{out}}}{M_{\mathrm{env}}}~=~(1-\epsilon_{sf})\frac{t_{\mathrm{dyn}}}{\tau_{\mathrm{Class\,0+I}}}\,.
\label{E:moutmenv}
\end{equation}

\noindent The ratio of $t_{\mathrm{dyn}}$ to $\tau_{\mathrm{Class\,0+I}}$ effectively expresses the duty cycle of the outflow.

For $\tau_{\mathrm{Class\,0/I}}\approx$0.5\,Myr \citep{Dunham2015,Heiderman2015,Carney2016} and a typical dynamical time for the outflow of approximately 10$^{4}$\,yrs, the ratio of outflow to envelope mass has a value of $\sim$0.01 if the $\epsilon_{sf}$=0.5. Thus, while certainly missing some details, and being affected by variation from source to source and with time, the fact that we find median value for $M_{\mathrm{out}}/M_{\mathrm{env}}$ of approximately 1$\%$ is consistent with protostellar outflows having an approximately constant entrainment efficiency per unit length, a core-to-star formation efficiency of approximately 50$\%$, and an outflow duty cycle of order $\sim$5$\%$.

\subsection{HIFI water and mid-$J$ CO emission}
\label{S:outflow_hifi}

\begin{figure*}
\begin{center}
\includegraphics[width=0.33\textwidth]{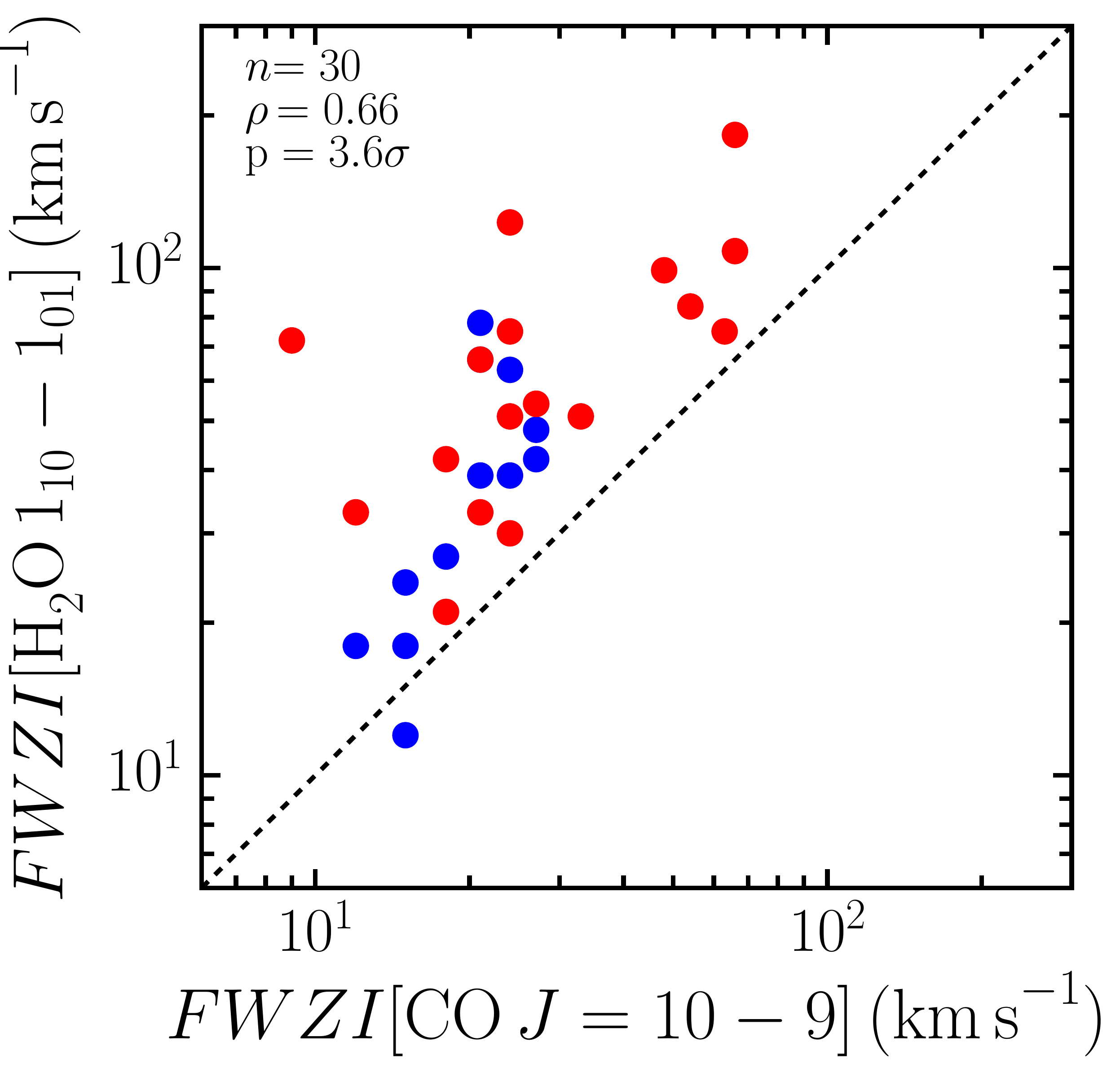}
\includegraphics[width=0.33\textwidth]{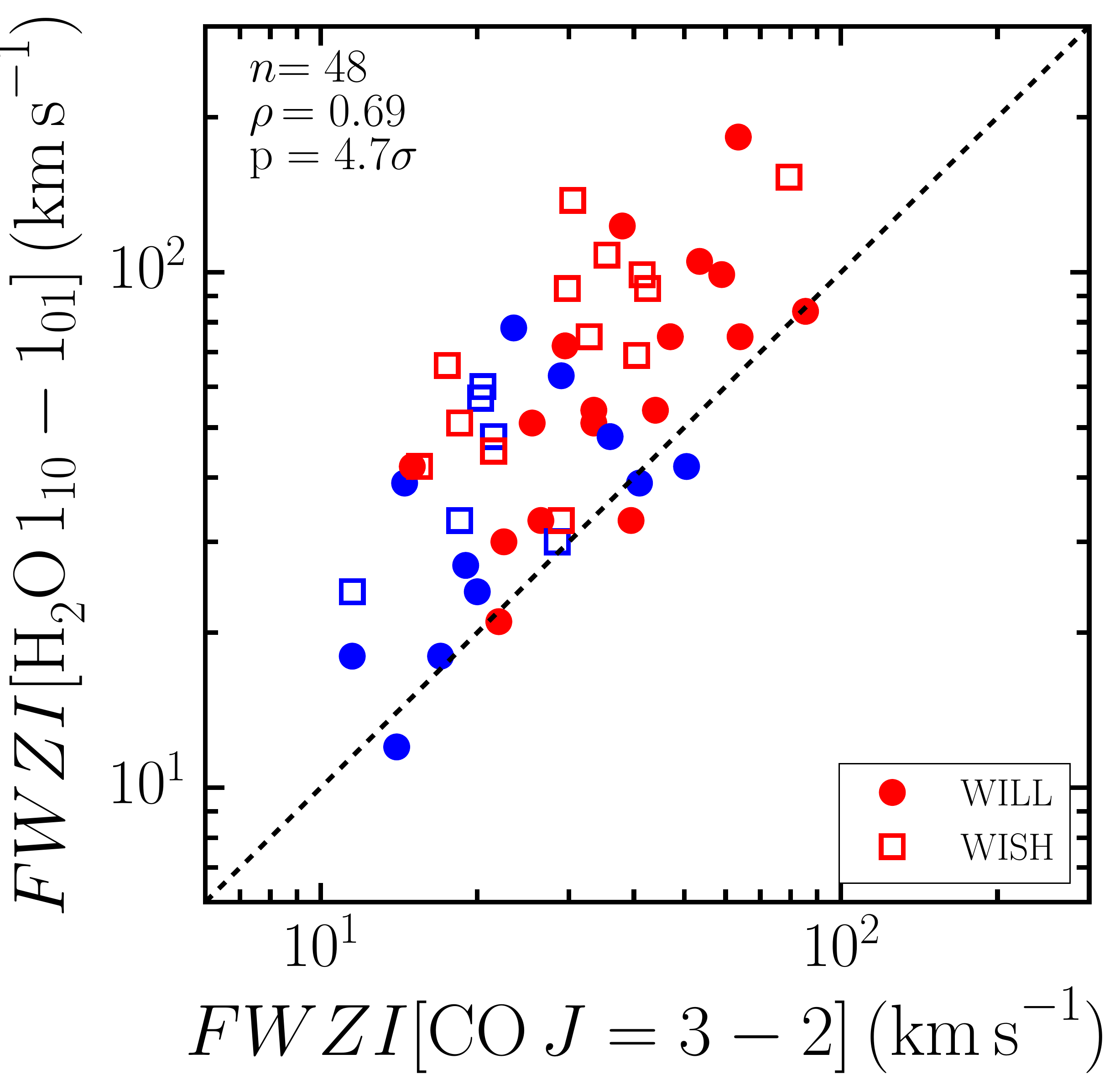}
\includegraphics[width=0.33\textwidth]{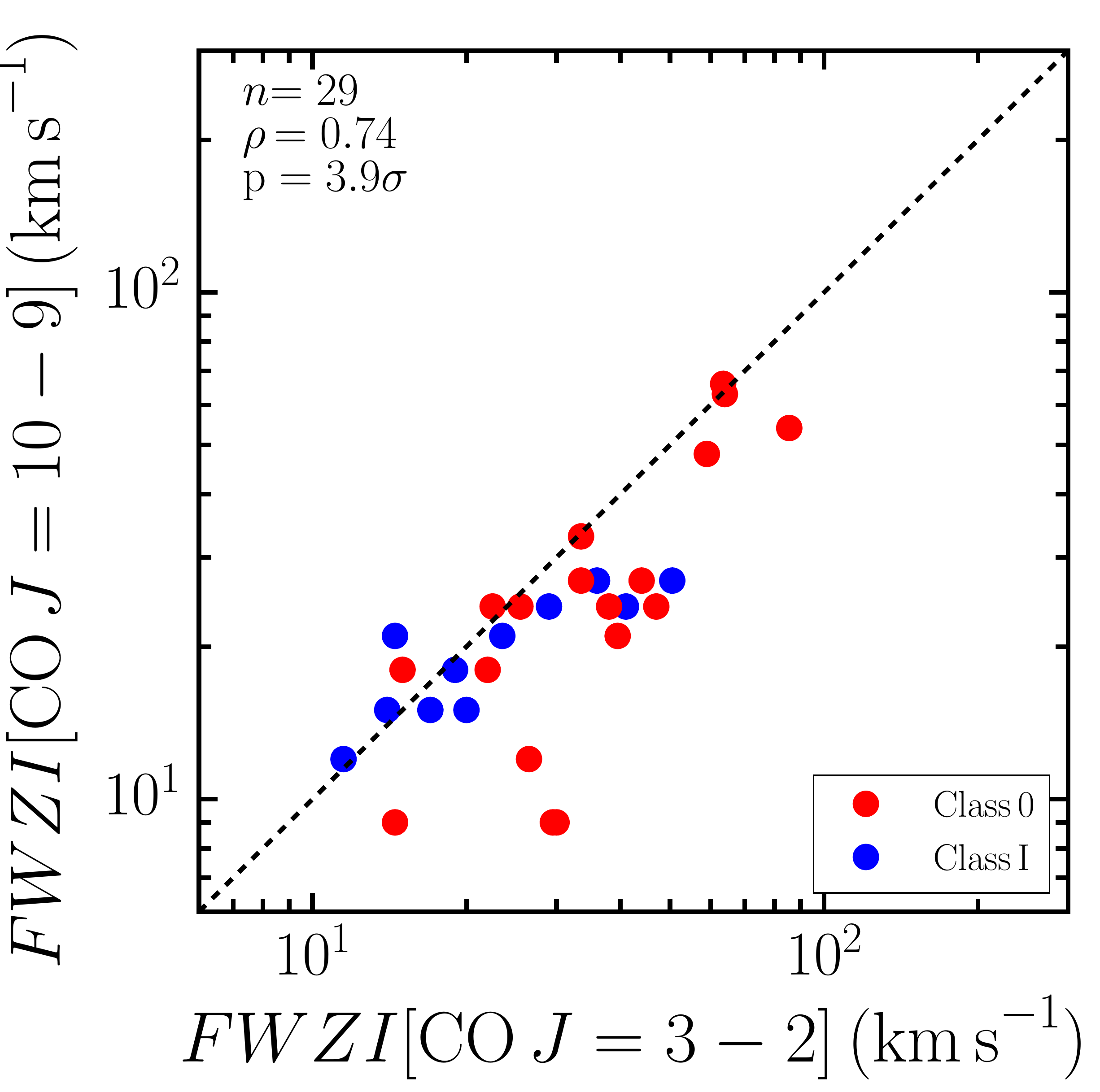}
\caption{Comparison of the full-width at zero intensity widths of H$_{2}$O 1$_{10}-$1$_{01}$, CO $J$=10$-$9 and 3$-$2. The dashed black lines indicate the line of equality.}
\label{F:outflow_hifico}
\end{center}
\end{figure*}

The water and CO $J$=10$-$9 spectra of many of the WILL sources show broad line wings, indicative of outflow emission, as seen in previous WISH observations \citep[see e.g. Fig.~\ref{F:hifi_557} and][]{Kristensen2012,SanJoseGarcia2013,Mottram2014,SanJoseGarcia2016}. The FWZI of H$_{2}$O is strongly correlated with CO $J$=10$-$9 and $J$=3$-$2 (see Fig.~\ref{F:outflow_hifico}), with H$_{2}$O consistently tracing faster material than these CO transitions, typically by a factor of $\sim$2. The narrower line-widths of CO $J$=10$-$9 with respect to CO $J$=3$-$2 are likely because the CO $J$=3$-$2 FWZI is calculated as the difference of the maximum red and blue velocity offsets anywhere in the 2\arcmin{}$\times$2\arcmin{} maps while for CO $J$=10$-$9 this is measured from a single HIFI spectrum with a 18.4\arcsec{} beam centred at the source position, that is, over a smaller region. 

\begin{figure*}
\begin{center}
\includegraphics[width=0.9\textwidth]{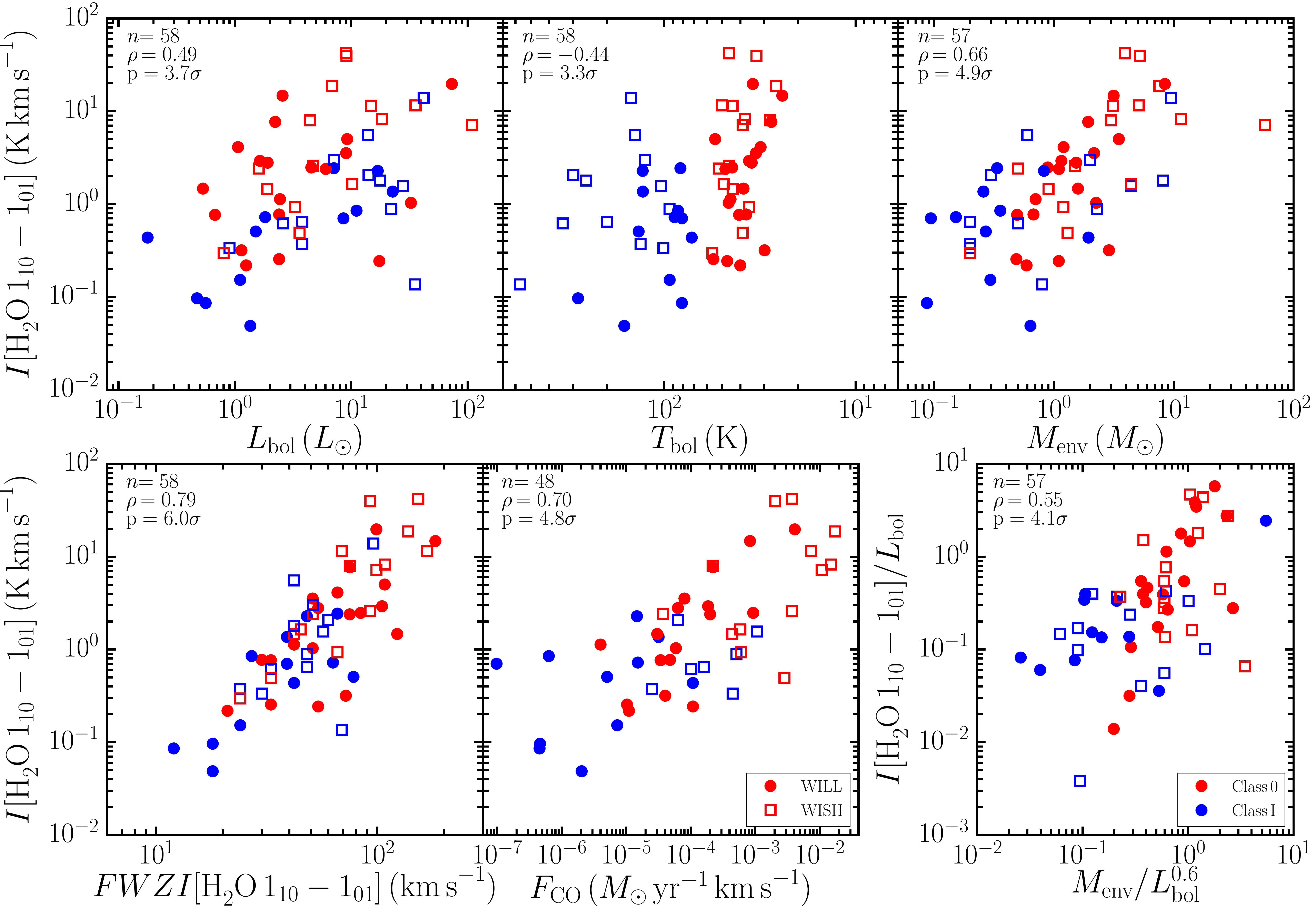}
\caption{Comparison of the integrated intensity of H$_{2}$O 1$_{10}-$1$_{01}$ linearly scaled to a distance of 200\,pc with various source (top) and outflow-related (bottom left and middle) properties, as well as the integrated intensity normalised by $L_{\mathrm{bol}}$ vs. the evolutionary indicator $M_{\mathrm{env}}/L_{\mathrm{bol}}^{0.6}$ (bottom right) proposed by \citet{Bontemps1996}. The number of sources, correlation coefficient and probability that the correlation is not just due to random distributions in the variables are shown in the upper-left of each panel.}
\label{F:outflow_hificorrelations}
\end{center}
\end{figure*}

Figure~\ref{F:outflow_hificorrelations} shows a comparison of various source and outflow-related properties with the integrated intensity of the H$_{2}$O 1$_{10}-$1$_{01}$ (557\,GHz) line after scaling to a common distance of 200\,pc. A linear scaling is used because the emission is dominated by outflows, which likely fill the beam along the outflow axis but not perpendicular to it \citep[see][ for more details]{Mottram2014}. We are able to confirm the strong correlation found by \citet{Kristensen2012} for the WISH sample alone between the integrated intensity of the water line with its FWZI (at 6.0$\sigma$) and $M_{\mathrm{env}}$ (4.9$\sigma$). A new correlation is also found with $F_{\mathrm{CO}}$ (4.8$\sigma$), firmly showing that water emission is related to, though not tracing the same material as, the entrained molecular outflow. Furthermore, shallower trends of water line intensity with $L_{\mathrm{bol}}$ and inversely with $T_{\mathrm{bol}}$, hinted at but not significant in the WISH sample \citep[e.g. see Fig.~6 of][]{Kristensen2012}, are now confirmed as statistically significant at 3.7 and 3.3$\sigma$ respectively. 

To further examine the variation of water emission with source evolution, the lower-right panel of Fig.~\ref{F:outflow_hificorrelations} shows the integrated intensity normalised by the source bolometric luminosity (thus minimising the contribution due to source brightness) vs. $M_{\mathrm{env}}/L_{\mathrm{bol}}^{0.6}$, which was proposed by \citet{Bontemps1996} as an evolutionary indicator. The clear positive correlation (4.1$\sigma$) seen in this panel reinforces the finding that the intensity of water emission decreases as sources evolve, independent of the relationship between integrated intensity and $L_{\mathrm{bol}}$. 

The WILL observations therefore reinforce and confirm the results from WISH: H$_{2}$O traces a warmer and faster component of protostellar outflow than the cold entrained molecular outflowing material traced by low-$J$ CO \citep[e.g.][]{Nisini2010,Kristensen2012,Karska2013,Santangelo2013,Busquet2014,Mottram2014,SanJoseGarcia2016}. In addition, they confirm that the intensity of H$_{2}$O is related to envelope mass and the strength of the entrained molecular outflow, and is higher for younger and/or more luminous sources.

\subsection{[O\,{\sc i}] emission}
\label{S:outflow_O1}

It has been suggested for some time that emission from [O\,{\sc i}] is a good alternative tracer of the mass loss from protostellar systems \citep[e.g.][]{Hollenbach1985,Giannini2001}. In Class 0/I protostars it is thought to primarily trace the atomic/ionised wind, because most PACS observations are spectrally unresolved and those few sources that do show velocity-resolved emission \citep[e.g. see][]{Nisini2015} are dominated by the unresolved ($\lesssim$100\kms{}) component. While there may be a contribution on-source from the disk, as in more evolved sources \citep[see e.g][]{Howard2013}, [O\,{\sc i}] emission in Class 0/I sources is often spatially extended and only fainter off-source by a factor of $\sim$2 compared to the peak position, so the wind likely still dominates.

The first comprehensive surveys of the [O\,{\sc i}] 63\micron{} transition towards samples of YSOs, observed in an 80\arcsec{} beam with the Infrared Space Observatory Long Wavelength Spectrometer \citep[ISO-LWS][]{Swinyard1998}, suggested a link between the mass loss in [O\,{\sc i}] and that in CO for Class 0 sources \citep{Giannini2001}. Only a marginal difference was seen in [O\,{\sc i}] luminosity between Class 0 and I sources \citep[][]{Nisini2002}, in contrast to the trend in CO. More recent studies by \citet{Podio2012} and \citet{Watson2015} with PACS on \textit{Herschel} at 9\arcsec{} resolution have used [O\,{\sc i}] observations to claim trends of decreasing mass loss in the wind between Classes 0, I and II. However, both suffered from low number statistics, and \citet{Podio2012} mixed the same ISO results where no trend was found with early detections from \textit{Herschel} PACS, which have significantly different beam sizes and observing methods that could induce such changes. For example, the chopping as part of PACS observations can cancel out up to 80$\%$ of the large-scale emission that is still detected by ISO \citep[see Appendix E of][]{Karska2013}. The combined WILL, WISH and DIGIT dataset, with consistent observations of a large number of YSOs is ideally placed to help solve this issue. 

\begin{figure*}
\begin{center}
\includegraphics[width=0.9\textwidth]{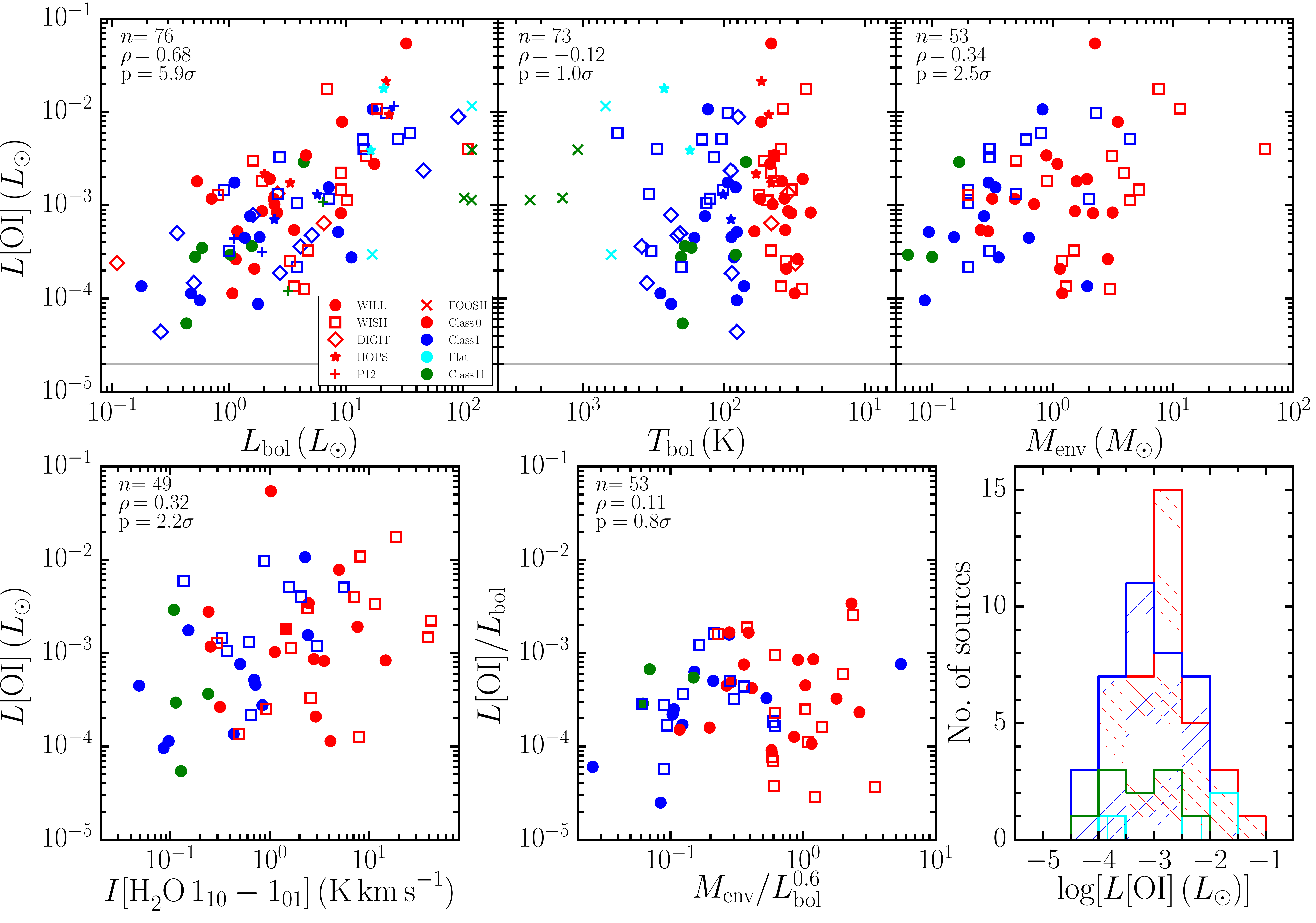}
\caption{[O\,{\sc i}] luminosity vs. $L_{\mathrm{bol}}$ (top-left), $T_{\mathrm{bol}}$ (top-middle), $M_{\mathrm{env}}$ (top-right) and integrated intensity in the H$_{2}$O 1$_{10}-$1$_{01}$ transition (bottom-left). Bottom-middle: [O\,{\sc i}] luminosity normalised by $L_{\mathrm{bol}}$ vs. the evolutionary indicator $M_{\mathrm{env}}/L_{\mathrm{bol}}^{0.6}$ proposed by \citet{Bontemps1996}. The plots include data from the WILL, WISH and DIGIT samples, as well as from the literature from \citet{Podio2012} and the HOPS \citep[][]{Watson2015} and FOOSH \citep[][]{Green2013b} surveys where available. The number of sources and correlation statistics in the upper-right of each panel include only Class 0/I sources so as to be conservative. The horizontal grey line in the top panels indicates the upper limit for disk emission from \citet[][]{Howard2013}. Bottom-right: Histogram of $L$[O\,{\sc i}] as a function of spectral type, including sources from all surveys.}
\label{F:outflow_O1_o1lbol}
\end{center}
\end{figure*}

Figure~\ref{F:outflow_O1_o1lbol} shows the distribution of [O\,{\sc i}] luminosity in the 63\micron{} line, integrated over the PACS spaxels associated with source outflows, and how this varies with various source parameters for the WILL, WISH and DIGIT samples (see Table~\ref{T:properties_mdot} for $L$[O\,{\sc i}] values). Also shown are the measurements from \textit{Herschel} studies of a number of Class I/II sources in Taurus \citep{Podio2012}, the ``\textit{Herschel} Orion Protostars'' survey \citep[HOPS][]{Watson2015}, and the ``FU Orionis Objects Surveyed with \textit{Herschel}'' survey \citep[FOOSH][]{Green2013b} which targeted a number of Flat spectrum and Class II sources that show evidence of FU Ori-type luminosity outbursts. None of the detected sources have a line luminosity below the upper limit for disk sources found by \citet[][]{Howard2013} towards sources in Taurus (4$\times$10$^{17}$\,Wm$^{-2}$, corresponding to $\sim$2$\times$10$^{-5}$\lsol{} assuming a distance of 140\,pc). 

Two primary results stand out from Fig.~\ref{F:outflow_O1_o1lbol}. First, $L$[O\,{\sc i}] is strongly correlated with $L_{\mathrm{bol}}$ but not with $M_{\mathrm{env}}$, with sources of all evolutionary classification following the overall trend. This is essentially the reverse of the situation found with low-$J$ CO, where the correlation is weak with $L_{\mathrm{bol}}$ and strong with $M_{\mathrm{env}}$ (see e.g. Fig.~\ref{F:outflow_fout}). H$_{2}$O shows clear correlations with both $M_{\mathrm{env}}$ and $L_{\mathrm{bol}}$ (see \ref{F:outflow_hificorrelations}), though the relationship is slightly stronger with $M_{\mathrm{env}}$ than $L_{\mathrm{bol}}$, consistent with it tracing actively shocked outflow material between the entrained outflow, probed by low-$J$ CO, and the wind, probed by [O\,{\sc i}]. 

Second, there is no statistically significant variation in $L$[O\,{\sc i}] with evolutionary stage, either when considering the flat distribution between $L$[O\,{\sc i}] and $T_{\mathrm{bol}}$ or the histogram of $L$[O\,{\sc i}], which shows remarkably similar distributions for Class 0, I or II sources. This is not due to an evolutionary trend being masked by the correlation with $L_{\mathrm{bol}}$, as shown by the flat distribution in $L$[O\,{\sc i}]$/L_{\mathrm{bol}}$ vs. $M_{\mathrm{env}}/L_{\mathrm{bol}}^{0.6}$. There is also no statistically significant correlation with envelope mass or integrated intensity in the H$_{2}$O 1$_{10}-$1$_{01}$, which is dominated by the fast, actively shocked component of the molecular outflow.

This apparent contradiction between the evolutionary behaviour of mass-loss indicators, that is, the decrease of CO and H$_{2}$O velocity, intensity etc. as sources evolve compared to the invariance of [O\,{\sc i}], will be explored and discussed in more detail in the following subsections. It is interesting to note that the FOOSH sources, which are all known to be undergoing luminosity outbursts, are on the upper end of, but consistent with, the distribution of other sources in terms of $L$[O\,{\sc i}] vs. $L_{\mathrm{bol}}$. Thus, [O\,\textsc{i}] must react relatively quickly to variations in the mass accretion rate, which has a significant contribution to the observed source luminosity.

\subsection{Mass accretion vs. loss}
\label{S:outflow_mdot}

The balance of mass loss vs. accretion is important in revealing the rate at which the central protostar gains mass, as well as what fraction of the initial envelope will become part of the central source, that is, the core to star efficiency of star formation. 

Direct measurement of the mass accretion rate is extremely challenging for embedded protostars because the UV, optical and near-IR continuum and lines typically used to do this in more evolved T-Tauri stars \citep[e.g.][]{Ingleby2013} are too heavily extincted. An approximate estimate can be obtained, however, by rearranging the equation for accretion luminosity, that is,

\begin{equation}
\dot{M}_{\mathrm{acc}}~=~\frac{L_{\mathrm{acc}}R_{*}}{GM_{*}}\,,
\label{E:mdotacc}
\end{equation}

\noindent with the aid of a number of empirically constrained assumptions. 

\begin{figure*}
\begin{center}
\includegraphics[width=0.9\textwidth]{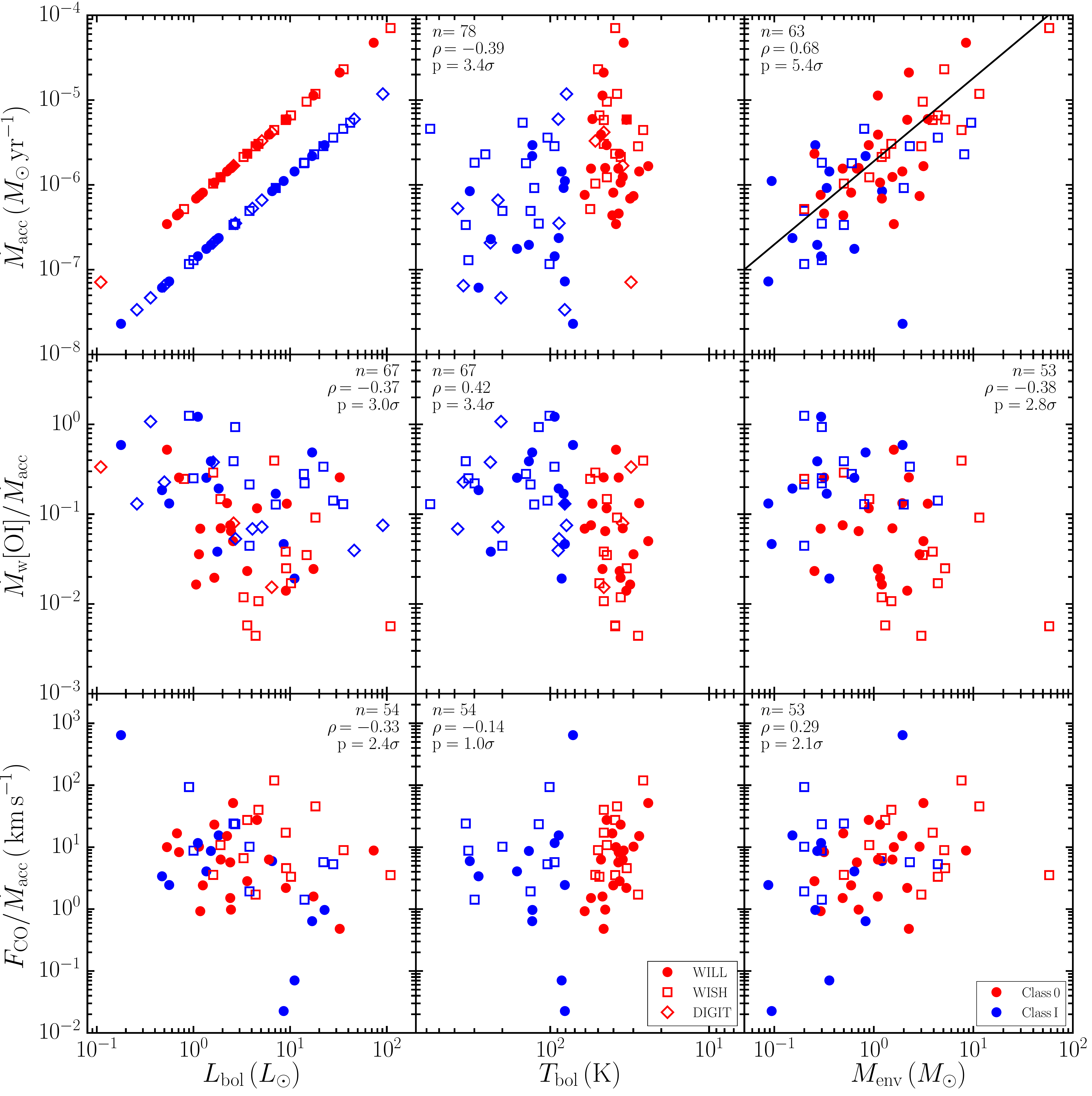}
\caption{Mass accretion rate ($\dot{M}_{\mathrm{acc}}$, top), the ratio of mass-loss rate in the wind from [O\,{\sc i}] to mass accretion rate (middle), and the ratio of outflow force from CO $J$=3$-$2 to mass accretion rate (bottom), vs. $L_{\mathrm{bol}}$ (left), $T_{\mathrm{bol}}$ (middle) and $M_{\mathrm{env}}$ (right). The solid line in the upper-right panel indicates the relationship between $\dot{M}_{\mathrm{acc}}$ and $M_{\mathrm{env}}$ from \citet{Duarte-Cabral2013}, which is part of the evolutionary models shown in Fig.~\ref{F:lboltbol}}
\label{F:outflow_mdots}
\end{center}
\end{figure*}

Firstly, accretion is assumed to generate all the observed bolometric luminosity for Class 0 sources and 50$\%$ for Class I sources, in keeping with the range observed in the few cases where this could be measured \citep[$L_{\mathrm{acc}}$/$L_{\mathrm{bol}}$ = 0.1 $-$ 0.8:][]{Nisini2005, Antoniucci2008,CarattioGaratti2012}. Next, a typical stellar mass ($M_{*}$) of 0.5\msol{} for Class I sources is assumed and 0.2\msol{} for Class 0 sources as they are still gaining mass, as in \citet{Nisini2015}. The chosen values are for sources that will end up slightly more massive than the peak of the IMF \citep[$\sim$0.2\msol{},][]{Chabrier2005}. However, as already discussed in Section~\ref{S:sample_evaluation}, our sample is biased towards slightly higher luminosities, and thus presumably stellar masses, than the global distribution, so this assumption is probably not far off. Indeed, these stellar masses are broadly in keeping with several recent mass determinations for similar embedded protostars from disk studies \citep{Tobin2012,Murillo2013,Harsono2014,Codella2014}. Finally, we assume a stellar radius ($R_{*}$) of 4\rsol{}. The calculated values are given in Table~\ref{T:properties_mdot} and shown vs. $L_{\mathrm{bol}}$, $T_{\mathrm{bol}}$ and $M_{\mathrm{env}}$ in the top panels of Fig.~\ref{F:outflow_mdots}. The solid line in the upper-left panel shows the relation assumed in the evolutionary models of \citet{Duarte-Cabral2013}, that is, Eqn.~\ref{E:dcmdotacc} with $\tau$=3$\times$10$^{5}$\,yrs.

\citet{Hollenbach1985} noted a simple scaling between the [O\,{\sc i}] line luminosity at 63\micron{} and the total mass-flux through the dissociative shock(s) producing it, given by:

\begin{equation}
\dot{M}_{\mathrm{s}}~=~10^{-4}~L\mathrm{[O\mathsc{i}\,63\mu m]}\,.
\label{E:mdotOI}
\end{equation}

\noindent For shocks generated by the wind, as is most likely the case for the emission probed by [O\,{\sc i}], the mass flux through the shock(s), $\dot{M}_{\mathrm{s}}$, is related to the wind mass-flux, $\dot{M}_{\mathrm{w}}$, by the general formula \citep[see][]{Dougados2010}:

\begin{equation}
\dot{M}_{\mathrm{s}}~=~N_{\mathrm{s}}\,\frac{\varv_{\mathrm{s}}}{\varv_{\mathrm{w}}\mathrm{cos}(\theta)}\dot{M}_{\mathrm{w}}\,,
\end{equation}

\noindent where $N_{\mathrm{s}}$ is the number of shocks in the beam, $\varv_{\mathrm{s}}$ is the shock speed, and $\theta$ is the angle between the normal to the shock front and the wind direction (the 1/cos($\theta$) term then accounts for the ratio of the shock area to the wind cross section). It may be seen that $\dot{M}_{\mathrm{s}}$=$\dot{M}_{\mathrm{w}}$ in the simple case considered by \citet{Hollenbach1985} if we are observing a static terminal shock where the wind is stopped against a much denser ambient medium; in this case, $N_{\mathrm{s}}$=1 and $\varv_{\mathrm{s}}$=$\varv_{\mathrm{w}}$cos($\theta$). This remains valid if the wind is not isotropic but collimated into a jet. 

If we are instead observing weaker internal shocks travelling along the jet/wind, then $\varv_{\mathrm{s}}$$<<\varv_{\mathrm{w}}$cos($\theta$) but this will tend to be compensated for by the presence of several shocks in the beam (i.e. $N_{\mathrm{s}}>$1), as suggested by the chains of closely spaced internal knots seen in optical jets. 

An alternative method for obtaining the average $\dot{M}_{\mathrm{w}}$ in this case is to consider that the [O\,{\sc i}] emission is approximately uniform along the flow within the aperture, and to divide the emitting gas mass by the aperture crossing time. The derivation of emitting mass requires assumptions on the temperature and electron density, which are somewhat uncertain without also having observations of the [O\,{\sc i}] 145\micron{} transition. However, \citet{Nisini2015} found that the differences are small between this alternative per-unit-length calculation and the \citet{Hollenbach1985} formulation for a terminal static wind shock (i.e. $\dot{M}_{\mathrm{s}}$=$\dot{M}_{\mathrm{w}}$). Hence, although we note that there are some uncertainties involved, we adopt $\dot{M}_{\mathrm{w}}$[O\,{\sc i}]=$\dot{M}_{\mathrm{s}}$ as given by Eqn.~\ref{E:mdotOI} to estimate the wind mass-flux from $L\mathrm{[O\mathsc{i}]}$ for our targets. The calculated values are given in Table~\ref{T:properties_mdot}.

The ratio of the mass-loss rate in the wind as measured from [O\,{\sc i}] using Eqn.~\ref{E:mdotOI} to the mass accretion rate (i.e. $\dot{M}_{\mathrm{w}}$[O\,{\sc i}]/$\dot{M}_{\mathrm{acc}}$) is compared to $L_{\mathrm{bol}}$, $T_{\mathrm{bol}}$ and $M_{\mathrm{env}}$ in the middle panels of Fig.~\ref{F:outflow_mdots}. $\dot{M}_{\mathrm{w}}$[O\,{\sc i}]/$\dot{M}_{\mathrm{acc}}$ varies from approximately 0.1$\%$ to 100$\%$ with a median of 13$\%$, in agreement with previous determinations \citep[e.g.][]{Cabrit2009,Ellerbroek2013} and in line with theoretical predictions \citep[e.g.][]{Konigl2000,Ferreira2006}. However, approximately two-thirds of all Class 0 sources lie below 10$\%$.

The lower panels of Fig.~\ref{F:outflow_mdots} show similar comparison using the outflow force as measured from CO $J$=3$-$2. Assuming the entrainment process is momentum conserving:

\begin{equation}
F_{\mathrm{CO}}=\dot{M}_{\mathrm{w}}\,\varv_{\mathrm{w}}\epsilon_{\mathrm{ent}}\,,
\end{equation}

\noindent where $\epsilon_{\mathrm{ent}}$ is the entrainment efficiency. The ratio with the mass accretion rate is then:

\begin{equation}
\frac{F_{\mathrm{CO}}}{\dot{M}_{\mathrm{acc}}}=\frac{\dot{M}_{\mathrm{w}}}{\dot{M}_{\mathrm{acc}}}\,\varv_{\mathrm{w}}\epsilon_{\mathrm{ent}}\,.
\end{equation}

\noindent $\frac{\dot{M}_{\mathrm{w}}}{\dot{M}_{\mathrm{acc}}}\,\varv_{\mathrm{w}}$ is expected to be approximately constant due to conservation of angular momentum, with a value close to the Keplerian velocity of the disk at the launching radius \citep{Duarte-Cabral2013}.

We find that $F_{\mathrm{CO}}/\dot{M}_{\mathrm{acc}}$ is relatively invariant with $L_{\mathrm{bol}}$, $T_{\mathrm{bol}}$ and $M_{\mathrm{env}}$, as shown by values of the Pearson coefficient $\rho$ consistent with 0 (i.e. $p<$3$\sigma$). Taken together, this suggests that the efficiency of entrainment, $\epsilon_{\mathrm{ent}}$, is not dependent on source properties. The Keplerian velocity for a disk around a 0.2 or 0.5\msol{} source is approximately 10$-$20\kms{} at 1AU, which, for a median value of $F_{\mathrm{CO}}/\dot{M}_{\mathrm{acc}}$ of 6.3\kms{}, suggests values for $\epsilon_{\mathrm{ent}}$ of approximately 0.3$-$0.6. If the wind is launched at larger radii then $\epsilon_{\mathrm{ent}}$ could be closer to 1.

$L$[O\,{\sc i}] does not vary with $T_{\mathrm{bol}}$, $M_{\mathrm{env}}$ or evolutionary stage (see Sect.~\ref{S:outflow_O1} and Fig.~\ref{F:outflow_O1_o1lbol}), so the increase of $\dot{M}_{\mathrm{w}}$[O\,{\sc i}]/$\dot{M}_{\mathrm{acc}}$ between Class 0 and I with increasing $T_{\mathrm{bol}}$ and with decreasing $M_{\mathrm{env}}$ is caused by the decrease in $\dot{M}_{\mathrm{acc}}$ , while $\dot{M}_{\mathrm{w}}$[O\,{\sc i}] remains relatively constant. In contrast, the invariance of $F_{\mathrm{CO}}/\dot{M}_{\mathrm{acc}}$ is caused by both $\dot{M}_{\mathrm{acc}}$ and $\dot{M}_{\mathrm{w}}$ decreasing with increasing $T_{\mathrm{bol}}$ and decreasing $M_{\mathrm{env}}$ (see the lower panels of Fig.~\ref{F:outflow_fout} for variation of $F_{\mathrm{CO}}$ with $T_{\mathrm{bol}}$ and $M_{\mathrm{env}}$). The reason for the difference in behaviour between these two measures of the ratio of mass loss to mass accretion is discussed further in the following section.

\subsection{On the difference between [O\,{\sc i}] and CO}
\label{S:outflow_o1co}

\begin{figure}
\begin{center}
\includegraphics[width=0.35\textwidth]{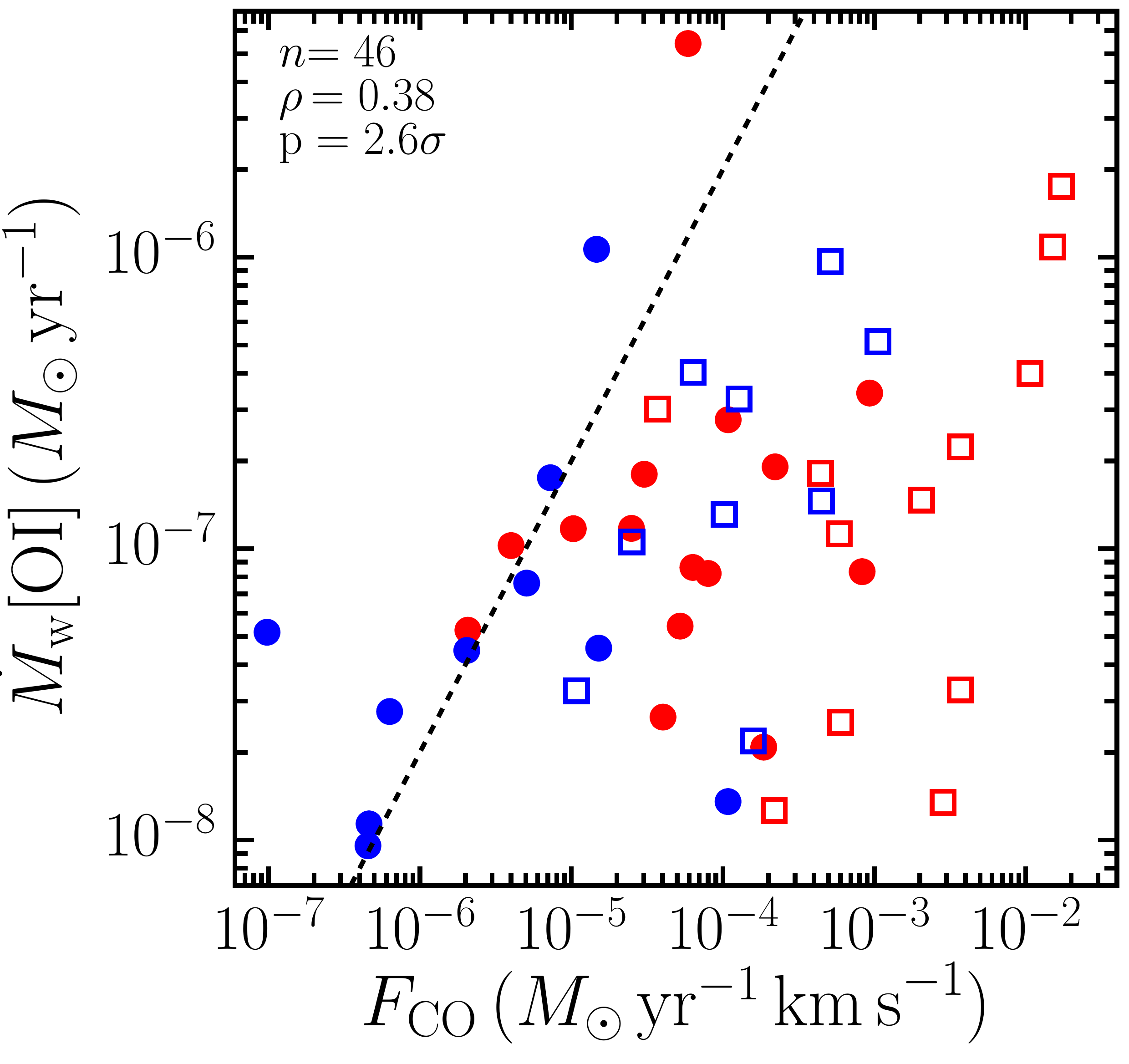}
\caption{Ratio of mass-loss rate in the wind from [O\,{\sc i}] to the outflow force from CO $J$=3$-$2. The dashed line indicates the expected locus if both trace the mass-loss rate in the wind, $\varv_{\mathrm{w}}$=100\kms{} and $\epsilon_{\mathrm{ent}}$=0.5. Lower values of $\varv_{\mathrm{w}}$ and/or $\epsilon_{\mathrm{ent}}$ move this line to the left. The symbols and colours have the same meaning as in Fig.~\ref{F:outflow_mdots}.}
\label{F:outflow_o1co}
\end{center}
\end{figure}

The difference in behaviour between the atomic component of the wind (as traced by [O\,{\sc i}]) and the entrained molecular outflow (as traced by low-$J$ CO) might seem to be in contradiction with models where the wind is the driving agent of the outflow \citep[see e.g.][]{Arce2007}. Indeed a direct comparison, shown in Fig.~\ref{F:outflow_o1co}, suggests that either the wind and outflow are not linked, [O\,{\sc i}] is under-estimating the mass loss rate in the wind or $F_{\mathrm{CO}}$ is overestimated. However, there are several factors relating to what component of the system each tracer probes that argue against rushing to such a conclusion.

First, [O\,{\sc i}] only traces the atomic component of the wind and/or jet. Jets in Class 0 protostars are known to have a significant molecular component, as identified from high-velocity features \citep[detected in e.g. CO, SiO and/or H$_{2}$O,][]{Bachiller1991,Tafalla2010,Kristensen2011} with typical mass-loss rates of approximately 10$^{-7}-$10$^{-5}$\msol{}\,yr$^{-1}$ \citep[e.g.][]{Santiago-Garcia2009,Lee2010b}. These are typically approximately ten times higher than measured from [O\,{\sc i}], but the molecular jet component disappears in older sources. This suggests an evolution in composition from molecular to atomic/ionised \citep[see][for a detailed discussion]{Nisini2015}, most likely due to increasing temperature of the protostar and decreasing density, and thus shielding, in the jet. Such arguments also hold for any wide-angle wind that could be present and contributing to driving the entrained CO outflow. Therefore, while the mass-loss rate due to the wind as a whole will decrease as the source evolves, in line with the decrease in the average mass accretion rate, the mass loss in the atomic component may remain approximately constant due to the shift in the composition of the wind.

Next, the optical depth of the continuum at 63\micron{} is likely considerable in the inner envelope in Class 0 sources \citep[see e.g.][]{Kristensen2012}, so the observed [O\,{\sc i}] flux may be significantly lower than the `true' emission. The continuum optical depth will decrease as the source evolves and $M_{\mathrm{env}}$ decreases, which may also act to counteract the evolution in the mass loss in the wind. However, such an effect should also cause the ratio of the 63\micron{} to 145\micron{} [O\,{\sc i}] lines to vary with continuum optical depth of the source, and a wavelength-dependent deficit in CO and H$_{2}$O transitions. Neither is clearly seen in PACS observations \citep[see e.g.][]{Karska2013}. This is therefore likely a minor effect dominating only for sight-lines directly towards the protostar through the disk.

Finally, there is increasing evidence that episodic or time-variable accretion is important in embedded protostars from the very earliest phases of their evolution \citep[see][for recent reviews]{Dunham2014a,Audard2014}. Accretion variability provides a consistent explanation for very low luminosity objects \citep[e.g.][]{Dunham2006}, the observed spread and trends in protostellar \citep[e.g.][]{Dunham2010} and outflow related \citep[e.g.][]{Duarte-Cabral2013} properties, and luminosity bursts, brighter by at least a factor  of ten, have now been observed in at least two embedded sources \citep{CarattioGaratti2011,Fischer2012,Safron2015}. Chains of high-velocity molecular knots or `bullets' observed in Class 0 outflows and jets, with typical spacings of 1000$-$10000\,AU between minor and major episodes, respectively \citep[e.g.][]{Santiago-Garcia2009,Lee2015}, past heating of CO$_{2}$ ice \citep[e.g.][]{Kim2012} and the difference between the expected and observed CO snow surface in a number of protostars \citep{Visser2015,Jorgensen2015} also provide indirect evidence of outbursts.

The imprint of time-variable accretion will be different for the molecular outflow and atomic wind, leading to differences in their properties. The luminosity will react quickly to any changes in the accretion rate \citep{Johnstone2013}, and thus traces the current or instantaneous activity. Since [O\,{\sc i}] is dominated by the wind, it traces material that is closely related to the current accretion state and thus is correlated with luminosity regardless of whether the source is in outburst (e.g. the FOOSH sources) or not (see Fig.~\ref{F:outflow_O1_o1lbol}). 

In contrast, the entrained molecular outflow traced by low-$J$ CO, particularly when measured over the full extent of the outflow, is an average of the ejection activity over at least 10$^{3}-$10$^{5}$\,yrs. Indeed, the highest intensity in the entrained outflow as traced by low-$J$ CO is usually offset from the central source. If these spots represent major ejections triggered by accretion bursts, then such episodes should have occurred approximately hundreds to thousands of years ago, leaving enough time for the luminosity and circumstellar material to cool back to pre-burst levels \citep[e.g.][]{Arce2001,Arce2013}. Thus, the mass loss in the molecular outflow is related to the time-averaged mass-accretion rate and may be dominated by any periods of high accretion/ejection during outbursts \citep[see also e.g.][]{Dunham2006,Lee2010a}. 

The decrease of $F_{\mathrm{CO}}$ between Class 0 and I (see Fig.~\ref{F:outflow_fout}) therefore shows that the average mass accretion rate decreases as sources evolve, as originally proposed by \citet{Bontemps1996}. The combination of decreasing mass accretion rates and episodic accretion was shown by \citet{Duarte-Cabral2013} to be consistent with the observed relationships between, and spread of, $L_{\mathrm{bol}}$, $M_{\mathrm{env}}$ and $F_{\mathrm{CO}}$. In particular, variation of the mass-accretion rate on shorter timescales than the dynamical timescale of the outflow helps to explain why outflow properties are less correlated with $L_{\mathrm{bol}}$ than with $M_{\mathrm{env}}$ (see Fig.~\ref{F:outflow_fout} and Section~\ref{S:outflow_jcmt}). Those sources that show particularly high outflow forces and/or peak emission close to the source position may therefore have recently finished such a burst, or have a higher duty cycle of outburst to quiescent accretion. Thus, the mass-loss rate in the wind from [O\,{\sc i}] and in the outflow force measured by low-$J$ CO are not directly correlated because the relationship between the current and time-averaged mass-accretion rate will be different for each source based on a complex combination of the source age, properties and mass-accretion history.

Some combination of the effects discussed above therefore causes the observed lack of correlation between CO and [O\,{\sc i}]. As such, [O\,{\sc i}] is not necessarily a direct alternative to CO for tracing mass loss and/or entrainment due to the jet/wind/outflow system in protostars, in contradiction to the early findings of \citet{Giannini2001}.

\section{Envelope}
\label{S:envelope}

\subsection{HCO$^{+}$ vs. H$_{2}$O}
\label{S:envelope_hifi}

\begin{figure*}
\begin{center}
\includegraphics[width=0.33\textwidth]{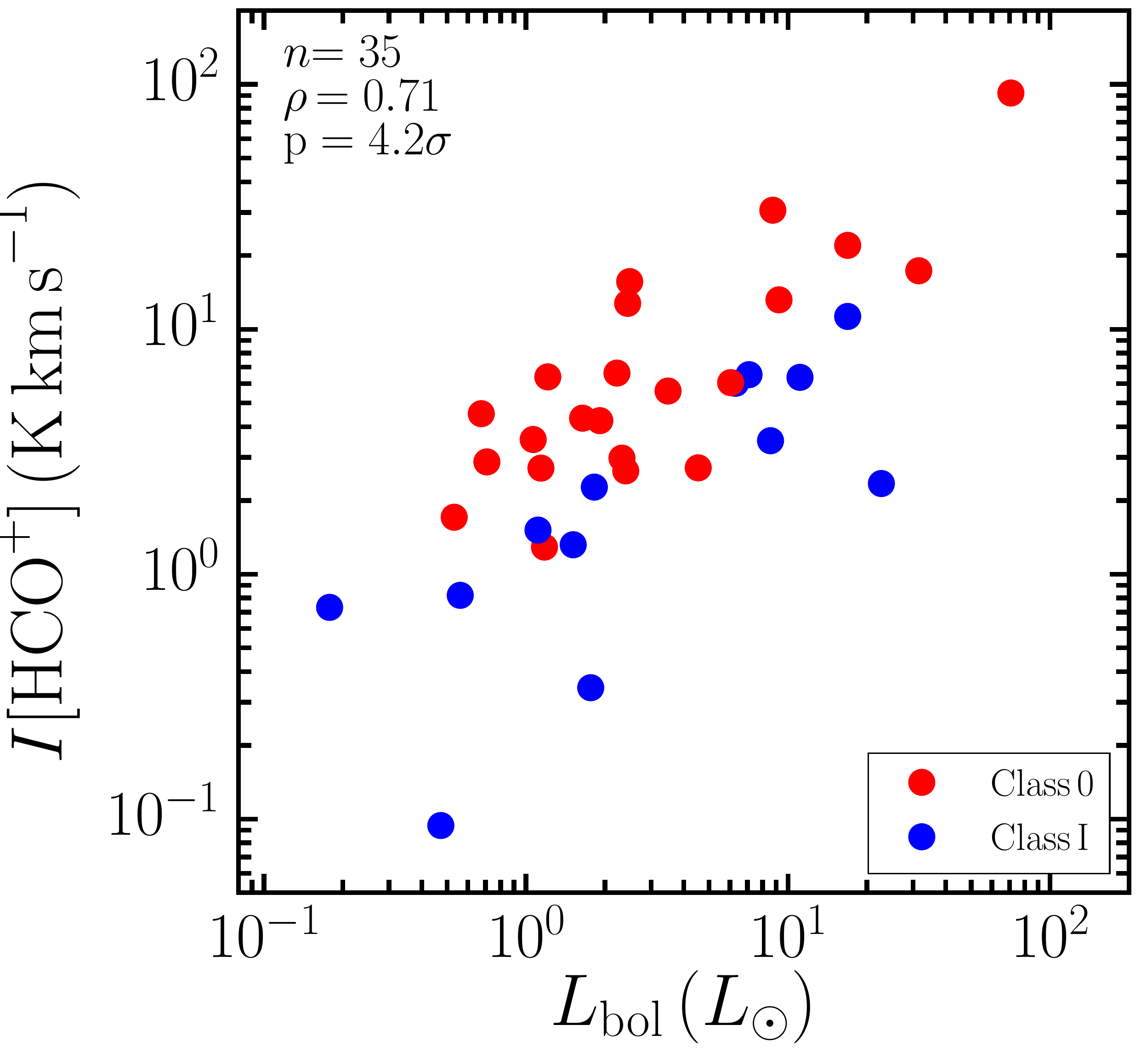}
\includegraphics[width=0.34\textwidth]{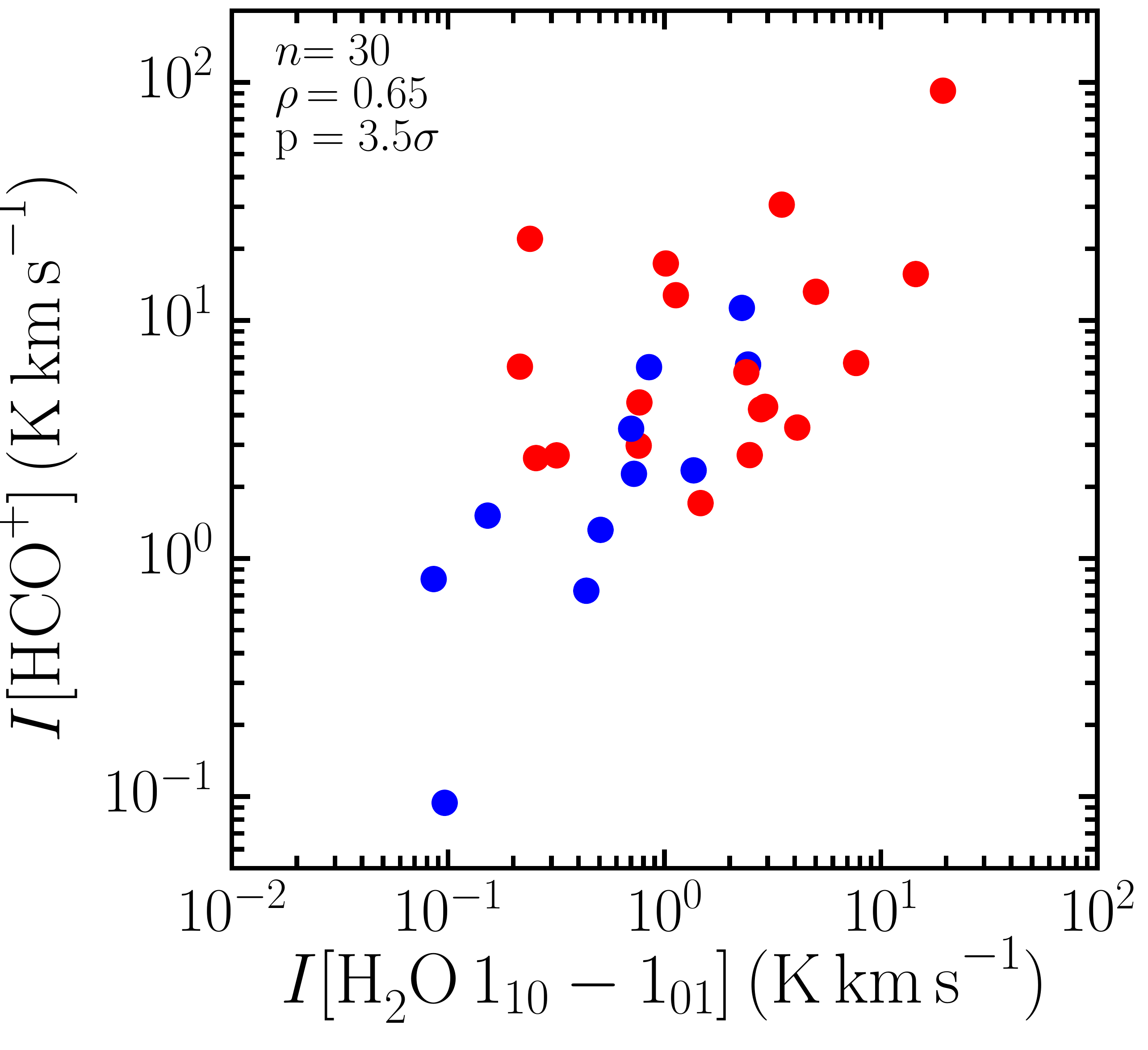}
\includegraphics[width=0.32\textwidth]{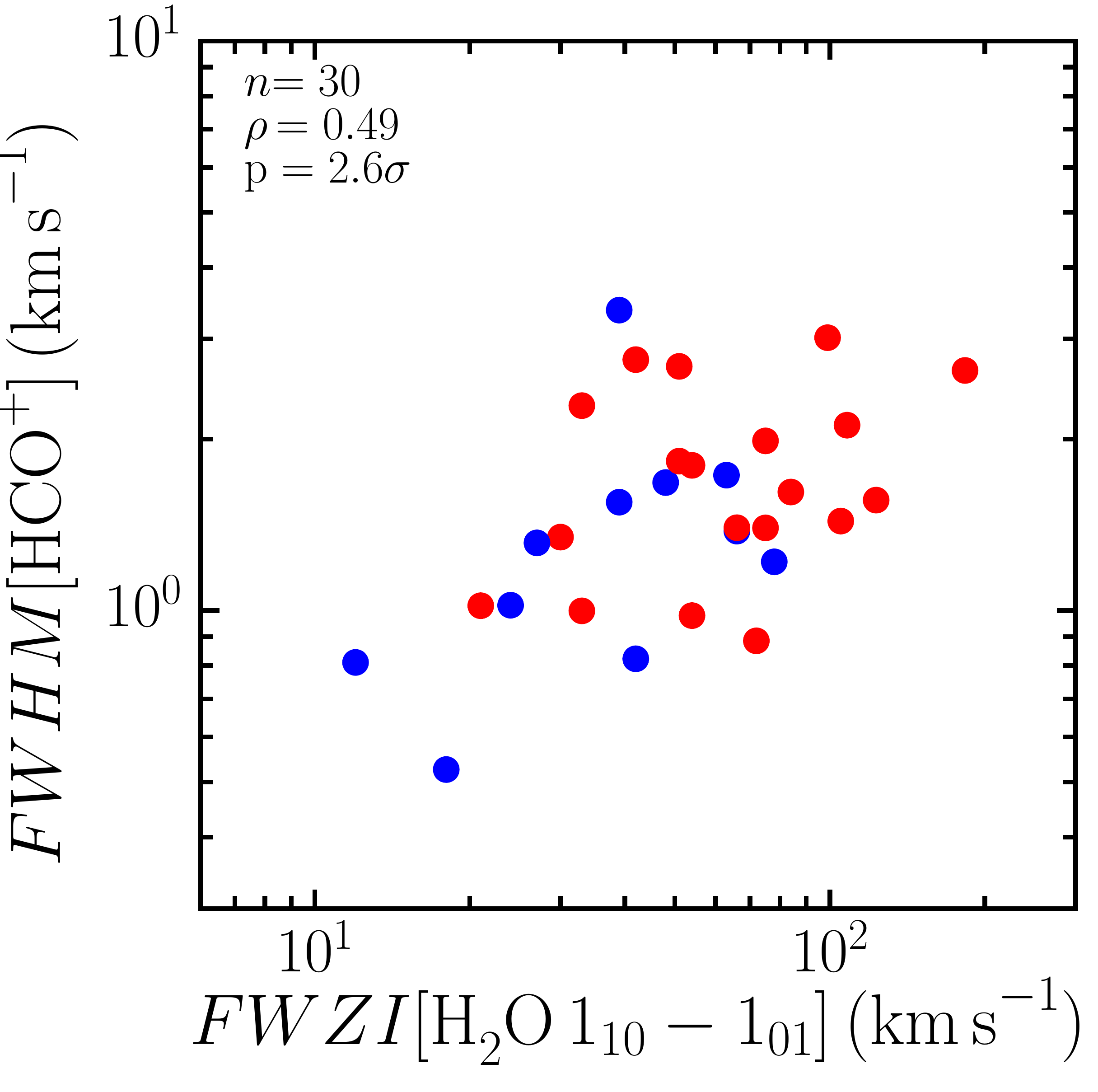}
\caption{Comparison of the integrated intensity of HCO$^{+}$ $J$=4$-$3 with $L_{\mathrm{bol}}$ (left) and the integrated intensity of H$_{2}$O 1$_{10}-$1$_{01}$ scaled to a distance of 200\,pc (middle), as well as the FWHM of HCO$^{+}$ vs. the FWZI of H$_{2}$O 1$_{10}-$1$_{01}$ (right).  The number of sources, correlation coefficient and probability that the correlation is not just due to random distributions in the variables are shown in the upper-left of each panel.}
\label{F:envelope_hcopwater}
\end{center}
\end{figure*}

Instead of outflows, HCO$^{+}$ $J$=4$-$3 emission primarily traces cool, high-density envelope material \citep[$T\sim$40\,K, $n_{\mathrm{cr}}$=2$\times$10$^{7}$\,cm$^{-3}$ though the effective density for optically thick emission could be as low as 10$^{4}$\,cm$^{-3}$, see][]{Shirley2015},  and so is a relatively clean discriminator between young, embedded protostars and pre-stellar or more evolved disk sources \citep[][]{vanKempen2009,Carney2016}. Spatially compact detections are found in this line towards most of the WILL sample sources, confirming them to be genuine embedded Class 0/I sources, while pre-stellar and Class II sources are either non-detections or show extended emission with no clear peak at the source position (see Appendix~\ref{S:properties_evolution} for details). 

H$_{2}$O emission is a good tracer of warm, relatively dense material in shocks related to protostellar outflows \citep[$T\gtrsim$300\,K, $n$=10$^{5}-$10$^{8}$\,cm$^{-3}$,][]{Kristensen2013,Mottram2014}. Sources with higher luminosities typically have stronger outflows (and thus stronger H$_{2}$O emission) and will lead, all other things being equal, to more mass at a given temperature in their envelopes and thus higher intensity in molecular tracers such as HCO$^{+}$. It is therefore not unreasonable to expect that the emission in these two tracers may be related in Class 0/I sources. 

\citet{Carney2016} compared the nature of HCO$^{+}$ emission (compact, confused or extended/not detected) and the detection of water in the WILL PACS 179\micron{} observations. There are 18 sources in common between their sample and WILL: 13 are classified as Class 0/I (i.e. compact HCO$^{+}$ $J$=4$-$3), 3 as confused and 2 as Class II (i.e. extended and/or non-detections in HCO$^{+}$). Most sources (14/18) have detected 179\micron{} water emission, with four Class 0/I sources and one confused source showing extended emission. As both Class II sources were detected, while three Class 0/I sources and one confused source was not, \citet{Carney2016} did not find a clear relationship with evolution between the spatial distributions and detection of HCO$^{+}$ and H$_{2}$O.

Considering the H$_{2}$O 1$_{10}-$1$_{01}$ HIFI observations (see Fig.~\ref{F:envelope_hcopwater}), for the two Class II sources, TAU\,07 is not detected at the 3$\sigma$ level and TAU\,09 shows very weak, narrow emission with an integrated intensity of 0.16\,K\kms{}, equivalent to 10$^{-18}$ W\,m$^{-2}$, which could include a contribution from the disk \citep[c.f.][]{Podio2012,Podio2013,Fedele2013}. Thus, while H$_{2}$O may be detected in either Class 0/I or II, the origin and intensity of the water emission changes as the source evolves. Strong detection of either HCO$^{+}$ or water is therefore still a good indication of the youth of a protostar.

For the Class 0/I sources, the full WILL dataset enables the relationship between these species to be probed further by comparing the integrated intensity and FWHM of HCO$^{+}$ with the integrated intensity and FWZI of the H$_{2}$O 1$_{10}-$1$_{01}$ HIFI observations (see Fig.~\ref{F:envelope_hcopwater}). HCO$^{+}$ intensity is correlated both with $L_{\mathrm{bol}}$ and the intensity of the water line at 4.2 and 3.5$\sigma$ significance, respectively. There is not a strong, statistically significant relationship between the kinematics of the two lines and no significant line-wings are seen in the HCO$^{+}$ spectra (see Fig.~\ref{F:hcop_c18o}). HCO$^{+}$ $J$=4$-$3 therefore seems to primarily trace parts of the envelope that are further from, and thus less disturbed by, the outflow. HCO$^{+}$ can be destroyed through reactions with H$_{2}$O \citep[e.g.][]{Jorgensen2013}, so the higher abundance of H$_{2}$O in the outflow \citep[$X\lbrack\mathrm{H}_{2}\mathrm{O}\rbrack$ or approximately 10$^{-5}-$10$^{-7}$:][]{Tafalla2013,Santangelo2013,Kristensen2017} compared to the envelope \citep[10$^{-8}-$10$^{-11}$:][]{Mottram2013,Schmalzl2014} could be suppressing the HCO$^{+}$ abundance in the outflow. This would explain why it is a poorer tracer of the outflow than might otherwise be expected. The correlation between the intensity of the two lines is therefore likely due to the relation between emission in each line and the source luminosity and structure, assisted by their tracing similar densities. 

\subsection{Infall signatures}
\label{S:envelope_infall}

\begin{figure}
\begin{center}
\includegraphics[width=0.35\textwidth]{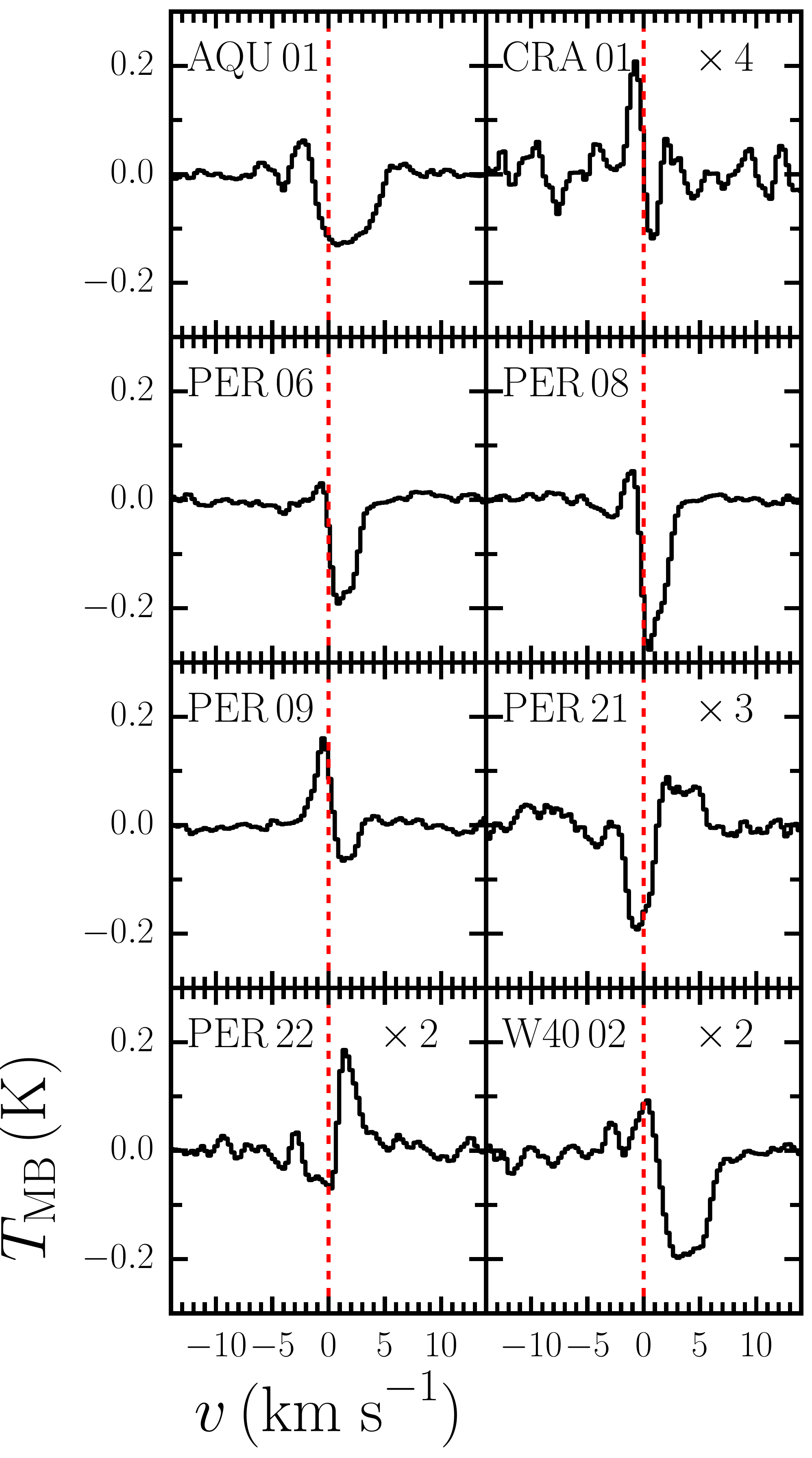}
\caption{Outflow-subtracted H$_{2}$O 1$_{10}-$1$_{01}$ residual line profiles for those sources showing either regular (PER\,21 and PER\,22) or inverse (all other) P-Cygni line profiles. All have been recentred so that the source velocity is at zero. The number in the upper-right corner of each panel indicates what factor the spectra have been multiplied by to aid visibility.}
\label{F:envelope_h2oipcrpc}
\end{center}
\end{figure}

The bulk of the H$_{2}$O emission comes from outflow-related shocks that have Gaussian-like profiles in velocity-resolved spectra. Once this contribution is removed, the residual profiles show the remaining water emission and/or absorption associated with the envelope. For the WISH sample, this process revealed seven sources with inverse P-Cygni line profiles indicative of infall and five with regular P-Cygni line profiles indicative of expansion motions in the envelope \citep{Kristensen2012,Mottram2013}. When the same procedure is performed for the WILL H$_{2}$O 1$_{10}-$1$_{01}$ observations, removing the shock emission using the Gaussian decomposition of the profiles from \citet[][]{SanJoseGarciaThesis}, six sources (3 Class 0 and 3 Class I) show inverse P-Cygni and two sources (both Class 0) show regular P-Cygni line profiles (see Fig.~\ref{F:envelope_h2oipcrpc}). 

The two WILL sources in Serpens South have broad water absorption features, but they are not offset enough to be considered inverse P-Cygni (see Fig.~\ref{F:hifi_557}). These may trace the large-scale cloud collision identified by \citet{Kirk2013}, similar to that in the Serpens Main cloud first identified by \citet{Duarte-Cabral2011} and revealed to be the origin of the strong inverse P-Cygni line profile in water in Serpens-SMM4 by \citet{Mottram2013}.

\begin{figure}
\begin{center}
\includegraphics[width=0.35\textwidth]{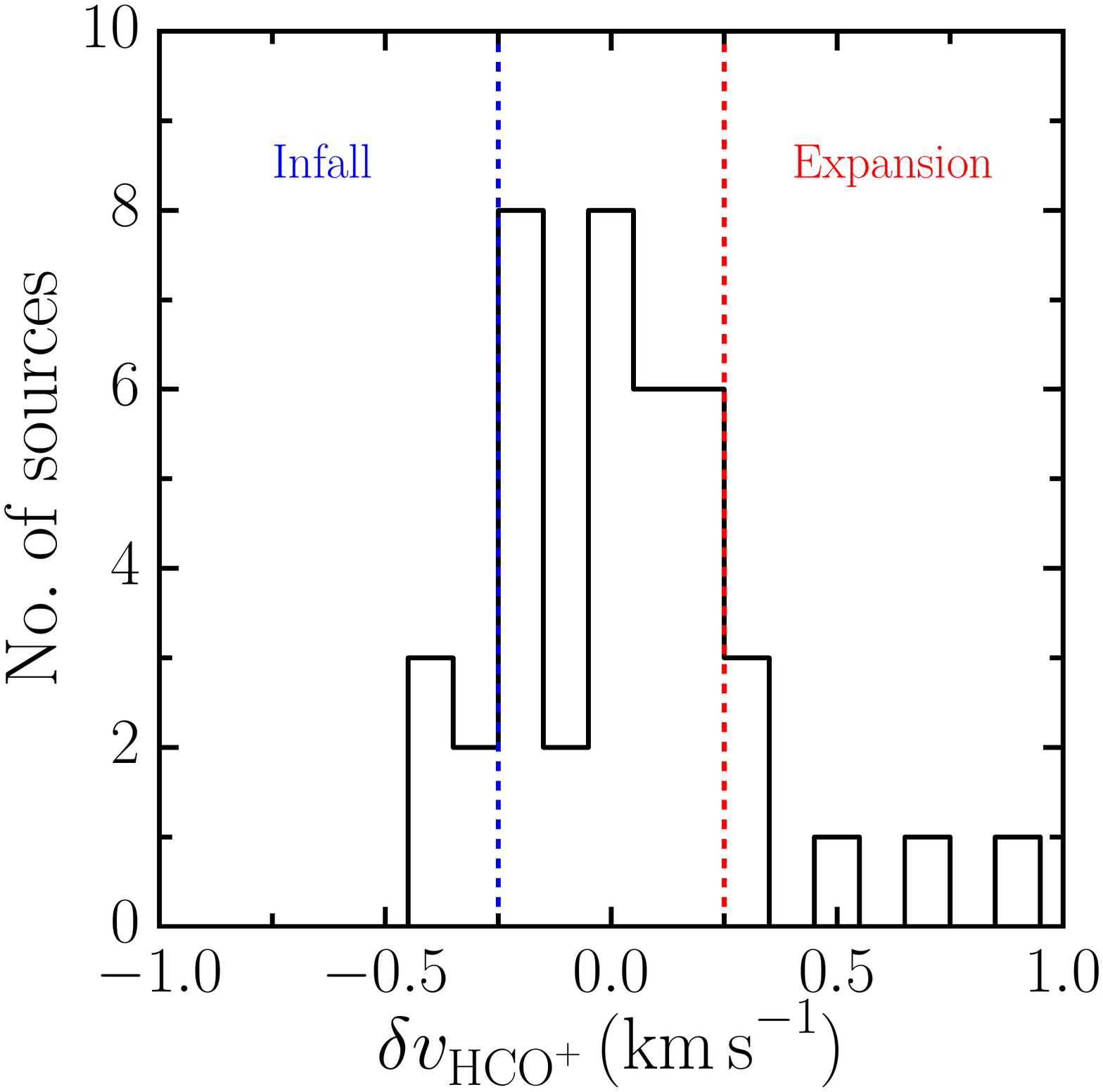}
\caption{Histogram of the normalised offset of the peak of HCO$^{+}$ 4$-$3 with respect to C$^{18}$O 3$-$2. The red and blue dashed lines indicate the boundaries outside which the offset is considered significant \citep[see][]{Mardones1997}.}
\label{F:envelope_hcopinfall}
\end{center}
\end{figure}

As an optically thick, higher density tracer, HCO$^{+}$ $J$=4$-$3 is also sensitive to infall and expansion motions in protostellar envelopes \citep[e.g.][]{Gregersen1997,Myers2000}. Though single-dish observations do not show absorption below the continuum, they can exhibit asymmetric line profiles, with the peak shifted to either the blue (infall) or red (expansion). The asymmetry between the red and blue peaks can be quantified using the $\delta\varv$ parameterisation suggested by \citet[][see Sect.~\ref{S:properties_ground} for more details]{Mardones1997}:

\begin{equation}
\delta\varv=\frac{\varv_{\mathrm{thick}}-\varv_{\mathrm{LSR}}}{\mathrm{FWHM}_{\mathrm{thin}}}\,,
\label{E:dv}
\end{equation}

\noindent where $\varv_{\mathrm{thick}}$ is the velocity of the peak emission in an optically thick tracer (in this case HCO$^{+}$) and FWHM$_{\mathrm{thin}}$ is the line width of an optically thin tracer (in this case C$^{18}$O). Values above or below 0.25 indicate a shift in the optically thick line of more than a quarter of the optically thin line width and so are considered significant. Thus, values above 0.25 indicate expansion motions while those below $-$0.25 indicate infall. Values between $-$0.25 and 0.25 are consistent with the optically thick and thin tracers being in agreement. A histogram of the values calculated for the WILL sample is shown in Fig.~\ref{F:envelope_hcopinfall}, with five sources showing blue asymmetry (i.e. $\delta\varv<-$0.25, 3 Class 0 and 2 Class I) and six showing red asymmetry (i.e. $\delta\varv>$0.25, 3 Class 0 and 3 Class I). 

Only one source (PER\,08) exhibits non-static line signatures in both water and HCO$^{+}$, however they conflict as the water shows an inverse P-Cygni line profile and the HCO$^{+}$ a red asymmetry, leaving the status of this source uncertain. It is likely that these two tracers probe different radii, and so perhaps infall and expansion dominate in different parts of the envelope.

What is possibly more puzzling is that the vast majority of sources do not show indications of either infall or expansion in either tracer. Two-sample Kolmogorov-Smirnov (K-S) tests were performed comparing the cumulative distributions of source properties (e.g. $L_{\mathrm{bol}}$, $T_{\mathrm{bol}}$ etc.) for sources that show infall or expansion motions with respect to those that do not, in order to see if any source properties correlate with the detection of infall or expansion signatures. Only the integrated intensity of C$^{18}$O 3$-$2 shows a statistically significant difference ($\leq$1$\%$ chance of being drawn from the same distribution)  between sources with either an infall or expansion signature and those that do not: sources with higher C$^{18}$O line intensity are more likely to show signs of infall. 

Sources with clear infall motions are statistically more likely to have higher FWZI in $^{12}$CO 3$-$2, that is, broader outflow line-wings, than those showing no or expansion envelope motions (0.9$\%$ likelihood of coming from the same distribution). However, there are some sources that have high FWZI but no indication of radial envelope motions. One other result worth noting is that the presence of envelope motions is not more likely for certain values of outflow inclination \citep[see][and Sect.~\ref{S:properties_outflow} for details of how these were determined]{vanderMarel2013}, suggesting that the orientation of the protostellar system is not the overriding cause of not detecting infall or expansion in our observations.

Infall must take place in all protostars, at least in the early phases, and at later times it seems unlikely that expansion of the envelope is restricted to a few select sources. Thus the low detection rate of such signatures in both tracers and the lack of a consistent trend with evolutionary Class is puzzling. It may well be that this is an observational effect, caused by small infall motions being lost in the general turbulent field on the large spatial scales that dominate single-dish observations. Mapping the velocity field inside protostellar envelopes with interferometers is likely needed to conclusively understand how material moves radially, and how this varies between different sources and over time \citep[e.g.][]{Yen2014,Aso2015,Evans2015}. 

\section{Evolution of water line profile components from Class 0 to Class I}
\label{S:evolution}

The intensity and line-width of water emission decreases for WISH sources between Class 0 and I \citep{Kristensen2012,Mottram2014}, while the rotational temperature of mid-$J$ CO and water excitation conditions do not \citep{Karska2013,Mottram2014}. \citet{Mottram2014} therefore suggested that the observed evolution in water line intensity and line-width from Class 0 to I was caused by a decrease in the velocity of the wind driving the outflow, due to the increase of the outflow cavity opening angle as proposed in the models of \citet{Panoglou2012}, for example, rather than a decrease in density in the H$_{2}$O emitting gas. However, the small sample size of the WISH survey when broken down into evolutionary classifications meant that some trends, while hinted at by the data, were not statistically significant.

\begin{figure}
\begin{center}
\includegraphics[width=0.40\textwidth]{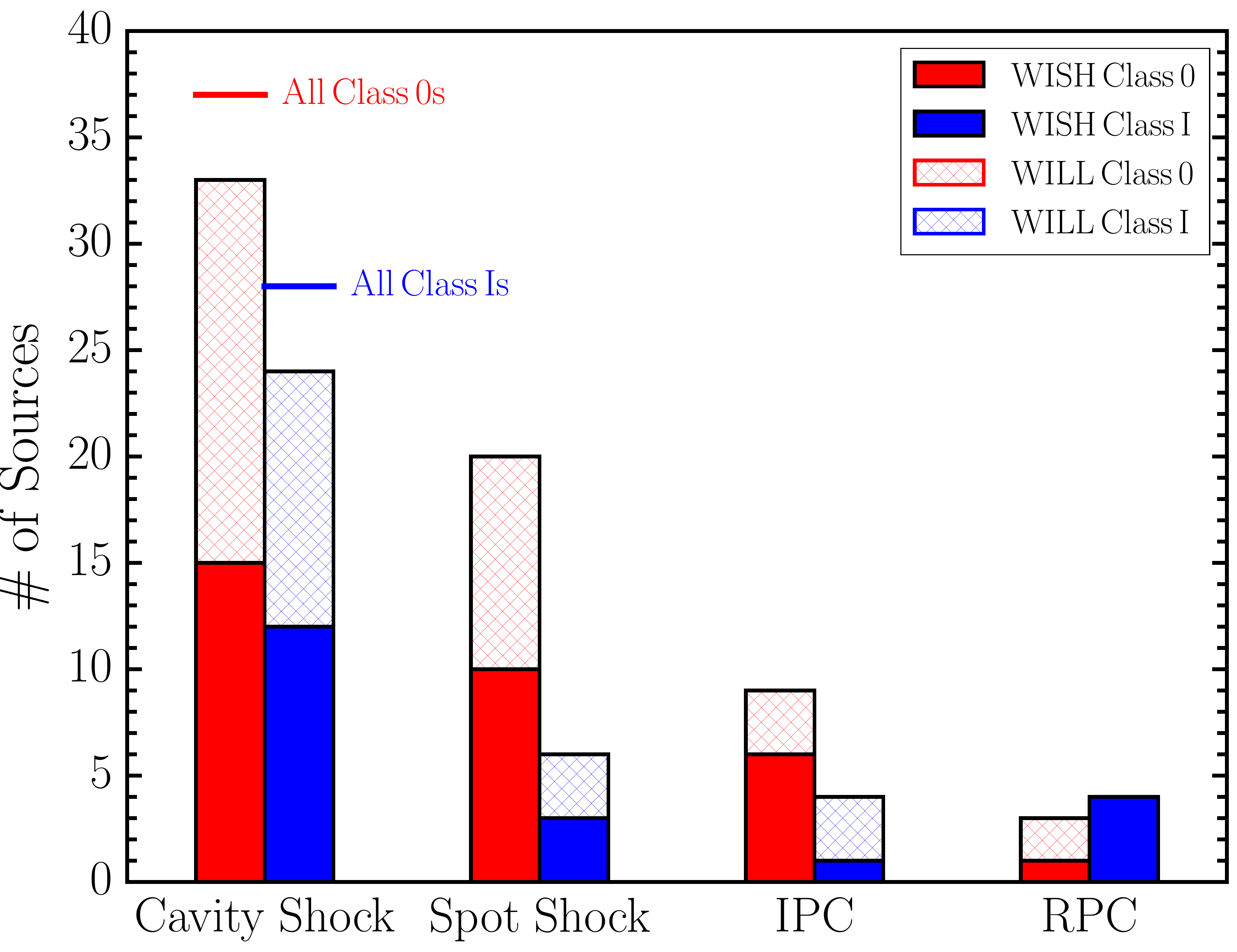}
\caption{Bar chart of the number of shock and inverse/regular P-Cygni envelope components seen in water in the WISH (solid) and WILL (hatched) surveys for Class 0 (red) and I (blue) sources. The horizontal red and blue lines indicate the total number of Class 0 and I sources across both samples respectively.}
\label{F:evolution_h2ohist}
\end{center}
\end{figure}

Inclusion of the WILL sample helps to resolve this issue. For example, Fig.~\ref{F:evolution_h2ohist} shows a histogram of the components found in HIFI water spectra, updated from that shown by \citet{Kristensen2012} to include the WILL survey sources and now using the more physically motivated nomenclature introduced by \citet{Mottram2014}. A broad cavity shock component is observed in almost all Class 0 and I sources. \citet{Mottram2014} argued that this is caused by C-type shocks in the outflow cavity wall, though \citet{Yvart2016} suggest an alternate explanation where this component is formed in a dusty disk wind. Spot shock components, associated with offset J-type shocks either in bullets along the jet or at the base of the outflow \citep{Kristensen2013,Mottram2014}, are far more likely to be detected in Class 0 than Class I sources. Inverse P-Cygni line profiles associated with infall are more common for younger sources, though the inclusion of the WILL sample means that expansion motions traced by regular P-Cygni line profiles are now approximately equally common in both Class 0 and Class I sources. With the exception of the regular P-Cygni profiles, the evolution of water line profile components found for the combined WISH and WILL samples confirms the conclusion of \citet{Kristensen2012} and \citet{Mottram2014} that the outflows of young Class 0 sources are more energetic and their envelopes more infall-dominated than their more evolved Class I counterparts.

Two-sample K-S tests show that there is less than 2$\%$ chance of the Class 0 and I distributions of $M_{\mathrm{env}}$, $M_{\mathrm{out}}$, $F_{\mathrm{CO}}$, $\dot{M}_{\mathrm{out}}$, the integrated intensity of HCO$^{+}$ and H$_{2}$O 1$_{10}-$1$_{01}$, and the FWZI of CO $J$=3$-$2 and H$_{2}$O 1$_{10}-$1$_{01}$ being drawn from the same distribution, with the values for Class 0 sources being larger on average than those of Class I sources. In particular, this confirms the decrease in $F_{\mathrm{CO}}$ and both the line-width and intensity of water emission with evolution of the central source, reinforcing the direct relation between water emission and outflow/shock activity.

One caveat is that the Class 0 and I sources in WISH and WILL sources are not evenly drawn from the sampled star-forming regions. For example, only Class I or II sources are included in the Taurus star-formation region, while many of the Class 0 sources are in Perseus, which is a much more active star-forming complex. The observed differences and trends between Class 0 and I may therefore be accentuated by environmental differences. However, all the evidence suggests that Class I sources have slower, less powerful outflows and show less sign of strong infall motions in their envelopes than Class 0 sources.

\section{Summary and conclusions}
\label{S:conclusions}

This paper has presented a set of \textit{Herschel} and ground-based follow-up observations, characterisation and initial analysis of a flux-limited sample of Class 0/I YSOs in the Gould Belt. From this comprehensive dataset, combined with observations from the WISH and DIGIT surveys, we are able to conclude that:

\begin{itemize}

\item Water line profiles are dominated by emission from the actively shocked regions in outflows, the activity of which decreases in strength (i.e. has lower intensity and FWZI, see Fig.~\ref{F:outflow_hificorrelations}) and has fewer J-type shocks (i.e. fewer spot-shock components, see Fig.~\ref{F:evolution_h2ohist}) as sources evolve from Class 0 to I. We also confirm the decrease in the force of the cooler and slower entrained outflowing gas, measured from low-$J$ CO, as young embedded protostars evolve.

\item The ratio of mass in the entrained outflow to envelope mass (i.e. $M_{\mathrm{out}}/M_{\mathrm{env}}$) remains relatively constant between Class 0 and I with a median of approximately 1$\%$, consistent with a core-to-star formation efficiency of approximately 50$\%$ and an outflow duty cycle of approximately 5$\%$. 

\item $F_{\mathrm{CO}}/\dot{M}_{\mathrm{acc}}$ is relatively constant with $L_{\mathrm{bol}}$, $T_{\mathrm{bol}}$ and $M_{\mathrm{env}}$, suggesting that the entrainment efficiency is constant and independent of the power and evolution of the driving source of the flow. The constant value of $F_{\mathrm{CO}}/\dot{M}_{\mathrm{acc}}$ implies a median velocity at the wind launching radius of 6.3\kms{}. This in turn suggests an entrainment efficiency of approximately 30$-$60$\%$ if the wind is launched around 1AU, or close to 100$\%$ if it is launched at larger radii.

\item $L$[O\,{\sc i}] is strongly correlated with $L_{\mathrm{bol}}$ but not with $M_{\mathrm{env}}$, in contrast to low-$J$ CO, which is strongly correlated with $M_{\mathrm{env}}$ and more weakly related to $L_{\mathrm{bol}}$. This suggests that [O\,{\sc i}] is more closely related to the current accretion activity while low-$J$ CO traces the average activity over timescales of approximately 10$^{2}-$10$^{4}$\,yrs. H$_{2}$O is more strongly correlated with $L_{\mathrm{bol}}$ than $M_{\mathrm{env}}$, but with a smaller difference than for low-$J$ CO, consistent with it tracing actively shocked material between the wind and entrained outflow.

\item $L$[O\,{\sc i}] does not vary significantly between Class 0 and I, likely because the molecular to atomic ratio in the wind and jet decreases as the source evolves, as suggested by \citet{Nisini2015}. This could be caused by increased temperature and decreased density (and thus shielding) in more evolved sources. [O\,{\sc i}] is therefore a poor tracer of the time-averaged mass-loss rate, and thus a poor alternative to CO.

\item Infall signatures are predominantly seen in Class 0 sources in both H$_{2}$O and HCO$^{+}$ single-dish observations, but with little overlap in detections between the two tracers. However, infall signatures remain elusive in the majority of sources. Thus, while water is a good tracer of infall, it is by no means the universal tracer needed to understand how this proceeds in general.

\item The conclusions drawn from the WISH sample hold and become more statistically robust when the combined WISH+WILL+DIGIT sample is analysed.

\end{itemize}

Further exploitation of these data and this sample can be found for the HIFI data in \citet{SanJoseGarciaThesis} and will be presented for the PACS data in Karska et al., (in prep.). The use of PACS and HIFI in this way is cementing the unique legacy of \textit{Herschel} on the energetics of star formation and the origin of water in the interstellar medium.


\begin{acknowledgements}

We thank the anonymous referee for constructive comments that improved the clarity and content of the paper. JCM acknowledges support from grant 614.001.008 from the Netherlands Organisation for Scientific Research (NWO), from the European Union A-ERC grant 291141 CHEMPLAN, and from the European Research Council under the European Community’s Horizon 2020 framework program (2014-2020) via the ERC Consolidator grant `From Cloud to Star Formation (CSF)’ (project number 648505). Astrochemistry in Leiden is supported by the Netherlands Research School for Astronomy (NOVA), by a Spinoza grant, by a Royal Netherlands Academy of Arts and Sciences (KNAW) professor prize, and by the CHEMPLAN A-ERC grant. AK acknowledges support from the Foundation for Polish Science (FNP) and the Polish National Science Center grant 2013/11/N/ST9/00400. D.F. acknowledges support from the Italian Ministry of Education, Universities and Research project SIR (RBSI14ZRHR).

HIFI has been designed and built by a consortium of institutes and university departments from across Europe, Canada and the United States under the leadership of SRON Netherlands Institute for Space Research, Groningen, The Netherlands and with major contributions from Germany, France and the US. Consortium members are: Canada: CSA, U.Waterloo; France: CESR, LAB, LERMA, IRAM; Germany: KOSMA, MPIfR, MPS; Ireland, NUI Maynooth; Italy: ASI, IFSI-INAF, Osservatorio Astrofisico di Arcetri- INAF; Netherlands: SRON, TUD; Poland: CAMK, CBK; Spain: Observatorio Astron{\'o}mico Nacional (IGN), Centro de Astrobiolog{\'i}a (CSIC-INTA). Sweden: Chalmers University of Technology - MC2, RSS $\&$ GARD; Onsala Space Observatory; Swedish National Space Board, Stockholm University - Stockholm Observatory; Switzerland: ETH Zurich, FHNW; USA: Caltech, JPL, NHSC. PACS has been developed by a consortium of institutes led by MPE (Germany) and including UVIE (Austria); KU Leuven, CSL, IMEC (Belgium); CEA, LAM (France); MPIA (Germany); INAF-IFSI/OAA/OAP/OAT, LENS, SISSA (Italy); IAC (Spain). This development has been supported by the funding agencies BMVIT (Austria), ESA-PRODEX (Belgium), CEA/CNES (France), DLR (Germany), ASI/INAF (Italy), and CICYT/MCYT (Spain).

This research made use of \textsc{astropy}, a community-developed core \textsc{python} package for Astronomy \citep{Astropy2013}, and \textsc{aplpy}, an open-source package for plotting astronomical images with \textsc{python} hosted at http://aplpy.github.com.

\end{acknowledgements}

\bibliography{will_overviewbib}

\bibliographystyle{aa}

\appendix

\section{Property determination}
\label{S:properties}

This section presents the various properties of the WILL sources, and details of how they were determined.

\subsection{Spectral energy distributions}
\label{S:properties_seds}

In constructing the spectral energy distributions (SEDs) of the WILL sources, the photometric flux densities from the near-IR to 24\micron{} from 2MASS \citep{Skrutskie2006} and \textit{Spitzer} have been collected, where detected, from the latest determinations by \citet{Dunham2015}. Flux densities for detections at 450 and/or 850\micron{} with SCUBA on the JCMT were taken from the catalogue of \citet{DiFrancesco2008}, and the 1.3\,mm detections with MAMBO on the IRAM 30m telescope by \citet{Maury2011} for Aquila, Serpens South and W40 were also included.

All WILL sources lie within the \textit{Herschel} PACS and SPIRE photometric maps observed in parallel-mode at 70, 160, 250, 350 and 500\micron{} as part of the \textit{Herschel} Gould Belt Survey, with many also within the smaller regions observed in PACS-only mode at 100 and 160\micron{}. Where both parallel and PACS-only mode observations are available at 160\micron{} we use the data taken in PACS-only mode, as these were observed with a slower scanning speed and thus are of better quality than those data taken in parallel with SPIRE observations. Processed and calibrated mosaics were downloaded from the \textit{Herschel} Science Archive (HSA\footnote{http://www.cosmos.esa.int/web/herschel/science-archive/}, see also the \textit{Herschel} Gould Belt survey archive for further products\footnote{http://gouldbelt-herschel.cea.fr/archives}). The observation identification numbers for these maps are given in Table~\ref{T:obsids_gbs}.

Aperture photometry was then performed using \textsc{python} routines from the \textsc{astropy} package to extract flux densities for all sources at all available wavelengths. The starting value for the source aperture, inner and outer sky radii for each source at each wavelength was set at 6, 10 and 12 pixels, typically corresponding to 6, 10 and 12 times the beam size. Rings at these radii were then overlaid on images of the data and the radii adjusted to best encompass each source and local background respectively, while excluding nearby sources. Aperture correction factors were calculated and applied for each combination of aperture and sky annuli using the relevant PACS and SPIRE point source function images. 

In some cases, additional nearby sources are blended with the primary WILL source at longer wavelengths where the \textit{Herschel} beam becomes large. For these sources, we specify the longest wavelength where the sources are reliably separated. The flux densities at longer wavelengths are then scaled down by the ratio of the flux densities in the reliable image between the aperture used for that wavelength and the aperture used for the longer wavelength data. For example, if a source is blended at 500\micron{}, then the flux density at that wavelength is scaled down by the ratio of the flux density at 350\micron{} to the flux density in the 350\micron{} image within the region defined by the aperture used for the 500\micron{} data.

\begin{figure*}
\begin{center}
\includegraphics[width=0.9\textwidth]{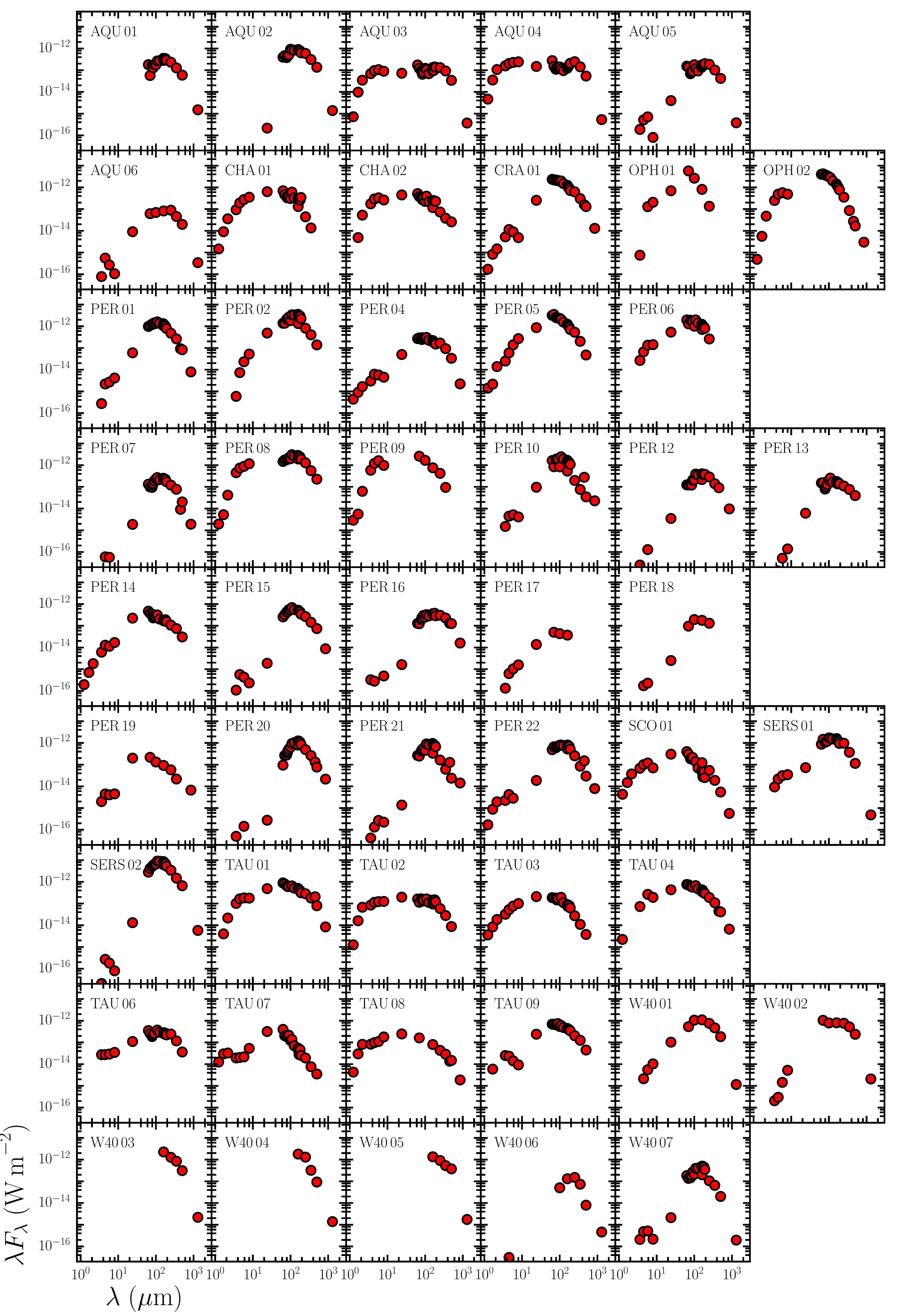}
\caption{SEDs for all WILL sources. Data points from the PACS spectra are only included when used in the $L_{\mathrm{bol}}$ calculations.}
\label{F:seds_overview}
\end{center}
\end{figure*}

Finally, continuum flux densities from the WILL PACS spectra were also obtained in the same manner as for the WISH sources \citep[see][for details]{Karska2013}. However, it is sometimes difficult in the PACS spectral maps to separate the emission related to the protostar from surrounding emission. Therefore, in cases where the PACS spectral continuum is significantly higher than obtained from the broad-band photometric maps, the latter is preferred and the PACS spectral continuum flux densities are not included in the SED. The broad-band photometric flux densities for all sources are given in Table~\ref{T:photometry_fluxes}, with the PACS spectral continuum flux densities given separately in Table \ref{T:pacs_fluxes}. The SEDs of all sources are shown in Fig.~\ref{F:seds_overview}.

For Aquila sources 01$-$04, the peak PACS continuum is offset from the coordinates in \citet{Maury2011}, and so extraction of the SED and ground-based molecular line emission is performed at the peak flux position.

\begin{table*}
\caption{SED photometric continuum fluxes. All numerical wavelengths in \micron{}.}
\begin{center}
\begin{tabular}{lcccccccc}
\hline \noalign {\smallskip}
Name & $J$ & $H$ & $K_{\mathrm{s}}$ & 3.6 & 4.5 & 5.8 & 8.0 & 24 \\
 & (mJy) & (mJy) & (mJy) & (mJy) & (mJy) & (mJy) & (mJy) & (mJy) \\
\hline\noalign {\smallskip}
AQU~01&$-$&$-$&$-$&$-$&$-$&$-$&$-$&$-$\\
AQU~02&$-$&$-$&$-$&$-$&$-$&$-$&$-$&\phantom{000}1.72$\pm$\phantom{000}0.60\\
AQU~03&\phantom{0}0.30$\pm$0.06&\phantom{0}5.39$\pm$0.17&25.00$\pm$0.60&\phantom{00}83.40$\pm$\phantom{00}4.17&\phantom{0}145.00$\pm$\phantom{00}6.99&\phantom{0}205.00$\pm$\phantom{00}9.76&\phantom{0}239.00$\pm$\phantom{0}11.40&\phantom{0}581.00$\pm$\phantom{00}54.10\\
AQU~04&\phantom{0}1.92$\pm$0.09&19.50$\pm$0.49&77.40$\pm$1.64&\phantom{0}190.00$\pm$\phantom{00}9.28&\phantom{0}306.00$\pm$\phantom{0}15.90&\phantom{0}441.00$\pm$\phantom{0}20.80&\phantom{0}650.00$\pm$\phantom{0}31.20&1190.00$\pm$\phantom{0}118.00\\
AQU~05&$-$&$-$&$-$&\phantom{000}0.22$\pm$\phantom{00}0.02&\phantom{000}0.77$\pm$\phantom{00}0.08&\phantom{000}1.37$\pm$\phantom{00}0.09&\phantom{000}0.21$\pm$\phantom{00}0.10&\phantom{00}31.90$\pm$\phantom{000}2.98\\
AQU~06&$-$&$-$&$-$&\phantom{000}0.09$\pm$\phantom{00}0.01&\phantom{000}0.83$\pm$\phantom{00}0.04&\phantom{000}0.52$\pm$\phantom{00}0.06&\phantom{000}0.28$\pm$\phantom{00}0.13&\phantom{00}72.60$\pm$\phantom{000}6.85\\
CHA~01&\phantom{0}0.60$\pm$0.06&\phantom{0}5.04$\pm$0.18&25.70$\pm$0.61&\phantom{0}112.00$\pm$\phantom{0}10.10&\phantom{0}287.00$\pm$\phantom{0}24.00&\phantom{0}533.00$\pm$\phantom{0}27.00&\phantom{0}939.00$\pm$\phantom{0}48.30&5080.00$\pm$\phantom{0}473.00\\
CHA~02&$-$&\phantom{0}2.70$\pm$0.12&38.00$\pm$1.10&\phantom{0}210.00$\pm$\phantom{0}10.00&\phantom{0}440.00$\pm$\phantom{0}10.00&\phantom{0}640.00$\pm$\phantom{0}31.00&\phantom{0}700.00$\pm$\phantom{0}34.00&3600.00$\pm$\phantom{0}340.00\\
CRA~01&\phantom{0}0.07$\pm$0.01&\phantom{0}0.46$\pm$0.05&\phantom{0}1.06$\pm$0.11&\phantom{000}6.30$\pm$\phantom{00}0.71&\phantom{00}17.20$\pm$\phantom{00}1.20&\phantom{00}17.30$\pm$\phantom{00}1.10&\phantom{00}12.90$\pm$\phantom{00}0.70&2040.00$\pm$\phantom{0}193.00\\
OPH~01&$-$&$-$&$-$&\phantom{000}0.90$\pm$\phantom{00}0.25&$-$&\phantom{0}250.00$\pm$\phantom{0}30.00&\phantom{0}550.00$\pm$130.00&5600.00$\pm$\phantom{0}840.00\\
OPH~02&\phantom{0}0.20$\pm$0.04&\phantom{0}3.10$\pm$0.28&34.00$\pm$2.10&\phantom{0}300.00$\pm$\phantom{0}24.00&\phantom{0}730.00$\pm$\phantom{0}48.00&1100.00$\pm$\phantom{0}55.00&1300.00$\pm$\phantom{0}65.00&$-$\\
PER~01&$-$&$-$&$-$&\phantom{000}0.33$\pm$\phantom{00}0.04&\phantom{000}3.30$\pm$\phantom{00}0.29&\phantom{000}5.20$\pm$\phantom{00}0.36&\phantom{00}11.00$\pm$\phantom{00}0.57&\phantom{0}480.00$\pm$\phantom{00}51.00\\
PER~02&$-$&$-$&$-$&\phantom{000}0.71$\pm$\phantom{00}0.15&\phantom{00}11.00$\pm$\phantom{00}0.74&\phantom{00}45.00$\pm$\phantom{00}2.20&\phantom{0}140.00$\pm$\phantom{00}7.10&3900.00$\pm$\phantom{0}420.00\\
PER~04&\phantom{0}0.18$\pm$0.01&\phantom{0}0.50$\pm$0.03&\phantom{0}1.20$\pm$0.08&\phantom{000}3.60$\pm$\phantom{00}0.29&\phantom{000}9.30$\pm$\phantom{00}0.65&\phantom{00}11.00$\pm$\phantom{00}0.77&\phantom{00}12.00$\pm$\phantom{00}0.82&\phantom{0}400.00$\pm$\phantom{00}42.00\\
PER~05&\phantom{0}0.58$\pm$0.03&\phantom{0}1.20$\pm$0.15&10.00$\pm$0.50&\phantom{00}30.00$\pm$\phantom{00}2.30&\phantom{00}89.00$\pm$\phantom{00}5.70&\phantom{0}270.00$\pm$\phantom{0}13.00&\phantom{0}720.00$\pm$\phantom{0}41.00&6900.00$\pm$1200.00\\
PER~06&$-$&$-$&$-$&\phantom{00}32.00$\pm$\phantom{00}2.20&\phantom{0}100.00$\pm$\phantom{00}6.00&\phantom{0}260.00$\pm$\phantom{0}13.00&\phantom{0}380.00$\pm$\phantom{0}23.00&4300.00$\pm$\phantom{0}460.00\\
PER~07&$-$&$-$&$-$&$-$&\phantom{000}0.09$\pm$\phantom{00}0.01&\phantom{000}0.11$\pm$\phantom{00}0.03&$-$&\phantom{00}15.00$\pm$\phantom{000}1.60\\
PER~08&\phantom{0}0.80$\pm$0.04&\phantom{0}2.90$\pm$0.25&30.00$\pm$1.80&\phantom{0}540.00$\pm$\phantom{0}46.00&1100.00$\pm$\phantom{0}82.00&1700.00$\pm$210.00&3100.00$\pm$190.00&$-$\\
PER~09&\phantom{0}1.20$\pm$0.06&\phantom{0}3.10$\pm$0.29&45.00$\pm$2.80&\phantom{0}700.00$\pm$\phantom{0}76.00&1800.00$\pm$170.00&3100.00$\pm$280.00&2600.00$\pm$250.00&$-$\\
PER~10&$-$&$-$&$-$&\phantom{000}1.80$\pm$\phantom{00}0.18&\phantom{000}6.70$\pm$\phantom{00}0.66&\phantom{00}10.00$\pm$\phantom{00}0.83&\phantom{00}11.00$\pm$\phantom{00}0.78&\phantom{0}770.00$\pm$\phantom{00}82.00\\
PER~12&$-$&$-$&$-$&\phantom{000}0.03$\pm$\phantom{00}0.01&$-$&\phantom{000}0.25$\pm$\phantom{00}0.04&$-$&\phantom{00}28.00$\pm$\phantom{000}3.00\\
PER~13&$-$&$-$&$-$&$-$&\phantom{000}0.02$\pm$\phantom{00}0.01&\phantom{000}0.10$\pm$\phantom{00}0.03&\phantom{000}0.37$\pm$\phantom{00}0.04&\phantom{00}49.00$\pm$\phantom{000}5.20\\
PER~14&\phantom{0}0.08$\pm$0.01&\phantom{0}0.39$\pm$0.02&\phantom{0}1.30$\pm$0.10&\phantom{000}7.30$\pm$\phantom{00}0.53&\phantom{00}19.00$\pm$\phantom{00}1.00&\phantom{00}22.00$\pm$\phantom{00}1.10&\phantom{00}45.00$\pm$\phantom{00}2.20&1800.00$\pm$\phantom{0}190.00\\
PER~15&$-$&$-$&$-$&\phantom{000}0.13$\pm$\phantom{00}0.01&\phantom{000}0.85$\pm$\phantom{00}0.07&\phantom{000}0.83$\pm$\phantom{00}0.07&\phantom{000}0.62$\pm$\phantom{00}0.06&\phantom{00}15.00$\pm$\phantom{000}1.50\\
PER~16&$-$&$-$&$-$&\phantom{000}0.39$\pm$\phantom{00}0.08&\phantom{000}0.42$\pm$\phantom{00}0.14&$-$&\phantom{000}1.30$\pm$\phantom{00}0.11&\phantom{00}13.00$\pm$\phantom{000}1.40\\
PER~17&$-$&$-$&$-$&\phantom{000}0.16$\pm$\phantom{00}0.01&\phantom{000}0.94$\pm$\phantom{00}0.07&\phantom{000}2.00$\pm$\phantom{00}0.15&\phantom{000}4.20$\pm$\phantom{00}0.30&\phantom{0}110.00$\pm$\phantom{00}12.00\\
PER~18&$-$&$-$&$-$&$-$&\phantom{000}0.26$\pm$\phantom{00}0.04&\phantom{000}0.44$\pm$\phantom{00}0.05&$-$&\phantom{00}20.00$\pm$\phantom{000}2.10\\
PER~19&$-$&$-$&$-$&\phantom{000}2.40$\pm$\phantom{00}0.17&\phantom{000}6.70$\pm$\phantom{00}0.34&\phantom{000}7.90$\pm$\phantom{00}0.40&\phantom{00}12.00$\pm$\phantom{00}0.60&1600.00$\pm$\phantom{0}170.00\\
PER~20&$-$&$-$&$-$&\phantom{000}0.06$\pm$\phantom{00}0.02&$-$&\phantom{000}0.28$\pm$\phantom{00}0.06&$-$&\phantom{000}2.20$\pm$\phantom{000}0.34\\
PER~21&$-$&$-$&$-$&\phantom{000}0.05$\pm$\phantom{00}0.01&\phantom{000}0.20$\pm$\phantom{00}0.06&\phantom{000}0.52$\pm$\phantom{00}0.06&\phantom{000}0.60$\pm$\phantom{00}0.12&\phantom{00}11.00$\pm$\phantom{000}1.20\\
PER~22&\phantom{0}0.07$\pm$0.01&\phantom{0}0.49$\pm$0.03&\phantom{0}1.40$\pm$0.16&\phantom{000}2.70$\pm$\phantom{00}0.21&\phantom{000}6.30$\pm$\phantom{00}0.66&\phantom{000}5.50$\pm$\phantom{00}0.32&$-$&\phantom{0}150.00$\pm$\phantom{00}20.00\\
SCO~01&\phantom{0}1.78$\pm$0.09&\phantom{0}8.33$\pm$0.34&26.80$\pm$0.79&\phantom{00}81.80$\pm$\phantom{00}4.33&\phantom{0}150.00$\pm$\phantom{00}7.85&\phantom{0}231.00$\pm$\phantom{0}10.90&\phantom{0}189.00$\pm$\phantom{0}11.60&2430.00$\pm$\phantom{0}248.00\\
SERS~01&$-$&$-$&$-$&\phantom{00}11.30$\pm$\phantom{00}0.56&\phantom{00}32.80$\pm$\phantom{00}1.62&\phantom{00}60.20$\pm$\phantom{00}2.84&\phantom{00}92.60$\pm$\phantom{00}4.56&\phantom{0}582.00$\pm$\phantom{0}105.00\\
SERS~02&$-$&$-$&$-$&\phantom{000}0.02$\pm$\phantom{00}0.01&\phantom{000}0.40$\pm$\phantom{00}0.05&\phantom{000}0.35$\pm$\phantom{00}0.06&\phantom{000}0.21$\pm$\phantom{00}0.10&\phantom{0}105.00$\pm$\phantom{00}11.90\\
TAU~01&$-$&\phantom{0}2.22$\pm$0.12&15.51$\pm$0.30&\phantom{0}121.49$\pm$\phantom{00}5.59&\phantom{0}248.06$\pm$\phantom{0}11.42&\phantom{0}350.51$\pm$\phantom{0}16.14&\phantom{0}473.22$\pm$\phantom{0}21.79&3887.74$\pm$\phantom{0}143.23\\
TAU~02&\phantom{0}0.52$\pm$0.04&\phantom{0}8.94$\pm$0.18&48.70$\pm$1.03&\phantom{0}101.99$\pm$\phantom{00}4.70&\phantom{0}170.04$\pm$\phantom{00}7.83&\phantom{0}238.07$\pm$\phantom{0}10.96&\phantom{0}333.47$\pm$\phantom{0}15.36&1576.52$\pm$\phantom{00}58.08\\
TAU~03&\phantom{0}1.48$\pm$0.11&\phantom{0}4.68$\pm$0.22&13.27$\pm$0.34&\phantom{00}38.42$\pm$\phantom{00}1.77&\phantom{00}78.44$\pm$\phantom{00}3.61&\phantom{0}147.47$\pm$\phantom{00}6.79&\phantom{0}267.34$\pm$\phantom{0}12.31&1666.00$\pm$\phantom{00}61.38\\
TAU~04&\phantom{0}0.92$\pm$0.06&$-$&$-$&\phantom{00}87.21$\pm$\phantom{00}4.02&$-$&\phantom{0}511.33$\pm$\phantom{0}23.55&\phantom{0}514.12$\pm$\phantom{0}23.68&3450.95$\pm$\phantom{0}128.24\\
TAU~06&$-$&$-$&$-$&\phantom{00}32.85$\pm$\phantom{00}1.82&\phantom{00}41.17$\pm$\phantom{00}1.90&\phantom{00}56.07$\pm$\phantom{00}2.58&\phantom{00}92.70$\pm$\phantom{00}4.27&\phantom{0}882.47$\pm$\phantom{00}32.51\\
TAU~07&\phantom{0}5.14$\pm$0.16&16.50$\pm$0.46&23.44$\pm$0.54&\phantom{00}23.15$\pm$\phantom{00}1.07&\phantom{00}30.38$\pm$\phantom{00}1.40&\phantom{00}41.75$\pm$\phantom{00}1.92&\phantom{0}141.54$\pm$\phantom{00}6.06&2545.06$\pm$\phantom{00}93.76\\
TAU~08&\phantom{0}1.77$\pm$0.07&16.50$\pm$0.53&58.55$\pm$1.19&\phantom{00}99.21$\pm$\phantom{00}4.57&\phantom{0}146.74$\pm$\phantom{00}6.76&\phantom{0}213.16$\pm$\phantom{00}9.82&\phantom{0}477.60$\pm$\phantom{0}21.99&1966.52$\pm$\phantom{00}72.44\\
TAU~09&$-$&\phantom{0}3.27$\pm$0.19&$-$&\phantom{00}29.96$\pm$\phantom{00}1.38&\phantom{00}34.88$\pm$\phantom{00}1.61&\phantom{00}27.33$\pm$\phantom{00}1.26&\phantom{00}24.61$\pm$\phantom{00}1.13&1912.93$\pm$\phantom{00}70.47\\
W40~01&$-$&$-$&$-$&$-$&\phantom{000}3.19$\pm$\phantom{00}0.18&\phantom{00}10.70$\pm$\phantom{00}0.84&\phantom{00}27.40$\pm$\phantom{00}3.78&\phantom{0}824.00$\pm$\phantom{00}78.40\\
W40~02&$-$&$-$&$-$&\phantom{000}0.25$\pm$\phantom{00}0.09&\phantom{000}0.45$\pm$\phantom{00}0.07&\phantom{000}2.80$\pm$\phantom{00}1.01&\phantom{00}13.90$\pm$\phantom{00}3.94&$-$\\
W40~03&$-$&$-$&$-$&$-$&$-$&$-$&$-$&$-$\\
W40~04&$-$&$-$&$-$&$-$&$-$&$-$&$-$&$-$\\
W40~05&$-$&$-$&$-$&$-$&$-$&$-$&$-$&$-$\\
W40~06&$-$&$-$&$-$&$-$&\phantom{000}0.05$\pm$\phantom{00}0.01&$-$&$-$&$-$\\
W40~07&$-$&$-$&$-$&\phantom{000}0.25$\pm$\phantom{00}0.03&\phantom{000}0.73$\pm$\phantom{00}0.05&\phantom{000}1.00$\pm$\phantom{00}0.07&\phantom{000}0.58$\pm$\phantom{00}0.06&\phantom{00}17.00$\pm$\phantom{000}1.58\\
\hline
\end{tabular}
\end{center}
\label{T:photometry_fluxes}
\end{table*}

\renewcommand{\thetable}{\thesection.\arabic{table} (Cont.)}
\addtocounter{table}{-1}

\begin{table*}
\caption{SED photometric continuum fluxes. All numerical wavelengths in \micron{}.}
\begin{center}
\begin{tabular}{lccccccccc}
\hline \noalign {\smallskip}
Name & 70 & 100 & 160 & 250 & 350 & 450\tablefootmark{a} & 500 & 850\tablefootmark{a} & 1300 \\
 & (Jy) & (Jy) & (Jy) & (Jy) & (Jy) & (Jy) & (Jy) & (Jy) & (Jy) \\
\hline\noalign {\smallskip}
AQU~01&\phantom{00}1.33$\pm$0.03&\phantom{00}5.89$\pm$0.08&\phantom{\tablefootmark{b}}\phantom{0}15.07$\pm$0.25\tablefootmark{b}&\phantom{\tablefootmark{c}}\phantom{0}19.43$\pm$0.67\tablefootmark{c}&\phantom{\tablefootmark{c}}\phantom{0}14.85$\pm$0.69\tablefootmark{c}&$-$&\phantom{\tablefootmark{c}}\phantom{00}9.87$\pm$0.52\tablefootmark{c}&$-$&\phantom{00}0.65$\pm$0.26\\
AQU~02&\phantom{0}10.88$\pm$0.20&\phantom{0}30.77$\pm$0.35&\phantom{\tablefootmark{b}}\phantom{0}47.80$\pm$1.01\tablefootmark{b}&\phantom{\tablefootmark{c}}\phantom{0}49.43$\pm$1.23\tablefootmark{c}&\phantom{\tablefootmark{c}}\phantom{0}36.19$\pm$1.12\tablefootmark{c}&$-$&\phantom{\tablefootmark{c}}\phantom{0}22.82$\pm$0.79\tablefootmark{c}&$-$&\phantom{00}0.60$\pm$0.19\\
AQU~03&\phantom{00}2.84$\pm$0.06&\phantom{00}4.09$\pm$0.05&\phantom{\tablefootmark{b}}\phantom{00}7.08$\pm$0.14\tablefootmark{b}&\phantom{\tablefootmark{c}}\phantom{0}10.94$\pm$0.41\tablefootmark{c}&\phantom{\tablefootmark{c}}\phantom{0}10.75$\pm$0.60\tablefootmark{c}&$-$&\phantom{\tablefootmark{c}}\phantom{00}5.69$\pm$0.48\tablefootmark{c}&$-$&\phantom{00}0.16$\pm$0.06\\
AQU~04&\phantom{00}3.56$\pm$0.09&\phantom{00}4.72$\pm$0.07&\phantom{\tablefootmark{b}}\phantom{00}6.67$\pm$0.17\tablefootmark{b}&\phantom{\tablefootmark{c}}\phantom{0}20.56$\pm$0.52\tablefootmark{c}&\phantom{\tablefootmark{c}}\phantom{0}16.71$\pm$0.66\tablefootmark{c}&$-$&\phantom{\tablefootmark{c}}\phantom{00}8.98$\pm$0.77\tablefootmark{c}&$-$&\phantom{00}0.23$\pm$0.10\\
AQU~05&\phantom{00}3.21$\pm$0.08&\phantom{00}5.95$\pm$0.08&\phantom{\tablefootmark{b}}\phantom{00}9.45$\pm$0.19\tablefootmark{b}&\phantom{\tablefootmark{c}}\phantom{0}15.47$\pm$0.49\tablefootmark{c}&\phantom{\tablefootmark{c}}\phantom{0}11.88$\pm$0.83\tablefootmark{c}&$-$&\phantom{\tablefootmark{c}}\phantom{00}6.98$\pm$0.60\tablefootmark{c}&$-$&\phantom{00}0.16$\pm$0.06\\
AQU~06&\phantom{00}1.44$\pm$0.06&\phantom{00}2.32$\pm$0.06&\phantom{\tablefootmark{b}}\phantom{00}4.38$\pm$0.24\tablefootmark{b}&\phantom{\tablefootmark{c}}\phantom{00}7.44$\pm$0.62\tablefootmark{c}&\phantom{\tablefootmark{c}}\phantom{00}5.33$\pm$0.62\tablefootmark{c}&$-$&\phantom{\tablefootmark{c}}\phantom{00}3.30$\pm$0.50\tablefootmark{c}&$-$&\phantom{00}0.15$\pm$0.06\\
CHA~01&\phantom{0}10.87$\pm$0.23&\phantom{0}10.23$\pm$0.14&\phantom{\tablefootmark{b}}\phantom{00}6.99$\pm$0.17\tablefootmark{b}&\phantom{\tablefootmark{c}}\phantom{00}3.67$\pm$0.25\tablefootmark{c}&\phantom{\tablefootmark{c}}\phantom{00}1.58$\pm$0.26\tablefootmark{c}&$-$&$-$&$-$&$-$\\
CHA~02&\phantom{00}8.87$\pm$0.17&\phantom{00}7.48$\pm$0.10&\phantom{\tablefootmark{b}}\phantom{00}6.25$\pm$0.13\tablefootmark{b}&\phantom{\tablefootmark{c}}\phantom{00}6.11$\pm$0.43\tablefootmark{c}&\phantom{\tablefootmark{c}}\phantom{00}4.42$\pm$0.54\tablefootmark{c}&$-$&\phantom{\tablefootmark{c}}\phantom{00}4.29$\pm$0.47\tablefootmark{c}&$-$&$-$\\
CRA~01&\phantom{0}51.72$\pm$0.94&\phantom{0}64.45$\pm$0.65&\phantom{\tablefootmark{b}}\phantom{0}66.03$\pm$0.81\tablefootmark{b}&\phantom{\tablefootmark{c}}\phantom{0}51.78$\pm$0.89\tablefootmark{c}&\phantom{\tablefootmark{c}}\phantom{0}35.52$\pm$0.77\tablefootmark{c}&\phantom{0}23.84$\pm$4.77&\phantom{\tablefootmark{c}}\phantom{0}21.79$\pm$0.51\tablefootmark{c}&\phantom{00}3.68$\pm$0.74&$-$\\
OPH~01&134.00$\pm$1.60&\phantom{0}88.39$\pm$0.78&\phantom{\tablefootmark{b}}\phantom{0}43.25$\pm$1.18\tablefootmark{b}&\phantom{\tablefootmark{c}}\phantom{0}11.22$\pm$1.48\tablefootmark{c}&$-$&$-$&$-$&$-$&$-$\\
OPH~02&\phantom{0}98.24$\pm$1.82&103.55$\pm$1.18&\phantom{\tablefootmark{b}}\phantom{0}71.39$\pm$1.14\tablefootmark{b}&\phantom{\tablefootmark{c}}\phantom{0}29.44$\pm$0.96\tablefootmark{c}&\phantom{\tablefootmark{c}}\phantom{00}9.99$\pm$0.76\tablefootmark{c}&\phantom{00}4.11$\pm$0.82&\phantom{\tablefootmark{c}}\phantom{00}2.78$\pm$0.51\tablefootmark{c}&\phantom{00}0.85$\pm$0.17&$-$\\
PER~01&\phantom{0}25.57$\pm$0.51&\phantom{0}44.87$\pm$0.55&\phantom{\tablefootmark{b}}\phantom{0}55.21$\pm$0.90\tablefootmark{b}&\phantom{\tablefootmark{c}}\phantom{0}40.01$\pm$0.91\tablefootmark{c}&\phantom{\tablefootmark{c}}\phantom{0}30.67$\pm$0.85\tablefootmark{c}&\phantom{0}13.83$\pm$2.77&\phantom{\tablefootmark{c}}\phantom{0}13.83$\pm$0.56\tablefootmark{c}&\phantom{00}2.24$\pm$0.45&$-$\\
PER~02&\phantom{0}32.16$\pm$0.72&\phantom{0}60.76$\pm$0.97&\phantom{\tablefootmark{b}}\phantom{0}70.74$\pm$2.59\tablefootmark{b}&\phantom{\tablefootmark{c}}\phantom{0}67.82$\pm$2.70\tablefootmark{c}&\phantom{\tablefootmark{c}}\phantom{0}47.30$\pm$2.46\tablefootmark{c}&$-$&\phantom{\tablefootmark{c}}\phantom{0}23.14$\pm$1.44\tablefootmark{c}&$-$&$-$\\
PER~04&\phantom{00}6.80$\pm$0.13&\phantom{00}9.36$\pm$0.12&\phantom{\tablefootmark{b}}\phantom{0}11.72$\pm$0.16\tablefootmark{b}&\phantom{\tablefootmark{c}}\phantom{0}14.12$\pm$0.20\tablefootmark{c}&\phantom{\tablefootmark{c}}\phantom{0}10.89$\pm$0.21\tablefootmark{c}&$-$&\phantom{\tablefootmark{c}}\phantom{00}5.60$\pm$0.13\tablefootmark{c}&\phantom{00}0.63$\pm$0.13&$-$\\
PER~05&\phantom{0}81.94$\pm$1.59&\phantom{0}77.36$\pm$0.92&\phantom{\tablefootmark{b}}\phantom{0}66.27$\pm$1.03\tablefootmark{b}&\phantom{\tablefootmark{c}}\phantom{0}44.23$\pm$0.76\tablefootmark{c}&\phantom{\tablefootmark{c}}\phantom{0}23.63$\pm$0.61\tablefootmark{c}&$-$&\phantom{\tablefootmark{c}}\phantom{00}7.93$\pm$0.35\tablefootmark{c}&$-$&$-$\\
PER~06&\phantom{0}30.24$\pm$0.84&\phantom{0}33.65$\pm$0.63&\phantom{\tablefootmark{b}}\phantom{0}38.17$\pm$0.72\tablefootmark{b}&\phantom{\tablefootmark{c}}\phantom{0}21.75$\pm$1.24\tablefootmark{c}&$-$&$-$&$-$&$-$&$-$\\
PER~07&\phantom{00}2.30$\pm$0.05&\phantom{00}7.75$\pm$0.13&\phantom{\tablefootmark{b}}\phantom{0}11.47$\pm$0.26\tablefootmark{b}&\phantom{\tablefootmark{c}}\phantom{0}10.05$\pm$0.32\tablefootmark{c}&\phantom{\tablefootmark{c}}\phantom{00}9.05$\pm$0.43\tablefootmark{c}&\phantom{00}1.38$\pm$0.28&\phantom{\tablefootmark{c}}\phantom{00}3.38$\pm$0.36\tablefootmark{c}&\phantom{00}0.54$\pm$0.11&$-$\\
PER~08&\phantom{0}40.55$\pm$0.62&\phantom{0}66.15$\pm$0.56&\phantom{\tablefootmark{b}}102.63$\pm$1.46\tablefootmark{b}&\phantom{\tablefootmark{c}}106.89$\pm$2.08\tablefootmark{c}&\phantom{\tablefootmark{c}}\phantom{0}64.92$\pm$1.97\tablefootmark{c}&$-$&\phantom{\tablefootmark{c}}\phantom{0}37.75$\pm$1.27\tablefootmark{c}&$-$&$-$\\
PER~09&\phantom{0}60.37$\pm$2.08&\phantom{0}55.78$\pm$1.24&\phantom{\tablefootmark{b}}\phantom{0}40.46$\pm$1.89\tablefootmark{b}&\phantom{\tablefootmark{c}}\phantom{0}34.99$\pm$2.77\tablefootmark{c}&\phantom{\tablefootmark{c}}\phantom{0}11.04$\pm$1.34\tablefootmark{c}&$-$&$-$&$-$&$-$\\
PER~10&\phantom{0}19.80$\pm$0.48&\phantom{0}27.83$\pm$0.49&\phantom{\tablefootmark{b}}\phantom{0}28.43$\pm$1.21\tablefootmark{b}&\phantom{\tablefootmark{c}}\phantom{0}16.50$\pm$1.31\tablefootmark{c}&\phantom{\tablefootmark{c}}\phantom{00}8.86$\pm$1.20\tablefootmark{c}&\phantom{0}41.51$\pm$8.30&\phantom{\tablefootmark{c}}\phantom{00}5.77$\pm$0.77\tablefootmark{c}&\phantom{00}6.46$\pm$1.29&$-$\\
PER~12&\phantom{00}2.87$\pm$0.07&\phantom{00}6.96$\pm$0.13&\phantom{\tablefootmark{b}}\phantom{0}11.60$\pm$0.32\tablefootmark{b}&\phantom{\tablefootmark{c}}\phantom{0}23.94$\pm$0.60\tablefootmark{c}&\phantom{\tablefootmark{c}}\phantom{0}16.26$\pm$1.09\tablefootmark{c}&\phantom{0}13.63$\pm$2.73&$-$&\phantom{00}2.74$\pm$0.55&$-$\\
PER~13&\phantom{00}3.63$\pm$0.07&\phantom{00}6.81$\pm$0.10&\phantom{\tablefootmark{b}}\phantom{00}9.47$\pm$0.16\tablefootmark{b}&\phantom{\tablefootmark{c}}\phantom{00}9.07$\pm$0.26\tablefootmark{c}&\phantom{\tablefootmark{c}}\phantom{00}8.87$\pm$0.31\tablefootmark{c}&$-$&\phantom{\tablefootmark{c}}\phantom{00}6.61$\pm$0.24\tablefootmark{c}&$-$&$-$\\
PER~14&\phantom{00}8.76$\pm$0.17&\phantom{00}8.65$\pm$0.11&\phantom{\tablefootmark{b}}\phantom{00}9.73$\pm$0.13\tablefootmark{b}&\phantom{\tablefootmark{c}}\phantom{00}8.88$\pm$0.22\tablefootmark{c}&\phantom{\tablefootmark{c}}\phantom{00}8.65$\pm$0.29\tablefootmark{c}&$-$&\phantom{\tablefootmark{c}}\phantom{00}5.08$\pm$0.21\tablefootmark{c}&$-$&$-$\\
PER~15&\phantom{00}7.53$\pm$0.14&\phantom{0}18.50$\pm$0.24&\phantom{\tablefootmark{b}}\phantom{0}26.54$\pm$0.42\tablefootmark{b}&\phantom{\tablefootmark{c}}\phantom{0}22.04$\pm$0.46\tablefootmark{c}&\phantom{\tablefootmark{c}}\phantom{0}16.89$\pm$0.46\tablefootmark{c}&$-$&\phantom{\tablefootmark{c}}\phantom{0}12.37$\pm$0.32\tablefootmark{c}&\phantom{00}2.47$\pm$0.49&$-$\\
PER~16&\phantom{00}2.90$\pm$0.06&\phantom{00}8.34$\pm$0.11&\phantom{\tablefootmark{b}}\phantom{0}19.21$\pm$0.30\tablefootmark{b}&\phantom{\tablefootmark{c}}\phantom{0}25.29$\pm$0.52\tablefootmark{c}&\phantom{\tablefootmark{c}}\phantom{0}26.48$\pm$0.64\tablefootmark{c}&\phantom{0}19.55$\pm$3.91&\phantom{\tablefootmark{c}}\phantom{0}20.97$\pm$0.48\tablefootmark{c}&\phantom{00}4.53$\pm$0.91&$-$\\
PER~17&\phantom{00}1.17$\pm$0.03&\phantom{00}1.45$\pm$0.03&\phantom{\tablefootmark{b}}\phantom{00}1.98$\pm$0.11\tablefootmark{b}&$-$&$-$&$-$&$-$&$-$&$-$\\
PER~18&\phantom{00}2.24$\pm$0.05&\phantom{00}6.44$\pm$0.09&\phantom{\tablefootmark{b}}\phantom{00}9.50$\pm$0.24\tablefootmark{b}&\phantom{\tablefootmark{c}}\phantom{0}10.88$\pm$0.65\tablefootmark{c}&$-$&$-$&$-$&$-$&$-$\\
PER~19&\phantom{00}5.08$\pm$0.10&\phantom{00}4.43$\pm$0.07&\phantom{\tablefootmark{b}}\phantom{00}4.90$\pm$0.12\tablefootmark{b}&\phantom{\tablefootmark{c}}\phantom{00}4.80$\pm$0.29\tablefootmark{c}&\phantom{\tablefootmark{c}}\phantom{00}2.55$\pm$0.41\tablefootmark{c}&$-$&$-$&\phantom{00}1.90$\pm$0.38&$-$\\
PER~20&\phantom{00}6.18$\pm$0.13&\phantom{0}21.36$\pm$0.27&\phantom{\tablefootmark{b}}\phantom{0}42.63$\pm$0.76\tablefootmark{b}&\phantom{\tablefootmark{c}}\phantom{0}41.10$\pm$1.49\tablefootmark{c}&\phantom{\tablefootmark{c}}\phantom{0}30.42$\pm$1.35\tablefootmark{c}&\phantom{0}19.88$\pm$3.98&\phantom{\tablefootmark{c}}\phantom{0}12.88$\pm$0.77\tablefootmark{c}&\phantom{00}6.13$\pm$1.23&$-$\\
PER~21&\phantom{00}5.82$\pm$0.15&\phantom{0}15.16$\pm$0.28&\phantom{\tablefootmark{b}}\phantom{0}17.89$\pm$0.71\tablefootmark{b}&\phantom{\tablefootmark{c}}\phantom{0}13.60$\pm$0.89\tablefootmark{c}&\phantom{\tablefootmark{c}}\phantom{00}7.16$\pm$0.84\tablefootmark{c}&\phantom{0}18.99$\pm$3.80&\phantom{\tablefootmark{c}}\phantom{00}3.96$\pm$0.61\tablefootmark{c}&\phantom{00}4.04$\pm$0.81&$-$\\
PER~22&\phantom{0}15.46$\pm$0.31&\phantom{0}26.24$\pm$0.40&\phantom{\tablefootmark{b}}\phantom{0}28.13$\pm$0.83\tablefootmark{b}&\phantom{\tablefootmark{c}}\phantom{0}21.27$\pm$0.80\tablefootmark{c}&\phantom{\tablefootmark{c}}\phantom{00}9.88$\pm$0.62\tablefootmark{c}&\phantom{0}22.52$\pm$4.50&\phantom{\tablefootmark{c}}\phantom{00}4.96$\pm$0.36\tablefootmark{c}&\phantom{00}2.24$\pm$0.45&$-$\\
SCO~01&\phantom{00}7.16$\pm$0.15&$-$&\phantom{\tablefootmark{b}}\phantom{00}6.57$\pm$0.11\tablefootmark{b}&\phantom{\tablefootmark{c}}\phantom{00}4.52$\pm$0.15\tablefootmark{c}&\phantom{\tablefootmark{c}}\phantom{00}2.21$\pm$0.15\tablefootmark{c}&$-$&\phantom{\tablefootmark{c}}\phantom{00}0.92$\pm$0.11\tablefootmark{c}&\phantom{00}0.16$\pm$0.03&$-$\\
SERS~01&\phantom{0}34.27$\pm$0.49&\phantom{0}56.60$\pm$0.36&\phantom{\tablefootmark{b}}\phantom{0}84.69$\pm$0.73\tablefootmark{b}&\phantom{\tablefootmark{c}}\phantom{0}80.67$\pm$1.28\tablefootmark{c}&\phantom{\tablefootmark{c}}\phantom{0}43.57$\pm$1.30\tablefootmark{c}&$-$&\phantom{\tablefootmark{c}}\phantom{0}19.09$\pm$0.92\tablefootmark{c}&$-$&\phantom{00}0.21$\pm$0.05\\
SERS~02&\phantom{0}96.64$\pm$1.39&219.00$\pm$1.75&\phantom{\tablefootmark{b}}327.90$\pm$4.45\tablefootmark{b}&\phantom{\tablefootmark{c}}283.98$\pm$8.56\tablefootmark{c}&\phantom{\tablefootmark{c}}172.40$\pm$6.58\tablefootmark{c}&$-$&\phantom{\tablefootmark{c}}109.94$\pm$3.96\tablefootmark{c}&$-$&\phantom{00}2.50$\pm$0.31\\
TAU~01&\phantom{0}18.67$\pm$0.38&$-$&\phantom{\tablefootmark{b}}\phantom{0}24.18$\pm$0.29\tablefootmark{b}&\phantom{\tablefootmark{c}}\phantom{0}23.52$\pm$0.39\tablefootmark{c}&\phantom{\tablefootmark{c}}\phantom{0}20.99$\pm$0.43\tablefootmark{c}&\phantom{0}30.74$\pm$6.15&\phantom{\tablefootmark{c}}\phantom{0}13.20$\pm$0.31\tablefootmark{c}&\phantom{00}2.38$\pm$0.48&$-$\\
TAU~02&\phantom{00}2.67$\pm$0.05&$-$&\phantom{\tablefootmark{b}}\phantom{00}5.09$\pm$0.06\tablefootmark{b}&\phantom{\tablefootmark{c}}\phantom{00}4.92$\pm$0.20\tablefootmark{c}&\phantom{\tablefootmark{c}}\phantom{00}3.28$\pm$0.23\tablefootmark{c}&$-$&\phantom{\tablefootmark{c}}\phantom{00}1.46$\pm$0.16\tablefootmark{c}&$-$&$-$\\
TAU~03&\phantom{00}4.21$\pm$0.09&$-$&\phantom{\tablefootmark{b}}\phantom{00}4.77$\pm$0.09\tablefootmark{b}&\phantom{\tablefootmark{c}}\phantom{00}2.27$\pm$0.13\tablefootmark{c}&\phantom{\tablefootmark{c}}\phantom{00}1.27$\pm$0.13\tablefootmark{c}&$-$&\phantom{\tablefootmark{c}}\phantom{00}0.63$\pm$0.09\tablefootmark{c}&$-$&$-$\\
TAU~04&\phantom{0}17.04$\pm$0.37&$-$&\phantom{\tablefootmark{b}}\phantom{0}18.77$\pm$0.33\tablefootmark{b}&\phantom{\tablefootmark{c}}\phantom{0}15.82$\pm$0.32\tablefootmark{c}&\phantom{\tablefootmark{c}}\phantom{0}12.66$\pm$0.36\tablefootmark{c}&\phantom{00}6.62$\pm$1.32&\phantom{\tablefootmark{c}}\phantom{00}6.86$\pm$0.24\tablefootmark{c}&\phantom{00}1.87$\pm$0.37&$-$\\
TAU~06&\phantom{00}5.96$\pm$0.12&$-$&\phantom{\tablefootmark{b}}\phantom{0}14.23$\pm$0.13\tablefootmark{b}&\phantom{\tablefootmark{c}}\phantom{0}20.28$\pm$0.28\tablefootmark{c}&\phantom{\tablefootmark{c}}\phantom{0}13.72$\pm$0.32\tablefootmark{c}&$-$&\phantom{\tablefootmark{c}}\phantom{00}6.05$\pm$0.21\tablefootmark{c}&$-$&$-$\\
TAU~07&\phantom{00}4.77$\pm$0.09&\phantom{00}4.10$\pm$0.05&\phantom{\tablefootmark{b}}\phantom{00}2.88$\pm$0.06\tablefootmark{b}&\phantom{\tablefootmark{c}}\phantom{00}1.60$\pm$0.10\tablefootmark{c}&\phantom{\tablefootmark{c}}\phantom{00}0.89$\pm$0.15\tablefootmark{c}&$-$&\phantom{\tablefootmark{c}}\phantom{00}0.59$\pm$0.13\tablefootmark{c}&$-$&$-$\\
TAU~08&\phantom{00}3.76$\pm$0.08&$-$&\phantom{\tablefootmark{b}}\phantom{00}4.28$\pm$0.07\tablefootmark{b}&\phantom{\tablefootmark{c}}\phantom{00}3.59$\pm$0.14\tablefootmark{c}&\phantom{\tablefootmark{c}}\phantom{00}3.35$\pm$0.17\tablefootmark{c}&\phantom{00}2.05$\pm$0.41&\phantom{\tablefootmark{c}}\phantom{00}2.47$\pm$0.14\tablefootmark{c}&\phantom{00}0.53$\pm$0.11&$-$\\
TAU~09&\phantom{0}15.73$\pm$0.26&\phantom{0}18.11$\pm$0.17&\phantom{\tablefootmark{b}}\phantom{0}23.20$\pm$0.24\tablefootmark{b}&\phantom{\tablefootmark{c}}\phantom{0}17.00$\pm$0.36\tablefootmark{c}&\phantom{\tablefootmark{c}}\phantom{0}14.60$\pm$0.43\tablefootmark{c}&$-$&\phantom{\tablefootmark{c}}\phantom{00}7.46$\pm$0.46\tablefootmark{c}&$-$&$-$\\
W40~01&\phantom{0}12.42$\pm$0.95&\phantom{0}35.37$\pm$0.88&\phantom{\tablefootmark{b}}\phantom{0}58.60$\pm$2.35\tablefootmark{b}&\phantom{\tablefootmark{c}}\phantom{0}62.61$\pm$3.19\tablefootmark{c}&\phantom{\tablefootmark{c}}\phantom{0}54.51$\pm$2.87\tablefootmark{c}&$-$&\phantom{\tablefootmark{c}}\phantom{0}30.67$\pm$1.76\tablefootmark{c}&$-$&\phantom{00}0.50$\pm$0.08\\
W40~02&\phantom{0}24.33$\pm$3.36&\phantom{0}26.28$\pm$1.73&\phantom{\tablefootmark{b}}\phantom{0}42.81$\pm$2.71\tablefootmark{b}&\phantom{\tablefootmark{c}}\phantom{0}63.53$\pm$3.16\tablefootmark{c}&\phantom{\tablefootmark{c}}\phantom{0}60.78$\pm$2.87\tablefootmark{c}&$-$&\phantom{\tablefootmark{c}}\phantom{0}39.55$\pm$1.80\tablefootmark{c}&$-$&\phantom{00}0.90$\pm$0.24\\
W40~03&$-$&$-$&\phantom{\tablefootmark{b}}120.48$\pm$3.55\tablefootmark{b}&\phantom{\tablefootmark{c}}106.22$\pm$3.83\tablefootmark{c}&\phantom{\tablefootmark{c}}\phantom{0}98.41$\pm$3.34\tablefootmark{c}&$-$&\phantom{\tablefootmark{c}}\phantom{0}51.79$\pm$1.94\tablefootmark{c}&$-$&\phantom{00}0.95$\pm$0.29\\
W40~04&$-$&$-$&\phantom{\tablefootmark{b}}\phantom{0}95.75$\pm$3.06\tablefootmark{b}&\phantom{\tablefootmark{c}}108.24$\pm$3.46\tablefootmark{c}&\phantom{\tablefootmark{c}}\phantom{0}37.51$\pm$3.02\tablefootmark{c}&$-$&\phantom{\tablefootmark{c}}\phantom{0}15.47$\pm$1.72\tablefootmark{c}&$-$&\phantom{00}0.60$\pm$0.14\\
W40~05&$-$&$-$&\phantom{\tablefootmark{b}}\phantom{0}71.84$\pm$2.85\tablefootmark{b}&\phantom{\tablefootmark{c}}\phantom{0}75.62$\pm$3.70\tablefootmark{c}&\phantom{\tablefootmark{c}}\phantom{0}61.72$\pm$5.65\tablefootmark{c}&$-$&\phantom{\tablefootmark{c}}\phantom{0}62.30$\pm$7.37\tablefootmark{c}&$-$&\phantom{00}0.76$\pm$0.23\\
W40~06&$-$&\phantom{00}1.68$\pm$0.09&\phantom{\tablefootmark{b}}\phantom{00}7.01$\pm$0.40\tablefootmark{b}&\phantom{\tablefootmark{c}}\phantom{0}12.62$\pm$0.91\tablefootmark{c}&\phantom{\tablefootmark{c}}\phantom{00}8.58$\pm$0.85\tablefootmark{c}&$-$&\phantom{\tablefootmark{c}}\phantom{00}1.33$\pm$0.77\tablefootmark{c}&$-$&\phantom{00}0.20$\pm$0.06\\
W40~07&\phantom{00}3.18$\pm$0.07&\phantom{00}7.01$\pm$0.10&\phantom{\tablefootmark{b}}\phantom{0}10.72$\pm$0.28\tablefootmark{b}&\phantom{\tablefootmark{c}}\phantom{00}8.76$\pm$0.53\tablefootmark{c}&\phantom{\tablefootmark{c}}\phantom{00}7.58$\pm$0.95\tablefootmark{c}&$-$&\phantom{\tablefootmark{c}}\phantom{00}3.35$\pm$0.49\tablefootmark{c}&$-$&\phantom{00}0.09$\pm$0.03\\
\hline
\end{tabular}
\tablefoot{\tablefoottext{a}{A 20$\%$ uncertainty is assumed for the SCUBA fluxes.} \tablefoottext{b}{Wavelength used to correct for contamination of nearby sources in longer-wavelength \textit{Herschel} fluxes.} \tablefoottext{c}{Flux corrected for contamination of nearby sources using lower-wavelength \textit{Herschel} map}}
\end{center}
\end{table*}

\renewcommand{\thetable}{\thesection.\arabic{table}}

\begin{table*}
\caption{Pacs spectral continuum fluxes. All fluxes are given in Jy, all numerical wavelengths in \micron{}.}
\begin{center}
\begin{tabular}{lcccccccccccc}
\hline \noalign {\smallskip}
Name & 63.2 & 79.2 & 81.8 & 84.6 & 90.0 & 108.8 & 125.4 & 157.7 & 164.0 & 169.1 & 179.5 & 190.0 \\
\hline\noalign {\smallskip}
AQU~01&\phantom{00}3.8&\phantom{00}4.5&\phantom{00}3.8&\phantom{00}3.6&\phantom{00}5.5&\phantom{00}9.7&\phantom{0}11.3&\phantom{0}18.5&\phantom{0}18.2&\phantom{0}17.5&\phantom{0}20.4&\phantom{0}18.5\\
AQU~02&\phantom{00}8.4&\phantom{0}10.1&\phantom{0}12.2&\phantom{0}13.1&\phantom{0}15.1&\phantom{0}31.3&\phantom{0}34.9&\phantom{0}46.5&\phantom{0}46.6&\phantom{0}44.4&\phantom{0}46.3&\phantom{0}39.9\\
AQU~03&\phantom{00}3.7&\phantom{00}2.5&\phantom{00}2.2&\phantom{00}1.8&\phantom{00}2.4&\phantom{00}3.8&\phantom{00}2.9&\phantom{00}4.8&\phantom{00}6.2&\phantom{00}5.4&\phantom{00}8.5&\phantom{00}8.6\\
AQU~04&\phantom{00}5.8&\phantom{00}2.9&\phantom{00}3.6&\phantom{00}3.9&\phantom{00}4.0&\phantom{00}4.1&\phantom{00}4.0&\phantom{00}6.4&\phantom{00}7.4&\phantom{00}8.6&\phantom{0}10.5&\phantom{0}13.2\\
AQU~05&\phantom{00}3.2&\phantom{00}1.8&\phantom{00}2.3&\phantom{00}2.2&\phantom{00}2.7&\phantom{00}4.3&\phantom{00}3.9&\phantom{00}7.2&\phantom{00}8.7&\phantom{00}9.1&\phantom{0}11.2&\phantom{0}12.7\\
AQU~06\\
CHA~01&\phantom{0}14.7&\phantom{00}9.7&\phantom{00}9.0&\phantom{0}10.2&\phantom{0}11.2&\phantom{0}22.0&\phantom{0}13.8&\phantom{0}16.6&\phantom{0}16.6&\phantom{0}15.8&\phantom{0}15.4&\phantom{0}21.7\\
CHA~02&\phantom{0}10.9&\phantom{00}9.4&\phantom{00}8.3&\phantom{00}7.6&\phantom{00}9.3&\phantom{0}14.3&\phantom{00}9.3&\phantom{0}13.0&\phantom{0}13.6&\phantom{0}13.8&\phantom{0}13.8&\phantom{0}14.5\\
CRA~01&\phantom{0}49.4&\phantom{0}57.7&\phantom{0}56.9&\phantom{0}58.4&\phantom{0}62.5&\phantom{0}71.5&\phantom{0}62.9&\phantom{0}65.2&\phantom{0}60.7&\phantom{0}55.9&\phantom{0}54.5&\phantom{0}45.5\\
OPH~01\tablefootmark{a}&139.1&157.8&161.3&161.5&159.7&161.0&137.4&111.0&102.6&\phantom{0}94.0&\phantom{0}86.7&\phantom{0}73.4\\
OPH~02&\phantom{0}84.7&\phantom{0}98.7&\phantom{0}99.3&102.2&101.9&102.5&\phantom{0}84.6&\phantom{0}75.8&\phantom{0}69.4&\phantom{0}64.3&\phantom{0}59.1&\phantom{0}49.1\\
PER~01&\phantom{0}20.7&\phantom{0}31.7&\phantom{0}35.3&\phantom{0}38.0&\phantom{0}42.8&\phantom{0}57.7&\phantom{0}58.1&\phantom{0}68.6&\phantom{0}65.9&\phantom{0}62.4&\phantom{0}60.4&\phantom{0}50.2\\
PER~02&\phantom{0}29.1&\phantom{0}55.7&\phantom{0}60.1&\phantom{0}66.4&\phantom{0}78.6&122.1&139.4&186.0&182.9&177.1&175.8&141.7\\
PER~04&\phantom{00}5.7&\phantom{00}7.1&\phantom{00}6.7&\phantom{00}8.4&\phantom{00}8.4&\phantom{0}11.1&\phantom{00}9.5&\phantom{0}11.6&\phantom{0}11.1&\phantom{0}10.1&\phantom{0}10.6&\phantom{00}9.6\\
PER~05&\phantom{0}67.1&\phantom{0}67.8&\phantom{0}67.3&\phantom{0}70.0&\phantom{0}71.7&\phantom{0}75.3&\phantom{0}65.2&\phantom{0}65.3&\phantom{0}62.0&\phantom{0}57.5&\phantom{0}54.0&\phantom{0}45.0\\
PER~06&\phantom{0}41.4&\phantom{0}50.2&\phantom{0}48.9&\phantom{0}48.1&\phantom{0}49.3&\phantom{0}71.2&\phantom{0}59.3&\phantom{0}65.1&\phantom{0}62.0&\phantom{0}56.6&\phantom{0}56.5&\phantom{0}51.2\\
PER~07&\phantom{00}2.8&\phantom{00}2.5&\phantom{00}3.2&\phantom{00}4.0&\phantom{00}5.7&\phantom{00}9.7&\phantom{00}9.0&\phantom{0}13.1&\phantom{0}13.1&\phantom{0}11.9&\phantom{0}13.2&\phantom{0}10.9\\
PER~08&\phantom{0}31.2&\phantom{0}47.0&\phantom{0}51.0&\phantom{0}55.2&\phantom{0}65.6&109.9&110.8&147.7&146.5&144.2&141.9&120.3\\
PER~09\tablefootmark{a}&242.5&291.1&301.7&308.8&324.4&398.7&357.1&310.7&281.6&260.8&227.9&190.6\\
PER~10&\phantom{0}34.4&\phantom{0}47.8&\phantom{0}52.3&\phantom{0}53.3&\phantom{0}58.4&\phantom{0}88.3&\phantom{0}80.9&\phantom{0}88.4&\phantom{0}84.3&\phantom{0}80.6&\phantom{0}78.6&\phantom{0}68.9\\
PER~12&\phantom{00}2.6&\phantom{00}3.6&\phantom{00}3.3&\phantom{00}3.4&\phantom{00}6.1&\phantom{0}14.1&\phantom{0}14.6&\phantom{0}21.0&\phantom{0}20.8&\phantom{0}20.5&\phantom{0}22.8&\phantom{0}24.0\\
PER~13&\phantom{00}3.2&\phantom{00}2.1&\phantom{00}2.7&\phantom{00}2.9&\phantom{00}3.9&\phantom{00}9.2&\phantom{00}7.0&\phantom{00}8.9&\phantom{00}8.9&\phantom{00}7.8&\phantom{00}9.3&\phantom{00}8.8\\
PER~14&\phantom{00}9.9&\phantom{00}9.0&\phantom{00}7.3&\phantom{00}6.5&\phantom{00}7.9&\phantom{0}11.4&\phantom{00}8.5&\phantom{00}9.1&\phantom{0}10.0&\phantom{00}9.3&\phantom{0}11.3&\phantom{0}10.0\\
PER~15&\phantom{00}5.4&\phantom{0}11.4&\phantom{0}11.9&\phantom{0}12.6&\phantom{0}16.6&\phantom{0}24.9&\phantom{0}23.0&\phantom{0}27.5&\phantom{0}27.7&\phantom{0}26.5&\phantom{0}26.0&\phantom{0}21.7\\
PER~16&\phantom{00}2.7&\phantom{00}5.1&\phantom{00}6.4&\phantom{00}6.4&\phantom{00}8.9&\phantom{0}11.8&\phantom{0}11.9&\phantom{0}18.6&\phantom{0}19.5&\phantom{0}18.3&\phantom{0}22.5&\phantom{0}18.2\\
PER~17\tablefootmark{a}&\phantom{00}2.7&\phantom{00}3.1&\phantom{00}1.9&\phantom{00}2.3&\phantom{00}2.8&\phantom{00}5.7&\phantom{00}6.2&\phantom{0}14.1&\phantom{0}14.1&\phantom{0}15.3&\phantom{0}16.2&\phantom{0}16.4\\
PER~18\tablefootmark{a}&\phantom{00}3.0&\phantom{00}5.1&\phantom{00}5.3&\phantom{00}5.5&\phantom{00}8.1&\phantom{0}11.5&\phantom{0}12.0&\phantom{0}20.3&\phantom{0}19.6&\phantom{0}20.0&\phantom{0}20.4&\phantom{0}20.5\\
PER~19\tablefootmark{a}&\phantom{00}0.9&\phantom{00}3.2&\phantom{00}2.5&\phantom{00}3.1&\phantom{00}3.5&\phantom{00}6.7&\phantom{00}4.5&\phantom{00}7.9&\phantom{00}7.0&\phantom{00}6.8&\phantom{00}8.4&\phantom{0}10.8\\
PER~20&\phantom{00}2.0&\phantom{00}7.3&\phantom{00}9.5&\phantom{0}11.9&\phantom{0}15.6&\phantom{0}33.9&\phantom{0}43.2&\phantom{0}65.0&\phantom{0}66.8&\phantom{0}63.6&\phantom{0}64.6&\phantom{0}54.3\\
PER~21&\phantom{00}5.6&\phantom{0}11.1&\phantom{0}13.0&\phantom{0}13.4&\phantom{0}17.0&\phantom{0}32.8&\phantom{0}34.8&\phantom{0}49.8&\phantom{0}49.0&\phantom{0}46.7&\phantom{0}46.9&\phantom{0}42.8\\
PER~22&\phantom{0}10.3&\phantom{0}15.8&\phantom{0}18.2&\phantom{0}19.2&\phantom{0}21.5&\phantom{0}29.2&\phantom{0}31.5&\phantom{0}40.9&\phantom{0}43.5&\phantom{0}40.8&\phantom{0}41.2&\phantom{0}32.7\\
SCO~01&\phantom{00}8.4&\phantom{00}4.9&\phantom{00}5.0&\phantom{00}5.9&\phantom{00}6.5&\phantom{00}5.0&\phantom{00}2.9&\phantom{00}2.8&\phantom{00}2.3&\phantom{00}1.5&\phantom{00}2.8&\phantom{00}1.6\\
SERS~01&\phantom{0}18.4&\phantom{0}28.6&\phantom{0}29.9&\phantom{0}32.6&\phantom{0}35.7&\phantom{0}59.8&\phantom{0}63.9&\phantom{0}77.5&\phantom{0}74.9&\phantom{0}72.2&\phantom{0}70.8&\phantom{0}60.4\\
SERS~02&\phantom{0}58.8&134.4&148.5&163.4&193.3&327.2&370.8&452.4&440.3&425.2&411.8&333.8\\
TAU~01&\phantom{0}18.8&\phantom{0}17.4&\phantom{0}17.4&\phantom{0}17.0&\phantom{0}18.2&\phantom{0}23.6&\phantom{0}21.4&\phantom{0}26.4&\phantom{0}25.5&\phantom{0}24.1&\phantom{0}23.7&\phantom{0}19.9\\
TAU~02&\phantom{00}3.4&\phantom{00}4.4&\phantom{00}3.8&\phantom{00}3.6&\phantom{00}4.4&\phantom{00}5.9&\phantom{00}4.7&\phantom{00}7.6&\phantom{00}6.6&\phantom{00}7.0&\phantom{00}8.8&\phantom{00}7.7\\
TAU~03&\phantom{00}3.9&\phantom{00}4.5&\phantom{00}4.8&\phantom{00}4.1&\phantom{00}5.2&\phantom{00}7.0&\phantom{00}4.6&\phantom{00}4.6&\phantom{00}4.0&\phantom{00}4.5&\phantom{00}4.8&\phantom{00}4.0\\
TAU~04&\phantom{0}15.9&\phantom{0}16.9&\phantom{0}16.6&\phantom{0}17.0&\phantom{0}17.1&\phantom{0}24.2&\phantom{0}19.9&\phantom{0}22.7&\phantom{0}23.0&\phantom{0}21.1&\phantom{0}19.7&\phantom{0}18.3\\
TAU~06&\phantom{00}7.4&\phantom{00}4.9&\phantom{00}5.8&\phantom{00}6.5&\phantom{00}7.8&\phantom{0}13.4&\phantom{0}12.2&\phantom{0}14.6&\phantom{0}15.3&\phantom{0}13.7&\phantom{0}15.7&\phantom{0}14.1\\
TAU~07&\phantom{00}8.4&\phantom{00}4.9&\phantom{00}5.0&\phantom{00}5.9&\phantom{00}6.5&\phantom{00}5.0&\phantom{00}2.9&\phantom{00}2.8&\phantom{00}2.3&\phantom{00}1.5&\phantom{00}2.8&\phantom{00}1.6\\
TAU~08\\
TAU~09&\phantom{0}14.7&\phantom{0}17.3&\phantom{0}19.5&\phantom{0}19.9&\phantom{0}22.0&\phantom{0}22.0&\phantom{0}20.4&\phantom{0}23.3&\phantom{0}22.8&\phantom{0}22.6&\phantom{0}23.1&\phantom{0}21.7\\
W40~01\tablefootmark{a}&\phantom{0}36.7&\phantom{0}67.3&\phantom{0}77.2&\phantom{0}84.7&103.8&193.8&226.6&274.0&267.7&252.9&240.7&193.0\\
W40~02\tablefootmark{a}&537.6&455.5&454.6&446.9&431.4&400.4&324.3&267.7&247.2&230.2&209.5&180.1\\
W40~03\tablefootmark{a}&128.8&204.8&219.6&230.9&251.7&352.6&353.5&363.1&339.4&316.0&288.7&238.3\\
W40~04\tablefootmark{a}&208.7&303.3&323.7&340.3&372.4&496.3&465.4&429.8&395.4&363.5&324.5&268.3\\
W40~05\tablefootmark{a}&106.8&173.7&187.8&200.3&228.2&343.1&346.4&362.9&335.7&316.1&291.6&238.9\\
W40~06\tablefootmark{a}&$-$&$-$&$-$&$-$&$-$&$-$&$-$&$-$&$-$&\phantom{00}0.1&\phantom{00}6.3&\phantom{00}8.5\\
W40~07&\phantom{00}3.9&\phantom{00}3.8&\phantom{00}5.0&\phantom{00}5.3&\phantom{00}8.0&\phantom{0}14.4&\phantom{0}17.0&\phantom{0}25.9&\phantom{0}27.4&\phantom{0}24.8&\phantom{0}27.4&\phantom{0}22.2\\
\hline
\end{tabular}
\tablefoot{\tablefoottext{a}{Data not included in SED.}}
\end{center}
\label{T:pacs_fluxes}
\end{table*}

\subsection{Continuum properties}
\label{S:properties_lboltbol}

A number of properties are often calculated from YSO SEDs in order to characterise them. While previous determinations have been made for the better-studied members of the WILL sample, variations in method and implementation could lead to biases or systematic effects between sources in different regions. In addition, \textit{Herschel} data provide significantly higher-resolution images for flux density determination than were previously available between 70 and 250\micron{} (e.g. \textit{Herschel} has a FWHM of $\sim$8\arcsec{} at 100\micron{} compared to 3\arcmin{}$\times$5\arcmin{} for \textit{IRAS}). In order to be consistent with other \textit{Herschel} surveys, most notably the WISH and DIGIT surveys, we follow the definitions in \citet{Dunham2010} and calculate the bolometric luminosity ($L_{\mathrm{bol}}$) using:

\begin{equation}
L_{\mathrm{bol}}~=~4\pi d^{2} \int_{0}^{\infty}F_{\nu}\mathrm{d}\nu\,,
\end{equation}

\noindent the sub-mm luminosity using:

\begin{equation}
L_{\mathrm{smm}}~=~4\pi d^{2} \int_{0}^{\nu=c/350\mu\mathrm{m}}F_{\nu}\mathrm{d}\nu\,,
\end{equation}

\noindent and bolometric temperature using:

\begin{equation}
T_{\mathrm{bol}}~=~1.25\times10^{-11}\frac{\int_{0}^{\infty}\nu F_{\nu}\mathrm{d}\nu}{\int_{0}^{\infty}F_{\nu}\mathrm{d}\nu}\,.
\end{equation}

\noindent In these equations, $F_{\nu}$ is the flux density at frequency $\nu$, and $d$ is the distance to the source. The integrals were calculated over the available SED flux densities using trapezium integration, which was found by \citet{Karska2013} to provide the most consistent results for the WISH survey sources.

The infrared spectral index \citep[see][]{Lada1987} is also calculated for those sources where the source is detected in at least three wavelengths between 2 and 25\micron{} using a least-squares linear fit to the logarithms of wavelength and flux density such that:

\begin{equation}
\alpha_{\mathrm{IR}}~=~\frac{d\,log_{10}(\lambda F_{\lambda})}{d\,log_{10}(\lambda)}\,,
\end{equation}

\noindent where $F_{\lambda}$ is the flux density at wavelength $\lambda$.

Finally, for the sources in the Aquila Rift we take the envelope mass ($M_{\mathrm{env}}$) from \citet{Maury2011}, calculated from the integrated intensity at 1.2\,mm, corrected to the updated distance for that region. In other regions we calculate $M_{\mathrm{env}}$ for those sources with SCUBA 850\micron{} observations using Eqn.~1 from \citet{Jorgensen2009}:

\begin{equation}
M_{\mathrm{env}}\,=\,0.44\,\left(\frac{L_{\mathrm{bol}}}{1\,L_{\odot}}\right)^{-0.36}\left(\frac{S_{15^{\prime\prime},\,850\mu\mathrm{m}}}{1\,\mathrm{Jy\,beam}^{-1}}\right)^{1.2}\left(\frac{d}{125\,\mathrm{pc}}\right)^{1.2}~M_{\odot}\,,
\end{equation}

\noindent where $S_{15^{\prime\prime},\,850\mu\mathrm{m}}$ is the peak SCUBA 850\micron{} flux density in a 15\arcsec{} pixel. This empirical relation was derived from comparison of observed source properties with the results of dust radiative transfer models. The results of these calculations are summarised in Table~\ref{T:properties}.

\subsection{PACS line properties and detection statistics}
\label{S:properties_pacs}

\begin{table*}
\begin{center}
\caption[]{Principle lines observed with PACS.}
\vspace{-2mm}
\begin{tabular}{lcccccccc}
\hline \noalign {\smallskip}
Species & Transition & Rest Frequency\tablefootmark{a} & Wavelength & E$_{\mathrm{u}}$/$k_{\mathrm{b}}$ & $A_{\mathrm{ul}}$\tablefootmark{b} & $n_{\mathrm{cr}}$ & $\theta_{mb}$\tablefootmark{c} &  Obs. Time\tablefootmark{d}   \\
& & (GHz) & (\micron{}) & (K) & (s$^{-1}$) & (cm$^{-3}$) & (\arcsec{})  & (min) \\
\hline\noalign {\smallskip}
o-H$_{2}$O & 2$_{12}-$1$_{01}$ & 1669.90477 & 179.527 & \phantom{0}114.4 & 5.59$\times$10$^{-2}$ & 3$\times$10$^{8\phantom{0}}$ & 12.7 & 36 \\
& 2$_{21}-$2$_{12}$ & 1661.00764 & 180.488 & \phantom{0}194.1 & 3.06$\times$10$^{-2}$ & 1$\times$10$^{8\phantom{0}}$ & 12.8 & 36 \\
& 2$_{21}-$1$_{10}$ & 2773.97659 & 108.073 & \phantom{0}194.1 & 2.56$\times$10$^{-1}$ & 1$\times$10$^{9\phantom{0}}$ & \phantom{0}7.6 & 17 \\
& 4$_{23}-$3$_{12}$ & 3807.25841 & \phantom{0}78.742 & \phantom{0}432.2 & 4.84$\times$10$^{-1}$ & 2$\times$10$^{9\phantom{0}}$ & \phantom{0}5.6 & 36 \\
& 6$_{16}-$5$_{05}$ & 3654.60328 & \phantom{0}82.031 & \phantom{0}643.5 & 7.49$\times$10$^{-1}$ & 4$\times$10$^{9\phantom{0}}$ & \phantom{0}5.8 & 36 \\
& 7$_{16}-$7$_{07}$ & 3536.66681 & \phantom{0}84.770 & 1013.2 & 2.16$\times$10$^{-1}$ & 1$\times$10$^{9\phantom{0}}$ & \phantom{0}6.0 & 36 \\
& 8$_{18}-$7$_{07}$ & 4734.29617 & \phantom{0}63.323 & 1070.7 & 1.75$\times$10$^{0\phantom{-}}$ & 1$\times$10$^{10}$ & \phantom{0}4.5 & 17 \\
\hline\noalign {\smallskip}
p-H$_{2}$O & 3$_{22}-$2$_{11}$ & 3331.45838 & \phantom{0}89.988 & \phantom{0}296.8 & 3.52$\times$10$^{-1}$ & 1$\times$10$^{9\phantom{0}}$ & \phantom{0}6.4 & 36 \\
& 4$_{04}-$3$_{13}$ & 2391.57263 & 125.354 & \phantom{0}319.5 & 1.73$\times$10$^{-1}$ & 9$\times$10$^{8\phantom{0}}$ & \phantom{0}8.9 & 36 \\
& 6$_{15}-$5$_{24}$ & 3798.28164  & \phantom{0}78.928 & \phantom{0}781.1 & 4.52$\times$10$^{-1}$ & 2$\times$10$^{9\phantom{0}}$ & \phantom{0}5.6 & 36 \\
\hline\noalign {\smallskip}
OH & $^{1}/_{2}$,$^{1}/_{2}-^{3}/_{2}$,$^{3}/_{2}$ & 3786.16998 & \phantom{0}79.181 & \phantom{0}181.7 & 3.52$\times$10$^{-2}$ & 1$\times$10$^{8\phantom{0}}$ & \phantom{0}5.6 & 36 \\
 & $^{1}/_{2}$,$^{1}/_{2}-^{3}/_{2}$,$^{3}/_{2}$ & 3789.17979 & \phantom{0}79.118 & \phantom{0}181.9 & 3.49$\times$10$^{-2}$ & 8$\times$10$^{7\phantom{0}}$ & \phantom{0}5.6 & 36 \\
 & $^{3}/_{2}$,$^{1}/_{2}-^{1}/_{2}$,$^{1}/_{2}$ & 1834.74735 & 163.397 & \phantom{0}269.8 & 6.37$\times$10$^{-2}$ & 2$\times$10$^{8\phantom{0}}$ & 11.6 & 36 \\
 & $^{3}/_{2}$,$^{1}/_{2}-^{1}/_{2}$,$^{1}/_{2}$ & 1837.81682 & 163.124 & \phantom{0}270.2 & 6.40$\times$10$^{-2}$ & 2$\times$10$^{8\phantom{0}}$ & 11.5 & 36 \\
 & $^{7}/_{2}$,$^{3}/_{2}-^{5}/_{2}$,$^{3}/_{2}$ & 3543.77937 & \phantom{0}84.597 & \phantom{0}290.5 & 5.13$\times$10$^{-1}$ & 2$\times$10$^{9\phantom{0}}$ & \phantom{0}6.0 & 36 \\
 & $^{7}/_{2}$,$^{3}/_{2}-^{5}/_{2}$,$^{3}/_{2}$ & 3551.18525 & \phantom{0}84.420 & \phantom{0}291.2 & 5.16$\times$10$^{-1}$ & 2$\times$10$^{9\phantom{0}}$ & \phantom{0}6.0 & 36 \\
\hline\noalign {\smallskip}
[O\,{\sc i}] & $^{3}P_{1}-^{3}P_{2}$ & 4744.77749 & \phantom{0}63.184 & \phantom{0}227.7 & 8.91$\times$10$^{-5}$ & 5$\times$10$^{5\phantom{0}}$ & \phantom{0}4.5 & 17 \\
\hline\noalign {\smallskip}
CO & 16$-$15 & 1841.34551 & 162.812  & \phantom{0}751.7 & 4.05$\times$10$^{-4}$ & 1$\times$10$^{6\phantom{0}}$ & 11.5 & 36 \\
 & 21$-$20 & 2413.91711 & 124.193  & 1276.1 & 8.83$\times$10$^{-4}$ & 2$\times$10$^{6\phantom{0}}$ & \phantom{0}8.8 & 36 \\
 & 24$-$23 & 2756.38758 & 108.763 & 1656.5 & 1.28$\times$10$^{-3}$ & 3$\times$10$^{6\phantom{0}}$ & \phantom{0}7.7 & 17 \\
& 29$-$28 & 3325.00528 & \phantom{0}90.163 & 2399.8 & 2.13$\times$10$^{-3}$ & 6$\times$10$^{6\phantom{0}}$ & \phantom{0}6.4 & 36 \\
& 31$-$30 & 3551.59236 & \phantom{0}84.411 & 2735.3 & 2.52$\times$10$^{-3}$ & 7$\times$10$^{6\phantom{0}}$ & \phantom{0}6.0 & 36 \\
& 32$-$31 & 3664.68418 & \phantom{0}81.806 & 2911.2 & 2.74$\times$10$^{-3}$ & 7$\times$10$^{6\phantom{0}}$ & \phantom{0}5.8 & 36 \\
& 33$-$32 & 3777.63573 & \phantom{0}79.360 & 3092.5 & 2.95$\times$10$^{-3}$ & 8$\times$10$^{6\phantom{0}}$ & \phantom{0}5.6 & 36 \\
\hline\noalign {\smallskip}
[C\,{\sc ii}] & $^{2}P_{^{3}/_{2}}-^{2}P_{^{1}/_{2}}$ & 1900.53690 & 157.741 & \phantom{00}91.2 & 2.32$\times$10$^{-6}$ & 3$\times$10$^{3\phantom{0}}$ & 11.2 & 36 \\
\noalign {\smallskip}\hline\noalign {\smallskip}
\label{T:observations_pacs_lines}
\end{tabular}
\vspace{-5mm}
\tablefoot{\tablefoottext{a}{Taken from the JPL database \citep{Pickett2010} for H$_{2}$O, OH and CO, \citet{Zink1991} for [O\,{\sc i}] and CDMS \citep{Muller2001,Muller2005} for [C\,{\sc ii}].} \tablefoottext{b}{Taken from \citet{Daniel2011} and \citet{Dubernet2009} for H$_{2}$O, the JPL database \citep{Pickett2010} for OH and CO, \citet{Fischer1983} for [O\,{\sc i}] and CDMS \citep{Muller2001,Muller2005} for [C{\sc ii}].} \tablefoottext{c}{Calculated using equation 3 from \citet{Roelfsema2012}.} \tablefoottext{d}{Total time including on+off source and overheads.}}
\vspace{-2mm}
\end{center}
\end{table*}

Table~\ref{T:observations_pacs_settings} gives the wavelength ranges covered by the two PACS settings used for the WILL survey. Table~\ref{T:observations_pacs_lines} summarises the properties of the principle transitions observed with PACS towards the WILL sample, while Table~\ref{T:pacs_detections} indicates which of these are detected towards each WILL source. The overall fraction of sources detected in each transition is also given in the bottom row of Table~\ref{T:pacs_detections}. The spectral resolution of PACS is not sufficient to separate the CO $J$=31$-$30 transition at 84.41\micron{} and the OH 84.42\micron{} transition. Emission at this wavelength is therefore marked as a detection for both lines but could be in only CO or OH. The detection (or not) of neighbouring transitions is likely a reasonable estimate of whether the detected emission is from CO, OH or a blend of the two.

\begin{table*}
\caption{PACS line detections.}
\begin{center}
\vspace{-2mm}
\begin{tabular}{l|ccccccc|ccc|c}
\hline \noalign {\smallskip}
Name & \multicolumn{7}{|c|}{o-H$_{2}$O} & \multicolumn{3}{|c|}{p-H$_{2}$O} & [O\,{\sc i}] \\
&$2_{12}-1_{01}$&$2_{21}-2_{12}$&$2_{21}-1_{10}$&$4_{23}-3_{12}$&$6_{16}-5_{05}$&$7_{16}-7_{07}$&$8_{18}-7_{07}$&$3_{22}-2_{11}$&$4_{04}-3_{13}$&$6_{15}-5_{24}$&$^{3}P_{1}-^{3}P_{2}$\\
\hline\noalign {\smallskip}
AQU\,01&Y&Y&Y&N&N&N&N&N&Y&N&Y\\
AQU\,02&Y&Y&Y&N&N&N&N&Y&Y&N&Y\\
AQU\,03&N&N&N&N&N&N&N&N&N&N&N\\
AQU\,04&N&N&N&N&N&N&N&N&N&N&N\\
AQU\,05&Y&N&Y&N&N&N&N&N&N&N&N\\
AQU\,06&N&N&N&N&N&N&N&N&N&N&N\\
CHA\,01&Y&N&Y&Y&Y&N&N&Y&Y&N&Y\\
CHA\,02&N&N&Y&Y&Y&N&N&N&N&Y&Y\\
CRA\,01&Y&N&Y&Y&N&N&N&Y&Y&N&Y\\
OPH\,01&Y&N&N&N&N&N&N&N&N&N&Y\\
OPH\,02&Y&Y&Y&Y&Y&N&Y&Y&Y&Y&Y\\
PER\,01&Y&Y&Y&Y&Y&N&Y&Y&Y&Y&Y\\
PER\,02&Y&Y&Y&Y&Y&N&Y&Y&Y&Y&Y\\
PER\,04&N&N&N&N&N&N&N&N&N&N&Y\\
PER\,05&Y&Y&Y&Y&Y&N&N&Y&Y&Y&Y\\
PER\,06&Y&N&Y&Y&Y&N&Y&N&Y&N&Y\\
PER\,07&N&N&N&N&N&N&N&N&N&N&Y\\
PER\,08&Y&N&Y&Y&Y&N&Y&Y&Y&N&Y\\
PER\,09&Y&Y&Y&Y&Y&N&Y&Y&Y&Y&Y\\
PER\,10&Y&N&Y&N&N&N&N&N&Y&N&Y\\
PER\,12&N&N&N&N&N&N&N&N&N&N&Y\\
PER\,13&N&N&Y&N&N&N&N&N&Y&N&N\\
PER\,14&Y&Y&Y&Y&Y&N&Y&Y&Y&N&Y\\
PER\,15&Y&N&Y&N&Y&N&N&N&Y&N&Y\\
PER\,16&N&N&N&N&N&N&N&N&N&N&Y\\
PER\,17&N&N&N&N&N&N&N&N&N&N&Y\\
PER\,18&Y&Y&Y&N&N&N&N&N&N&N&Y\\
PER\,19&Y&Y&Y&N&Y&N&N&N&Y&N&Y\\
PER\,20&Y&Y&Y&Y&N&N&N&N&Y&N&Y\\
PER\,21&Y&Y&Y&Y&Y&N&N&Y&Y&N&Y\\
PER\,22&Y&Y&Y&Y&Y&N&N&N&Y&N&Y\\
SCO\,01&Y&N&Y&Y&Y&N&Y&N&Y&N&Y\\
SERS\,01&N&N&N&N&N&N&N&N&N&N&Y\\
SERS\,02&Y&Y&Y&Y&N&N&Y&Y&Y&N&Y\\
TAU\,01&Y&N&N&N&N&N&N&N&Y&N&Y\\
TAU\,02&N&N&Y&Y&Y&N&Y&N&N&N&Y\\
TAU\,03&Y&N&Y&N&Y&N&Y&Y&N&N&Y\\
TAU\,04&N&N&N&N&N&N&N&N&N&N&Y\\
TAU\,06&N&N&N&N&N&N&N&N&N&N&Y\\
TAU\,07&Y&N&Y&Y&Y&N&Y&N&Y&N&Y\\
TAU\,08\tablefootmark{a}&-&-&-&-&-&-&-&-&-&-&-\\
TAU\,09&Y&N&Y&Y&N&N&N&Y&N&Y&Y\\
W40\,01&N&N&Y&N&N&N&N&N&N&N&Y\\
W40\,02&N&N&Y&N&N&N&N&N&Y&N&Y\\
W40\,03&N&N&N&N&N&N&N&N&N&N&Y\\
W40\,04&N&N&N&N&N&N&N&N&N&N&Y\\
W40\,05&N&N&N&N&N&N&N&N&N&N&Y\\
W40\,06&N&N&N&N&N&N&N&N&N&N&\phantom{\tablefootmark{b}}N\tablefootmark{b}\\
W40\,07&N&N&N&N&N&N&N&N&N&N&Y\\
\hline\noalign{\smallskip}
Total&27/48&14/48&30/48&19/48&18/48&\phantom{0}0/48&12/48&14/48&24/48&\phantom{0}7/48&42/48\\
\hline\noalign {\smallskip}
\end{tabular}
\vspace{-2mm}
\tablefoot{\tablefoottext{a}{Source not observed with PACS due to end of \textit{Herschel} mission.} \tablefoottext{b}{In absorption due to contamination of the off-position.} \tablefoottext{c}{The CO $J$=31$-$30 transition at 84.41\micron{} and the OH 84.42\micron{} transition are blended.}}
\end{center}
\label{T:pacs_detections}
\end{table*}
\renewcommand{\thetable}{\thesection.\arabic{table} (Cont.)}
\addtocounter{table}{-1}
\begin{table*}
\caption{PACS line detections.}
\begin{center}
\vspace{-2mm}
\begin{tabular}{l|ccccccc|cccccc}
\hline \noalign {\smallskip}
Name & \multicolumn{7}{|c|}{CO} & \multicolumn{6}{|c}{OH} \\
&16$-$14&21$-$20&24$-$23&29$-$28&31$-$30\tablefootmark{c}&32$-$31&33$-$32&79.18\micron{}&79.12\micron{}&163.40\micron{}&163.12\micron{}&84.60\micron{}&84.42\micron{}\tablefootmark{c}\\
\hline\noalign {\smallskip}
AQU\,01&Y&Y&Y&N&N&N&N&N&N&N&N&N&N\\
AQU\,02&Y&Y&Y&Y&Y&N&N&N&N&N&N&Y&Y\\
AQU\,03&N&N&N&N&N&N&N&N&N&N&N&N&N\\
AQU\,04&N&N&N&N&N&N&N&N&N&N&N&N&N\\
AQU\,05&Y&Y&N&N&N&N&N&N&N&N&N&N&N\\
AQU\,06&N&N&N&N&N&N&N&N&N&N&N&N&N\\
CHA\,01&Y&Y&Y&Y&Y&N&N&Y&Y&Y&Y&Y&Y\\
CHA\,02&Y&Y&N&N&Y&N&N&Y&Y&N&N&Y&Y\\
CRA\,01&Y&Y&Y&Y&Y&N&N&Y&Y&Y&Y&Y&Y\\
OPH\,01&Y&N&N&N&N&N&N&N&N&N&N&N&N\\
OPH\,02&Y&Y&Y&Y&Y&Y&N&Y&Y&Y&Y&Y&Y\\
PER\,01&Y&Y&Y&Y&Y&Y&Y&Y&Y&Y&Y&Y&Y\\
PER\,02&Y&Y&Y&Y&Y&Y&Y&Y&Y&Y&Y&Y&Y\\
PER\,04&N&N&N&N&N&N&N&N&N&N&N&N&N\\
PER\,05&Y&Y&Y&Y&Y&N&Y&Y&Y&N&N&Y&Y\\
PER\,06&Y&Y&Y&N&Y&Y&N&Y&Y&Y&Y&Y&Y\\
PER\,07&N&N&N&N&N&N&N&N&N&N&N&N&N\\
PER\,08&Y&Y&Y&Y&Y&Y&N&Y&Y&Y&Y&Y&Y\\
PER\,09&Y&Y&Y&Y&Y&Y&Y&Y&Y&Y&Y&Y&Y\\
PER\,10&Y&Y&Y&Y&Y&N&N&Y&Y&Y&Y&Y&Y\\
PER\,12&N&N&N&N&N&N&N&N&N&N&N&N&N\\
PER\,13&N&N&N&N&N&N&N&N&N&N&N&N&N\\
PER\,14&Y&Y&Y&Y&Y&Y&Y&Y&Y&Y&Y&Y&Y\\
PER\,15&Y&Y&Y&N&Y&N&N&N&N&Y&N&Y&Y\\
PER\,16&Y&Y&N&N&N&N&N&N&N&N&N&N&N\\
PER\,17&N&N&N&N&N&N&N&N&N&N&N&N&N\\
PER\,18&Y&Y&Y&N&N&N&N&N&N&Y&Y&N&N\\
PER\,19&Y&Y&Y&N&Y&N&N&Y&Y&Y&Y&Y&Y\\
PER\,20&Y&Y&Y&N&Y&N&N&Y&Y&Y&Y&Y&Y\\
PER\,21&Y&Y&Y&Y&Y&N&N&Y&Y&Y&N&Y&Y\\
PER\,22&Y&Y&Y&N&Y&N&N&Y&Y&N&Y&Y&Y\\
SCO\,01&Y&Y&Y&Y&Y&N&N&Y&Y&Y&Y&Y&Y\\
SERS\,01&N&N&N&N&N&N&N&N&N&N&N&N&N\\
SERS\,02&Y&Y&Y&Y&Y&N&N&N&N&N&N&Y&Y\\
TAU\,01&Y&Y&N&N&Y&N&N&Y&Y&N&N&Y&Y\\
TAU\,02&Y&Y&Y&N&Y&N&N&Y&Y&Y&Y&Y&Y\\
TAU\,03&N&Y&Y&N&N&N&N&Y&Y&N&N&Y&N\\
TAU\,04&Y&Y&N&Y&Y&N&N&Y&Y&N&N&Y&Y\\
TAU\,06&N&N&N&N&N&N&N&N&N&N&N&N&N\\
TAU\,07&Y&Y&Y&Y&Y&N&N&Y&Y&Y&Y&Y&Y\\
TAU\,08\tablefootmark{a}&-&-&-&-&-&-&-&-&-&-&-&-&-\\
TAU\,09&Y&Y&Y&Y&Y&N&N&N&N&Y&N&Y&Y\\
W40\,01&Y&Y&N&N&N&N&N&N&N&N&N&N&N\\
W40\,02&Y&Y&Y&N&Y&N&N&N&N&Y&Y&Y&Y\\
W40\,03&N&N&N&N&N&N&N&N&N&N&N&N&N\\
W40\,04&N&N&N&N&N&N&N&N&N&N&N&N&N\\
W40\,05&N&N&N&N&N&N&N&N&N&N&N&N&N\\
W40\,06&N&N&N&N&N&N&N&N&N&N&N&N&N\\
W40\,07&N&N&N&N&N&N&N&N&N&N&N&N&N\\
\hline\noalign{\smallskip}
Total&32/48&32/48&26/48&17/48&26/48&\phantom{0}7/48&\phantom{0}5/48&22/48&22/48&20/48&18/48&27/48&26/48\\
\hline\noalign {\smallskip}
\end{tabular}
\vspace{-2mm}
\tablefoot{\tablefoottext{a}{Source not observed with PACS due to end of \textit{Herschel} mission.} \tablefoottext{b}{In absorption due to contamination of the off-position.} \tablefoottext{c}{The CO $J$=31$-$30 transition at 84.41\micron{} and the OH 84.42\micron{} transition are blended.}}
\end{center}
\end{table*}
\renewcommand{\thetable}{\thesection.\arabic{table}}

\subsection{Entrained outflow}
\label{S:properties_outflow}

Molecular outflows, usually detected through observations of low-$J$ $^{12}$CO, are a ubiquitous signpost of ongoing star formation. In order to identify and quantify the properties of the entrained outflowing material associated with WILL sources from $^{12}$CO $J$=3$-$2 maps, a number of steps were taken, following \citet{vanderMarel2013}:

\begin{enumerate}

\item The data were resampled to 0.5\kms{} to improve the sensitivity to line-wing emission.

\item Maps of maximum red and blue-shifted velocities were identified in all spectra as the channel where the emission first reaches 1$\sigma_{\mathrm{rms}}$.

\item The outer velocity ($\varv_{\mathrm{out}}$) of each outflow lobe was defined from the maximum velocity maps as the most offset value from the source velocity. The maximum velocity ($\varv_{\mathrm{max}}$) is the absolute difference between $\varv_{\mathrm{out}}$ and $\varv_{\mathrm{LSR}}$.

\item The inner velocity ($\varv_{\mathrm{in}}$) for each lobe was defined by the same approach using a spectrum without any outflow emission, so as to mask out cloud emission.

\item Integrated maps for the red and blue outflow lobes were created by integrating between the minimum and maximum velocities. Visual inspection and comparison of these maps with the continuum emission was then used to identify the spectra associated with an outflow from the source. Excluding low-velocity emission will lead to an underestimate in the mass and related properties by factors of a few \citep{Downes2007}. However, this is generally preferable to performing an incorrect correction based on poor knowledge of the contribution from the envelope \citep{Cabrit1990}.

\item Each source was assigned an inclination ($i$) of 10, 30, 50 or 70\deg{} between the outflow axis and the line of sight, such that $i$=0$^{\circ}$ is pole-on, through visual inspection of the overlap between the red and blue lobes with each other and the source position.

\item The radius associated with each outflow lobe ($R_{\mathrm{CO}}$) was defined as the distance between the source position and furthest pixel containing outflow emission.

\item The mass in each channel of each pixel ($m_{ij}$) was calculated assuming an excitation temperature ($T_{\mathrm{ex}}$) of 75\,K, $\mu$=2.8 \citep{Kauffmann2008} and a CO abundance relative to H$_{2}$ of 1.2$\times$10$^{-4}$ \citep{Frerking1982}, consistent with the outflow properties determined by \citet{Yildiz2015} for the WISH sample. Changing $T_{\mathrm{ex}}$ to 100\, or 50\,K would only raise or lower the mass by a factor of 1.2. No correction for $\tau$ is performed, as the optically thick parts of the line near the line centre are excluded and the line wings are typically optically thin in outflows associated with low-mass protostars \citep[see][]{vanderMarel2013,Yildiz2015}.

\item The physical properties of the outflow were calculated using the separation method \citep[M7 in][]{vanderMarel2013}, where the mass, momentum and energy are calculated separately for each lobe on a per-channel basis for each spectrum, then summed over all channels and spectra, while the maximum velocity is used to calculate the dynamical time of the flow, that is:

\begin{equation}
M_{\mathrm{out}}\,=\,\sum_{j=1}^{\mathrm{npix}}\sum_{\varv_{\mathrm{in}}}^{\varv_{\mathrm{out}}}m_{i,j}\,dv\,,
\end{equation}

\begin{equation}
P_{\mathrm{out}}\,=\,\sum_{j=1}^{\mathrm{npix}}\sum_{\varv_{\mathrm{in}}}^{\varv_{\mathrm{out}}}m_{i,j}\mid\varv_{i}-\varv_{\mathrm{LSR}}\mid\,dv\,,
\end{equation}

\begin{equation}
E_{\mathrm{out}}\,=\,\sum_{j=1}^{\mathrm{npix}}\sum_{\varv_{\mathrm{in}}}^{\varv_{\mathrm{out}}}m_{i,j}\mid\varv_{i}-\varv_{\mathrm{LSR}}\mid^{2}\,dv\,,
\end{equation}

\begin{equation}
t_{\mathrm{dyn}}\,=\,\frac{R_{\mathrm{CO}}}{\varv_{\mathrm{max}}}\,,
\end{equation}

\begin{equation}
\dot{M}_{\mathrm{out}}\,=\,\frac{c_{\mathrm{f}}\,M_{\mathrm{out}}}{t_{\mathrm{dyn}}}\,,
\end{equation}

\noindent and

\begin{equation}
F_{\mathrm{CO}}\,=\,\frac{c_{\mathrm{f}}\,P_{\mathrm{out}}}{t_{\mathrm{dyn}}}\,,
\end{equation}

\noindent where c$_{\mathrm{f}}$ is a correction factor to account for inclination, given in Table~\ref{T:outflow_correction} and derived from the models of \citet{Downes2007} by \citet{vanderMarel2013}.

\end{enumerate}

\begin{table}
\begin{center}
\caption[]{Correction factors for line-of-sight outflow inclination.}
\vspace{-2mm}
\begin{tabular}{lcccc}
\hline \noalign {\smallskip}
$i$ & 10\deg{} & 30\deg{} & 50\deg{} & 70\deg{} \\
\hline\noalign {\smallskip}
c$_{\mathrm{f}}$ & 1.2 & 2.8 & 4.4 & 7.1 \\
\hline\noalign {\smallskip}
\label{T:outflow_correction}
\end{tabular}
\vspace{-5mm}
\end{center}
\vspace{-2mm}
\end{table}

\begin{figure*}
\begin{center}
\includegraphics[width=0.95\textwidth]{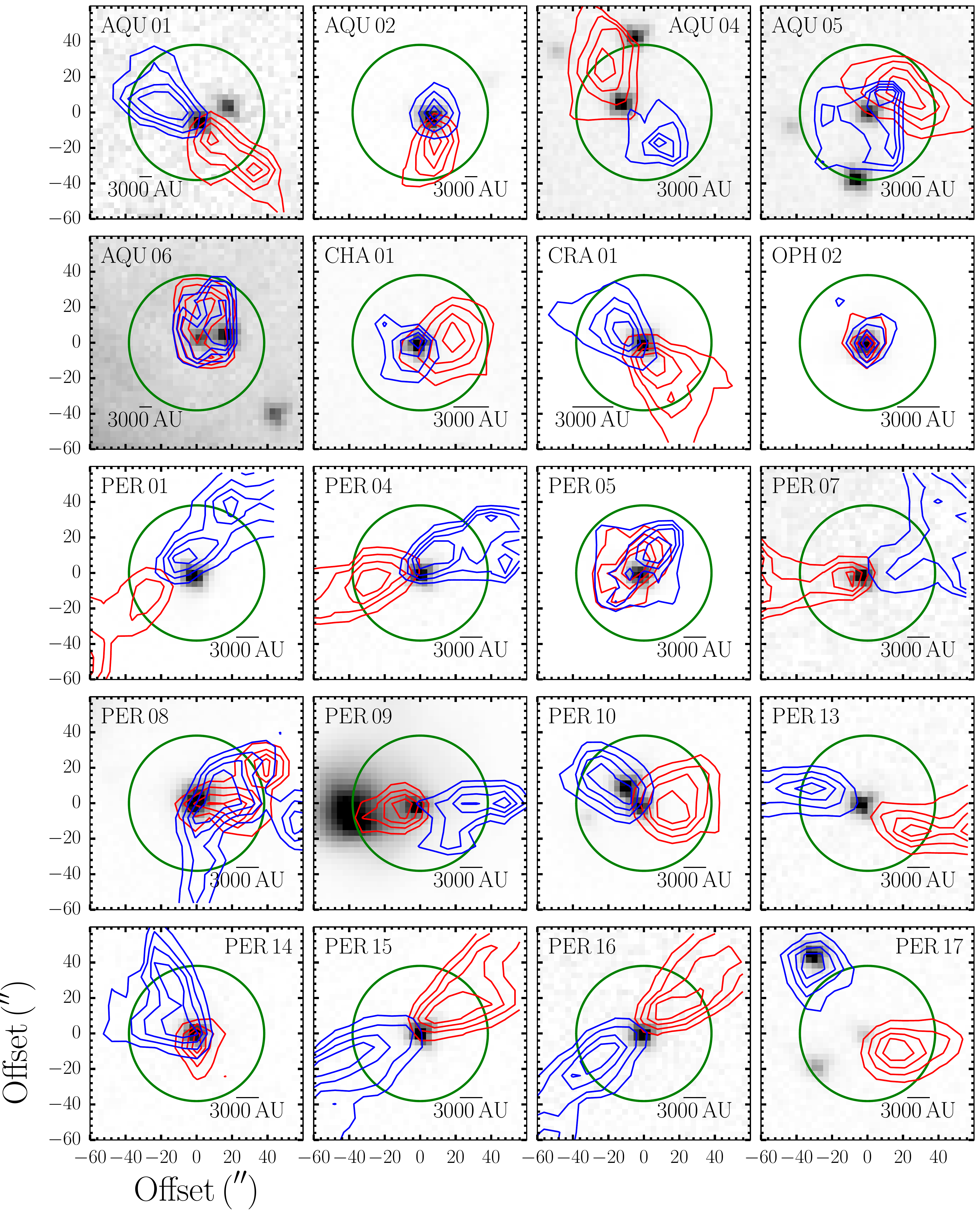}
\caption{Outflow maps. The grey-scale images show the 70\micron{} continuum emission from \textit{Herschel}, while the red and blue contours show the outflow lobes detected in $^{12}$CO $J$=3$-$2. The levels for the contours are at 20, 40, 60 and 80$\%$ of the maximum velocity-integrated emission. The green circle indicates the HIFI beam for the H$_{2}$O 1$_{10}-$1$_{01}$ transition. All maps show a region of 2\arcmin{}\,$\times$2\arcmin{} centred on the source position. The black scale-bar in the lower panel of each figure indicates 3000\,AU at the distance of the source.}
\label{F:outflow_maps}
\end{center}
\end{figure*}

Table~\ref{T:outflow_properties} gives the calculated outflow properties for the red and blue lobes separately, including the velocity limits, outflow mass ($M_{\mathrm{out}}$), momentum ($P_{\mathrm{out}}$), kinetic energy ($E_{\mathrm{out}}$), radius, inclination, dynamical time ($t_{\mathrm{dyn}}$), force ($F_{\mathrm{CO}}$) and mass outflow rate ($\dot{M}_{\mathrm{out}}$). In some cases, the outflow may extend beyond the 2\arcmin{}$\times$2\arcmin{} coverage of the observations, so these values may be lower limits to the total value. Figure~\ref{F:outflow_maps} shows the outflow lobes of the observed outflows associated with the WILL sources overlaid on the \textit{Herschel} PACS 70\,micron{} maps.

\renewcommand{\thefigure}{\thesection.\arabic{figure} (Cont.)}
\addtocounter{figure}{-1}

\begin{figure*}
\begin{center}
\includegraphics[width=0.95\textwidth]{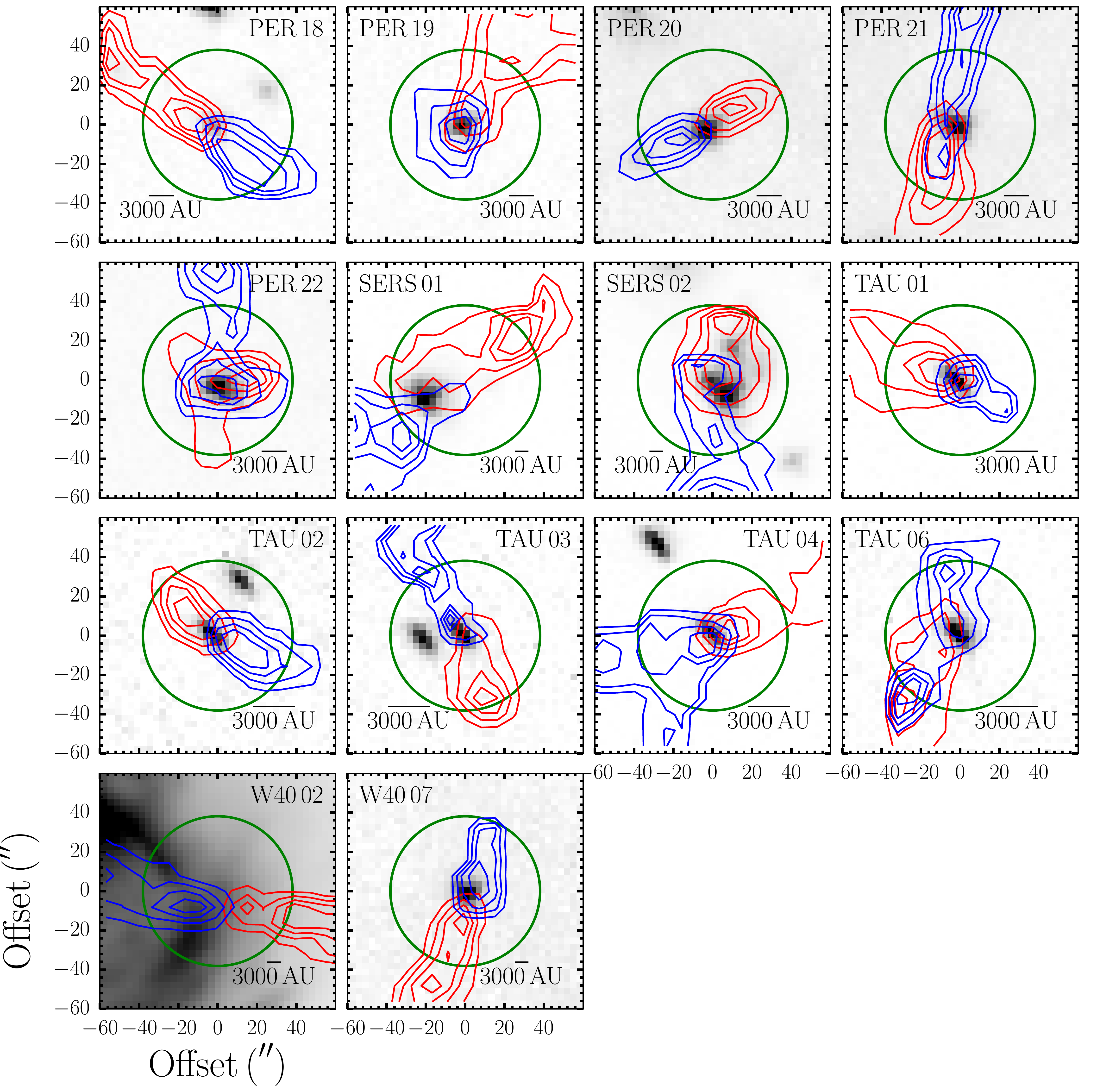}
\caption{Outflow maps. The grey-scale images show the 70\micron{} continuum emission from \textit{Herschel}, while the red and blue contours show the outflow lobes detected in $^{12}$CO $J$=3$-$2. The levels for the contours are at 20, 40, 60 and 80$\%$ of the maximum velocity-integrated emission. The green circle indicates the HIFI beam for the H$_{2}$O 1$_{10}-$1$_{01}$ transition. All maps show a region of 2\arcmin{}\,$\times$2\arcmin{} centred on the source position. The black scale-bar in the lower panel of each figure indicates 3000\,AU at the distance of the source.}
\end{center}
\end{figure*}

\renewcommand{\thefigure}{\thesection.\arabic{figure}}

\begin{table*}
\caption{Outflow properties.}
\begin{center}
\vspace{-2mm}
\begin{tabular}{lcccccccccccc}
\hline \noalign {\smallskip}
Name & $i$ & Lobe & $\varv_{in}$\tablefootmark{a} & $\varv_{out}$\tablefootmark{a}  & $\mid\varv_{max}\mid$\tablefootmark{a} & $M_{\mathrm{out}}$\tablefootmark{a} & $P_{\mathrm{out}}$\tablefootmark{a} & $E_{\mathrm{out}}$\tablefootmark{a} & $R_{\mathrm{CO}}$\tablefootmark{a} & $t_{\mathrm{dyn}}$\tablefootmark{a} & $F_{\mathrm{CO}}$\tablefootmark{b} & $\dot{M}_{\mathrm{out}}$\tablefootmark{b}\\
& (\deg{}) & & (km\,s$^{-1}$) & (km\,s$^{-1}$) & (km\,s$^{-1}$) & (\msol{}) & (\msol{}\,km\,s$^{-1}$) & (ergs) & (10$^{3}$\,AU) & (10$^{3}$\,yrs) & (\msol{}\,yr$^{-1}$\,km\,s$^{-1}$) & (\msol{}\,yr$^{-1}$) \\
\hline\noalign {\smallskip}
AQU\,01&70&R&11.0&33.5&26.1&1.6$\times$10$^{-02}$&1.5$\times$10$^{-01}$&1.8$\times$10$^{+43}$&15.0&\phantom{0}2.7&3.9$\times$10$^{-04}$&4.2$\times$10$^{-05}$\\
&&B&\phantom{0}3.0&$-$30.0\phantom{$-$}&37.4&2.2$\times$10$^{-02}$&2.2$\times$10$^{-01}$&3.3$\times$10$^{+43}$&28.1&\phantom{0}3.6&4.5$\times$10$^{-04}$&4.3$\times$10$^{-05}$\\
\hline\noalign {\smallskip}
AQU\,02&30&R&10.5&25.0&17.5&7.4$\times$10$^{-03}$&4.2$\times$10$^{-02}$&3.1$\times$10$^{+42}$&16.4&\phantom{0}4.4&2.7$\times$10$^{-05}$&4.6$\times$10$^{-06}$\\
&&B&\phantom{0}4.5&\phantom{0}$-$8.5\phantom{$-$}&16.0&4.8$\times$10$^{-03}$&3.1$\times$10$^{-02}$&2.6$\times$10$^{+42}$&\phantom{0}5.5&\phantom{0}1.6&5.4$\times$10$^{-05}$&8.2$\times$10$^{-06}$\\
\hline\noalign {\smallskip}
AQU\,03&$-$\tablefootmark{c}&$-$&$-$&$-$&$-$&$-$&$-$&$-$&$-$&$-$&$-$&$-$\\
\hline\noalign {\smallskip}
AQU\,04&70&R&11.0&17.5& 9.9&7.7$\times$10$^{-03}$&3.7$\times$10$^{-02}$&1.9$\times$10$^{+42}$&29.2&14.0&1.9$\times$10$^{-05}$&3.9$\times$10$^{-06}$\\
&&B&\phantom{0}2.5&\phantom{0}$-$1.0\phantom{$-$}&\phantom{0}8.6&8.7$\times$10$^{-04}$&5.4$\times$10$^{-03}$&3.4$\times$10$^{+41}$&10.0&\phantom{0}5.5&6.9$\times$10$^{-06}$&1.1$\times$10$^{-06}$\\
\hline\noalign {\smallskip}
AQU\,05&30&R&11.0&18.5&11.2&1.9$\times$10$^{-02}$&9.8$\times$10$^{-02}$&5.5$\times$10$^{+42}$&15.1&\phantom{0}6.4&4.3$\times$10$^{-05}$&8.2$\times$10$^{-06}$\\
&&B&\phantom{0}2.5&\phantom{0}$-$4.0\phantom{$-$}&11.3&1.5$\times$10$^{-03}$&1.2$\times$10$^{-02}$&9.8$\times$10$^{+41}$&16.4&\phantom{0}6.8&4.8$\times$10$^{-06}$&6.1$\times$10$^{-07}$\\
\hline\noalign {\smallskip}
AQU\,06&10&R&11.0&18.5&10.2&2.7$\times$10$^{-03}$&1.2$\times$10$^{-02}$&6.5$\times$10$^{+41}$&\phantom{0}5.3&\phantom{0}2.5&6.0$\times$10$^{-06}$&1.3$\times$10$^{-06}$\\
&&B&\phantom{0}3.0&\phantom{0}$-$3.5\phantom{$-$}&11.8&1.1$\times$10$^{-03}$&9.1$\times$10$^{-03}$&7.7$\times$10$^{+41}$&\phantom{0}5.3&\phantom{0}2.2&5.0$\times$10$^{-06}$&6.3$\times$10$^{-07}$\\
\hline\noalign {\smallskip}
CHA\,01&30&R&\phantom{0}7.5&12.0&\phantom{0}7.1&9.7$\times$10$^{-04}$&3.7$\times$10$^{-03}$&1.5$\times$10$^{+41}$&10.7&\phantom{0}7.1&1.4$\times$10$^{-06}$&3.8$\times$10$^{-07}$\\
&&B&\phantom{0}3.0&\phantom{0}2.0&\phantom{0}2.9&8.8$\times$10$^{-05}$&2.0$\times$10$^{-04}$&4.8$\times$10$^{+39}$&\phantom{0}4.8&\phantom{0}7.8&7.3$\times$10$^{-08}$&3.2$\times$10$^{-08}$\\
\hline\noalign {\smallskip}
CHA\,02&$-$\tablefootmark{c}&$-$&$-$&$-$&$-$&$-$&$-$&$-$&$-$&$-$&$-$&$-$\\
\hline\noalign {\smallskip}
CRA\,01&50&R&\phantom{0}7.3&22.8&17.1&1.5$\times$10$^{-03}$&6.8$\times$10$^{-03}$&4.8$\times$10$^{+41}$&18.7&\phantom{0}5.2&5.7$\times$10$^{-06}$&1.3$\times$10$^{-06}$\\
&&B&\phantom{0}3.8&$-$16.7\phantom{$-$}&22.4&1.4$\times$10$^{-03}$&6.2$\times$10$^{-03}$&4.2$\times$10$^{+41}$&27.9&\phantom{0}5.9&4.6$\times$10$^{-06}$&1.0$\times$10$^{-06}$\\
\hline\noalign {\smallskip}
OPH\,01&$-$\tablefootmark{c}&$-$&$-$&$-$&$-$&$-$&$-$&$-$&$-$&$-$&$-$&$-$\\
\hline\noalign {\smallskip}
OPH\,02&10&R&\phantom{0}7.0&10.5&\phantom{0}6.3&5.1$\times$10$^{-05}$&1.9$\times$10$^{-04}$&7.3$\times$10$^{+39}$&\phantom{0}9.2&\phantom{0}7.0&3.2$\times$10$^{-08}$&8.7$\times$10$^{-09}$\\
&&B&\phantom{0}1.0&\phantom{0}$-$4.0\phantom{$-$}&\phantom{0}8.2&9.3$\times$10$^{-05}$&3.7$\times$10$^{-04}$&1.6$\times$10$^{+40}$&11.8&\phantom{0}6.8&6.6$\times$10$^{-08}$&1.6$\times$10$^{-08}$\\
\hline\noalign {\smallskip}
PER\,01&70&R&\phantom{0}7.2&49.2&45.2&2.4$\times$10$^{-02}$&2.1$\times$10$^{-01}$&3.2$\times$10$^{+43}$&41.9&\phantom{0}4.4&3.4$\times$10$^{-04}$&3.9$\times$10$^{-05}$\\
&&B&\phantom{0}$-$2.3\phantom{$-$}&$-$36.3\phantom{$-$}&40.3&1.0$\times$10$^{-02}$&1.5$\times$10$^{-01}$&2.9$\times$10$^{+43}$&15.2&\phantom{0}1.8&5.9$\times$10$^{-04}$&4.2$\times$10$^{-05}$\\
\hline\noalign {\smallskip}
PER\,02&$-$\tablefootmark{d}&$-$&$-$&$-$&$-$&$-$&$-$&$-$&$-$&$-$&$-$&$-$\\
\hline\noalign {\smallskip}
PER\,04&50&R&\phantom{0}6.2&12.7&\phantom{0}7.5&2.3$\times$10$^{-03}$&5.9$\times$10$^{-03}$&2.0$\times$10$^{+41}$&27.9&17.6&1.5$\times$10$^{-06}$&5.8$\times$10$^{-07}$\\
&&B&\phantom{0}1.7&\phantom{0}$-$1.8\phantom{$-$}&\phantom{0}7.0&2.8$\times$10$^{-04}$&1.3$\times$10$^{-03}$&6.2$\times$10$^{+40}$&14.2& 9.6&5.9$\times$10$^{-07}$&1.3$\times$10$^{-07}$\\
\hline\noalign {\smallskip}
PER\,05&10&R&12.7&17.7&10.5&2.5$\times$10$^{-04}$&1.8$\times$10$^{-03}$&1.4$\times$10$^{+41}$&13.9&\phantom{0}6.3&3.5$\times$10$^{-07}$&4.8$\times$10$^{-08}$\\
&&B&\phantom{0}3.2&\phantom{0}$-$1.3\phantom{$-$}&\phantom{0}8.5&3.4$\times$10$^{-04}$&1.8$\times$10$^{-03}$&1.0$\times$10$^{+41}$&13.9&\phantom{0}7.7&2.9$\times$10$^{-07}$&5.4$\times$10$^{-08}$\\
\hline\noalign {\smallskip}
PER\,06&$-$\tablefootmark{d}&$-$&$-$&$-$&$-$&$-$&$-$&$-$&$-$&$-$&$-$&$-$\\
\hline\noalign {\smallskip}
PER\,07&70&R&11.2&20.7&13.4&1.7$\times$10$^{-03}$&1.1$\times$10$^{-02}$&7.3$\times$10$^{+41}$&29.2&10.4&7.2$\times$10$^{-06}$&1.2$\times$10$^{-06}$\\
&&B&\phantom{0}2.2&\phantom{0}$-$5.3\phantom{$-$}&12.6&2.4$\times$10$^{-03}$&1.6$\times$10$^{-02}$&1.2$\times$10$^{+42}$&17.4&\phantom{0}6.6&1.8$\times$10$^{-05}$&2.6$\times$10$^{-06}$\\
\hline\noalign {\smallskip}
PER\,08&10&R&15.5&21.5&13.8&5.4$\times$10$^{-04}$&5.1$\times$10$^{-03}$&5.0$\times$10$^{+41}$&12.7&\phantom{0}4.4&1.4$\times$10$^{-06}$&1.5$\times$10$^{-07}$\\
&&B&\phantom{0}$-$0.5\phantom{$-$}&$-$14.5\phantom{$-$}&22.2&3.4$\times$10$^{-03}$&3.6$\times$10$^{-02}$&4.1$\times$10$^{+42}$&15.1&\phantom{0}3.2&1.3$\times$10$^{-05}$&1.2$\times$10$^{-06}$\\
\hline\noalign {\smallskip}
PER\,09&70&R&13.0&25.0&17.5&5.3$\times$10$^{-04}$&4.9$\times$10$^{-03}$&5.1$\times$10$^{+41}$&14.6&\phantom{0}4.0&8.8$\times$10$^{-06}$&9.6$\times$10$^{-07}$\\
&&B&\phantom{0}0.5&$-$16.0\phantom{$-$}&23.5&7.6$\times$10$^{-04}$&9.3$\times$10$^{-03}$&1.3$\times$10$^{+42}$&14.1&\phantom{0}2.8&2.3$\times$10$^{-05}$&1.9$\times$10$^{-06}$\\
\hline\noalign {\smallskip}
PER\,10&50&R&13.2&32.2&23.6&1.3$\times$10$^{-02}$&9.6$\times$10$^{-02}$&8.7$\times$10$^{+42}$&14.2&\phantom{0}2.9&1.5$\times$10$^{-04}$&1.9$\times$10$^{-05}$\\
&&B&\phantom{0}2.2&$-$14.8\phantom{$-$}&23.4&6.1$\times$10$^{-03}$&6.1$\times$10$^{-02}$&6.8$\times$10$^{+42}$&24.9&\phantom{0}5.0&5.3$\times$10$^{-05}$&5.4$\times$10$^{-06}$\\
\hline\noalign {\smallskip}
PER\,12&$-$\tablefootmark{c}&$-$&$-$&$-$&$-$&$-$&$-$&$-$&$-$&$-$&$-$&$-$\\
\hline\noalign {\smallskip}
PER\,13&70&R&10.7&22.2&14.3&4.1$\times$10$^{-03}$&1.9$\times$10$^{-02}$&1.1$\times$10$^{+42}$&14.2&\phantom{0}4.7&2.9$\times$10$^{-05}$&6.1$\times$10$^{-06}$\\
&&B&\phantom{0}5.7&\phantom{0}$-$4.3\phantom{$-$}&12.2&1.8$\times$10$^{-03}$&7.9$\times$10$^{-03}$&4.8$\times$10$^{+41}$&27.0&10.4&5.4$\times$10$^{-06}$&1.2$\times$10$^{-06}$\\
\hline\noalign {\smallskip}
PER\,14&50&R&\phantom{0}9.0&20.0&13.8&3.0$\times$10$^{-04}$&1.7$\times$10$^{-03}$&1.1$\times$10$^{+41}$&\phantom{0}4.0&\phantom{0}1.4&5.3$\times$10$^{-06}$&9.7$\times$10$^{-07}$\\
&&B&\phantom{0}4.5&\phantom{0}$-$9.0\phantom{$-$}&15.2&5.3$\times$10$^{-03}$&2.0$\times$10$^{-02}$&1.0$\times$10$^{+42}$&27.9&\phantom{0}8.7&9.9$\times$10$^{-06}$&2.7$\times$10$^{-06}$\\
\hline\noalign {\smallskip}
PER\,15&70&R& 9.7&28.2&21.2&1.0$\times$10$^{-02}$&6.4$\times$10$^{-02}$&5.4$\times$10$^{+42}$&17.4&\phantom{0}3.9&1.2$\times$10$^{-04}$&1.8$\times$10$^{-05}$\\
&&B&\phantom{0}1.2&$-$25.3\phantom{$-$}&32.3&4.9$\times$10$^{-03}$&5.3$\times$10$^{-02}$&7.2$\times$10$^{+42}$&37.0&\phantom{0}5.4&6.9$\times$10$^{-05}$&6.4$\times$10$^{-06}$\\
\hline\noalign {\smallskip}
PER\,16&70&R&\phantom{0}9.2&23.2&16.2&4.6$\times$10$^{-03}$&2.3$\times$10$^{-02}$&1.7$\times$10$^{+42}$&17.4&\phantom{0}5.1&3.3$\times$10$^{-05}$&6.4$\times$10$^{-06}$\\
&&B&\phantom{0}2.7&\phantom{0}$-$6.3\phantom{$-$}&13.3&2.2$\times$10$^{-03}$&1.4$\times$10$^{-02}$&1.1$\times$10$^{+42}$&37.0&13.2&7.8$\times$10$^{-06}$&1.2$\times$10$^{-06}$\\
\hline\noalign {\smallskip}
PER\,17&50&R&10.8&30.3&23.7&5.2$\times$10$^{-03}$&3.8$\times$10$^{-02}$&3.4$\times$10$^{+42}$&12.8&\phantom{0}2.6&6.6$\times$10$^{-05}$&8.9$\times$10$^{-06}$\\
&&B&\phantom{0}2.8&$-$20.2\phantom{$-$}&26.8&7.2$\times$10$^{-03}$&5.2$\times$10$^{-02}$&4.9$\times$10$^{+42}$&30.1&\phantom{0}5.3&4.3$\times$10$^{-05}$&5.9$\times$10$^{-06}$\\
\hline\noalign {\smallskip}
PER\,18&70&R&10.8&32.3&25.6&2.1$\times$10$^{-03}$&2.1$\times$10$^{-02}$&2.7$\times$10$^{+42}$&32.4&\phantom{0}6.0&2.5$\times$10$^{-05}$&2.4$\times$10$^{-06}$\\
&&B&\phantom{0}2.8&\phantom{0}$-$5.7\phantom{$-$}&12.4&7.0$\times$10$^{-04}$&3.9$\times$10$^{-03}$&2.5$\times$10$^{+41}$&12.7&\phantom{0}4.9&5.7$\times$10$^{-06}$&1.0$\times$10$^{-06}$\\
\hline\noalign {\smallskip}
PER\,19&50&R& 9.7&15.7&\phantom{0}9.0&2.4$\times$10$^{-03}$&9.3$\times$10$^{-03}$&3.9$\times$10$^{+41}$&17.4&\phantom{0}9.2&4.5$\times$10$^{-06}$&1.2$\times$10$^{-06}$\\
&&B&\phantom{0}2.7&\phantom{0}$-$4.3\phantom{$-$}&11.0&6.0$\times$10$^{-04}$&3.2$\times$10$^{-03}$&1.9$\times$10$^{+41}$&11.8&\phantom{0}5.1&2.8$\times$10$^{-06}$&5.2$\times$10$^{-07}$\\
\hline\noalign {\smallskip}
PER\,20&50&R&11.2&50.2&41.3&5.6$\times$10$^{-03}$&6.0$\times$10$^{-02}$&1.2$\times$10$^{+43}$&11.8&\phantom{0}1.4&1.9$\times$10$^{-04}$&1.8$\times$10$^{-05}$\\
&&B&\phantom{0}6.2&$-$13.8\phantom{$-$}&22.7&4.7$\times$10$^{-03}$&3.8$\times$10$^{-02}$&4.2$\times$10$^{+42}$&28.1&\phantom{0}5.9&2.8$\times$10$^{-05}$&3.5$\times$10$^{-06}$\\
\hline\noalign {\smallskip}
\end{tabular}
\vspace{-2mm}
\tablefoot{\tablefoottext{a}{Measured values, i.e. not corrected for inclination.} \tablefoottext{b}{Corrected for inclination as discussed in detail in \citet{vanderMarel2013} and \citet{Yildiz2015}.} \tablefoottext{c}{No outflow associated with this source.} \tablefoottext{d}{An outflow is detected but is too contaminated by outflows from others sources for meaningful determinations to be made. See Sect.~\ref{S:cases} for details.}}
\end{center}
\label{T:outflow_properties}
\end{table*}
\renewcommand{\thetable}{\thesection.\arabic{table} (Cont.)}
\addtocounter{table}{-1}
\begin{table*}
\caption{Outflow properties.}
\begin{center}
\vspace{-2mm}
\begin{tabular}{lcccccccccccc}
\hline \noalign {\smallskip}
Name & $i$ & Lobe & $\varv_{in}$\tablefootmark{a} & $\varv_{out}$\tablefootmark{a}  & $\mid\varv_{max}\mid$\tablefootmark{a} & $M_{\mathrm{out}}$\tablefootmark{a} & $P_{\mathrm{out}}$\tablefootmark{a} & $E_{\mathrm{out}}$\tablefootmark{a} & $R_{\mathrm{CO}}$\tablefootmark{a} & $t_{\mathrm{dyn}}$\tablefootmark{a} & $F_{\mathrm{CO}}$\tablefootmark{b} & $\dot{M}_{\mathrm{out}}$\tablefootmark{b}\\
& (\deg{}) & & (km\,s$^{-1}$) & (km\,s$^{-1}$) & (km\,s$^{-1}$) & (\msol{}) & (\msol{}\,km\,s$^{-1}$) & (ergs) & (10$^{3}$\,AU) & (10$^{3}$\,yrs) & (\msol{}\,yr$^{-1}$\,km\,s$^{-1}$) & (\msol{}\,yr$^{-1}$) \\
\hline\noalign {\smallskip}
PER\,21&50&R&11.2&28.2&19.5&4.0$\times$10$^{-03}$&2.5$\times$10$^{-02}$&2.3$\times$10$^{+42}$&26.4&\phantom{0}6.4&1.7$\times$10$^{-05}$&2.7$\times$10$^{-06}$\\
&&B&\phantom{0}4.7&$-$15.8\phantom{$-$}&24.5&2.4$\times$10$^{-03}$&2.2$\times$10$^{-02}$&2.6$\times$10$^{+42}$&10.8&\phantom{0}2.1&4.6$\times$10$^{-05}$&5.1$\times$10$^{-06}$\\
\hline\noalign {\smallskip}
PER\,22&30&R&11.2&18.2&\phantom{0}8.5&5.4$\times$10$^{-03}$&1.2$\times$10$^{-02}$&3.2$\times$10$^{+41}$&21.9&12.3&2.7$\times$10$^{-06}$&1.2$\times$10$^{-06}$\\
&&B&\phantom{0}6.2&\phantom{0}3.2&\phantom{0}6.5&9.0$\times$10$^{-04}$&3.6$\times$10$^{-03}$&1.5$\times$10$^{+41}$&10.8&\phantom{0}7.8&1.3$\times$10$^{-06}$&3.2$\times$10$^{-07}$\\
\hline\noalign {\smallskip}
SCO\,01&$-$\tablefootmark{c}&$-$&$-$&$-$&$-$&$-$&$-$&$-$&$-$&$-$&$-$&$-$\\
\hline\noalign {\smallskip}
SERS\,01&50&R&11.0&18.5&10.3&1.5$\times$10$^{-02}$&6.4$\times$10$^{-02}$&3.1$\times$10$^{+42}$&19.9&\phantom{0}9.1&3.1$\times$10$^{-05}$&7.2$\times$10$^{-06}$\\
&&B&\phantom{0}3.5&$-$15.0\phantom{$-$}&23.2&1.6$\times$10$^{-02}$&1.2$\times$10$^{-01}$&1.0$\times$10$^{+43}$&32.4&\phantom{0}6.6&7.8$\times$10$^{-05}$&1.1$\times$10$^{-05}$\\
\hline\noalign {\smallskip}
SERS\,02&50&R&13.5&40.5&32.7&1.0$\times$10$^{-01}$&9.9$\times$10$^{-01}$&1.2$\times$10$^{+44}$&\phantom{0}7.9&\phantom{0}1.1&3.8$\times$10$^{-03}$&3.9$\times$10$^{-04}$\\
&&B&\phantom{0}1.0&$-$18.5\phantom{$-$}&26.3&3.0$\times$10$^{-02}$&3.2$\times$10$^{-01}$&3.9$\times$10$^{+43}$&26.4&\phantom{0}4.8&2.9$\times$10$^{-04}$&2.8$\times$10$^{-05}$\\
\hline\noalign {\smallskip}
TAU\,01&50&R&\phantom{0}8.3&16.8&10.0&2.1$\times$10$^{-03}$&5.4$\times$10$^{-03}$&1.8$\times$10$^{+41}$&26.4&12.6&1.9$\times$10$^{-06}$&7.5$\times$10$^{-07}$\\
&&B&\phantom{0}4.8&\phantom{0}$-$6.7\phantom{$-$}&13.5&2.6$\times$10$^{-04}$&8.5$\times$10$^{-04}$&3.9$\times$10$^{+40}$&\phantom{0}3.4&\phantom{0}1.2&3.2$\times$10$^{-06}$&9.5$\times$10$^{-07}$\\
\hline\noalign {\smallskip}
TAU\,02&50&R&\phantom{0}8.0&11.5&\phantom{0}4.9&3.1$\times$10$^{-04}$&5.6$\times$10$^{-04}$&1.2$\times$10$^{+40}$&18.5&17.9&1.4$\times$10$^{-07}$&7.7$\times$10$^{-08}$\\
&&B&\phantom{0}5.0&\phantom{0}$-$0.0\phantom{$-$}&\phantom{0}6.6&2.5$\times$10$^{-04}$&6.2$\times$10$^{-04}$&1.9$\times$10$^{+40}$&11.8&\phantom{0}8.5&3.2$\times$10$^{-07}$&1.3$\times$10$^{-07}$\\
\hline\noalign {\smallskip}
TAU\,03&50&R&\phantom{0}8.5&14.0&\phantom{0}6.6&6.3$\times$10$^{-04}$&1.3$\times$10$^{-03}$&3.4$\times$10$^{+40}$&11.8&\phantom{0}8.5&6.5$\times$10$^{-07}$&3.2$\times$10$^{-07}$\\
&&B&\phantom{0}4.5&\phantom{0}3.0&\phantom{0}4.4&3.5$\times$10$^{-05}$&1.1$\times$10$^{-04}$&3.5$\times$10$^{+39}$&28.1&30.2&1.6$\times$10$^{-08}$&5.0$\times$10$^{-09}$\\
\hline\noalign {\smallskip}
TAU\,04&50&R&\phantom{0}7.5&18.0&11.7&9.5$\times$10$^{-04}$&2.9$\times$10$^{-03}$&1.3$\times$10$^{+41}$&16.2&\phantom{0}6.6&1.9$\times$10$^{-06}$&6.3$\times$10$^{-07}$\\
&&B&\phantom{0}5.5&\phantom{0}1.0&\phantom{0}5.3&4.7$\times$10$^{-04}$&6.5$\times$10$^{-04}$&1.2$\times$10$^{+40}$&27.9&24.8&1.2$\times$10$^{-07}$&8.3$\times$10$^{-08}$\\
\hline\noalign {\smallskip}
TAU\,06&30&R&\phantom{0}9.0&14.0&\phantom{0}6.8&5.4$\times$10$^{-04}$&1.5$\times$10$^{-03}$&4.6$\times$10$^{+40}$&26.4&18.4&2.2$\times$10$^{-07}$&8.3$\times$10$^{-08}$\\
&&B&\phantom{0}5.5&\phantom{0}0.0&\phantom{0}7.2&5.3$\times$10$^{-04}$&1.3$\times$10$^{-03}$&3.7$\times$10$^{+40}$&23.6&15.5&2.3$\times$10$^{-07}$&9.5$\times$10$^{-08}$\\
\hline\noalign {\smallskip}
TAU\,07&$-$\tablefootmark{c}&$-$&$-$&$-$&$-$&$-$&$-$&$-$&$-$&$-$&$-$&$-$\\
\hline\noalign {\smallskip}
TAU\,08&$-$\tablefootmark{c}&$-$&$-$&$-$&$-$&$-$&$-$&$-$&$-$&$-$&$-$&$-$\\
\hline\noalign {\smallskip}
TAU\,09&$-$\tablefootmark{c}&$-$&$-$&$-$&$-$&$-$&$-$&$-$&$-$&$-$&$-$&$-$\\
\hline\noalign {\smallskip}
W40\,01&$-$\tablefootmark{c}&$-$&$-$&$-$&$-$&$-$&$-$&$-$&$-$&$-$&$-$&$-$\\
\hline\noalign {\smallskip}
W40\,02&70&R&11.5&19.0&14.2&2.4$\times$10$^{-03}$&2.1$\times$10$^{-02}$&1.9$\times$10$^{+42}$&15.8&\phantom{0}5.3&2.8$\times$10$^{-05}$&3.2$\times$10$^{-06}$\\
&&B&\phantom{0}2.5&\phantom{0}$-$6.5\phantom{$-$}&11.3&1.1$\times$10$^{-02}$&4.8$\times$10$^{-02}$&2.5$\times$10$^{+42}$&26.4&11.0&3.1$\times$10$^{-05}$&6.9$\times$10$^{-06}$\\
\hline\noalign {\smallskip}
W40\,03&$-$\tablefootmark{c}&$-$&$-$&$-$&$-$&$-$&$-$&$-$&$-$&$-$&$-$&$-$\\
\hline\noalign {\smallskip}
W40\,04&$-$\tablefootmark{c}&$-$&$-$&$-$&$-$&$-$&$-$&$-$&$-$&$-$&$-$&$-$\\
\hline\noalign {\smallskip}
W40\,05&$-$\tablefootmark{c}&$-$&$-$&$-$&$-$&$-$&$-$&$-$&$-$&$-$&$-$&$-$\\
\hline\noalign {\smallskip}
W40\,06&$-$\tablefootmark{c}&$-$&$-$&$-$&$-$&$-$&$-$&$-$&$-$&$-$&$-$&$-$\\
\hline\noalign {\smallskip}
W40\,07&50&R&10.5&26.0&18.6&5.4$\times$10$^{-03}$&4.6$\times$10$^{-02}$&5.1$\times$10$^{+42}$&26.4&\phantom{0}6.7&3.0$\times$10$^{-05}$&3.5$\times$10$^{-06}$\\
&&B&\phantom{0}3.0&\phantom{0}$-$4.0\phantom{$-$}&11.4&1.2$\times$10$^{-03}$&8.7$\times$10$^{-03}$&6.9$\times$10$^{+41}$&\phantom{0}4.2&\phantom{0}1.7&2.2$\times$10$^{-05}$&3.0$\times$10$^{-06}$\\
\hline\noalign {\smallskip}
\end{tabular}
\vspace{-2mm}
\tablefoot{\tablefoottext{a}{Measured values, that is, not corrected for inclination.} \tablefoottext{b}{Corrected for inclination as discussed in detail in \citet{vanderMarel2013} and \citet{Yildiz2015}.} \tablefoottext{c}{No outflow associated with this source.} \tablefoottext{d}{An outflow is detected but is too contaminated by outflows from others sources for meaningful determinations to be made. See Sect.~\ref{S:cases} for details.}}
\end{center}
\end{table*}

\renewcommand{\thetable}{\thesection.\arabic{table}}

\subsection{Mass accretion and loss}
\label{S:properties_mdot}

Table~\ref{T:properties_mdot} presents the calculated mass accretion rates for all WILL, WISH and DIGIT sources, calculated as discussed in Sect.~\ref{S:outflow_mdot} using Eqn.~\ref{E:mdotacc}, as well as the observed luminosity in the [O\,{\sc i}] 63\micron{} transition and the mass-loss rate in the wind derived using the relation from \citet{Hollenbach1985}, given in Eqn.~\ref{E:mdotOI}. $L\mathrm{[OI\,63\mu m]}$ is not given for OPH\,01, W40\,01 and W40\,03$-$06 as the detections in [O\,{\sc i}] towards these sources are almost certainly due to PDR emission.

\begin{table*}
\caption{Mass-accretion rate, the luminosity of [O\,{\sc i}] in the 63\micron{} line, and mass-loss rate calculated from it.}
\begin{center}
\vspace{-2mm}
\begin{tabular}{lcccclccc}
\hline \noalign {\smallskip}
Name & $\dot{M}_{\mathrm{acc}}$ & $L$[O\,{\sc i}~63$\mu$m] & $\dot{M}_{\mathrm{w}}$ &~~~& Name & $\dot{M}_{\mathrm{acc}}$ & $L$[O\,{\sc i}~63$\mu$m] & $\dot{M}_{\mathrm{w}}$ \\
 & (\msol{}\,yr$^{-1}$) & (\lsol{}) & (\msol{}\,yr$^{-1}$) && & (\msol{}\,yr$^{-1}$) & (\lsol{}) & (\msol{}\,yr$^{-1}$) \\ 
\hline\noalign {\smallskip}
AQU\,01&1.7$\times$10$^{-06}$&8.3$\times$10$^{-04}$&8.3$\times$10$^{-08}$&&AQU\,02&5.9$\times$10$^{-06}$&8.2$\times$10$^{-04}$&8.2$\times$10$^{-08}$\\
AQU\,03&\phantom{\tablefootmark{a}}$-$\tablefootmark{a}&\phantom{\tablefootmark{b}}$-$\tablefootmark{b}&$-$&&AQU\,04&8.4$\times$10$^{-07}$&\phantom{\tablefootmark{b}}$-$\tablefootmark{b}&$-$\\
AQU\,05&1.6$\times$10$^{-06}$&\phantom{\tablefootmark{b}}$-$\tablefootmark{b}&$-$&&AQU\,06&8.1$\times$10$^{-07}$&\phantom{\tablefootmark{b}}$-$\tablefootmark{b}&$-$\\
CHA\,01&\phantom{\tablefootmark{a}}$-$\tablefootmark{a}&3.7$\times$10$^{-04}$&3.7$\times$10$^{-08}$&&CHA\,02&2.3$\times$10$^{-07}$&8.7$\times$10$^{-05}$&8.7$\times$10$^{-09}$\\
CRA\,01&1.6$\times$10$^{-06}$&1.2$\times$10$^{-03}$&1.2$\times$10$^{-07}$&&OPH\,01&\phantom{\tablefootmark{a}}$-$\tablefootmark{a}&\phantom{\tablefootmark{c}}2.9$\times$10$^{-03}$\tablefootmark{c}&$-$\\
OPH\,02&1.1$\times$10$^{-06}$&5.2$\times$10$^{-04}$&5.2$\times$10$^{-08}$&&PER\,01&2.9$\times$10$^{-06}$&3.4$\times$10$^{-03}$&3.4$\times$10$^{-07}$\\
PER\,02&6.0$\times$10$^{-06}$&7.8$\times$10$^{-03}$&7.8$\times$10$^{-07}$&&PER\,04&7.6$\times$10$^{-07}$&5.3$\times$10$^{-04}$&5.3$\times$10$^{-08}$\\
PER\,05&1.4$\times$10$^{-06}$&2.8$\times$10$^{-04}$&2.8$\times$10$^{-08}$&&PER\,06&9.2$\times$10$^{-07}$&1.6$\times$10$^{-03}$&1.6$\times$10$^{-07}$\\
PER\,07&4.6$\times$10$^{-07}$&1.2$\times$10$^{-03}$&1.2$\times$10$^{-07}$&&PER\,08&2.2$\times$10$^{-06}$&1.1$\times$10$^{-02}$&1.1$\times$10$^{-06}$\\
PER\,09&2.9$\times$10$^{-06}$&\phantom{\tablefootmark{e}}$-$\tablefootmark{e}&$-$&&PER\,10&3.9$\times$10$^{-06}$&\phantom{\tablefootmark{e}}$-$\tablefootmark{e}&$-$\\
PER\,12&6.9$\times$10$^{-07}$&1.1$\times$10$^{-04}$&1.1$\times$10$^{-08}$&&PER\,13&4.4$\times$10$^{-07}$&\phantom{\tablefootmark{b}}$-$\tablefootmark{b}&$-$\\
PER\,14&2.4$\times$10$^{-07}$&4.6$\times$10$^{-04}$&4.6$\times$10$^{-08}$&&PER\,15&1.1$\times$10$^{-06}$&2.1$\times$10$^{-04}$&2.1$\times$10$^{-08}$\\
PER\,16&7.4$\times$10$^{-07}$&2.6$\times$10$^{-04}$&2.6$\times$10$^{-08}$&&PER\,17&2.3$\times$10$^{-08}$&1.4$\times$10$^{-04}$&1.4$\times$10$^{-08}$\\
PER\,18&3.4$\times$10$^{-07}$&1.8$\times$10$^{-03}$&1.8$\times$10$^{-07}$&&PER\,19&1.4$\times$10$^{-07}$&1.8$\times$10$^{-03}$&1.8$\times$10$^{-07}$\\
PER\,20&1.4$\times$10$^{-06}$&1.9$\times$10$^{-03}$&1.9$\times$10$^{-07}$&&PER\,21&1.2$\times$10$^{-06}$&8.6$\times$10$^{-04}$&8.6$\times$10$^{-08}$\\
PER\,22&1.6$\times$10$^{-06}$&1.0$\times$10$^{-03}$&1.0$\times$10$^{-07}$&&SCO\,01&\phantom{\tablefootmark{a}}$-$\tablefootmark{a}&2.8$\times$10$^{-04}$&2.8$\times$10$^{-08}$\\
SERS\,01&1.1$\times$10$^{-05}$&2.8$\times$10$^{-03}$&2.8$\times$10$^{-07}$&&SERS\,02&4.7$\times$10$^{-05}$&\phantom{\tablefootmark{e}}$-$\tablefootmark{e}&$-$\\
TAU\,01&2.0$\times$10$^{-07}$&7.6$\times$10$^{-04}$&7.6$\times$10$^{-08}$&&TAU\,02&6.1$\times$10$^{-08}$&1.1$\times$10$^{-04}$&1.1$\times$10$^{-08}$\\
TAU\,03&\phantom{\tablefootmark{a}}$-$\tablefootmark{a}&5.4$\times$10$^{-05}$&5.4$\times$10$^{-09}$&&TAU\,04&1.8$\times$10$^{-07}$&4.5$\times$10$^{-04}$&4.5$\times$10$^{-08}$\\
TAU\,06&7.3$\times$10$^{-08}$&9.6$\times$10$^{-05}$&9.6$\times$10$^{-09}$&&TAU\,07&\phantom{\tablefootmark{a}}$-$\tablefootmark{a}&3.5$\times$10$^{-04}$&3.5$\times$10$^{-08}$\\
TAU\,08&\phantom{\tablefootmark{a}}$-$\tablefootmark{a}&\phantom{\tablefootmark{f}}$-$\tablefootmark{f}&$-$&&TAU\,09&\phantom{\tablefootmark{a}}$-$\tablefootmark{a}&2.9$\times$10$^{-04}$&2.9$\times$10$^{-08}$\\
W40\,01&\phantom{\tablefootmark{a}}$-$\tablefootmark{a}&\phantom{\tablefootmark{c}}3.1$\times$10$^{-04}$\tablefootmark{c}&$-$&&W40\,02&2.1$\times$10$^{-05}$&5.4$\times$10$^{-02}$&5.4$\times$10$^{-06}$\\
W40\,03&\phantom{\tablefootmark{a}}$-$\tablefootmark{a}&\phantom{\tablefootmark{c}}1.5$\times$10$^{-01}$\tablefootmark{c}&$-$&&W40\,04&\phantom{\tablefootmark{a}}$-$\tablefootmark{a}&\phantom{\tablefootmark{c}}1.1$\times$10$^{-01}$\tablefootmark{c}&$-$\\
W40\,05&\phantom{\tablefootmark{a}}$-$\tablefootmark{a}&\phantom{\tablefootmark{c}}7.8$\times$10$^{-02}$\tablefootmark{c}&$-$&&W40\,06&\phantom{\tablefootmark{a}}$-$\tablefootmark{a}&\phantom{\tablefootmark{c,d}}0.0$\times$10$^{+00}$\tablefootmark{c,d}&$-$\\
W40\,07&2.3$\times$10$^{-06}$&5.4$\times$10$^{-04}$&5.4$\times$10$^{-08}$&&L1448-MM&5.8$\times$10$^{-06}$&2.2$\times$10$^{-03}$&2.2$\times$10$^{-07}$\\
IRAS03245+3002&4.2$\times$10$^{-06}$&6.4$\times$10$^{-04}$&6.4$\times$10$^{-08}$&&L1455-IRS3&4.7$\times$10$^{-08}$&5.0$\times$10$^{-04}$&5.0$\times$10$^{-08}$\\
NGC1333-IRAS2A&2.3$\times$10$^{-05}$&\phantom{\tablefootmark{b}}$-$\tablefootmark{b}&$-$&&NGC1333-IRAS3A&5.4$\times$10$^{-06}$&\phantom{\tablefootmark{f}}$-$\tablefootmark{f}&$-$\\
NGC1333-IRAS4A&5.9$\times$10$^{-06}$&\phantom{\tablefootmark{g}}1.5$\times$10$^{-03}$\tablefootmark{g}&$-$&&NGC1333-IRAS4B&2.9$\times$10$^{-06}$&1.3$\times$10$^{-04}$&1.3$\times$10$^{-08}$\\
IRAS03301+3111&5.3$\times$10$^{-07}$&3.6$\times$10$^{-04}$&3.6$\times$10$^{-08}$&&B1\,a&3.5$\times$10$^{-07}$&1.9$\times$10$^{-04}$&1.9$\times$10$^{-08}$\\
B1\,c&3.3$\times$10$^{-06}$&\phantom{\tablefootmark{b}}$-$\tablefootmark{b}&$-$&&L1489&4.9$\times$10$^{-07}$&2.2$\times$10$^{-04}$&2.2$\times$10$^{-08}$\\
L1551-IRS5&2.9$\times$10$^{-06}$&9.7$\times$10$^{-03}$&9.7$\times$10$^{-07}$&&TMR1&4.9$\times$10$^{-07}$&1.1$\times$10$^{-03}$&1.1$\times$10$^{-07}$\\
TMC1A&3.5$\times$10$^{-07}$&3.3$\times$10$^{-03}$&3.3$\times$10$^{-07}$&&L1527&1.2$\times$10$^{-06}$&1.8$\times$10$^{-03}$&1.8$\times$10$^{-07}$\\
TMC1&1.2$\times$10$^{-07}$&1.5$\times$10$^{-03}$&1.5$\times$10$^{-07}$&&IRAM04191&7.1$\times$10$^{-08}$&2.4$\times$10$^{-04}$&2.4$\times$10$^{-08}$\\
HH46&3.6$\times$10$^{-06}$&5.1$\times$10$^{-03}$&5.1$\times$10$^{-07}$&&Ced110-IRS4&5.2$\times$10$^{-07}$&1.3$\times$10$^{-03}$&1.3$\times$10$^{-07}$\\
BHR71&9.6$\times$10$^{-06}$&3.4$\times$10$^{-03}$&3.4$\times$10$^{-07}$&&IRAS12496&4.6$\times$10$^{-06}$&5.9$\times$10$^{-03}$&5.9$\times$10$^{-07}$\\
IRAS15398&1.0$\times$10$^{-06}$&3.0$\times$10$^{-03}$&3.0$\times$10$^{-07}$&&GSS30&1.8$\times$10$^{-06}$&5.1$\times$10$^{-03}$&5.1$\times$10$^{-07}$\\
VLA1623-243&1.7$\times$10$^{-06}$&1.3$\times$10$^{-03}$&1.3$\times$10$^{-07}$&&WL12&2.1$\times$10$^{-07}$&7.9$\times$10$^{-04}$&7.9$\times$10$^{-08}$\\
Elias29&1.8$\times$10$^{-06}$&4.0$\times$10$^{-03}$&4.0$\times$10$^{-07}$&&IRS44&6.6$\times$10$^{-07}$&4.8$\times$10$^{-04}$&4.8$\times$10$^{-08}$\\
IRS46&6.5$\times$10$^{-08}$&1.5$\times$10$^{-04}$&1.5$\times$10$^{-08}$&&IRS63&1.3$\times$10$^{-07}$&3.3$\times$10$^{-04}$&3.3$\times$10$^{-08}$\\
RNO91&3.4$\times$10$^{-07}$&1.3$\times$10$^{-03}$&1.3$\times$10$^{-07}$&&L483&6.6$\times$10$^{-06}$&1.1$\times$10$^{-03}$&1.1$\times$10$^{-07}$\\
Ser-SMM1&7.1$\times$10$^{-05}$&4.0$\times$10$^{-03}$&4.0$\times$10$^{-07}$&&Ser-SMM4&4.4$\times$10$^{-06}$&1.8$\times$10$^{-02}$&1.8$\times$10$^{-06}$\\
Ser-SMM3&1.2$\times$10$^{-05}$&1.1$\times$10$^{-02}$&1.1$\times$10$^{-06}$&&RCrA-IRS5A&9.2$\times$10$^{-07}$&1.2$\times$10$^{-03}$&1.2$\times$10$^{-07}$\\
RCrA\,I7A&1.2$\times$10$^{-05}$&8.9$\times$10$^{-03}$&8.9$\times$10$^{-07}$&&RCrA\,I7B&6.0$\times$10$^{-06}$&2.4$\times$10$^{-03}$&2.4$\times$10$^{-07}$\\
HH100&2.3$\times$10$^{-06}$&\phantom{\tablefootmark{f}}$-$\tablefootmark{f}&$-$&&L723&2.3$\times$10$^{-06}$&1.3$\times$10$^{-04}$&1.3$\times$10$^{-08}$\\
B335&2.1$\times$10$^{-06}$&2.5$\times$10$^{-04}$&2.5$\times$10$^{-08}$&&L1157&3.0$\times$10$^{-06}$&3.3$\times$10$^{-04}$&3.3$\times$10$^{-08}$\\
\hline\noalign {\smallskip}
\end{tabular}
\vspace{-2mm}
\tablefoot{\tablefoottext{a}{Not Class 0/I.} \tablefoottext{b}{Non-detection.} \tablefoottext{c}{Dominated by PDR.} \tablefoottext{d}{Line in absorption.} \tablefoottext{e}{Dominated by neighbouring source.} \tablefoottext{f}{Source not observed with PACS.} \tablefoottext{g}{Only one off position was used due to possible contamination but this flux may still be underestimated.}}
\end{center}
\label{T:properties_mdot}
\end{table*}

\subsection{Ground-based line fitting}
\label{S:properties_ground}

\begin{table*}
\begin{center}
\caption[]{Species and transitions targeted during the ground-based spectral follow-up.}
\vspace{-2mm}
\begin{tabular}{lcccccccccc}
\hline \noalign {\smallskip}
Species & Transition & Rest Frequency\tablefootmark{a}& E$_{\mathrm{u}}$/$k_{\mathrm{b}}$ & $A_{\mathrm{ul}}$ & $n_{\mathrm{cr}}$ & JCMT $\eta_{mb}$\tablefootmark{b} & APEX $\eta_{mb}$\tablefootmark{c} & JCMT $\theta_{mb}$ & APEX $\theta_{mb}$ & Median\,$\sigma_{\mathrm{rms}}$\tablefootmark{d} \\
& & (GHz) & (K) & (s$^{-1}$) & (cm$^{-3}$) & & & (\arcsec{}) & (\arcsec{}) & (K) \\
\hline\noalign {\smallskip}
CO & 2$-$1 & 230.53800 & 16.60 & 6.91$\times$10$^{-7}$ & 6$\times$10$^{3}$ & $-$ & 0.75 & $-$ & 27.3 & 0.41\\
$^{13}$CO & 2$-$1 & 220.39868 & 15.87 & 6.04$\times$10$^{-7}$ & 5$\times$10$^{3}$ & $-$ & 0.75 & $-$ & 28.5 & 0.08\\
C$^{18}$O & 2$-$1 & 219.56035 & 15.81 & 6.01$\times$10$^{-7}$ & 5$\times$10$^{3}$ & $-$ & 0.75 & $-$ & 28.6 & 0.08\\
C$^{17}$O & 2$-$1 & 224.71439 & 16.18 & 6.43$\times$10$^{-7}$ & 6$\times$10$^{3}$ & $-$ & 0.75 & $-$ & 28.0 & 0.08\\
CO & 3$-$2 & 345.79599 & 33.19 & 2.50$\times$10$^{-6}$ & 2$\times$10$^{4}$ & 0.63 & 0.73 & 14.5 & 18.2 & 0.09\\
$^{13}$CO & 3$-$2 & 330.58797 & 31.73 & 2.18$\times$10$^{-6}$ & 1$\times$10$^{4}$ & 0.63 & 0.73 & 15.2 & 19.0 & 0.22\\
C$^{18}$O & 3$-$2 & 329.33055 & 31.61 & 2.17$\times$10$^{-6}$ & 1$\times$10$^{4}$ & 0.63 & 0.73 & 15.3 & 19.1 & 0.30\\
C$^{17}$O & 3$-$2 & 337.06113 & 32.35 & 2.32$\times$10$^{-6}$ & 1$\times$10$^{4}$ & $-$ & 0.73 & $-$ & 18.7 & 0.09\\
CO & 4$-$3 & 461.04077 & 55.32 & 6.13$\times$10$^{-6}$ & 3$\times$10$^{4}$ & $-$ & 0.60 & $-$ & 13.6 & 0.12 \\
\hline\noalign {\smallskip}
HCO$^{+}$ & 4$-$3 & 356.73429 & 42.80 & 3.63$\times$10$^{-3}$ & 2$\times$10$^{7}$ & 0.63 & 0.73 & 14.1 & 17.6 & 0.28\\
H$^{13}$CO$^{+}$ & 4$-$3 & 346.99834 & 41.63 & 3.29$\times$10$^{-3}$ & 2$\times$10$^{7}$ & 0.63 & 0.73 & 14.5 & 18.2 & 0.13\\
\hline\noalign {\smallskip}
\end{tabular}
\tablefoot{\tablefoottext{a}{Taken from the JPL database \citep{Pickett2010}.} \tablefoottext{b}{Historical value from long-term monitoring of standards used for consistency with WISH results \citep[e.g.][]{Yildiz2015}} \tablefoottext{c}{http://www.apex-telescope.org/telescope/efficiency/} \tablefoottext{d}{In 0.2\kms{} channels except for CO $J$=3$-$2, which is in 0.5\kms{} channels.}}
\end{center}
\vspace{-2mm}
\label{T:observations_jcmtapex}
\end{table*}

Table~\ref{T:observations_jcmtapex} presents the basic properties of all transitions observed with the JCMT and APEX. Gaussian fitting was performed for the central spectra of all sources for $^{13}$CO, C$^{18}$O, HCO$^{+}$ and H$^{13}$CO$^{+}$ in order to determine line-widths and central velocities. The $\varv_{\mathrm{LSR}}$ for each source is defined as the peak position of the fit to the C$^{18}$O $J$=3$-$2 central spectrum. For the four sources where this is not detected, three (SCO\,01, TAU\,03, TAU\,07) were also observed and detected in $^{13}$CO $J$=3$-$2, so the fit to this line is used instead. For the remaining source (TAU\,08) we use the velocity derived by \cite{Caselli2002} from N$_{2}$H$^{+}$ $J$=1$-$0 observations. The integrated intensity of the lines is measured over a window of $\pm$3\,FWHM. 

For the optically thick HCO$^{+}$ and $^{13}$CO, we also quantify any blue/red asymmetry by calculating $\delta\varv$ as defined by \citet[][]{Mardones1997} using Eqn.~\ref{E:dv}. For most sources, we use the peak of the Gaussian fit to the line, but in the case of five sources (AQU\,02, PER\,04, SERS\,01, SERS\,02 and TAU\,09), the position of the maximum intensity is used in this calculation for HCO$^{+}$ because strong self-absorption or a broad, fainter second component skews the Gaussian fit away from the maximum intensity. 

\begin{figure*}
\begin{center}
\includegraphics[width=0.9\textwidth]{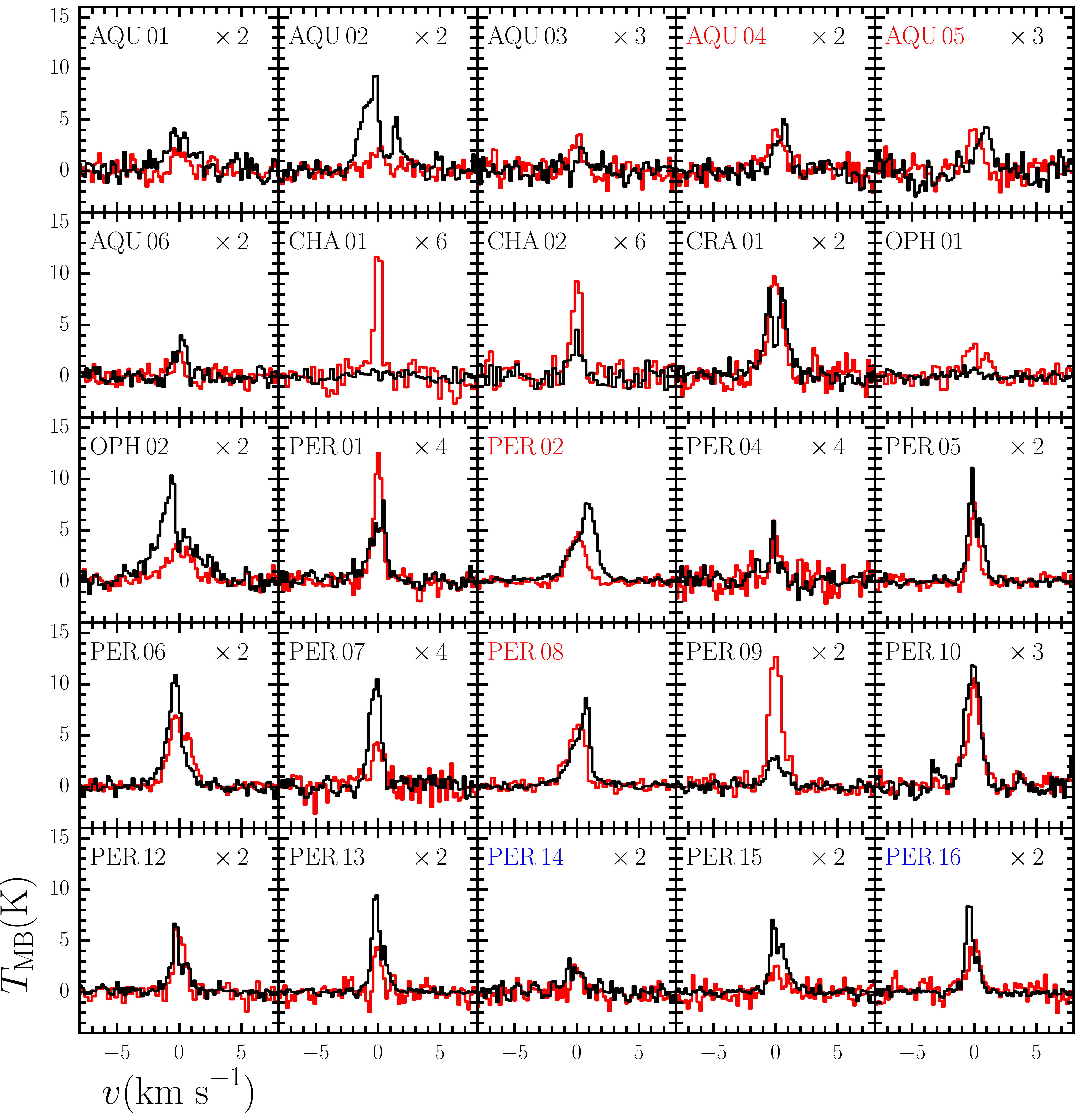}
\caption{Overview of the central HCO$^{+}$ $J$=4$-$3 and C$^{18}$O $J$=3$-$2 spectra shown in black and red respectively. All have been recentred so that the source velocity is at zero.  Sources that are considered blue-skewed (i.e. $\delta\varv<-$0.25) have their names in blue, while those considered red-skewed (i.e. $\delta\varv>$0.25) have their names in red. The number in the upper-right corner of each panel indicates what factor the spectra have been multiplied by.}
\label{F:hcop_c18o}
\end{center}
\end{figure*}

The results for all sources are presented in Table~\ref{T:ground_spectra}, and Fig.~\ref{F:hcop_c18o} shows the central HCO$^{+}$ and C$^{18}$O spectra for all sources.

\renewcommand{\thefigure}{\thesection.\arabic{figure} (Cont.)}
\addtocounter{figure}{-1}

\begin{figure*}
\begin{center}
\includegraphics[width=0.9\textwidth]{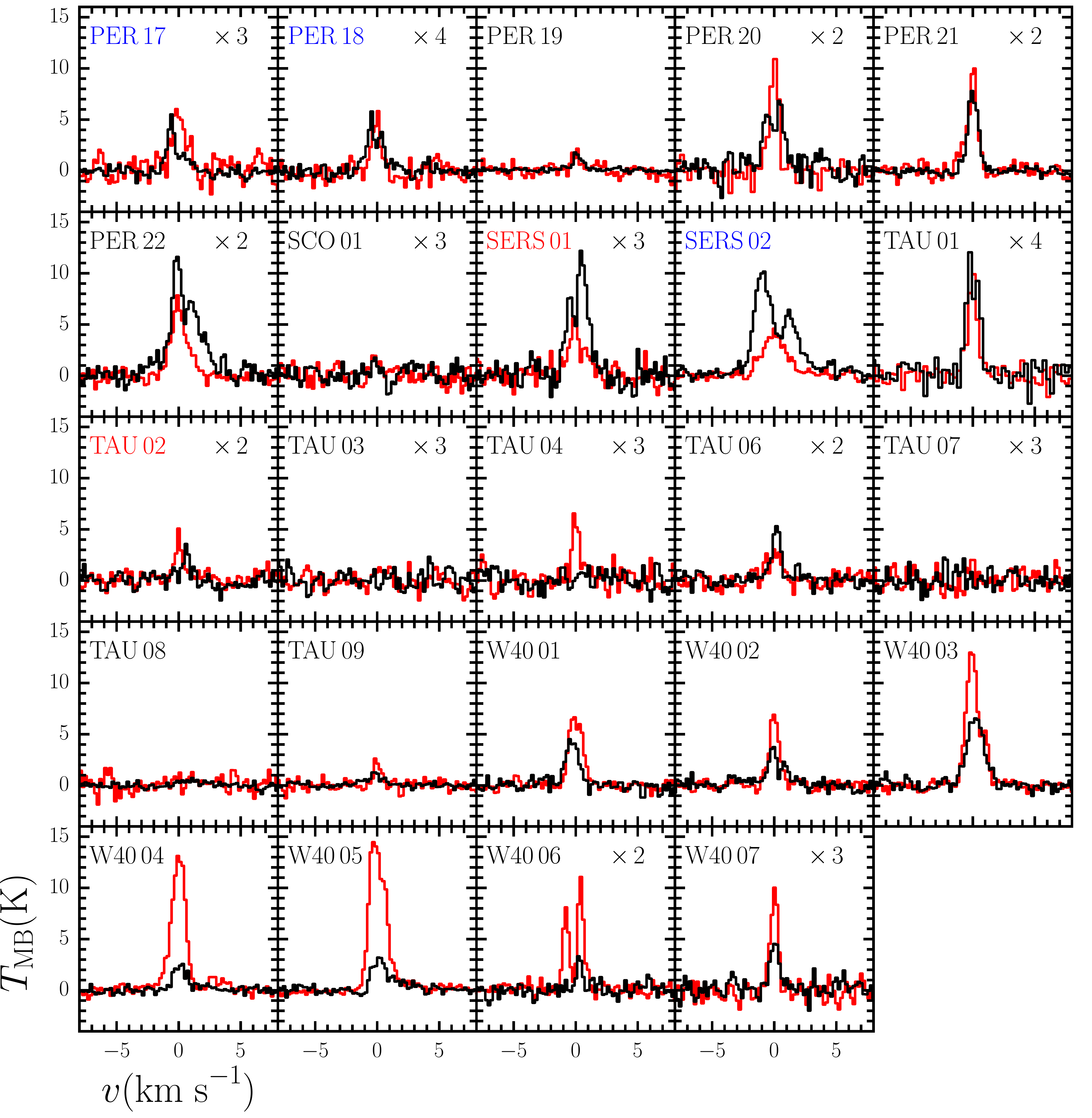}
\caption{Overview of the central HCO$^{+}$ $J$=4$-$3 and C$^{18}$O $J$=3$-$2 spectra shown in black and red respectively. All have been recentred so that the source velocity is at zero.  Sources that are considered blue-skewed (i.e. $\delta\varv<-$0.25) have their names in blue, while those considered red-skewed (i.e. $\delta\varv>$0.25) have their names in red. The number in the upper-right corner of each panel indicates what factor the spectra have been multiplied by.}
\end{center}
\end{figure*}

\renewcommand{\thefigure}{\thesection.\arabic{figure}}

\begin{table*}
\caption{Ground-based spectral results.}
\begin{center}
\vspace{-2mm}
\begin{tabular}{llccccccccc}
\hline \noalign {\smallskip}
Name & Molecule & Transition & $\sigma_{\mathrm{rms}}$ & $\varv_{\mathrm{max}}$ & T$_{\mathrm{max}}$ & $\varv_{\mathrm{peak}}$\tablefootmark{a} & T$_{\mathrm{peak}}$\tablefootmark{a} & FWHM\tablefootmark{a} & $\int T_{\mathrm{MB}}\,\mathrm{d}\varv$\tablefootmark{b} & $\delta\varv$\tablefootmark{c} \\
 & & & (K) & (km\,s$^{-1}$) & (K) & (km\,s$^{-1}$) & (K) & (km\,s$^{-1}$) & (K\kms{}) & (km\,s$^{-1}$) \\ 
\hline\noalign {\smallskip}
AQU\,01&$^{13}$CO&3$-$2&0.25&7.2&\phantom{0}2.33&7.3$\pm$0.1&\phantom{0}2.1$\pm$0.2&1.9$\pm$0.2&\phantom{0}4.23$\pm$0.57&$-$0.07$\pm$0.03\phantom{$-$}\\
&C$^{18}$O&3$-$2&0.32&7.4&\phantom{0}1.48&7.4$\pm$0.2&\phantom{0}1.2$\pm$0.4&1.1$\pm$0.4&\phantom{0}1.34$\pm$0.74&0.00$\pm$0.01\\
&HCO$^{+}$&4$-$3&0.39&7.0&\phantom{0}2.08&7.4$\pm$0.2&\phantom{0}1.6$\pm$0.2&2.6$\pm$0.4&\phantom{0}4.56$\pm$0.89&$-$0.04$\pm$0.01\phantom{$-$}\\
&H$^{13}$CO$^{+}$&4$-$3&0.18&$-$&$<$0.55&$-$&$-$&$-$&$-$&$-$\\
AQU\,02&$^{13}$CO&3$-$2&0.21&7.6&\phantom{0}4.47&7.5$\pm$0.1&\phantom{0}4.2$\pm$0.2&1.8$\pm$0.1&\phantom{0}7.95$\pm$0.48&$-$0.01$\pm$0.01\phantom{$-$}\\
&C$^{18}$O&3$-$2&0.26&7.8&\phantom{0}1.19&7.5$\pm$0.2&\phantom{0}1.0$\pm$0.3&1.2$\pm$0.4&\phantom{0}1.38$\pm$0.60&0.00$\pm$0.01\\
&HCO$^{+}$\,\tablefootmark{d}&4$-$3&0.30&7.4&\phantom{0}4.63&7.6$\pm$0.1&\phantom{0}4.7$\pm$0.3&2.7$\pm$0.1&13.28$\pm$0.94&\phantom{\,\tablefootmark{e}}$-$0.09$\pm$0.03\,\tablefootmark{e}\phantom{$-$}\\
AQU\,03&$^{13}$CO&3$-$2&0.21&7.4&\phantom{0}2.33&7.1$\pm$0.1&\phantom{0}2.2$\pm$0.2&1.4$\pm$0.2&\phantom{0}3.27$\pm$0.49&0.02$\pm$0.01\\
&C$^{18}$O&3$-$2&0.28&7.4&\phantom{0}1.20&7.1$\pm$0.2&\phantom{0}1.1$\pm$0.4&1.1$\pm$0.4&\phantom{0}1.22$\pm$0.65&0.00$\pm$0.01\\
&HCO$^{+}$&4$-$3&0.28&$-$&$<$0.85&$-$&$-$&$-$&$-$&$-$\\
&H$^{13}$CO$^{+}$&4$-$3&0.25&$-$&$<$0.75&$-$&$-$&$-$&$-$&$-$\\
AQU\,04&$^{13}$CO&3$-$2&0.22&7.4&\phantom{0}3.90&7.5$\pm$0.1&\phantom{0}3.2$\pm$0.1&2.8$\pm$0.1&\phantom{0}9.29$\pm$0.55&$-$0.01$\pm$0.01\phantom{$-$}\\
&C$^{18}$O&3$-$2&0.36&7.6&\phantom{0}2.03&7.6$\pm$0.1&\phantom{0}1.8$\pm$0.3&1.6$\pm$0.3&\phantom{0}3.07$\pm$0.83&0.00$\pm$0.01\\
&HCO$^{+}$&4$-$3&0.33&8.2&\phantom{0}2.52&8.1$\pm$0.1&\phantom{0}2.1$\pm$0.4&1.3$\pm$0.3&\phantom{0}2.81$\pm$0.78&0.33$\pm$0.07\\
&H$^{13}$CO$^{+}$&4$-$3&0.19&$-$&$<$0.56&$-$&$-$&$-$&$-$&$-$\\
AQU\,05&$^{13}$CO&3$-$2&0.27&7.0&\phantom{0}2.79&7.4$\pm$0.1&\phantom{0}1.9$\pm$0.1&2.9$\pm$0.2&\phantom{0}5.91$\pm$0.66&0.10$\pm$0.03\\
&C$^{18}$O&3$-$2&0.30&7.4&\phantom{0}1.36&7.3$\pm$0.2&\phantom{0}1.4$\pm$0.4&1.2$\pm$0.4&\phantom{0}1.72$\pm$0.69&0.00$\pm$0.01\\
&HCO$^{+}$&4$-$3&0.31&8.2&\phantom{0}1.44&8.2$\pm$0.2&\phantom{0}1.4$\pm$0.3&1.3$\pm$0.4&\phantom{0}1.98$\pm$0.71&0.70$\pm$0.21\\
&H$^{13}$CO$^{+}$&4$-$3&0.14&$-$&$<$0.43&$-$&$-$&$-$&$-$&$-$\\
AQU\,06&$^{13}$CO&3$-$2&0.25&8.4&\phantom{0}2.52&8.3$\pm$0.1&\phantom{0}2.0$\pm$0.2&1.9$\pm$0.2&\phantom{0}4.11$\pm$0.57&0.11$\pm$0.07\\
&C$^{18}$O&3$-$2&0.33&8.4&\phantom{0}1.19&8.3$\pm$0.2&\phantom{0}1.3$\pm$0.7&0.6$\pm$0.4&\phantom{0}0.90$\pm$0.76&0.00$\pm$0.01\\
&HCO$^{+}$&4$-$3&0.29&8.4&\phantom{0}2.02&8.4$\pm$0.1&\phantom{0}1.8$\pm$0.4&1.0$\pm$0.3&\phantom{0}1.96$\pm$0.67&0.22$\pm$0.14\\
&H$^{13}$CO$^{+}$&4$-$3&0.16&$-$&$<$0.47&$-$&$-$&$-$&$-$&$-$\\
CHA\,01&$^{13}$CO&2$-$1&0.08&5.2&\phantom{0}6.63&4.9$\pm$0.1&\phantom{0}5.9$\pm$0.2&1.5$\pm$0.1&\phantom{0}9.62$\pm$0.43&$-$0.01$\pm$0.01\phantom{$-$}\\
&C$^{18}$O&2$-$1&0.08&4.7&\phantom{0}4.44&4.7$\pm$0.1&\phantom{0}4.8$\pm$0.3&0.7$\pm$0.1&\phantom{0}3.63$\pm$0.39&0.00$\pm$0.01\\
&C$^{17}$O&2$-$1&0.08&5.2&\phantom{0}1.47&5.2$\pm$0.1&\phantom{0}1.4$\pm$0.3&0.7$\pm$0.2&\phantom{0}1.09$\pm$0.37&0.00$\pm$0.01\\
&$^{13}$CO&3$-$2&0.12&5.1&\phantom{0}4.04&4.8$\pm$0.1&\phantom{0}3.7$\pm$0.2&1.6$\pm$0.1&\phantom{0}6.43$\pm$0.55&$-$0.13$\pm$0.05\phantom{$-$}\\
&C$^{18}$O&3$-$2&0.16&4.8&\phantom{0}1.94&4.9$\pm$0.1&\phantom{0}2.2$\pm$0.7&0.6$\pm$0.2&\phantom{0}1.49$\pm$0.77&0.00$\pm$0.01\\
&C$^{17}$O&3$-$2&0.08&5.1&\phantom{0}0.45&4.8$\pm$0.2&\phantom{0}0.4$\pm$0.2&0.9$\pm$0.6&\phantom{0}0.43$\pm$0.35&0.00$\pm$0.01\\
&HCO$^{+}$&4$-$3&0.07&$-$&$<$0.20&$-$&$-$&$-$&$-$&$-$\\
&H$^{13}$CO$^{+}$&4$-$3&0.06&$-$&$<$0.18&$-$&$-$&$-$&$-$&$-$\\
CHA\,02&$^{13}$CO&2$-$1&0.09&3.5&\phantom{0}4.66&3.2$\pm$0.1&\phantom{0}4.7$\pm$0.2&1.3$\pm$0.1&\phantom{0}6.71$\pm$0.47&0.18$\pm$0.07\\
&C$^{18}$O&2$-$1&0.08&3.1&\phantom{0}3.60&3.0$\pm$0.1&\phantom{0}3.7$\pm$0.3&0.8$\pm$0.1&\phantom{0}2.98$\pm$0.43&0.00$\pm$0.01\\
&C$^{17}$O&2$-$1&0.08&3.5&\phantom{0}1.36&3.5$\pm$0.1&\phantom{0}1.2$\pm$0.3&0.8$\pm$0.2&\phantom{0}1.06$\pm$0.37&0.00$\pm$0.01\\
&$^{13}$CO&3$-$2&0.13&3.6&\phantom{0}2.24&3.0$\pm$0.1&\phantom{0}2.0$\pm$0.2&1.8$\pm$0.2&\phantom{0}3.84$\pm$0.59&$-$0.01$\pm$0.01\phantom{$-$}\\
&C$^{18}$O&3$-$2&0.17&3.0&\phantom{0}1.54&3.0$\pm$0.1&\phantom{0}1.6$\pm$0.5&0.9$\pm$0.3&\phantom{0}1.47$\pm$0.75&0.00$\pm$0.01\\
&C$^{17}$O&3$-$2&0.10&3.3&\phantom{0}0.40&3.2$\pm$0.3&\phantom{0}0.4$\pm$0.3&0.8$\pm$0.8&\phantom{0}0.31$\pm$0.43&0.00$\pm$0.01\\
&HCO$^{+}$&4$-$3&0.11&3.0&\phantom{0}0.76&3.0$\pm$0.2&\phantom{0}0.6$\pm$0.3&1.0$\pm$0.6&\phantom{0}0.65$\pm$0.49&$-$0.04$\pm$0.02\phantom{$-$}\\
&H$^{13}$CO$^{+}$&4$-$3&0.10&$-$&$<$0.29&$-$&$-$&$-$&$-$&$-$\\
CRA\,01&$^{13}$CO&3$-$2&0.43&5.1&\phantom{0}8.82&5.1$\pm$0.1&\phantom{0}7.2$\pm$0.5&1.2$\pm$0.1&\phantom{0}9.22$\pm$1.05&$-$0.41$\pm$0.05\phantom{$-$}\\
&C$^{18}$O&3$-$2&0.51&5.5&\phantom{0}4.89&5.6$\pm$0.1&\phantom{0}4.7$\pm$0.5&1.4$\pm$0.2&\phantom{0}6.82$\pm$1.18&0.00$\pm$0.01\\
&HCO$^{+}$&4$-$3&0.31&6.1&\phantom{0}4.32&5.7$\pm$0.1&\phantom{0}3.2$\pm$0.2&2.3$\pm$0.2&\phantom{0}7.69$\pm$0.76&0.04$\pm$0.01\\
OPH\,01&$^{13}$CO&3$-$2&0.50&3.0&\phantom{0}6.79&3.7$\pm$0.1&\phantom{0}6.9$\pm$0.4&2.4$\pm$0.2&17.87$\pm$1.49&$-$0.07$\pm$0.02\phantom{$-$}\\
&C$^{18}$O&3$-$2&0.61&3.9&\phantom{0}3.18&3.8$\pm$0.2&\phantom{0}2.6$\pm$0.5&2.1$\pm$0.5&\phantom{0}5.86$\pm$1.73&0.00$\pm$0.01\\
&HCO$^{+}$&4$-$3&0.34&$-$&$<$1.01&$-$&$-$&$-$&$-$&$-$\\
&H$^{13}$CO$^{+}$&4$-$3&0.15&$-$&$<$0.46&$-$&$-$&$-$&$-$&$-$\\
OPH\,02&$^{13}$CO&3$-$2&0.22&4.6&\phantom{0}7.84&4.4$\pm$0.1&\phantom{0}6.2$\pm$0.2&2.5$\pm$0.1&16.68$\pm$0.62&0.08$\pm$0.01\\
&C$^{18}$O&3$-$2&0.27&4.0&\phantom{0}1.82&4.2$\pm$0.1&\phantom{0}1.5$\pm$0.1&2.6$\pm$0.3&\phantom{0}4.25$\pm$0.61&0.00$\pm$0.01\\
&HCO$^{+}$&4$-$3&0.32&3.6&\phantom{0}5.16&3.6$\pm$0.1&\phantom{0}3.4$\pm$0.1&3.4$\pm$0.2&12.22$\pm$0.80&$-$0.22$\pm$0.03\phantom{$-$}\\
&H$^{13}$CO$^{+}$&4$-$3&0.13&$-$&$<$0.40&$-$&$-$&$-$&$-$&$-$\\
PER\,01&$^{13}$CO&3$-$2&0.16&3.9&\phantom{0}5.03&4.1$\pm$0.1&\phantom{0}4.9$\pm$0.2&1.6$\pm$0.1&\phantom{0}8.24$\pm$0.41&-0.00$\pm$0.01\\
&C$^{18}$O&3$-$2&0.15&4.1&\phantom{0}3.13&4.1$\pm$0.1&\phantom{0}3.0$\pm$0.3&0.8$\pm$0.1&\phantom{0}2.62$\pm$0.36&0.00$\pm$0.01\\
&HCO$^{+}$&4$-$3&0.15&4.5&\phantom{0}1.97&4.1$\pm$0.1&\phantom{0}1.5$\pm$0.1&1.6$\pm$0.2&\phantom{0}2.52$\pm$0.34&0.06$\pm$0.01\\
PER\,02&$^{13}$CO&3$-$2&0.20&4.5&\phantom{0}7.89&4.6$\pm$0.1&\phantom{0}7.9$\pm$0.1&2.2$\pm$0.1&18.84$\pm$0.48&0.06$\pm$0.01\\
&C$^{18}$O&3$-$2&0.31&4.7&\phantom{0}4.82&4.5$\pm$0.1&\phantom{0}4.5$\pm$0.3&1.6$\pm$0.1&\phantom{0}7.60$\pm$0.70&0.00$\pm$0.01\\
&HCO$^{+}$&4$-$3&0.16&5.3&\phantom{0}7.60&5.3$\pm$0.1&\phantom{0}6.7$\pm$0.1&2.1$\pm$0.1&15.10$\pm$0.48&0.46$\pm$0.03\\
&H$^{13}$CO$^{+}$&4$-$3&0.16&$-$&$<$0.47&$-$&$-$&$-$&$-$&$-$\\
PER\,04&C$^{18}$O&3$-$2&0.26&5.1&\phantom{0}1.48&5.2$\pm$0.2&\phantom{0}1.2$\pm$0.5&0.8$\pm$0.4&\phantom{0}1.04$\pm$0.61&0.00$\pm$0.01\\
&HCO$^{+}$&4$-$3&0.16&5.1&\phantom{0}1.48&4.9$\pm$0.1&\phantom{0}0.8$\pm$0.1&1.6$\pm$0.3&\phantom{0}1.29$\pm$0.38&\phantom{\,\tablefootmark{e}}$-$0.17$\pm$0.08\,\tablefootmark{e}\phantom{$-$}\\
&H$^{13}$CO$^{+}$&4$-$3&0.12&$-$&$<$0.36&$-$&$-$&$-$&$-$&$-$\\
PER\,05&$^{13}$CO&3$-$2&0.24&7.1&\phantom{0}7.29&7.3$\pm$0.1&\phantom{0}6.2$\pm$0.3&1.4$\pm$0.1&\phantom{0}9.07$\pm$0.59&0.01$\pm$0.01\\
&C$^{18}$O&3$-$2&0.22&7.3&\phantom{0}3.85&7.3$\pm$0.1&\phantom{0}3.6$\pm$0.4&0.8$\pm$0.1&\phantom{0}3.11$\pm$0.50&0.00$\pm$0.01\\
\hline
\end{tabular}
\vspace{-2mm}
\tablefoot{\tablefoottext{a}{From Gaussian fits.} \tablefoottext{b}{Calculated by integrating over $\varv_{\rm LSR}\pm$3\,FWHM.} \tablefoottext{c}{Calculated using Eqn.\,1 from \citet{Mardones1997}.} \tablefoottext{d}{Central absorption masked in Gaussian fit.} \tablefoottext{e}{Position of maximum of optically thick line used due to non-Gaussian line shape.} \tablefoottext{f}{Resampled to 0.3\,\kms{} to fit weak line.}}
\end{center}
\label{T:ground_spectra}
\end{table*}
\renewcommand{\thetable}{\thesection.\arabic{table} (Cont.)}
\addtocounter{table}{-1}
\begin{table*}
\caption{Ground-based spectral results.}
\begin{center}
\vspace{-2mm}
\begin{tabular}{llccccccccc}
\hline \noalign {\smallskip}
Name & Molecule & Transition & $\sigma_{\mathrm{rms}}$ & $\varv_{\mathrm{max}}$ & T$_{\mathrm{max}}$ & $\varv_{\mathrm{peak}}$\tablefootmark{a} & T$_{\mathrm{peak}}$\tablefootmark{a} & FWHM\tablefootmark{a} & $\int T_{\mathrm{MB}}\,\mathrm{d}\varv$\tablefootmark{b}  & $\delta\varv$\tablefootmark{c} \\
 & & & (K) & (km\,s$^{-1}$) & (K) & (km\,s$^{-1}$) & (K) & (km\,s$^{-1}$) & (K\kms{}) & (km\,s$^{-1}$) \\ 
\hline\noalign {\smallskip}
&HCO$^{+}$&4$-$3&0.15&7.1&\phantom{0}5.54&7.3$\pm$0.1&\phantom{0}4.1$\pm$0.2&1.3$\pm$0.1&\phantom{0}5.70$\pm$0.40&$-$0.01$\pm$0.01\phantom{$-$}\\
&H$^{13}$CO$^{+}$&4$-$3&0.13&7.5&\phantom{0}0.70&7.4$\pm$0.1&\phantom{0}0.7$\pm$0.2&0.9$\pm$0.3&\phantom{0}0.65$\pm$0.31&0.00$\pm$0.01\\
PER\,06&$^{13}$CO&3$-$2&0.16&6.6&\phantom{0}8.39&7.2$\pm$0.1&\phantom{0}6.8$\pm$0.1&2.8$\pm$0.1&20.58$\pm$0.55&$-$0.08$\pm$0.01\phantom{$-$}\\
&C$^{18}$O&3$-$2&0.15&7.0&\phantom{0}3.46&7.3$\pm$0.1&\phantom{0}3.2$\pm$0.1&2.0$\pm$0.1&\phantom{0}6.84$\pm$0.36&0.00$\pm$0.01\\
&HCO$^{+}$&4$-$3&0.20&7.0&\phantom{0}5.44&7.0$\pm$0.1&\phantom{0}5.0$\pm$0.2&1.4$\pm$0.1&\phantom{0}7.33$\pm$0.47&$-$0.18$\pm$0.01\phantom{$-$}\\
&H$^{13}$CO$^{+}$&4$-$3&0.11&6.8&\phantom{0}0.44&7.0$\pm$0.2&\phantom{0}0.4$\pm$0.2&0.8$\pm$0.4&\phantom{0}0.36$\pm$0.26&0.00$\pm$0.01\\
PER\,07&$^{13}$CO&3$-$2&0.21&7.3&\phantom{0}5.34&7.2$\pm$0.1&\phantom{0}4.6$\pm$0.1&2.2$\pm$0.1&10.93$\pm$0.51&$-$0.12$\pm$0.04\phantom{$-$}\\
&C$^{18}$O&3$-$2&0.23&7.3&\phantom{0}1.08&7.4$\pm$0.2&\phantom{0}1.0$\pm$0.3&1.3$\pm$0.4&\phantom{0}1.31$\pm$0.53&0.00$\pm$0.01\\
&HCO$^{+}$&4$-$3&0.14&7.3&\phantom{0}2.63&7.1$\pm$0.1&\phantom{0}2.6$\pm$0.2&1.1$\pm$0.1&\phantom{0}2.96$\pm$0.32&$-$0.17$\pm$0.05\phantom{$-$}\\
&H$^{13}$CO$^{+}$&4$-$3&0.19&$-$&$<$0.58&$-$&$-$&$-$&$-$&$-$\\
PER\,08&$^{13}$CO&3$-$2&0.34&7.8&11.17&7.3$\pm$0.1&11.0$\pm$0.2&2.9$\pm$0.1&33.92$\pm$1.12&$-$0.22$\pm$0.02\phantom{$-$}\\
&C$^{18}$O&3$-$2&0.38&7.8&\phantom{0}6.04&7.7$\pm$0.1&\phantom{0}6.1$\pm$0.4&1.6$\pm$0.1&10.30$\pm$1.09&0.00$\pm$0.01\\
&HCO$^{+}$&4$-$3&0.19&8.4&\phantom{0}8.63&8.1$\pm$0.1&\phantom{0}6.9$\pm$0.2&1.7$\pm$0.1&12.31$\pm$0.60&0.29$\pm$0.02\\
&H$^{13}$CO$^{+}$&4$-$3&0.10&8.2&\phantom{0}1.20&8.3$\pm$0.1&\phantom{0}1.2$\pm$0.2&0.7$\pm$0.1&\phantom{0}0.90$\pm$0.24&0.00$\pm$0.01\\
PER\,09&$^{13}$CO&3$-$2&0.31&6.9&20.45&7.2$\pm$0.1&19.1$\pm$0.4&2.0$\pm$0.1&39.84$\pm$1.20&$-$0.24$\pm$0.02\phantom{$-$}\\
&C$^{18}$O&3$-$2&0.31&7.5&\phantom{0}6.33&7.5$\pm$0.1&\phantom{0}6.4$\pm$0.5&1.2$\pm$0.1&\phantom{0}8.28$\pm$0.89&0.00$\pm$0.01\\
&HCO$^{+}$&4$-$3&0.19&7.6&\phantom{0}1.50&7.5$\pm$0.1&\phantom{0}1.4$\pm$0.2&1.6$\pm$0.2&\phantom{0}2.23$\pm$0.44&$-$0.03$\pm$0.01\phantom{$-$}\\
&H$^{13}$CO$^{+}$&4$-$3&0.11&$-$&$<$0.34&$-$&$-$&$-$&$-$&$-$\\
PER\,10&$^{13}$CO&3$-$2&0.23&8.1&\phantom{0}8.26&8.1$\pm$0.1&\phantom{0}7.5$\pm$0.1&3.0$\pm$0.1&23.67$\pm$0.58&$-$0.38$\pm$0.03\phantom{$-$}\\
&C$^{18}$O&3$-$2&0.20&8.7&\phantom{0}3.53&8.7$\pm$0.1&\phantom{0}3.1$\pm$0.2&1.4$\pm$0.1&\phantom{0}4.50$\pm$0.48&0.00$\pm$0.01\\
&HCO$^{+}$&4$-$3&0.30&8.5&\phantom{0}3.94&8.5$\pm$0.1&\phantom{0}4.0$\pm$0.3&1.4$\pm$0.1&\phantom{0}5.89$\pm$0.71&$-$0.08$\pm$0.01\phantom{$-$}\\
&H$^{13}$CO$^{+}$&4$-$3&0.18&9.1&\phantom{0}0.89&9.0$\pm$0.1&\phantom{0}0.9$\pm$0.4&0.7$\pm$0.3&\phantom{0}0.63$\pm$0.42&0.00$\pm$0.01\\
PER\,12&$^{13}$CO&3$-$2&0.28&7.8&\phantom{0}5.45&7.8$\pm$0.1&\phantom{0}5.1$\pm$0.2&2.0$\pm$0.1&10.89$\pm$0.80&0.05$\pm$0.01\\
&C$^{18}$O&3$-$2&0.31&7.5&\phantom{0}3.05&7.8$\pm$0.1&\phantom{0}3.1$\pm$0.5&1.0$\pm$0.2&\phantom{0}3.24$\pm$0.88&0.00$\pm$0.01\\
&HCO$^{+}$&4$-$3&0.21&7.4&\phantom{0}3.33&7.6$\pm$0.1&\phantom{0}2.3$\pm$0.2&1.4$\pm$0.2&\phantom{0}3.47$\pm$0.52&$-$0.18$\pm$0.04\phantom{$-$}\\
&H$^{13}$CO$^{+}$&4$-$3&0.13&7.8&\phantom{0}0.59&8.0$\pm$0.2&\phantom{0}0.6$\pm$0.3&0.7$\pm$0.4&\phantom{0}0.43$\pm$0.30&0.00$\pm$0.01\\
PER\,13&C$^{18}$O&3$-$2&0.35&7.9&\phantom{0}2.19&8.0$\pm$0.1&\phantom{0}2.4$\pm$0.8&0.6$\pm$0.2&\phantom{0}1.57$\pm$0.82&0.00$\pm$0.01\\
&HCO$^{+}$&4$-$3&0.11&7.9&\phantom{0}4.70&7.9$\pm$0.1&\phantom{0}4.2$\pm$0.2&1.0$\pm$0.1&\phantom{0}4.44$\pm$0.30&$-$0.18$\pm$0.07\phantom{$-$}\\
&H$^{13}$CO$^{+}$&4$-$3&0.17&8.1&\phantom{0}1.42&8.1$\pm$0.1&\phantom{0}1.3$\pm$0.3&0.7$\pm$0.2&\phantom{0}0.94$\pm$0.40&0.00$\pm$0.01\\
PER\,14&$^{13}$CO&3$-$2&0.28&6.0&\phantom{0}5.09&6.0$\pm$0.1&\phantom{0}4.4$\pm$0.4&1.1$\pm$0.1&\phantom{0}5.12$\pm$0.67&$-$0.34$\pm$0.20\phantom{$-$}\\
&C$^{18}$O&3$-$2&0.35&6.0&\phantom{0}1.33&6.2$\pm$0.2&\phantom{0}1.3$\pm$0.7&0.7$\pm$0.4&\phantom{0}1.01$\pm$0.82&0.00$\pm$0.01\\
&HCO$^{+}$&4$-$3&0.24&5.6&\phantom{0}1.64&5.9$\pm$0.1&\phantom{0}1.1$\pm$0.2&1.7$\pm$0.3&\phantom{0}2.10$\pm$0.56&$-$0.43$\pm$0.26\phantom{$-$}\\
&H$^{13}$CO$^{+}$&4$-$3&0.13&$-$&$<$0.38&$-$&$-$&$-$&$-$&$-$\\
PER\,15&C$^{18}$O\,\tablefootmark{f}&3$-$2&0.30&6.9&\phantom{0}1.20&6.9$\pm$0.2&\phantom{0}1.0$\pm$0.4&1.4$\pm$0.6&\phantom{0}1.46$\pm$0.83&0.00$\pm$0.01\\
&HCO$^{+}$&4$-$3&0.12&6.7&\phantom{0}3.52&7.1$\pm$0.1&\phantom{0}2.8$\pm$0.1&1.4$\pm$0.1&\phantom{0}4.26$\pm$0.32&0.10$\pm$0.04\\
&H$^{13}$CO$^{+}$&4$-$3&0.16&7.1&\phantom{0}0.82&7.2$\pm$0.2&\phantom{0}0.6$\pm$0.3&0.7$\pm$0.4&\phantom{0}0.45$\pm$0.38&0.00$\pm$0.01\\
PER\,16&C$^{18}$O&3$-$2&0.41&7.1&\phantom{0}2.54&7.0$\pm$0.1&\phantom{0}2.2$\pm$0.5&1.2$\pm$0.3&\phantom{0}2.87$\pm$0.94&0.00$\pm$0.01\\
&HCO$^{+}$&4$-$3&0.15&6.5&\phantom{0}4.17&6.6$\pm$0.1&\phantom{0}3.9$\pm$0.3&0.9$\pm$0.1&\phantom{0}3.70$\pm$0.38&$-$0.29$\pm$0.07\phantom{$-$}\\
&H$^{13}$CO$^{+}$&4$-$3&0.16&7.3&\phantom{0}0.56&6.9$\pm$0.2&\phantom{0}0.5$\pm$0.2&1.0$\pm$0.5&\phantom{0}0.56$\pm$0.37&0.00$\pm$0.01\\
PER\,17&$^{13}$CO&3$-$2&0.19&6.0&\phantom{0}3.68&6.5$\pm$0.1&\phantom{0}3.4$\pm$0.1&2.3$\pm$0.1&\phantom{0}8.31$\pm$0.47&$-$0.09$\pm$0.02\phantom{$-$}\\
&C$^{18}$O&3$-$2&0.32&6.4&\phantom{0}2.01&6.6$\pm$0.1&\phantom{0}1.9$\pm$0.3&1.4$\pm$0.3&\phantom{0}2.69$\pm$0.74&0.00$\pm$0.01\\
&HCO$^{+}$&4$-$3&0.15&6.0&\phantom{0}1.84&6.0$\pm$0.1&\phantom{0}1.6$\pm$0.3&0.8$\pm$0.2&\phantom{0}1.39$\pm$0.35&$-$0.44$\pm$0.09\phantom{$-$}\\
&H$^{13}$CO$^{+}$&4$-$3&0.11&$-$&$<$0.32&$-$&$-$&$-$&$-$&$-$\\
PER\,18&$^{13}$CO&3$-$2&0.18&6.7&\phantom{0}3.86&6.6$\pm$0.1&\phantom{0}4.1$\pm$0.1&1.7$\pm$0.1&\phantom{0}7.42$\pm$0.41&$-$0.03$\pm$0.01\phantom{$-$}\\
&C$^{18}$O&3$-$2&0.20&6.7&\phantom{0}2.60&6.6$\pm$0.1&\phantom{0}2.7$\pm$0.3&0.9$\pm$0.1&\phantom{0}2.55$\pm$0.47&0.00$\pm$0.01\\
&HCO$^{+}$&4$-$3&0.13&6.2&\phantom{0}1.45&6.4$\pm$0.1&\phantom{0}1.0$\pm$0.1&1.6$\pm$0.2&\phantom{0}1.67$\pm$0.29&$-$0.28$\pm$0.04\phantom{$-$}\\
PER\,19&C$^{18}$O&3$-$2&0.51&6.9&\phantom{0}2.17&6.8$\pm$0.2&\phantom{0}1.8$\pm$1.0&0.7$\pm$0.5&\phantom{0}1.45$\pm$1.19&0.00$\pm$0.01\\
&HCO$^{+}$&4$-$3&0.18&6.7&\phantom{0}1.80&6.9$\pm$0.1&\phantom{0}1.5$\pm$0.2&1.0$\pm$0.2&\phantom{0}1.67$\pm$0.41&0.22$\pm$0.14\\
&H$^{13}$CO$^{+}$&4$-$3&0.15&$-$&$<$0.46&$-$&$-$&$-$&$-$&$-$\\
PER\,20&$^{13}$CO&3$-$2&0.28&8.7&\phantom{0}9.61&8.9$\pm$0.1&10.0$\pm$0.3&1.5$\pm$0.1&16.13$\pm$0.78&$-$0.01$\pm$0.01\phantom{$-$}\\
&C$^{18}$O&3$-$2&0.38&9.0&\phantom{0}5.46&8.9$\pm$0.1&\phantom{0}5.5$\pm$0.8&0.8$\pm$0.1&\phantom{0}4.72$\pm$1.08&0.00$\pm$0.01\\
&HCO$^{+}$&4$-$3&0.39&9.3&\phantom{0}3.42&8.9$\pm$0.1&\phantom{0}2.8$\pm$0.3&2.0$\pm$0.2&\phantom{0}5.93$\pm$0.93&0.01$\pm$0.01\\
&H$^{13}$CO$^{+}$&4$-$3&0.16&9.3&\phantom{0}1.09&9.2$\pm$0.1&\phantom{0}1.1$\pm$0.3&0.7$\pm$0.2&\phantom{0}0.76$\pm$0.37&0.00$\pm$0.01\\
PER\,21&$^{13}$CO&3$-$2&0.21&8.7&10.32&8.6$\pm$0.1&10.4$\pm$0.2&1.7$\pm$0.1&19.13$\pm$0.56&$-$0.12$\pm$0.01\phantom{$-$}\\
&C$^{18}$O&3$-$2&0.28&8.9&\phantom{0}5.00&8.8$\pm$0.1&\phantom{0}4.6$\pm$0.4&0.9$\pm$0.1&\phantom{0}4.51$\pm$0.67&0.00$\pm$0.01\\
&HCO$^{+}$&4$-$3&0.14&8.7&\phantom{0}3.90&8.8$\pm$0.1&\phantom{0}3.8$\pm$0.2&1.0$\pm$0.1&\phantom{0}4.01$\pm$0.33&0.02$\pm$0.01\\
&H$^{13}$CO$^{+}$&4$-$3&0.15&9.1&\phantom{0}0.71&9.0$\pm$0.1&\phantom{0}0.8$\pm$0.4&0.5$\pm$0.3&\phantom{0}0.42$\pm$0.35&0.00$\pm$0.01\\
PER\,22&$^{13}$CO&3$-$2&0.21&9.9&\phantom{0}8.59&9.7$\pm$0.1&\phantom{0}8.5$\pm$0.1&2.5$\pm$0.1&22.67$\pm$0.57&$-$0.06$\pm$0.01\phantom{$-$}\\
&C$^{18}$O&3$-$2&0.24&9.7&\phantom{0}3.92&9.8$\pm$0.1&\phantom{0}3.3$\pm$0.2&1.5$\pm$0.1&\phantom{0}5.30$\pm$0.55&0.00$\pm$0.01\\
&HCO$^{+}$&4$-$3&0.38&9.7&\phantom{0}5.80&10.2$\pm$0.1&\phantom{0}4.3$\pm$0.2&2.8$\pm$0.1&12.72$\pm$0.91&0.24$\pm$0.02\\
&H$^{13}$CO$^{+}$&4$-$3&0.15&$-$&$<$0.46&$-$&$-$&$-$&$-$&$-$\\
SCO\,01&$^{13}$CO&3$-$2&0.23&3.6&\phantom{0}4.02&3.6$\pm$0.1&\phantom{0}3.5$\pm$0.3&1.0$\pm$0.1&\phantom{0}3.61$\pm$0.53&0.00$\pm$0.01\\
\hline
\end{tabular}
\vspace{-2mm}
\tablefoot{\tablefoottext{a}{From Gaussian fits.} \tablefoottext{b}{Calculated by integrating over $\varv_{\rm LSR}\pm$3\,FWHM.} \tablefoottext{c}{Calculated using Eqn.\,1 from \citet{Mardones1997}.} \tablefoottext{d}{Central absorption masked in Gaussian fit.} \tablefoottext{e}{Position of maximum of optically thick line used due to non-Gaussian line shape.} \tablefoottext{f}{Resampled to 0.3\,\kms{} to fit weak line.}}
\end{center}
\end{table*}
\renewcommand{\thetable}{\thesection.\arabic{table} (Cont.)}
\addtocounter{table}{-1}
\begin{table*}
\caption{Ground-based spectral results.}
\begin{center}
\vspace{-2mm}
\begin{tabular}{llccccccccc}
\hline \noalign {\smallskip}
Name & Molecule & Transition & $\sigma_{\mathrm{rms}}$ & $\varv_{\mathrm{max}}$ & T$_{\mathrm{max}}$ & $\varv_{\mathrm{peak}}$\tablefootmark{a} & T$_{\mathrm{peak}}$\tablefootmark{a} & FWHM\tablefootmark{a} & $\int T_{\mathrm{MB}}\,\mathrm{d}\varv$\tablefootmark{b}  & $\delta\varv$\tablefootmark{c} \\
 & & & (K) & (km\,s$^{-1}$) & (K) & (km\,s$^{-1}$) & (K) & (km\,s$^{-1}$) & (K\kms{}) & (km\,s$^{-1}$) \\ 
\hline\noalign {\smallskip}
&C$^{18}$O&3$-$2&0.26&$-$&$<$0.79&$-$&$-$&$-$&$-$&$-$\\
&HCO$^{+}$&4$-$3&0.30&$-$&$<$0.90&$-$&$-$&$-$&$-$&$-$\\
&H$^{13}$CO$^{+}$&4$-$3&0.22&$-$&$<$0.65&$-$&$-$&$-$&$-$&$-$\\
SERS\,01&$^{13}$CO&3$-$2&0.23&7.4&\phantom{0}4.55&8.1$\pm$0.1&\phantom{0}3.7$\pm$0.2&2.4$\pm$0.1&\phantom{0}9.25$\pm$0.63&$-$0.02$\pm$0.01\phantom{$-$}\\
&C$^{18}$O&3$-$2&0.30&8.2&\phantom{0}1.88&8.2$\pm$0.1&\phantom{0}1.4$\pm$0.3&1.6$\pm$0.3&\phantom{0}2.45$\pm$0.68&0.00$\pm$0.01\\
&HCO$^{+}$&4$-$3&0.36&8.6&\phantom{0}4.07&8.5$\pm$0.1&\phantom{0}3.3$\pm$0.3&1.8$\pm$0.2&\phantom{0}6.30$\pm$0.85&\phantom{\,\tablefootmark{e}}0.26$\pm$0.06\,\tablefootmark{e}\\
SERS\,02&$^{13}$CO&3$-$2&0.26&7.2&\phantom{0}6.65&8.1$\pm$0.1&\phantom{0}5.4$\pm$0.1&5.0$\pm$0.1&29.00$\pm$0.85&0.12$\pm$0.01\\
&C$^{18}$O&3$-$2&0.33&7.8&\phantom{0}4.59&7.8$\pm$0.1&\phantom{0}4.0$\pm$0.2&2.3$\pm$0.1&\phantom{0}9.59$\pm$0.77&0.00$\pm$0.01\\
&HCO$^{+}$\,\tablefootmark{d}&4$-$3&0.38&7.0&10.17&7.7$\pm$0.1&11.5$\pm$0.4&3.0$\pm$0.1&36.94$\pm$1.61&\phantom{\,\tablefootmark{e}}$-$0.36$\pm$0.02\,\tablefootmark{e}\phantom{$-$}\\
TAU\,01&$^{13}$CO&3$-$2&0.13&7.2&\phantom{0}4.13&7.0$\pm$0.1&\phantom{0}4.0$\pm$0.1&1.8$\pm$0.1&\phantom{0}7.66$\pm$0.39&0.18$\pm$0.03\\
&C$^{18}$O&3$-$2&0.20&6.9&\phantom{0}2.48&6.8$\pm$0.1&\phantom{0}2.4$\pm$0.4&0.9$\pm$0.2&\phantom{0}2.31$\pm$0.55&0.00$\pm$0.01\\
&HCO$^{+}$&4$-$3&0.22&6.6&\phantom{0}3.01&6.8$\pm$0.1&\phantom{0}2.7$\pm$0.3&1.2$\pm$0.2&\phantom{0}3.49$\pm$0.65&$-$0.01$\pm$0.01\phantom{$-$}\\
&H$^{13}$CO$^{+}$&4$-$3&0.11&6.9&\phantom{0}0.46&6.9$\pm$0.2&\phantom{0}0.4$\pm$0.3&0.6$\pm$0.4&\phantom{0}0.26$\pm$0.28&0.00$\pm$0.01\\
TAU\,02&C$^{18}$O&3$-$2&0.29&6.6&\phantom{0}2.54&6.6$\pm$0.1&\phantom{0}2.1$\pm$0.6&0.7$\pm$0.2&\phantom{0}1.58$\pm$0.67&0.00$\pm$0.01\\
&HCO$^{+}$&4$-$3&0.34&7.2&\phantom{0}1.79&7.2$\pm$0.1&\phantom{0}1.7$\pm$0.9&0.5$\pm$0.3&\phantom{0}0.95$\pm$0.79&0.85$\pm$0.28\\
&H$^{13}$CO$^{+}$&4$-$3&0.15&$-$&$<$0.45&$-$&$-$&$-$&$-$&$-$\\
TAU\,03&$^{13}$CO&3$-$2&0.23&7.8&\phantom{0}2.26&7.4$\pm$0.1&\phantom{0}1.6$\pm$0.1&2.3$\pm$0.2&\phantom{0}3.84$\pm$0.55&0.00$\pm$0.01\\
&C$^{18}$O&3$-$2&0.27&$-$&$<$0.82&$-$&$-$&$-$&$-$&$-$\\
&HCO$^{+}$&4$-$3&0.29&$-$&$<$0.86&$-$&$-$&$-$&$-$&$-$\\
&H$^{13}$CO$^{+}$&4$-$3&0.13&$-$&$<$0.40&$-$&$-$&$-$&$-$&$-$\\
TAU\,04&$^{13}$CO&3$-$2&0.24&6.6&\phantom{0}5.29&6.5$\pm$0.1&\phantom{0}4.1$\pm$0.2&1.7$\pm$0.1&\phantom{0}7.39$\pm$0.58&0.16$\pm$0.06\\
&C$^{18}$O&3$-$2&0.31&6.2&\phantom{0}2.19&6.3$\pm$0.1&\phantom{0}2.1$\pm$0.6&0.7$\pm$0.2&\phantom{0}1.54$\pm$0.72&0.00$\pm$0.01\\
&HCO$^{+}$&4$-$3&0.28&$-$&$<$0.84&$-$&$-$&$-$&$-$&$-$\\
&H$^{13}$CO$^{+}$&4$-$3&0.15&$-$&$<$0.44&$-$&$-$&$-$&$-$&$-$\\
TAU\,06&C$^{18}$O&3$-$2&0.32&7.2&\phantom{0}1.53&7.2$\pm$0.2&\phantom{0}1.3$\pm$0.3&1.4$\pm$0.4&\phantom{0}1.93$\pm$0.73&0.00$\pm$0.01\\
&HCO$^{+}$&4$-$3&0.28&7.4&\phantom{0}2.66&7.4$\pm$0.1&\phantom{0}2.7$\pm$0.5&0.8$\pm$0.2&\phantom{0}2.29$\pm$0.66&0.12$\pm$0.03\\
TAU\,07&$^{13}$CO&3$-$2&0.21&5.8&\phantom{0}1.70&6.3$\pm$0.1&\phantom{0}1.5$\pm$0.2&2.0$\pm$0.2&\phantom{0}3.11$\pm$0.49&0.00$\pm$0.01\\
&C$^{18}$O&3$-$2&0.26&$-$&$<$0.78&$-$&$-$&$-$&$-$&$-$\\
&HCO$^{+}$&4$-$3&0.27&$-$&$<$0.82&$-$&$-$&$-$&$-$&$-$\\
&H$^{13}$CO$^{+}$&4$-$3&0.18&$-$&$<$0.54&$-$&$-$&$-$&$-$&$-$\\
TAU\,08&C$^{18}$O&3$-$2&0.80&$-$&$<$2.41&$-$&$-$&$-$&$-$&$-$\\
&HCO$^{+}$&4$-$3&0.25&$-$&$<$0.75&$-$&$-$&$-$&$-$&$-$\\
&H$^{13}$CO$^{+}$&4$-$3&0.17&$-$&$<$0.50&$-$&$-$&$-$&$-$&$-$\\
TAU\,09&C$^{18}$O&3$-$2&0.48&5.3&\phantom{0}2.63&5.5$\pm$0.1&\phantom{0}2.4$\pm$1.0&0.7$\pm$0.3&\phantom{0}1.76$\pm$1.10&0.00$\pm$0.01\\
&HCO$^{+}$&4$-$3&0.21&5.3&\phantom{0}1.27&5.3$\pm$0.1&\phantom{0}1.0$\pm$0.2&1.8$\pm$0.3&\phantom{0}1.89$\pm$0.47&\phantom{\,\tablefootmark{e}}$-$0.25$\pm$0.12\,\tablefootmark{e}\phantom{$-$}\\
&H$^{13}$CO$^{+}$&4$-$3&0.11&$-$&$<$0.32&$-$&$-$&$-$&$-$&$-$\\
W40\,01&$^{13}$CO&3$-$2&0.26&5.2&15.10&5.2$\pm$0.1&16.1$\pm$0.2&1.9$\pm$0.1&32.16$\pm$0.75&0.22$\pm$0.01\\
&C$^{18}$O&3$-$2&0.36&4.8&\phantom{0}6.65&4.9$\pm$0.1&\phantom{0}6.8$\pm$0.4&1.3$\pm$0.1&\phantom{0}9.36$\pm$0.82&0.00$\pm$0.01\\
&HCO$^{+}$&4$-$3&0.41&4.4&\phantom{0}4.52&4.6$\pm$0.1&\phantom{0}4.5$\pm$0.5&1.2$\pm$0.2&\phantom{0}5.78$\pm$0.97&$-$0.23$\pm$0.02\phantom{$-$}\\
&H$^{13}$CO$^{+}$&4$-$3&0.14&4.6&\phantom{0}0.84&4.7$\pm$0.1&\phantom{0}0.7$\pm$0.3&0.8$\pm$0.3&\phantom{0}0.58$\pm$0.33&0.00$\pm$0.01\\
W40\,02&$^{13}$CO&3$-$2&0.30&4.8&27.47&4.8$\pm$0.1&27.2$\pm$0.4&1.1$\pm$0.1&31.95$\pm$0.75&0.01$\pm$0.01\\
&C$^{18}$O&3$-$2&0.37&4.8&\phantom{0}6.90&4.8$\pm$0.1&\phantom{0}6.8$\pm$0.6&0.9$\pm$0.1&\phantom{0}6.85$\pm$0.87&0.00$\pm$0.01\\
&HCO$^{+}$&4$-$3&0.49&4.8&\phantom{0}3.74&4.9$\pm$0.1&\phantom{0}2.7$\pm$0.4&1.8$\pm$0.3&\phantom{0}5.18$\pm$1.15&0.10$\pm$0.01\\
&H$^{13}$CO$^{+}$&4$-$3&0.15&5.0&\phantom{0}0.63&5.1$\pm$0.2&\phantom{0}0.5$\pm$0.3&0.8$\pm$0.5&\phantom{0}0.42$\pm$0.35&0.00$\pm$0.01\\
W40\,03&$^{13}$CO&3$-$2&0.29&5.8&19.39&6.1$\pm$0.1&12.0$\pm$0.4&2.9$\pm$0.1&36.48$\pm$1.89&$-$0.20$\pm$0.01\phantom{$-$}\\
&C$^{18}$O&3$-$2&0.34&6.2&12.97&6.4$\pm$0.1&12.1$\pm$0.4&1.4$\pm$0.1&17.45$\pm$0.91&0.00$\pm$0.01\\
&HCO$^{+}$&4$-$3&0.48&6.6&\phantom{0}6.55&6.6$\pm$0.1&\phantom{0}6.6$\pm$0.4&1.6$\pm$0.1&11.43$\pm$1.10&0.17$\pm$0.01\\
W40\,04&$^{13}$CO&3$-$2&0.29&6.0&16.20&7.4$\pm$0.1&\phantom{0}8.7$\pm$0.3&5.6$\pm$0.2&51.37$\pm$2.27&0.52$\pm$0.02\\
&C$^{18}$O&3$-$2&0.35&6.6&13.12&6.7$\pm$0.1&13.2$\pm$0.4&1.3$\pm$0.1&18.29$\pm$0.85&0.00$\pm$0.01\\
&HCO$^{+}$&4$-$3&0.35&7.0&\phantom{0}2.57&6.8$\pm$0.1&\phantom{0}2.4$\pm$0.4&1.3$\pm$0.2&\phantom{0}3.32$\pm$0.80&0.09$\pm$0.01\\
&H$^{13}$CO$^{+}$&4$-$3&0.11&$-$&$<$0.32&$-$&$-$&$-$&$-$&$-$\\
W40\,05&$^{13}$CO&3$-$2&0.21&5.8&14.66&7.6$\pm$0.1&11.5$\pm$0.3&4.6$\pm$0.1&55.71$\pm$1.92&0.73$\pm$0.02\\
&C$^{18}$O&3$-$2&0.26&6.2&14.47&6.5$\pm$0.1&14.6$\pm$0.3&1.5$\pm$0.1&23.71$\pm$0.76&0.00$\pm$0.01\\
&HCO$^{+}$&4$-$3&0.28&6.6&\phantom{0}3.20&6.8$\pm$0.1&\phantom{0}3.0$\pm$0.3&1.5$\pm$0.2&\phantom{0}4.78$\pm$0.67&0.16$\pm$0.01\\
&H$^{13}$CO$^{+}$&4$-$3&0.11&6.2&\phantom{0}0.39&6.5$\pm$0.2&\phantom{0}0.4$\pm$0.2&0.7$\pm$0.5&\phantom{0}0.27$\pm$0.26&0.00$\pm$0.01\\
W40\,06&$^{13}$CO&3$-$2&0.20&5.8&12.81&6.0$\pm$0.1&11.8$\pm$0.3&1.9$\pm$0.1&23.49$\pm$0.91&$-$0.29$\pm$0.03\phantom{$-$}\\
&C$^{18}$O&3$-$2&0.23&7.0&\phantom{0}5.54&6.6$\pm$0.1&\phantom{0}3.0$\pm$0.2&1.9$\pm$0.2&\phantom{0}6.26$\pm$0.73&0.00$\pm$0.01\\
&HCO$^{+}$&4$-$3&0.29&6.8&\phantom{0}1.66&6.9$\pm$0.1&\phantom{0}1.7$\pm$0.9&0.5$\pm$0.3&\phantom{0}0.91$\pm$0.68&0.16$\pm$0.01\\
&H$^{13}$CO$^{+}$&4$-$3&0.12&$-$&$<$0.36&$-$&$-$&$-$&$-$&$-$\\
W40\,07&$^{13}$CO&3$-$2&0.20&6.8&\phantom{0}2.25&7.1$\pm$0.1&\phantom{0}1.7$\pm$0.1&2.2$\pm$0.2&\phantom{0}4.00$\pm$0.47&$-$0.38$\pm$0.07\phantom{$-$}\\
&C$^{18}$O&3$-$2&0.25&7.4&\phantom{0}3.34&7.4$\pm$0.1&\phantom{0}3.3$\pm$0.5&0.7$\pm$0.1&\phantom{0}2.49$\pm$0.57&0.00$\pm$0.01\\
&HCO$^{+}$&4$-$3&0.28&7.4&\phantom{0}1.51&7.4$\pm$0.1&\phantom{0}1.6$\pm$0.5&0.9$\pm$0.3&\phantom{0}1.45$\pm$0.65&0.06$\pm$0.01\\
&H$^{13}$CO$^{+}$&4$-$3&0.12&$-$&$<$0.35&$-$&$-$&$-$&$-$&$-$\\
\hline
\end{tabular}
\vspace{-2mm}
\tablefoot{\tablefoottext{a}{From Gaussian fits.} \tablefoottext{b}{Calculated by integrating over $\varv_{\rm LSR}\pm$3\,FWHM.} \tablefoottext{c}{Calculated using Eqn.\,1 from \citet{Mardones1997}.} \tablefoottext{d}{Central absorption masked in Gaussian fit.} \tablefoottext{e}{Position of maximum of optically thick line used due to non-Gaussian line shape.} \tablefoottext{f}{Resampled to 0.3\,\kms{} to fit weak line.}}
\end{center}
\end{table*}

\renewcommand{\thetable}{\thesection.\arabic{table}}

\subsection{Evolutionary classification}
\label{S:properties_evolution}

Table~\ref{T:properties_evolution} presents a summary of the various indicators used to reach the final classification of the evolutionary state of sources in the WILL sample. Firstly, we consider whether or not an entrained molecular outflow is associated with the source in CO $J$=3$-$2, and whether or not there are broad line-wings in the HIFI water and CO $J$=10$-$9 spectra. Sources with all three signatures are likely to be the youngest protostars, with strong, energetic and likely warm outflows. Those without any detected outflow signatures are likely either pre-stellar or more evolved (i.e. Class II) sources. 

Next, we follow the method of \citet{vanKempen2009} and \citet{Carney2016} using maps of the molecular emission in HCO$^{+}$ $J$=4$-$3 and C$^{18}$O $J$=3$-$2 to separate Class 0/I embedded protostars from edge-on Class II disk sources. If both transitions are strong and spatially concentrated then the source is most likely a genuine embedded (i.e. Class 0/I) YSO. If not, the source is either too cold (i.e. pre-stellar) or does not have a significant envelope and so is a more evolved disk source. In the W40 sources, the extended emission in both lines is likely due to the PDR and may mask the presence of compact emission associated with the sources, so they are designated as confused. A few other sources also receive this designation, in line with \citet{Carney2016}, if there are multiple sources in the field that cannot be disentangled at the resolution of the observations.

The presence of strong, compact sub-mm continuum emission is also indicative of a young (i.e. pre-stellar or embedded Class 0/I) source, allowing pre-stellar and Class II sources to be distinguished \citep[e.g.][]{Andre2000}. Finally, the T$_{bol}$ classification (i.e. T$_{bol}<$70\,K corresponds to Class 0, 70 $\leq T_{\mathrm{bol}} <$ 650\,K to Class I) can be used to separate Class 0 and I sources, though on its own this calculation is not always able to correctly separate other highly-reddened sources; for example edge-on Class II disks \citep[][]{Crapsi2008}.

Overall, the WILL sample consists of 23 Class 0, 14 Class I, 8 Class II and 4 potentially pre-stellar sources. Most of the Class II sources are in Taurus, while the pre-stellar sources are all in Aquila/W40. Six sources (one Class 0, one Class II and four pre-stellar) have narrow yet bright $^{12}$CO $J$=10$-$9 emission (see Fig.~\ref{F:hifi_co10-9}) suggestive of a PDR along the line of sight to the source. The details of some of the more complex determinations are discussed in Appendix~\ref{S:cases}.

\begin{table*}
\caption{Source evolution.}
\begin{center}
\vspace{-2mm}
\begin{tabular}{lcccccc}
\hline \noalign {\smallskip}
Name & Outflow & H$_{2}$O\tablefootmark{a} & mid-$J$ CO\tablefootmark{a} & Stage\tablefootmark{b} & $T_{\rm bol}$ Class & Final Class\tablefootmark{c} \\
\hline\noalign {\smallskip}
AQU\,01&Y&B&B&0/I&0&0\\
AQU\,02&Y&B&B&0/I&0&0\\
AQU\,03&N&E&E&II&I&II\\
AQU\,04&Y&E&N&0/I&I&I\\
AQU\,05&Y&B&B&0/I&0&0\\
AQU\,06&Y&E&N&0/I&0&0\\
CHA\,01&Y&E&E&II&I&\phantom{\tablefootmark{d}}II\tablefootmark{d}\\
CHA\,02&N&N&N&0/I&I&I\\
CRA\,01&Y&B&E&0/I&0&0\\
OPH\,01&N&E&E&II&I&\phantom{\tablefootmark{d}}II+PDR?\tablefootmark{d}\\
OPH\,02&Y&B&B&0/I&I&I\\
PER\,01&Y&B&B&0/I&0&0\\
PER\,02&Y&B&B&0/I&0&0\\
PER\,04&Y&N&E&0/I&0&0\\
PER\,05&Y&B&E&0/I&I&I\\
PER\,06&Y&B&B&0/I&I&I\\
PER\,07&Y&B&N&0/I&0&0\\
PER\,08&Y&B&B&C&I&I\\
PER\,09&Y&B&B&C&I&I\\
PER\,10&Y&B&B&0/I&0&0\\
PER\,12&N&B&E&0/I&0&0\\
PER\,13&Y&B&E&0/I&0&0\\
PER\,14&Y&B&B&0/I&I&I\\
PER\,15&Y&B&N&0/I&0&0\\
PER\,16&Y&E&E&0/I&0&0\\
PER\,17&Y&B&E&C&I&I\\
PER\,18&Y&B&B&0/I&0&0\\
PER\,19&Y&B&B&0/I&I&I\\
PER\,20&Y&B&B&0/I&0&0\\
PER\,21&Y&B&B&0/I&0&0\\
PER\,22&Y&B&B&0/I&0&0\\
SCO\,01&N&N&E&II&I&II\\
SERS\,01&Y&B&B&0/I&0&0\\
SERS\,02&Y&B&B&0/I&0&0\\
TAU\,01&Y&B&E&0/I&I&I\\
TAU\,02&Y&B&E&C&I&I\\
TAU\,03&Y&B&N&II&I&\phantom{\tablefootmark{d}}II\tablefootmark{d}\\
TAU\,04&Y&B&B&0/I&I&I\\
TAU\,06&Y&B&E&0/I&I&I\\
TAU\,07&N&E&E&II&I&II\\
TAU\,08&N&N&E&II&I&II\\
TAU\,09&N&E&E&II&I&II\\
W40\,01&N&B+P&P&C&0&\phantom{\tablefootmark{d}}0+PDR\tablefootmark{d}\\
W40\,02&Y&B&B&0/I&0&0\\
W40\,03&N&P&P&C&0&\phantom{\tablefootmark{d}}PS?+PDR\tablefootmark{d}\\
W40\,04&N&P&P&C&0&\phantom{\tablefootmark{d}}PS?+PDR\tablefootmark{d}\\
W40\,05&N&P&P&C&0&\phantom{\tablefootmark{d}}PS?+PDR\tablefootmark{d}\\
W40\,06&N&N&P&C&0&\phantom{\tablefootmark{d}}PS?+PDR\tablefootmark{d}\\
W40\,07&Y&N&E&C&0&0\\
\hline
\end{tabular}
\vspace{-2mm}
\tablefoot{\tablefoottext{a}{From HIFI observations: B=broad, E=envelope, P=PDR and N=non-detection.} \tablefoottext{b}{Based on the scheme of \citet{Carney2016}, C=confused.}  \tablefoottext{c}{Final classification, PS?=potentially pre-stellar, PDR=narrow, bright $^{12}$CO $J$=10$-$9 emission consistent with a photon-dominated region.} \tablefoottext{d}{See Appendix~\ref{S:cases} for more details.}}
\end{center}
\label{T:properties_evolution}
\end{table*}

\section{Observation IDs}
\label{S:obsids}

This section presents the \textit{Herschel} observation ID numbers for all WILL HIFI and PACS spectral observations (Table~\ref{T:obsids_will}), as well as those of the \textit{Herschel} PACS and SPIRE photometric maps used to extract source photometry (Table~\ref{T:obsids_gbs}).

\begin{table*}[ph!]
\caption[]{\textit{Herschel} observation identification numbers for WILL HIFI and PACS observations.}
\centering
\begin{tabular}{lccccccccc}
\hline \noalign {\smallskip}
\centering
Name &  H$_{2}$O 1$_{10}$-1$_{01}$\tablefootmark{a} & H$_{2}$O 3$_{12}$-2$_{21}$\tablefootmark{b} & H$_{2}$O 1$_{11}$-0$_{00}$\tablefootmark{c} & H$_{2}$O 2$_{02}$-1$_{11}$\tablefootmark{d} & PACS 1 & PACS 2 \\
\hline\noalign {\smallskip}
AQU~01&1342268681&1342268121&1342266489&1342268160&1342254232&1342254233\\
AQU~02&1342268682&1342268125&1342266488&1342268159&1342254270&1342254271\\
AQU~03&1342268683&1342268486&1342266501&1342268154&1342254226&1342254227\\
AQU~04&1342268684&1342268492&1342266499&1342268156&1342254225&1342254224\\
AQU~05&1342268462&1342268119&1342266500&1342268155&1342254229&1342254228\\
AQU~06&1342268463&1342268123&1342266498&1342268153&1342254272&1342254273\\
CHA~01&1342263152&1342254889&1342263403&1342257661&1342267618&1342267619\\
CHA~02&1342263153&1342266398&1342263404&1342267974&1342265695&1342265694\\
CRA~01&1342254318&1342254375&1342254377&1342254338&1342254254&1342254253\\
OPH~01&1342263422&1342263173&1342266509&1342266759&1342266926&1342266925\\
OPH~02&1342266423&1342263172&1342266508&1342266758&1342263470&1342263469\\
PER~01&1342263524&1342262778&1342263321&1342262807&1342263509&1342263508\\
PER~02&1342263526&1342262779&1342263322&1342262806&1342263507&1342263506\\
PER~03&1342263525&1342262777&1342263323&1342262808&1342263511&1342263510\\
PER~04&1342263523&1342262776&1342263324&1342262809&1342264251&1342264250\\
PER~05&1342263529&1342262784&1342263325&1342262795&1342264249&1342264248\\
PER~06&1342263530&1342262783&1342263326&1342262794&1342264246&1342264247\\
PER~07&1342263531&1342262782&1342263327&1342262802&1342264244&1342264245\\
PER~08&1342263535&1342262786&1342263328&1342262793&1342264242&1342264243\\
PER~09&1342263431&1342262787&1342263329&1342262792&1342267611&1342267612\\
PER~10&1342263534&1342262785&1342263330&1342262791&1342267615&1342267616\\
PER~11&1342263433&1342262780&1342263331&1342262803&1342267607&1342267608\\
PER~12&1342263532&1342262781&1342263332&1342262804&1342267609&1342267610\\
PER~13&1342263536&1342262788&1342263333&1342262790&1342267613&1342267614\\
PER~14&1342263537&1342262774&1342263334&1342266763&1342263512&1342263513\\
PER~15&1342263434&1342262775&1342263335&1342262810&1342263514&1342263515\\
PER~16&1342263538&1342263161&1342263336&1342266764&1342265447&1342265448\\
PER~17&1342263539&1342263163&1342263337&1342266765&1342263486&1342263487\\
PER~18&1342263540&1342263162&1342263338&1342266766&1342265449&1342265450\\
PER~19&1342263541&1342263164&1342263339&1342266767&1342265451&1342265452\\
PER~20&1342263542&1342263165&1342263340&1342266768&1342265453&1342265454\\
PER~21&1342263543&1342263166&1342263341&1342266769&1342265455&1342265456\\
PER~22&1342263544&1342263167&1342263342&1342266770&1342265701&1342265702\\
SCO~01&1342266428&1342263174&1342263319&1342266760&1342267175&1342267176\\
SERS~01&1342268464&1342268487&1342266502&1342268158&1342254231&1342254230\\
SERS~02&1342268465&1342268118&1342266497&1342268157&1342254223&1342254222\\
TAU~01&1342266913&1342268127&1342266486&1342266771&1342265458&1342265457\\
TAU~02&1342266931&1342268128&1342266485&1342266772&1342265460&1342265459\\
TAU~03&1342266932&1342268129&1342266484&1342266773&1342265462&1342265461\\
TAU~04&1342266933&1342268131&1342266482&1342266774&1342265464&1342265463\\
TAU~06&1342266934&1342268130&1342266483&1342266775&1342265465&1342265466\\
TAU~07&1342266935&1342268133&1342266481&1342266776&1342265467&1342265468\\
TAU~08&1342266930&1342268135&1342266477&1342266778&\phantom{\tablefootmark{e}}$-$\tablefootmark{e}&\phantom{\tablefootmark{e}}$-$\tablefootmark{e}\\
TAU~09&1342266936&1342268134&1342266478&1342268144&1342267856&1342267857\\
W40~01&1342268272&1342268488&1342266496&1342268152&1342254221&1342254220\\
W40~02&1342268273&1342268122&1342266495&1342268151&1342254269&1342254268\\
W40~03&1342268676&1342268489&1342266494&1342268150&1342254267&1342254266\\
W40~04&1342268677&1342268124&1342266493&1342268149&1342254261&1342254260\\
W40~05&1342268678&1342268490&1342266492&1342268148&1342254265&1342254264\\
W40~06&1342268679&1342268491&1342266491&1342268147&1342254258&1342254259\\
W40~07&1342268680&1342268120&1342266490&1342268146&1342254262&1342254263\\
\hline\noalign {\smallskip}
\end{tabular}
\label{T:obsids_will}
\tablefoot{\tablefoottext{a}{Observation also contains H$_{2}$$^{18}$O 1$_{10}-$1$_{01}$ in the other sideband.} \tablefoottext{b}{Observation also contains CO $J$=10$-$9 in the same sideband.} \tablefoottext{c}{Observation also contains H$_{2}$$^{18}$O 1$_{11}-$0$_{00}$ in other sideband and $^{13}$CO $J$=10$-$9 in the same sideband.} \tablefoottext{d}{Observation also contains C$^{18}$O $J$=9$-$8 in the same sideband.} \tablefoottext{e}{Observation scheduled but not successfully observed before the end of science operations.}}
\end{table*}

\begin{table*}[ph!]
\caption[]{\textit{Herschel} observation identification numbers for continuum PACS and SPIRE observations used to determine far-IR SED fluxes.}
\centering
\begin{tabular}{lcccc}
\hline \noalign {\smallskip}
\centering
Region & PACS 70\micron{} & PACS 100\micron{} & PACS 160\micron{} & SPIRE 250, 350 \& 500\micron{}\\
\hline\noalign {\smallskip}
Aquila, Serpens South \& W40 & 1342186277,1342186278 & 1342193534,1342193535 & 1342193534,1342193535 & 1342186277,1342186278 \\
Chameleon & 1342213178,1342213179 & 1342224782,1342224783 & 1342224782,1342224783 & 1342213178,1342213179 \\
& 1342213180,1342213181 & 1342212708,1342212709 & 1342212708,1342212709 & 1342213180,1342213181 \\
Corona Australis & 1342206677,1342206678  & 1342218806,1342218807 & 1342218806,1342218807 & 1342206677,1342206678 \\
Ophiuchus & 1342205093,1342205094 & 1342227148,1342227149 & 1342227148,1342227149 & 1342205093,1342205094 \\
& & 1342227148,1342227149 & 1342227148,1342227149 & \\
Perseus & 1342190326,1342190327 & 1342227103,1342227104 & 1342227103,1342227104 & 1342190326,1342190327 \\
& 1342214504,1342214505 & 1342216077,1342216078 & 1342216077,1342216078 & 1342214504,1342214505 \\
Scorpius & 1342267724,1342267725 & $-$ & 1342267724,1342267725 & \\
Taurus & 1342190616,1342190654 & 1342228005,1342228006 &1342190616,1342228005 & 1342190616,1342190652 \\
& 1342190655 & & 1342228006,1342190652 & 1342190653,1342190654 \\
& & & 1342190653 & 1342190655,1342202253 \\
& & & & 1342202254 \\
\hline\noalign {\smallskip}
\end{tabular}
\label{T:obsids_gbs}
\end{table*}

\section{Discussion of individual cases}
\label{S:cases}

This section presents notes on individual sources to explain oddities in the data that bear specific mention.

\textbf{\textit{AQU\,01}:} The additional emission component observed only in the H$_{2}$O 1$_{10}-$1$_{01}$ line towards this source (which has the largest beam) is almost certainly due to another source on the edge of the beam. This source is outside the beam for all other HIFI observations.

\textbf{\textit{CHA\,01}:} This source shows a relatively small and weak but detectible outflow in CO $J$=3$-$2, as well as a relatively narrow detection in H$_{2}$O with HIFI and detections in [O\,{\sc i}], H$_{2}$O, CO and OH with PACS. The principle reason this source is not designated as a Class I is the non-detection of HCO$^{+}$. The low velocity of the outflow suggests this may either be a remnant from the Class I phase, or a disk-wind such as recently seen by ALMA in HD 163296 \citep{Klaassen2013}.

\textbf{\textit{CHA\,02}:} The non-detection of water emission in the HIFI observations for this Class I source is unsurprising given that only a few lines are marginally detected in PACS and no outflow is detected in CO $J$=3$-$2. There is a detection in [O\,{\sc i}], suggesting that some form of jet or wind is present. This source is therefore probably nearing the Class II phase and simply has too tenuous an envelope for a significant outflow component to still be present and detectible.

\textbf{\textit{OPH\,01}:} This source does not show compact HCO$^{+}$ or C$^{18}$O emission, so is not Class 0/I. There is a cold, starless core to the north that increasingly dominates at longer wavelengths \citep[e.g.][]{Sadavoy2010} and is causing the source to appear more embedded than it really is. The strong, narrow CO $J$=10$-$9 emission profile, along with the PACS detection of CO $J$=16$-$15 are most likely from a PDR or bow-shock, possibly caused by interaction between the infrared source and the cold core if they are spatially associated. Given that the primary infrared source is visible in the mid-IR \citep[e.g.][]{Brown2013}, we classify it as Class II, whilst noting that there is also PDR-like emission.

\textbf{\textit{PER\,02}:} High-resolution BIMA observations of the CO outflows and \textit{Spitzer} IRAC scattered light observations show that the blue outflow lobe of this source is contaminated by the blue outflow lobe originating from L1448-MM to the south \citep{Kwon2006,Tobin2007}. We therefore do not report outflow properties as it is impossible to disentangle the two flows. However, the BIMA observations do show activity related to the targeted source, so it is still classified as having a detected outflow for the purposes of source evolution. The broad absorption in the blue outflow lobe in the ground-state water lines likely takes place against the outflow from L1448-MM, indicating that it is between this source and the observer. In particular, the saturated absorption feature below the continuum in H$_{2}$O 1$_{11}-$0$_{00}$ \citep[see][]{SanJoseGarciaThesis} rules out the possibility that this is instead caused by a combination of emission from the different outflows in the beam. Thus, the source is most likely further away than the outflow from L1448-MM and the two are probably not interacting directly.

\textbf{\textit{PER\,04}:} The non-detection of water for this Class 0 source is slightly surprising given the detection of an outflow in CO $J$=3$-$2. However, the low velocity of the outflow may mean that any shocks are not fast enough to sputter water from the grains or warm enough to lead to efficient gas-phase formation. This may indicate that this source is particularly young or has been in a lower-accretion phase for some time.

\textbf{\textit{PER\,06}:} While the CO $J$=3$-$2 data are consistent with an outflow originating from this source (NGC1333-IRAS2B), it is impossible to disentangle the contribution of NGC1333-IRAS2A, something even \citet{Plunkett2013} find impossible at three times higher spatial resolution with CARMA, so we follow those authors in not quoting any outflow property values.

\textbf{\textit{PER\,08}:} At least part of the strange outflow morphology from this source may be from another nearby source or indicate that it is a multiple. However, it is a single star in high-resolution VLA observations \citep{Tobin2016} and there are no other obvious infrared candidates nearby so for now we attribute all the outflow emission to this protostar.

\textbf{\textit{PER\,12}:} The red outflow lobe of NGC1333-IRAS4A passes through the H$_{2}$O 1$_{10}-$1$_{01}$ beam but not the other transitions that have smaller beam-sizes; hence the detection in only this transition. There is no evidence in the CO $J$=3$-$2 data of an outflow associated with this source. However, \citet{Tobin2015} note that \textit{Spitzer} images are suggestive of a jet or outflow related to the source (see their Fig.~19), so the non-detection is likely because the outflow is in the plane of the sky rather than because this source does not have an outflow.

\textbf{\textit{PER\,22}:} The morphology of the outflow in CO $J$=3$-$2 is suggestive of two outflows, particularly in the red lobe, with one being approximately north-south and the other east-west. \citet{Enoch2009} find another, more evolved Class I source (Per-emb 55) $\sim$9\arcsec{} away, which therefore lies within the \textit{Herschel} beam. VLA observations resolve this additional source into a binary \citep{Tobin2016}, suggesting that this is actually a triple if Per-emb 8 and 55 are spatially associated.

\textbf{\textit{SERS\,01}:} This source is significantly offset in both the continuum and molecular lines from the position given by \citet{Maury2011}, which is likely why it is relatively weak in the HIFI and PACS observations.

\textbf{\textit{TAU\,03}:} Most of the outflow near this source is likely due to a neighbouring, probably younger, source, though there does seem to be a weak flow from the primary target as well.

\textbf{\textit{TAU\,06}:} The outflow observed in CO $J$=3$-$2 seems to have two blue outflow lobes, and the red and blue outflow lobes are not well aligned. Given the low velocity of the outflow, it may be close to the plane of the sky, or there may be multiple outflows in the region. There is no other obvious driving source in the vicinity however, so we assign all the emission to the target.

\textbf{\textit{TAU\,07$-$09}:} These Class II sources exhibit some line broadening in CO $J$=3$-$2 around the source position. While the line-wings are not at high enough velocity offset (FWZI$\lesssim$10\kms{}) to be considered related to a true outflow, this is suggestive of either higher turbulence or a small disk wind \citep[c.f.][]{Klaassen2013} in these sources.

\textbf{\textit{W40\,01}:} The HIFI H$_{2}$O spectra show a broad line-wing, in addition to PDR-like absorption close to the $\varv_{\mathrm{LSR}}$, consistent with an outflow, though there is no outflow detected in low-$J$ CO. The source is compact and relatively bright at 70\micron{}, which led \citet{Konyves2015} to designate this source as protostellar, rather than prestellar, and it is detected in the mid-IR. It is therefore most likely a Class 0 source, though the presence of the W40 PDR complicates the detection and passage of the outflow.

\textbf{\textit{W40\,03$-$06}:} These sources show little or only very weak continuum emission at wavelengths shorter than 100\micron{}, while at longer wavelengths, the emission is not particularly compact. They also do not show any signs of outflow and the only line detections are related to emission from the W40 PDR. We therefore designate these sources as potentially prestellar, but note that the PDR makes a reliable classification more difficult.

\textbf{\textit{W40\,07}:} The outflow observed towards this source in CO $J$=3$-$2 is surprisingly strong given that there is no emission detected in H$_{2}$O and only a faint, narrow and tentative detection in CO $J$=10$-$9. The low fluxes in the mid-IR would seem to suggest that there is little warm gas around this source, but the sub-mm and mm continuum detections indicate a significant reservoir of cold dust. The bright and compact nature of the 70\micron{} PACS continuum emission and shape of the SED led \citet{Konyves2015} to designate this source as protostellar. It is therefore most likely a very young protostellar source where the outflow has not become fast and warm enough to be detected in H$_{2}$O and high-J CO.

\end{document}